# The Experimental Probe of Inflationary Cosmology

## *A Mission Concept Study for NASA's Einstein Inflation Probe*

EPIC Mission Study Team

Jet Propulsion Laboratory

February 2008

Members of the EPIC Mission Study Team

Science Team

| | |
|---|---|
| Alex Amblard | UC Irvine |
| Charles Beichman | JPL/IPAC |
| James Bock* (PI) | JPL/Caltech |
| Robert Caldwell | Dartmouth |
| John Carlstrom | U. Chicago |
| Sarah Church | Stanford U. |
| Asantha Cooray* | UC Irvine |
| Clive Dickenson | JPL |
| Scott Dodelson | Fermilab |
| Darren Dowell | JPL |
| Mark Dragovan | JPL |
| Todd Gaier | JPL |
| Ken Ganga | IAP, France |
| Walter Gear | U. Cardiff, UK |
| Jason Glenn | U. Colorado |
| Sunil Golwala | Caltech |
| Krzysztof Gorski | JPL/Caltech |
| Shaul Hanany* | U. Minnesota |
| Carl Heiles | UC Berkeley |
| Eric Hivon | IAP, France |
| Bill Holzapfel | UC Berkeley |
| Kent Irwin | NIST |
| Jeff Jewell | JPL |
| Marc Kamionkowski | Caltech |
| Manoj Kaplinghat | UC Irvine |
| Brian Keating* | UC San Diego |
| Lloyd Knox | UC Davis |
| Adrian Lee* | UC Berkeley/LBNL |
| Andrew Lange | Caltech/JPL |
| Charles Lawrence | JPL |
| Erik Leitch | JPL |
| Steven Levin | JPL |
| Tomotake Matsumura* | U. Minnesota |
| Michael Milligan* | U. Minnesota |
| Hien Nguyen | JPL |
| Tim Pearson | Caltech |
| Jeff Peterson | Carnegie Mellon U. |
| Nicolas Ponthieu* | IAS, France |
| Clem Pryke | U. Chicago |
| Jean-Loup Puget | IAS, France |
| Anthony Readhead | Caltech |
| Tom Renbarger* | UC San Diego |
| Paul Richards | UC Berkeley |
| Ron Ross | JPL |
| Mike Seiffert | JPL |
| Helmut Spieler | LBNL |
| Meir Shimon | UCSD |
| Huan Tran* | UC Berkeley/SSL |
| Martin White | UC Berkeley |
| Jonas Zmuidzinas | Caltech/JPL |

Technology Team

| | |
|---|---|
| Terry Cafferty* | TC Technology |
| Dustin Crumb* | ATK Space |
| Peter Day | JPL |
| Warren Holmes* | JPL |
| Bob Kinsey* | JPL |
| Rick LeDuc | JPL |
| Mark Lysek* | JPL |
| Sara MacLellan* | JPL |
| Aluizio Prata* | USC |
| Celeste Satter* | JPL |
| Hemali Vyas* | JPL |
| Brett Williams* | JPL |

*Science and Technology Working Group

## *Cover Art*

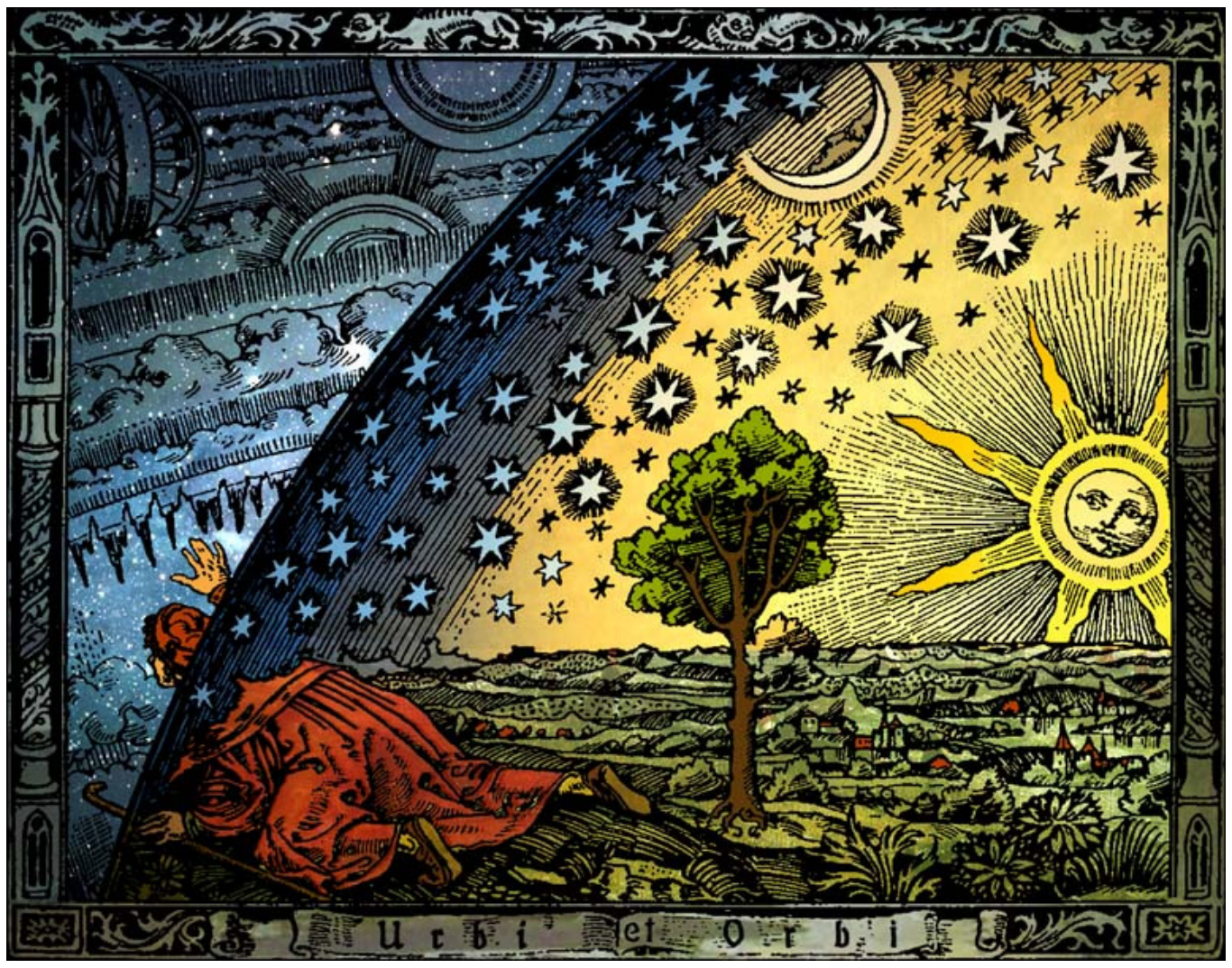

*What, then, is this blue [sky], which certainly does exist, and which veils from us the stars during the day?"* from Camille Flammarion from L'Atmosphere: Météorologie Populaire (1888).

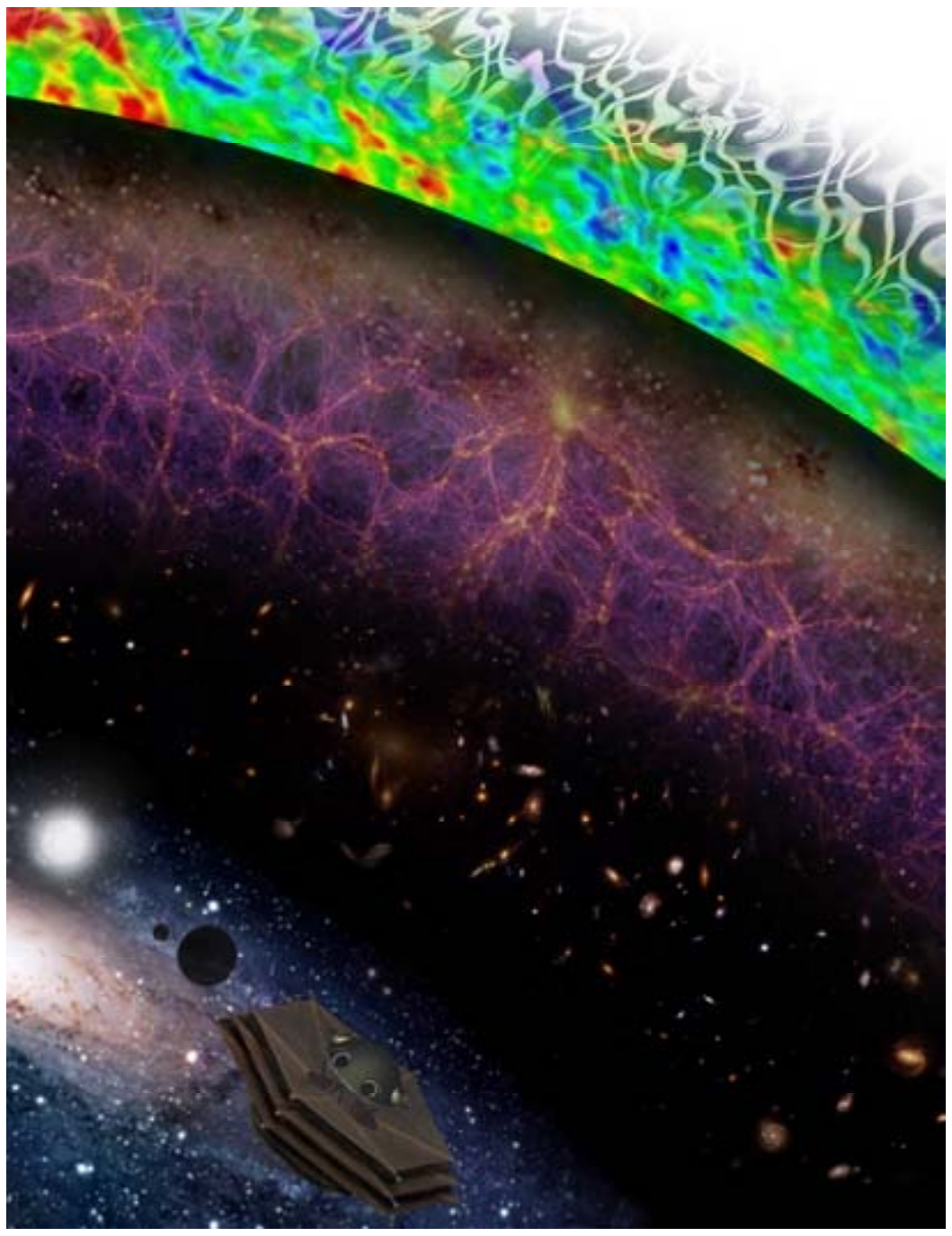

Cover Art: The cover for this report is modeled after Camille Flammarion's original woodcut where an astronomer peers outside of the orb of the night sky, to see what lies beyond. We show the EPIC satellite mission, at L2 orbit with the earth and moon in shadow, looking out through our own Galaxy and into the distant universe through successive orbs. In order of increasing distance, and further back in time, are first *galaxies and galaxy clusters*. These galaxies blend into *large scale structures* of galaxies seeded by dark matter. The onset of the formation of the *first stars and galaxies* follows from the gravitational collapse of the first structures after the *dark ages*, when the universe consisted largely of neutral hydrogen and helium. The *Cosmic Microwave Background (CMB)* originates from the photons last scattered on the surface of an ionized plasma, and has been rendered to show both temperature anisotropy and polarization. The temperature anisotropy and polarization grew out of *space-time fluctuations*, sourced by both matter/energy over-densities and gravitational waves, emerging after the period of *Inflation*.
(artwork by J. Park, JPL and Dustin Crumb, ATK Space).

# TABLE OF CONTENTS

## Executive Summary

What powered the Big Bang? Increasingly precise measurements of the Cosmic Microwave Background (CMB) support Inflation, a period of exponential expansion in the first moments after the Big Bang singularity. Yet in spite of mounting experimental evidence for the existence of Inflation, the physics causing Inflation remain a mystery. If Inflation is related to grand unification, as many theorists believe, the physical process would be well beyond the reach of modern terrestrial particle accelerators. Fortunately, the CMB provides a powerful test of Inflation, in the form of a polarization signal produced by a background of gravitational waves remaining from Inflation. This signal has unique vector properties and a distinctive power spectrum that allow it to be distinguished from both foregrounds and CMB polarization produced by more prosaic density fluctuations. NASA has envisioned the Inflation Probe, a moderate-cost Einstein Probe mission in the Beyond Einstein Program, to search for CMB polarization from Inflationary gravitational waves.

In this report, we describe a feasibility study for the Inflation Probe named the Experimental Probe of Inflationary Cosmology (EPIC). The starting point for EPIC is the Task Force for CMB Research (TFCR), a joint NSF/NASA/DOE report that describes the scientific goals of the Inflation Probe, and provides a technology roadmap leading to its realization [1]. EPIC takes full advantage of the unique advantages of a space-borne observation: all-sky coverage with a redundant scan strategy optimized for polarization, high sensitivity, multiple frequency bands for foreground removal, and rigorous control of systematic errors. EPIC uses high-sensitivity scan-modulated bolometer arrays in broad frequency bands ranging from 30 to 300 GHz. We studied two possible architectures, a *low-cost* option with six 30 cm telescopes targeted to search only for Inflationary polarization, and a *comprehensive-science* option with a single 4 m telescope with angular resolution sufficient to also study polarization from density fluctuations and gravitational lensing.

When we began our study we sought to answer five fundamental implementation questions: 1) can foregrounds be measured and subtracted to a sufficiently low level?; 2) can systematic errors be controlled?; 3) can we develop optics with sufficiently large throughput, low polarization, and frequency coverage from 30 to 300 GHz?; 4) is there a technical path to realizing the sensitivity and systematic error requirements?; and 5) what are the specific mission architecture parameters, including cost? Detailed answers to these questions are contained in this report. In brief, we find that EPIC, which assumes only modest development in focal plane technology, can indeed meet the sensitivity and band coverage requirements. We have developed new strategies to control systematic errors, and we find no fundamental problems to controlling systematics at the required level, although a full study is beyond the scope of this report. The removal of foregrounds can only be partially answered, since foregrounds are currently not well measured in polarization, particularly Galactic dust emission. However, we explored two techniques for removal, and found each gave sufficient subtraction (to $r \sim 0.01$) assuming a foreground model based on best current knowledge.

In summary, we see a clear path forward to the Inflation Probe, and this study shows such a mission can be modest and simple, with 30-cm telescopes, a commercial 3-axis spacecraft, a conventional liquid helium cryostat, and a single observing mode with low telemetry requirements. What developments that are needed are already proceeding rapidly, e.g. representative optics are already in use in sub-orbital experiments, larger focal plane arrays than needed for even the most ambitious version of EPIC will see first light this year, and high-quality measurements of polarized foregrounds will be available from sub-orbital experiments well before Planck data becomes public. Assuming NASA supports the technology development and mission planning activities described in the TFCR report, our study supports the feasibility of the TFCR mission timeline and a 2011 mission start.

## 1. Science

The wealth and quality of data from a suite of ground-breaking sub-orbital CMB experiments [1-8] and now WMAP [9] have unveiling an increasingly accurate description of the Universe's geometry, energy, and mass. We are now *confident* that the Universe is flat, while arguing about the second or third significant figures in the values of most cosmological parameters. Yet fundamental questions remain. Dark Matter and Dark Energy ("vacuum energy") dominate the composition of the Universe today, but their nature is unknown. The physical mechanism that laid down the primordial perturbations to the dark matter and photons eludes us, since it occurred at an energy scale far beyond the grasp of any terrestrial particle accelerator.

Inflation, the prevailing paradigm related to the origin of density perturbations, posits that an explosive ~$e^{60}$ expansion stretched space at super-luminal velocities in the first moments after the Big Bang. Although revolutionary, inflationary models have withstood a barrage of experimental tests, based entirely upon increasingly precise observations of the CMB, confirming all of the following predictions: 1) nearly scale-invariant spectrum on large angular scales; 2) a nearly flat geometry; 3) adiabatic fluctuations; 4) nearly perfectly Gaussian fluctuations; and 5) super-horizon fluctuations. Recently, WMAP reported a slight departure from a scale-invariant spectrum [10]. This result, assuming it holds up with further observations, may be the first data supporting a specific class of Inflationary models.

The Experimental Probe of Inflationary Cosmology (EPIC) will pursue the CMB-polarization signature of the Inflationary Gravitational Waves (IGWs) – a hallmark of inflation. A detection of the primordial gravitational wave background would be a truly spectacular achievement and will not only establish inflation as the source of density perturbations, but also allow a way to connect inflationary models to fundamental physics at a specific energy scale. The *low-cost* EPIC-*LC* scenario carries out a powerful search for IGW polarization in a modest mission configuration. The *comprehensive science* EPIC-*CS* mission has the ability to map the secondary polarization signal produced by the interaction of the CMB with intervening matter. These maps will be powerful new tools for cosmology, enabling us to precisely study neutrino masses and probe the equation of state of Dark Energy.

### 1.1 Inflationary Gravitational-Wave Background

Inflation, a period of accelerated expansion in the very early Universe, is driven by a form of "Dark Energy" associated with some high-energy phase transition. Inflation was postulated to solve [11-13] the horizon and magnetic-monopole problems, but remained speculative -- purely the realm of theorists -- until recently. Two of inflation's predictions, a nearly scale-invariant spectrum of primordial density perturbations and a flat Universe, have now been confirmed. BOOMERanG, DASI, and MAXIMA's discovery of multiple peaks in the CMB power spectrum verified gravitational amplification of primordial density perturbations as the origin of large-scale structure in the Universe today. The location of the first peak tells us that the Universe must be very close to, if not precisely, flat, exactly as inflation predicts. Finally, on angular scales of several degrees the anti-correlation between temperature and polarization patterns measured by WMAP provides evidence for modes that exited the horizon during an inflationary phase. These cannot be explained by post-inflation causal physics, but are a natural prediction of inflation.

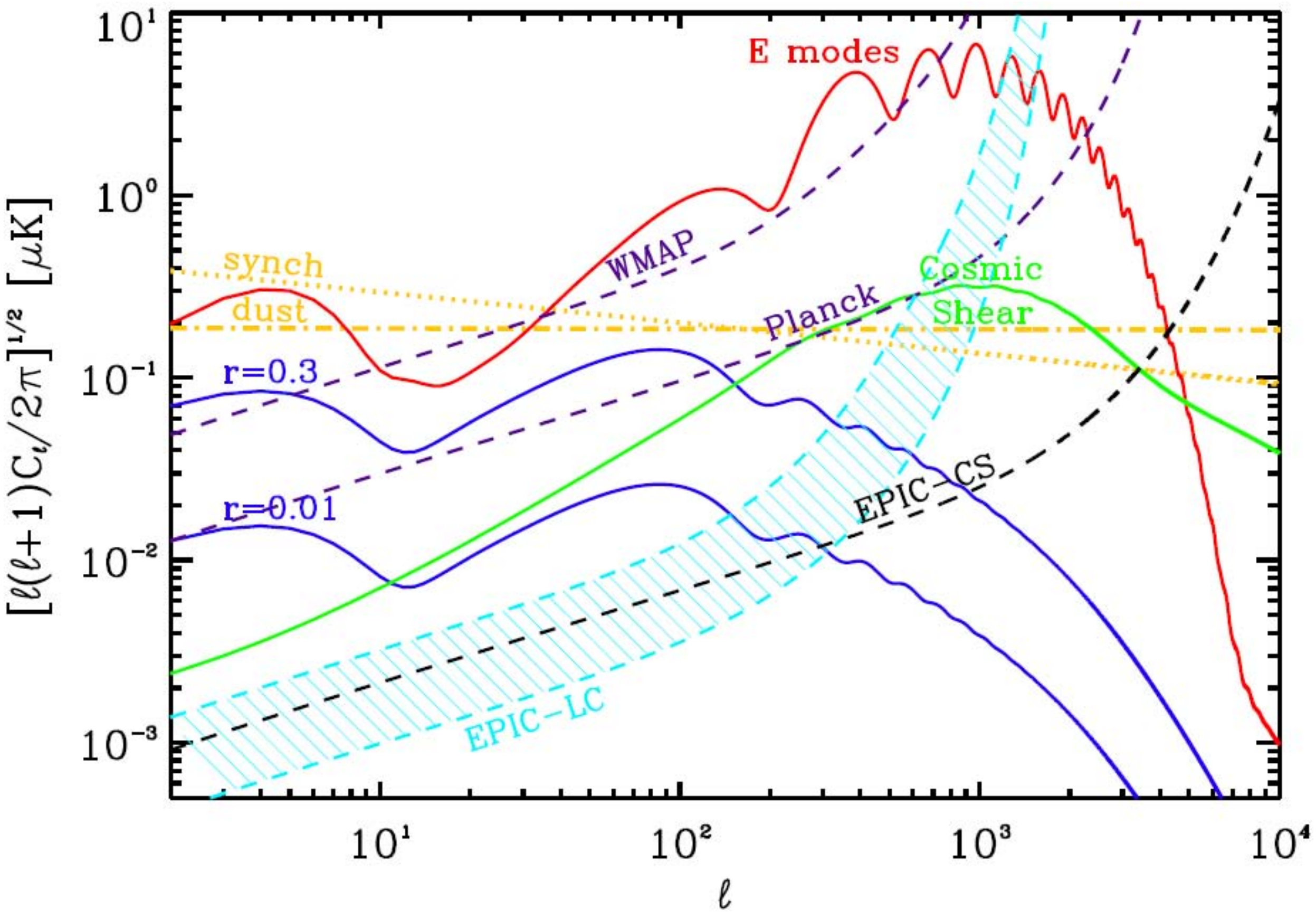

Fig. 1.1.1. The sensitivity of EPIC-LC, WMAP and Planck to CMB polarization anisotropy. E-mode polarization anisotropy from scalar perturbations are shown in red; B-mode from tensor perturbations are shown in blue for r = 0.3 and r = 0.01 Inflationary Gravitational Waves (IGWs); and B-mode polarization produced by lensing of the E-mode polarization is shown in green. The science goal of EPIC is to reach the level of r = 0.01 for the entire $\ell < 100$ multipole range after foreground subtraction. Expected B-mode foreground power spectra for polarized dust (orange dash-dotted) and synchrotron (orange dotted) at 70 GHz are determined by power-law models fits to the foreground power in a combination of WMAP 23 GHz polarization maps [29], low frequency radio maps [30], and 100 micron dust map for |b| > 20˚ [31] for a 65% sky cut. The sensitivity of EPIC-LC is given over a range from the required baseline sensitivity (top of the cyan band) and a 1-year mission to the design TES-option sensitivity and a 2-year mission (bottom of the cyan band). The sensitivity for EPIC-CS is taken from required mission parameters. WMAP assumes an 8-year mission life; Planck assumes 1.2 years at goal sensitivities for HFI. Note the sensitivity curves show band-combined sensitivities to $C_\ell$ is broad $\Delta\ell/\ell = 0.3$ bins in order to compare the full raw statistical power of the three experiments in the same manner. Final sensitivity to r after foregrounds removal will naturally be reduced.

The next step is to determine the new ultra-high-energy physics responsible for inflation. Inflation predicts a cosmological background of stochastic gravitational waves, produced during inflation through quantum-mechanical excitations of the gravitational field [14-15]. Since the production process is purely gravitational, the theory predicts that the amplitude of the gravitational-wave background depends only on the universal expansion rate -- or equivalently on the cosmological energy density, the age of the Universe, or the height of the inflaton potential -- during inflation. Since the cosmological energy density during inflation varies from one model of inflation to another, the amplitude of gravitational wave background cannot be predicted in a model-independent manner. Measuring the amplitude of the IGW background, however, provides a direct and a robust measurement of the energy scale of inflation. If the energy scale turns out to be $10^{16}$ GeV, then inflation was most likely associated with unification of the strong, weak, and electromagnetic interactions. If the energy scale is lower, then inflation may have had to do with Peccei-Quinn symmetry breaking, supersymmetry breaking [16], or some other new high-energy physics. Finally, while there are sources of gravitational waves within the horizon leading to sub-horizon wavelength waves, such as due to massive binary black holes [17], a primordial phenomenon such as inflation is required to produce super-horizon wavelength gravitational waves that can be detected with CMB polarization measurements. Thus, if inflationary gravitational waves can be detected, they provide an important probe of the physics related to cosmic inflation.

Currently favored inflation theories based on simple scale-field potentials suggest that the IGW amplitude when extrapolated to frequencies of a few mHz to a few Hz corresponding to the frequency range probed within LIGO/LISA bands will be below threshold for these experiments [18]. If the gravitational wave background is detectable at a tensor-to-scalar ratio above 0.001, the relic background present today may be detectable with a post-LISA experiment called Big Bang Observer, one of two Einstein vision missions in NASA's roadmap [19]. Before a direct detection of the relic background, CMB polarization field provides the most promising tool to probe the amplitude of inflationary gravitational waves with a clear signature of their presence in the form a unique "curl" pattern. The vector-like properties of the polarization allow it to be decomposed into curl ("B-mode") and curl-free grad ("E-mode") components [21-22]. Primordial density perturbations produce only a curl-free polarization pattern. However, gravitational waves induce a curl in the polarization of the CMB [23], producing a unique signature. The curl pattern does not correlate with either the temperature or the electric-type parity pattern, providing a way to distinguish the detection from any systematic effects. The power spectrum[1] for the curl component of the CMB polarization due to a background of inflationary gravitational waves is shown in Fig. 1.1.1.

While the expected amplitude of inflationary gravitational waves is highly uncertain, recent results from WMAP [10] provide some guidance. The perturbations generated by inflation for both density and gravitational waves (or tensors) have power-law power spectra with $P_s \propto k^{ns}$ and $P_t \propto k^{nt}$, as a function of the wave number k. These are in return related to the scalar-field potential $V(\phi)$ responsible for inflation as the field $\phi$ rolls down it and the derivatives of the potential with respect to the scalar field. In the standard slow-roll descriptions of inflation involving a single inflaton field, the tensor-to-scalar ratio involving the ratio of amplitudes between gravitational wave and density perturbation power spectra can be written as $r = 16\varepsilon$ where $\varepsilon$ is the first-order slow-roll parameter given as $\varepsilon = [M_{pl}V']^2/16\pi V^2$. With the second slow roll parameter $\eta = [M^2_{pl}V'']/8\pi V$, we can write the scalar spectral index as $n_s = 1 - 6\varepsilon + 2\eta$ [23]. With $n_s = 0.958 \pm 0.016$ from recent WMAP second data analysis [10], and if $\varepsilon \sim \eta$ in an optimistic description of the inflationary scenario, then we find that $r \sim 0.1\dot{6}$, which is within detection limits of Planck [24].

The true scenario, however, is likely to be more complex given that we have limited knowledge of inflationary physics and the shape of the inflaton potential. For analytical models of the inflationary potential involving models such as power-law with $V(\phi) \propto e^{\phi/\mu}$, chaotic inflation model with $V(\phi) \propto (\phi/\mu)^p$, and spontaneous symmetry-breaking potential with $V(\phi) \propto [1-(\phi/\nu)^2]^2$, the recent WMAP results guide towards a gravitational wave background with tensor-to-scalar ratio, in general, greater than 0.01 [25] (see Fig 1.1.2) given that in all these descriptions of inflation the behavior is such that as one moves away from $n_s = 1$, the tensor-to-scalar ratio increases. For example, in the case of the power-law inflationary potential, $r = 8(1-n_s)$ while with chaotic inflation this relation is modified as $r = 8p(1- n_s)/(p+2)$. The WMAP result that $n_s$ differs from 1 at the $2\sigma$ to $3\sigma$ level can then be interpreted as evidence for a detectable gravitational wave background for EPIC. Furthermore, the combined information of r and $n_s$ can be used to distinguish between inflationary models.

[1]In this report we define r to be the ratio of the initial tensor/scalar spectra (as used by the CAMB program) rather than the T/S ratio for the quadrupole (as used by the CMBFAST program). The result of this is that our values for achievable r should be divided by a factor of ~1.6 when comparing to the other convention.

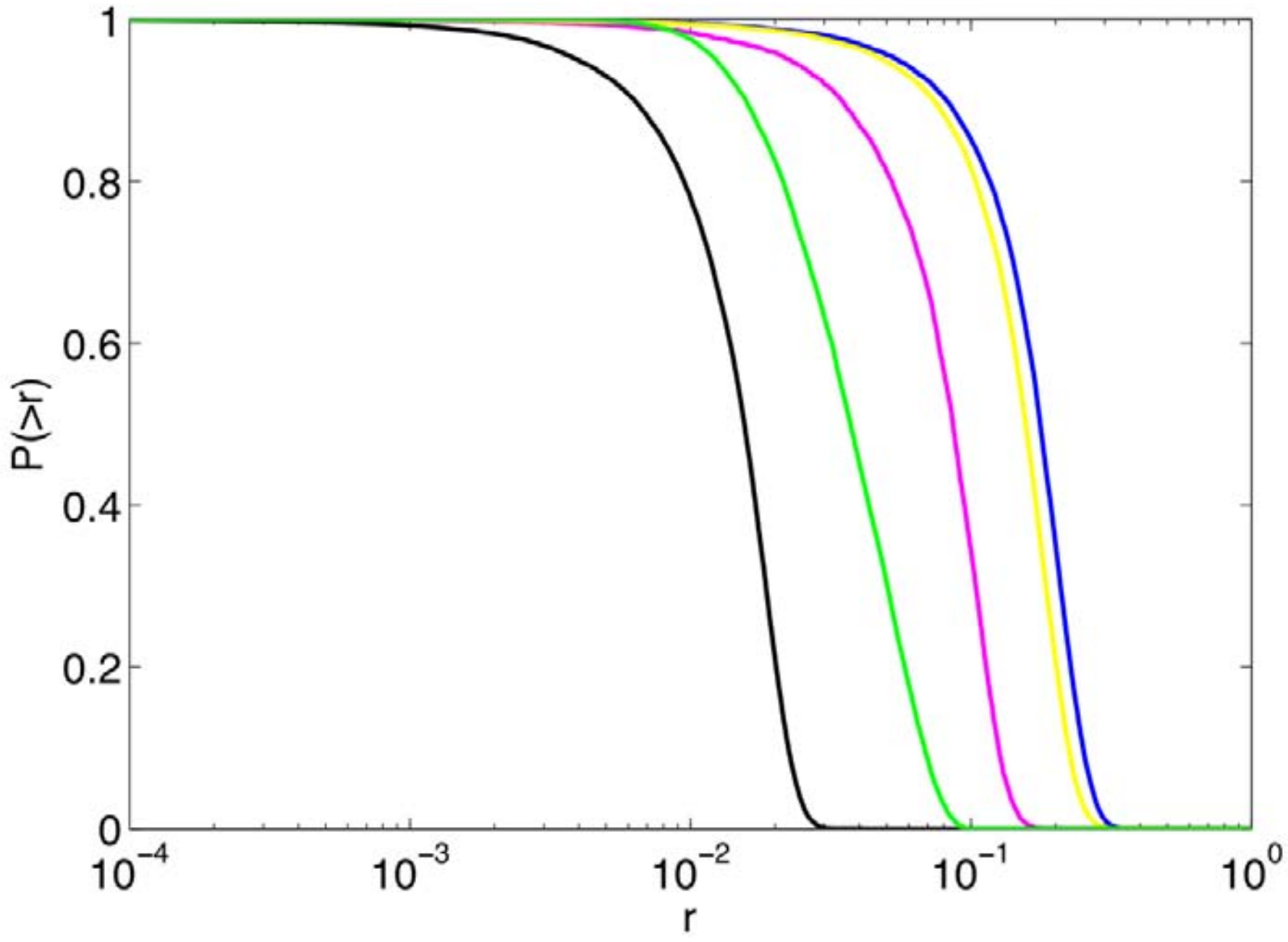

Fig. 1.1.2. Percentage of inflationary models with a tensor-to-scalar ratio above a threshold value of r, for the analytical potentials involving power-law (blue line), chaotic with p = 8 (yellow), chaotic with p = 1 (purple), spontaneous symmetry-breaking (green), and chaotic with p = 0.1 (black) models (see text for details). The figure is reproduced from Ref. [25].

While in Ref. [25] only analytical models are discussed, similar studies can be extended to consider numerically generated inflationary models that do not have to follow a specific analytical form. In Fig. 1.1.3 we show the expected distribution of a large number of single inflationary potentials that have an arbitrary shape for $V(\phi)$ as a function of the tensor-to-scalar ratio and the scalar spectral index. The potentials are generated with a large number of Monte-Carlo models of inflation under the Hamilton-Jacobi approach [26]. While these models span a large range of the parameter space including a wide distribution in the gravitational wave background amplitude, probing the tensor-to-scalar ratio down to 0.001 allows us to probe a category of potentials that are generally described as large-field models, in which the field moves a width $\Delta\phi$ of order $M_{pl}$ as the field rolls from scales corresponding to CMB to the end of inflation to inflate the Universe between 40 and 70 e-folds of expansion [27]. Arguments related to a large tensor-to-scalar ratio can also be made in terms of the number of degrees used to fine-tune the potential [28]. As summarized in Fig. 1.1.3 over the range of fine-tuning to more than 9 degrees of extra fine-tuning the tensor-to-scalar ratio is generally at the level between 0.01 and 0.1. Observations with EPIC will allow us to probe this interesting range of parameter space.

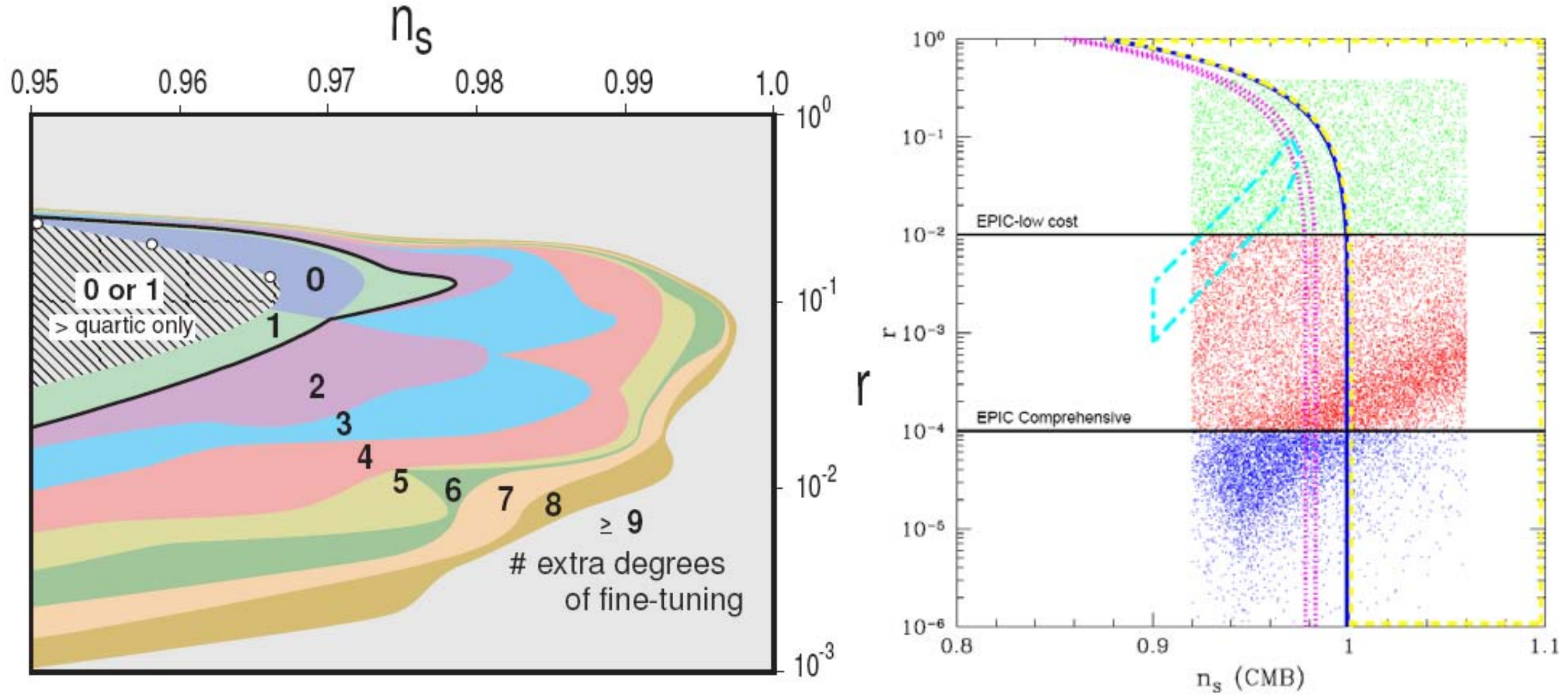

Fig. 1.1.3. Predictions for tensor-to-scalar ratio r vs. the scalar spectral index. In the left panel, we reproduce results from an analytical calculation on how the predictions vary with the degree to which the inflation potential is finely tuned [27]. In increasing order of fine-tuning, the potentials behave as monomial, quadratic, cubic, and quartic. The right panel shows the expected distribution of model points for general polynomial description of the inflaton potential with coefficients of the polynomial generated by numerical Monte-Carlo models of the Hamilton-Jacobi equation [28]. The hatched region in the left-hand figure is for inflaton potentials of the form $\phi^n$ where power-law n is more than 4, where as the whole region takes n >= 2. 0, 1 etc refers to the # of degrees of freedom in these models (where degrees of freedom are in terms of derivatives of the potential with $\phi$). While there is no specific region in the tensor-to-scalar ratio vs. scalar spectral index preferred by these generic scalar-field potentials models which can be generally described as large-field potentials will be probed with EPIC polarization measurements.

Of course, if the energy scale of inflation is low, < $10^{15}$ GeV, then the IGB could be undetectable by CMB polarization measurements. However, if inflation had something to do with grand unification, as many theorists believe, then the IGB amplitude will be detectable by EPIC, providing us a glimpse of the conditions in the Universe roughly $10^{-38}$ seconds after the Big Bang. And it would constitute perhaps the first detection of radiation produced by Hawking-like effects of quantum field theory in curved space-time. This IGW background thus provides an astonishing opportunity for NASA.

## 1.2 Precision CMB Polarimetry

EPIC will improve upon Planck's raw sensitivity by a factor of ~10. In addition to the measurements we review briefly here, the power of EPIC will open a discovery space for breakthroughs we cannot anticipate now.

*Scalar polarization:* EPIC will extract all of the information encoded in the CMB surface-of-last-scattering, and achieve the grad-mode cosmic-variance limit out the beam resolution. A measurement of the scalar power spectrum will enable new tests of the physics of recombination and probes for exotic phenomena [32].

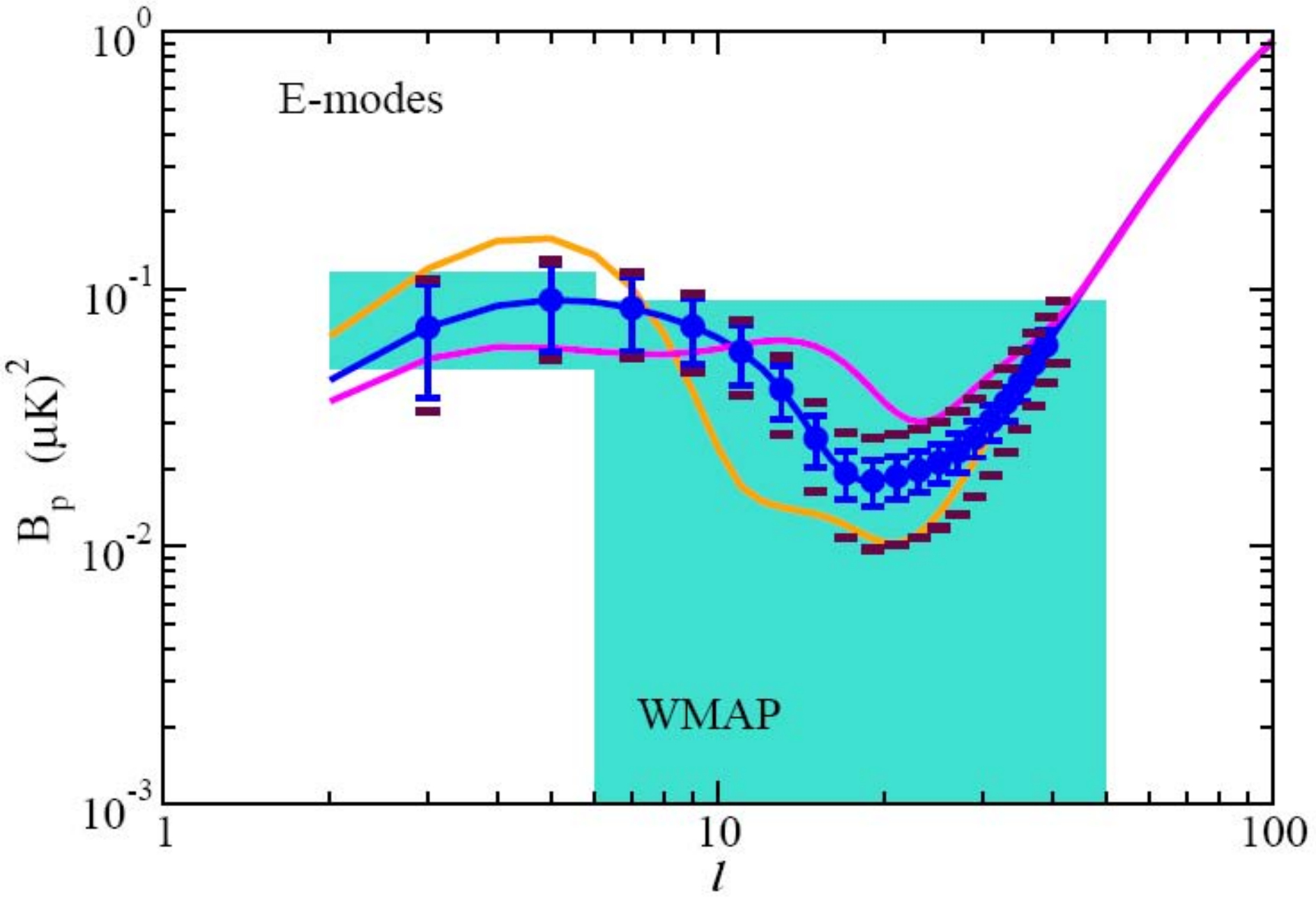

Fig. 1.2.1 The low-multipole (ℓ < 100) region of the CMB polarization E-mode angular power spectrum with a bump related to reionization such that the ionization fraction of electrons has variations at two redshifts with complete reionization at a redshift of 6.3 consistent with SDSS [32]. Between a redshift of 6.3 and $z_{ri}$, the ionization fraction has a varying value such that the total optical depth is still normalized to 0.1 consistent with WMAP [10]. Between 10 < ℓ < 40, from bottom, middle, and top curves are for $z_{ri}$ = 13, 30, and 50, respectively. The error bars show the cosmic variance limited errors in the E-mode spectrum possible with EPIC, while extended errors marked by horizontal lines show the errors expected from Planck. With the cosmic variance limited measurements available with EPIC, one can establish additional details of the reionization process beyond the integrated optical depth [34].

*Reionization:* The recent WMAP report of a large-angle polarization excess has indicated the possibility of reionization at a redshift *z*~10 [10, 35]. There is a large uncertainty, however, on both the integrated optical depth as well as the exact reionization history of the Universe. None of the ground-based experiments will improve this result to the limit allow by foregrounds as observations will be limited to a small area of the sky. EPIC will complete this task with a cosmic-variance limited measurement of the E-mode spectrum to extract all of the available information about the reionization process, including ways to address whether the Universe reionized once or twice [34].

*Cosmic Shear:* The tens of arcminute and finer angular scale CMB temperature and polarization anisotropy provide a unique probe of the integrated mass distribution along the line of sight through lensing modification to the anisotropy structure [36]. This secondary lensing signal can be studied through higher-order statistics leading a direct measurement of the integrated mass power spectrum [37-38]. In combination the power spectrum provides a measure of the neutrino mass since a massive neutrino affects the formation of small-scale structure [39-40]. In Fig. 1.2.2 we summarize our results related to neutrino masses. While existing cosmological studies limit the sum neutrino masses to be below about 0.66 eV (95% CL) [10], a combination of CMB lensing studies with Planck combined with all CMB information in the low-resolution version of EPIC can be used to probe a sum of the neutrino masses down to 0.15 eV (95% CL), while CMB lensing information in the EPIC high resolution version alone can extend this down to 0.05 eV.

These cosmological results expected from EPIC can be put in the context of neutrino experiments motivated by known particle physics. Neutrino oscillation experimental data fix the difference of neutrino mass squared between two states and for solar and atmospheric neutrinos the mass squared differences are $2.5\times10^{-3}$ $eV^2$ [41] and $8\times10^{-5}$ $eV^2$ [42], respectively. When combined, these estimates of mass-squared differences lead to two potential mass hierarchies shown in the inset of Fig.1.2.2 [43]. With lensing information from the high resolution version of EPIC, it is possible to estimate both the sum of neutrino masses, but also distinguish between the two options involving inverted and normal hierarchies.

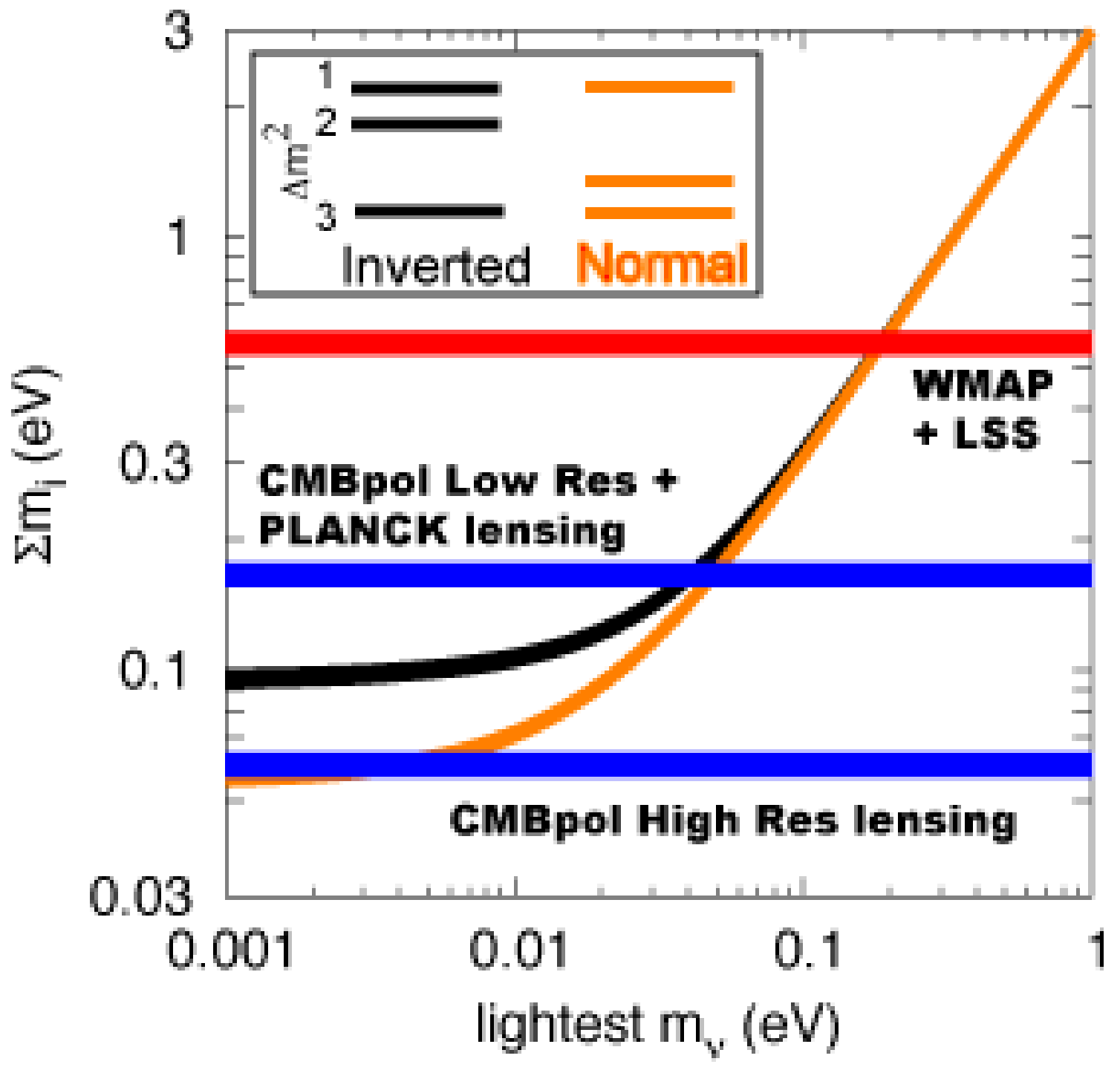

Fig. 1.2.2 The sum of neutrino masses as a function of the lightest neutrino mass by making use of atmospheric and solar neutrino oscillation data [41-42]. The two lines show the relation between sum of the neutrino masses and the mass of the lightest neutrino for the two possible mass hierarchies [43]. The horizontal lines show the limits reached with existing data (top line from [10]) and limits reachable with EPIC either in terms of the two low-resolution and high-resolution versions. This figure was adapted from Ref. [43].

*Secondary anisotropy and polarization*: While resolution is a limiting factor for secondary polarization studies with EPIC-LC, the high resolution provided by EPIC-CS will open up temperature and polarization maps for a variety of studies related to secondary polarization signals from the large-scale structure. In the case of temperature maps alone, EPIC-CS will improve the cluster detection through the SZ effect relative to the cluster catalog in Planck given the improvement in noise by at least an order of magnitude. This will allow detection of clusters with total mass below $10^{14}$ M_sun or at least a factor of 5 improvement in mass limit detectable with Planck while at the same time extending the redshift range of clusters to a higher redshift than Planck.

While the homogenous reionization signal peaks at large angular scales, density or ionization fraction modulation of reionization will lead to an additional polarization signal at small angular scales. Moreover, scattering of the temperature anisotropy quadrupole by electrons in galaxy clusters will generate another polarization signal. The cluster locations can be identified based on SZ detections in the temperature map and the cluster polarization detection can be optimized through known locations and depths of the SZ signal.

By averaging over large samples of clusters, one can determine the CMB quadrupole projected at various cluster locations in redshift space. The evolution of the mean cluster polarization with redshift reflects the growth of the quadrupole, from the integrated Sachs-Wolfe effect, and this depends on dark energy properties. This measurement will enable a unique measurement of the dark energy equation of state. The EPIC-CS mission can also lead to an additional measurement of the equation of state of dark energy using the evolution of the lensing power spectrum as can be extracted from EPIC-CS temperature and polarization maps through an analysis of the lensed CMB anisotropy.

*Non-Gaussianity:* To a first approximation, inflation predicts that primordial perturbations have a mostly Gaussian distribution. To next order, though, some very small degree of non-Gaussianity is to be generically expected [44]. Specific slow-roll inflationary models predict the amplitude and nature of that non-Gaussianity, although the complete range of predictions in current viable inflationary models is quite expected to be below the detectability level of CMB data alone. Alternatives to slow-roll inflation, such as under D-brane inflation motivated by string theory arguments generally however suggest a large non-Gaussianity level than single-field slow-roll models of inflation [45]. Thus, if detected, non-Gaussianity of primordial perturbations as seen by CMB would provide a unique avenue toward the new ultra-high-energy physics responsible for inflation.

In addition to non-Gaussianity associated with primary anisotropy, a large number of secondary effects in CMB data will generate non-Gaussian signals, especially at small angular scales that will be probed with the high-resolution version of EPIC [46]. These signals provide information related to growth of structures as well as cosmology and astrophysics during the reionization era and later.

*Interstellar Magnetic Fields:* The interstellar magnetic field, together with gravity and gas pressure, is one of the three major forces acting on interstellar gas. Although key to understanding interstellar medium physics, our current data on Galactic magnetic fields (Zeeman splitting, Faraday rotation, optical polarization) is quite limited. A sensitive, multi-frequency survey of diffuse linear polarization with arc-minute resolution will revolutionize our understanding of the diffuse interstellar magnetic field on length scales of a parsec.

### 1.3 Angular Resolution

The cosmic shear, scalar polarization and interstellar magnetic field science themes are especially dependent on the choice of angular resolution. While these themes are important, and robustly predicted by standard cosmology, they are outside the main science goal advocated by the TFCR, namely to probe the IGW B-mode signal to at least $r = 0.01$. A deep search for IGW B-mode polarization does not require high angular resolution, at least until confusion with cosmic shear B-mode polarization becomes problematic [47-48]. We therefore have split our study into two mission concepts, a *comprehensive-science* scenario with a 4-m telescope that measures scalar polarization to cosmic variance into the damping tail, and a *low-cost* mission with 30-cm telescopes that is designed solely to search for IGW B-mode polarization down to the cosmic shear confusion limit of r ~ 0.01 in both the reionization and recombination peaks of the B-mode power spectrum. A description of these science goals may be found in Table 4.1.1.

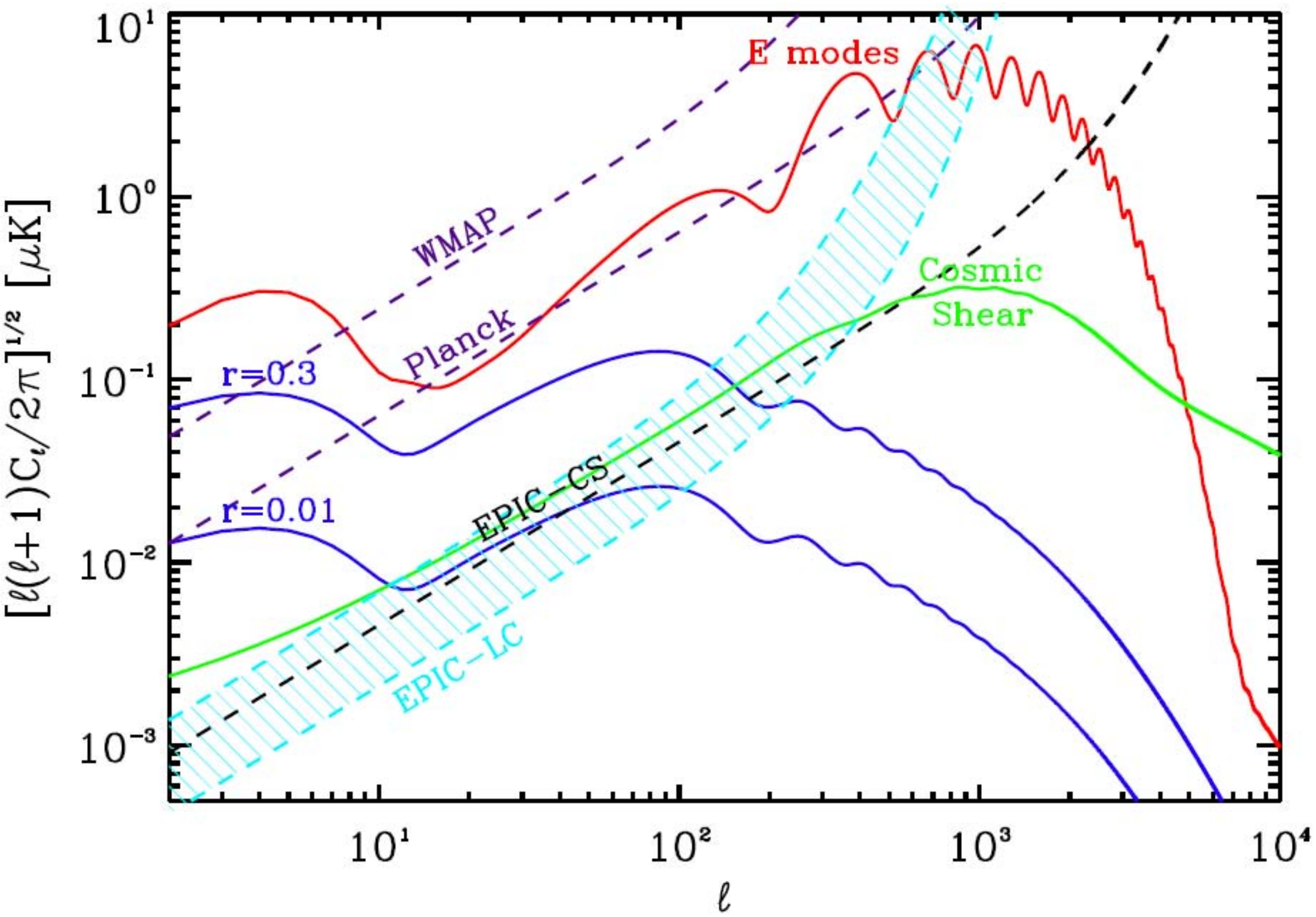

Fig. 1.3.1. The noise $C_\ell$s of EPIC-LC, WMAP and Planck, with curves same as Fig.1.1.1. A comparison of these noise power spectra and the signals, such as primordial B-mode spectra shown in blue for r = 0.01 and r = 0.3 reveals the angular scale, or the multipole moment, where cosmic variance of primordial signals dominate the measurement. As shown, the detection of low-multipole reionization bump is dominated by cosmic variance while for low r models, the recombination bump at degree angular scales is the transition between noise domination to cosmic variance domination.

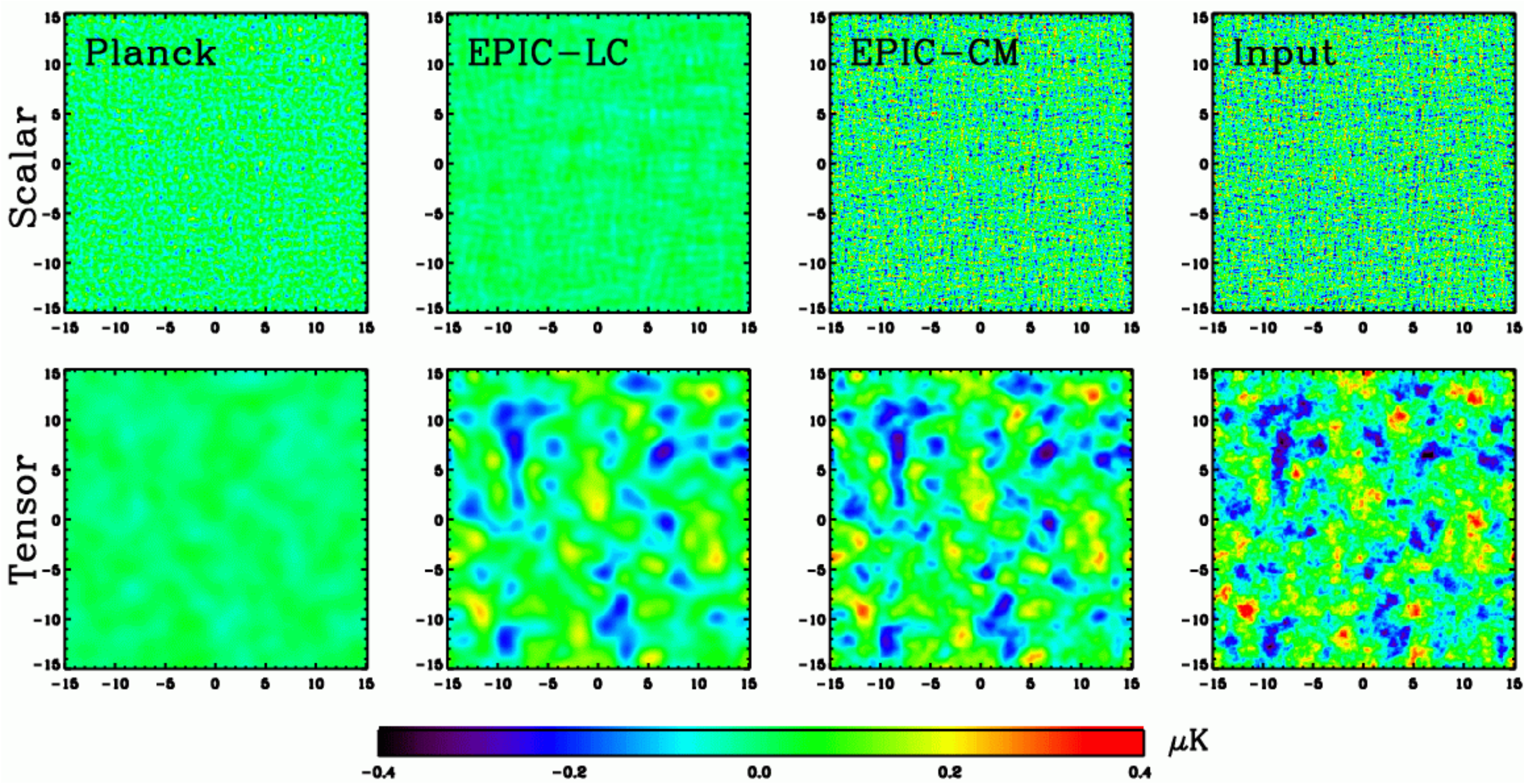

Fig. 1.3.2. Maps of the Q Stokes parameter for purely scalar perturbations (top row) and purely tensorial perturbation (bottom row), with a tensor to scalar ratio T/S = 0.3 and an optical depth to reionization of 0.088. The right-most panels show a 10x10 deg patch of a simulated CMB sky, assuming the cosmological parameters measured by WMAP [10]. The right 3 panels in both rows are Wiener filtered maps of the same patch as they would be observed by Planck-LC (NTD version) and EPIC-CS. (Note in this presentation, the effect of instrument noise is to reduce the amplitude of cosmological structure in the image, rather to add noise to the image).

## 2. Foregrounds

Polarized Galactic emission will likely set the practical limit to detecting primordial B-mode polarization. We have designed EPIC to have the best possible prospect of distinguishing the large angular scale E- and B-mode signals from Galactic emission.

There are two characteristic signals due to primordial B-modes. The first signature is due to rescattering of the primordial B-modes after reionization, and yields a peak at $\ell \approx 8$. The second, truly primordial, signature is a peak in the power spectrum at $\ell \approx 100$. The first signal is thus present on the largest scales while the second is present on small scales – of order 2˚. There are thus two very different regimes for estimating the foregrounds that may contaminate these signals. On large scales, Galactic emission is expected to be bright, roughly comparable to the largest expected IGW signal, and thus must be deeply subtracted. On degree scales however, one can restrict observations to very clean patches of sky, where the foregrounds may even be as faint as the r = 0.01 IGW B-mode signal, and still obtain sufficient cosmic variance precision to provide a good measurement.

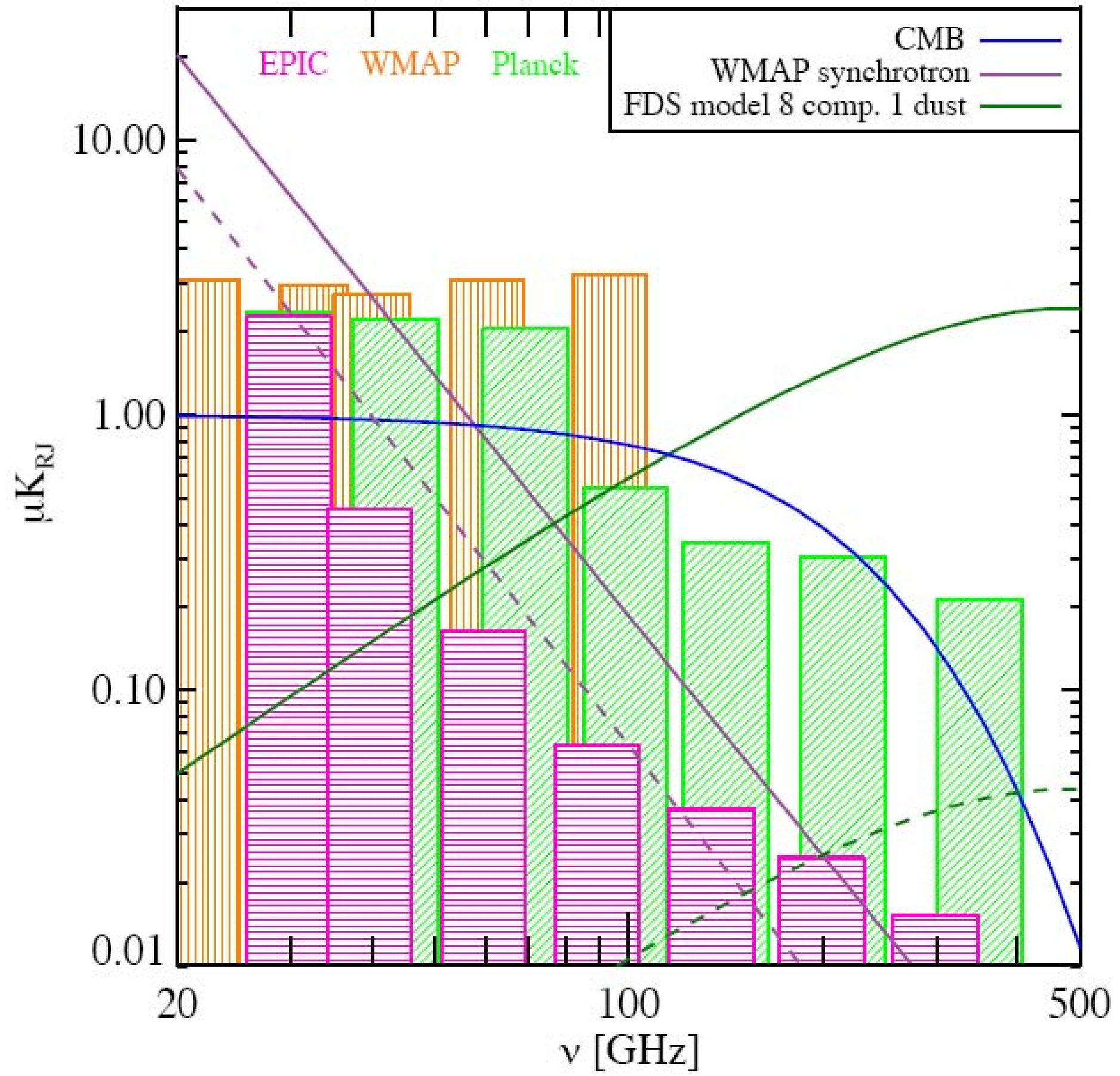

Fig. 2.0.1. Frequency spectra of galactic emission for 75% of the sky, shown by solid curves in dust (green) and synchrotron (purple), and a clean 2% patch, dashed curves, compared with the spectrum of the CMB (solid blue curve). Note that there is more variation in the dust than in the synchrotron, and that the minimum frequency changes depending on the region of sky. In bands we show the EPIC-LC required, WMAP, and Planck sensitivities, where the heights of the bands correspond to the per-pixel rms instrumental noise for 14′ pixels.

## 2.1 Foregrounds Taxonomy

### *2.1.1 Synchrotron Emission*

Synchrotron radiation is emitted by electrons spiraling in supernova remnant and galactic magnetic fields. The emission is an approximate power law in frequency with exponent (in Rayleigh-Jeans temperature) $\beta_s$ and polarization fraction $\Pi_s$, and both of these parameters may vary spatially. Geometrical projection suppresses polarization in a manner not necessarily correlated with $\beta_s$. WMAP provides the best measurement of polarized synchrotron over the full sky. Investigating small 900 $deg^2$ fields in the multipole range $\ell = 30 - 100$, we observe a factor of ~3000 variation in synchrotron brightness. In the cleanest portions of the sky, we expect foreground levels as low as $\Delta T_{rms} \sim 60\ nK_{RJ}$ at 100 GHz, which at a raw level is already close to the science goal of r = 0.01.

### *2.1.2 Thermal Dust Emission*

Dust grains emit blackbody radiation modified by a frequency-dependent emissivity, and becomes polarized because the grains preferentially align perpendicularly to magnetic fields.

There is good evidence that the frequency dependence of the emissivity is not well described by a power law [1,2], and that the dust temperature is not described by a single temperature component.

Finkbeiner *et al.* [2] ('FDS') fit a two-component dust model to unpolarized IRAS, DIRBE, and FIRAS data. FDS interpret the components as large (~100 nm) silicate grains with $\langle T_1 \rangle$ = 9.4 K and small (~10 nm) graphite grains with $\langle T_2 \rangle$ = 16.1 K. Although this model has some shortcomings, it does provide a useful starting point for modeling Galactic emission removal. Again for small fields of size 900 $deg^2$, we find significantly reduced dust emission levels in clean patches of the sky. A typical "clean" field has, according to the FDS maps, $\Delta T_{rms}$ ~ 10 $nK_{RJ}$ at 100 GHz.

Recently WMAP provided a measurement of dust polarization at 100 GHz [3]. The WMAP team constructs a template of *polarized* thermal dust emission based on FDS intensity and polarization orientations derived from observations of dust-polarized starlight. The mean observed polarization fraction of high latitude dust is 3.6 ± 1.1%. The majority (> 97%) of polarized foreground emission measured in WMAP is explained by a simple two-component model of thermal dust and synchrotron emission with a slowly spatially varying spectral index. The WMAP team argues there is no evidence for significant polarized emission from spinning dust grains, and any such component contributes < 1% of the total polarized signal variance at any frequency.

The foreground levels given above are shown in Fig. 2.0.1 for the two different scenarios: the cleanest 75% of the sky one would observe to detect the $\ell$ ~ 8 signal and the cleanest 2% of the sky one would use to detect the $\ell$ ~ 100 signal. In the former case, the sum of synchrotron and dust has a minimum in the range 60 to 100 GHz, while in the latter case, the greater reduction in dust emission pushes the optimal observing frequency up to roughly 150 GHz.

*2.1.3 Dust "Exotic" Emission*

There are many claimed detections of spinning dust grains, which emit via rotational and vibrational modes and produce a spectral bump at tens of GHz with a non-negligible tail extending past 100 GHz [4]. The WMAP team argues that such emission is a sufficiently sub-dominant contributor to temperature anisotropy that no template removal is necessary [5]. This, combined with the expectation of only a few% polarization [6,1] (vs. 50-75% for synchrotron), should put spinning dust emission below the required level.

WMAP synchrotron maps could possibly contain a hidden component from thermal magnetic dipole emission [7] (e.g., iron-containing) grains; and such emission can be polarized to 30% at 100 GHz [8]. However, this component is probably not significant because it does not match the measured spectral dependence [1].

*2.1.4 Sub-dominant Foregrounds*

Extragalactic radio and infrared compact sources are sufficiently diluted on the angular scales of interest that only the brightest sources need be removed. Tucci *et al.* [9] estimate that one can use the Planck compact source catalog to remove the brightest radio sources (> 200 mJy at 100 GHz) and leave a polarized contamination level of less than 10 nK at $\ell$ = 8. Infrared point sources are expected to be largely unpolarized.

Free-free emission is intrinsically unpolarized, though the edges of HII regions may appear polarized via the same effect that gives rise to E-mode polarization of the CMB [10]. The effect would be small compared to other galactic emission.

## 2.2 Foreground Removal Strategies

We have investigated two scenarios for foreground removal, one based in pixel space and one in Fourier space, with detailed calculations for the EPIC-LC scenario. The pixel-based technique is comparatively insensitive to the amplitude of the foregrounds, in that if the foregrounds are described by the model used to fit and remove them, the residual errors in the CMB do not depend on the foreground amplitude. However, this technique becomes less effective if the spectral indices have significant spatial variation, because more free parameters are required for removal. The spectral technique only assumes that the CMB spectrum is precisely known, and removes components that do not match this spectral template. The spectral technique is insensitive to variations in spectral index, but degrades if the foregrounds are larger in amplitude. To understand how each technique handles more complicated realistic sky models, numerical simulations are required.

### *2.2.1 Pixel-Based Foreground Removal*

We compare the two mission options for EPIC using a pixel-based foreground separation technique [11] to see how well the CMB can be reconstructed for each mission configuration. This Bayesian technique fits for parametric models of the individual foreground components using a MCMC algorithm to find the best-fitting parameters and errors for each pixel on the sky. We use a realistic model for the sky and fit for only the 2 dominant foregrounds expected in polarization (synchrotron and thermal dust). We model the spectra as simple power-laws as a function of frequency, fitting for both the amplitude and spectral index simultaneously, along with the CMB amplitude for the Stokes parameters I, Q and U. For this investigation, we fit for a given foreground model, and evaluate errors after 1000 realizations of CMB and noise.

The sky model consists of CMB and 4 foreground components (synchrotron, free-free, thermal dust and spinning dust emissions) as given in Table 2.2.1. The amplitudes and spectra were chosen based on our current best knowledge of foreground emissions from recent work [12, 13]. The foregrounds are known to vary considerably from pixel-to-pixel on the sky and the details of each foreground component are still not well characterized, particularly in polarization [1]. One example is the assumption that the synchrotron spectral index is constant with frequency. It is known to steepen with frequency due to spectral-ageing of the CR electrons [2]. However, most of the steepening occurs at lower frequencies than those considered for EPIC (< 30 GHz). The polarization fractions are typical values expected at high Galactic latitudes and position angles for each component (i.e. distribution of Stokes Q and U) were given a random distribution in each realization. Little is known about the "anomalous dust" component, which emits strongly at frequencies < 60 GHz. For this study, we chose to use a typical spinning dust model [4] and assumed a relatively low polarization fraction as expected from spinning dust grains [6]; see Table 2.2.1.

We then evaluated the removal of foregrounds assuming the band frequency coverage between 30 and 300 GHz shown in Fig. 2.0.1 and described in section 5.1. This analysis is confined to the EPIC-LC scenario with either NTD Ge detectors or TES arrays. We estimate the residual uncertainty in our measurement of the CMB emission in each pixel after foreground signals have been removed using the multiple bands. We follow schematically the technique of [10], first fitting the nonlinear model parameters (power-law indices and dust temperature) on large pixels, then smoothing the nonlinear parameter fields spatially and fixing them when fitting for the amplitude components on smaller pixels. The technique is particularly appropriate given that Galactic emission anisotropy power is primarily on large scales.

**Table 2.2.1 Definition of Input Sky Model Used in Pixel-Based Removal Scheme**

| **Component** | **Spectrum ($I \propto \nu^{\beta}$)** | **Amplitude (Stokes I)** | **Pol. fraction** |
|---|---|---|---|
| CMB | β=0 (thermal) | 70μK | 1% |
| Synchrotron | β=-3.0 | 40μK @ 23GHz | 10% |
| Free-free | β=-2.15 | 20μK @ 23GHz | 1% |
| Thermal dust | FDS model 8 (β~+1.7) | 10μK @ 94GHz | 5% |
| Spinning dust | DL98 (WNM) | 50μK @ 23GHz | 2% |

For brevity, we quote the average true error of the Stokes Q and U results from 1000 realizations in Table 2.2.2 (the Q and U values were almost identical in each case). This error includes a term for the CMB bias (i.e. the error in the fitted CMB value). An example of the fitted components over multiple simulations is shown in Fig. 2.2.1. The results indicate that foreground removal results in a slight degradation if the indices are fitted on large patches of sky. The most conservative case, assuming the indices are fitted in each pixel independently, results in a degradation of ~3 compared to the raw band-combined sensitivity. Even in this scenario, the full foreground-cleaned EPIC NTD-Ge scenario with required sensitivities marginally sufficient statistical sensitivity to provide a detection of an r = 0.01 IGW signal in both the recombination and the reionization peaks.

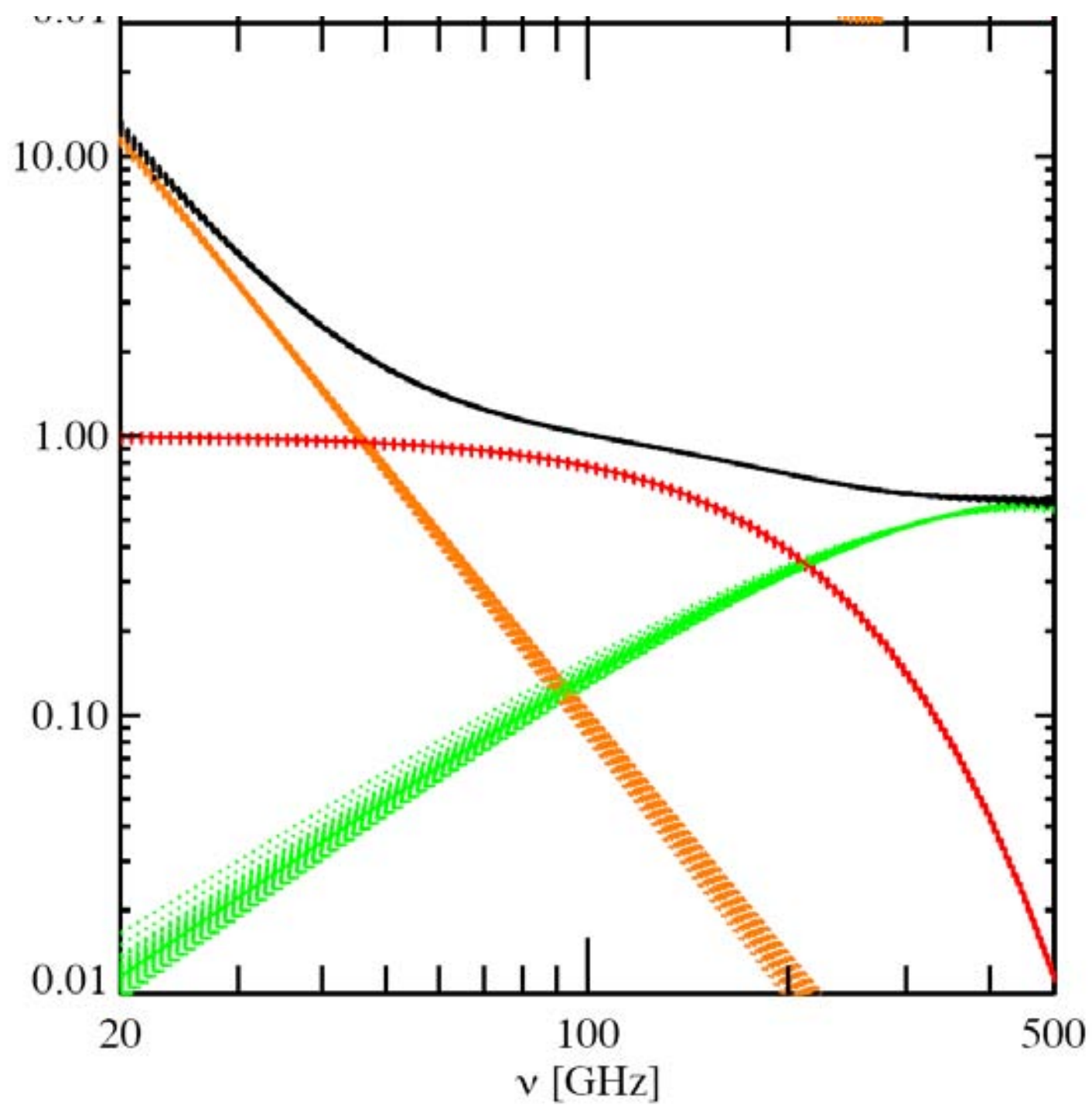

Fig. 2.2.1 Fitted models for 100 data simulations on 30˚ pixels, fitting for amplitudes, $\beta_s$ and $\beta_d$. The rms spread of the CMB fits (red curves) is the uncertainty on the CMB amplitude after foreground estimation. Also shown are errors in fitting synchrotron (orange) and dust (green) emission.

The technique described is linear, in that it does not depend on how bright the foregrounds are, assuming they are described by the model fits, but rather how many parameters are used to carry out the removal. However, fewer parameters can be used to describe foregrounds in regions where they are dim, as the errors in the parameters have less effect. Conversely, more parameters will be needed in bright regions. The major uncertainty with this analysis so far is how smoothly the spectral indices vary, as this determines how large a patch can be used to fix the indices.

We have also investigated the adding WMAP 8-year data at lower frequencies, K-band (22 GHz) and Ka-band (33 GHz). However, the sensitivity of WMAP and EPIC are sufficiently disparate that the WMAP data provides little benefit. We investigated replacing the 30 GHz EPIC band with additional pixels at 40 GHz, and find that there is no significant change to the results. This may indicate the band coverage can be reduced somewhat.

We note that a full-sky map-based foreground subtraction routine has been recently developed [14], and could be applied to the case of EPIC as a future project.

**Table 2.2.2 Estimated Sensitivities After Foreground Removal**

| Case | Planck | EPIC/NTD | EPIC/TES |
|---|---|---|---|
| No foregrounds | 325 | 35 | 11 |
| $\beta_s$ and $\beta_d$ fixed | 592 | 77 | 26 |
| $\beta_s$ and $\beta_d$ fitted in 15° pixels | 595 | 81 | 26 |
| $\beta_s$ and $\beta_d$ fitted in 10° pixels | 599 | 85 | 28 |
| $\beta_s$ and $\beta_d$ fitted in 5° pixels | 621 | 108 | 34 |
| $\beta_s$ and $\beta_d$ fitted in 2° pixels | 751 | 203 | 62 |

Note: expected per-pixel sensitivities in polarization in 2x2 degree pixels in $nK_{CMB}$, including foreground degradation, as explained in the text. We present numbers for both the EPIC-LC NTD Ge and TES sensitivity cases. The results are negligibly different for 75% sky and 2% sky in this model, since this subtraction technique is not sensitive to the amplitude of the foreground. A key question remaining is what level of bias is introduced by this technique in an all-sky measurement.

*2.2.2 Fourier-Space Removal*

In addition to pixel-based methods, foregrounds can also be removed in the harmonic space especially along the same manner that Tegmark *et al.* [15] used to produce a foreground-cleaned WMAP map (TOH map). Following Amblard *et al.* [16], we computed EPIC efficiency to remove foregrounds (we limited ourselves to the 2 dominant emissions: dust and synchrotron polarization), which combines optimally the $a_{lm}$ coefficients of the different frequencies to reduce the overall power spectrum while preserving the CMB signal:

$$a_{lm} = \sum_i w_l^i a_{lm}^i \qquad C_l = w_l^i C_l^{ij} w_l^j \qquad \sum_i w_l^i = 1$$

The CMB part of the correlation matrix $C^{ij}_l$ is determined with CAMB with standard cosmological parameters. The instrumental noise part of this matrix follows the NET and angular resolution values in Table 5.1.3. The dust and synchrotron correlation is obtained through simulated maps of these emissions. We simulated these foreground maps as observed by EPIC between 30 and 300 GHz using data from WMAP at 23 GHz [1,5]. Assuming this channel is dominated by synchrotron emission (we reduced $l > 40$ power to remove some noise), we extrapolated this map at higher frequencies using the software provided by the WOMBAT project to obtain our synchrotron maps. The WOMBAT project uses the spectral index $\beta$ obtained from combining the Rhodes/HartRAO 2326 MHz survey [17], the Stockert 21 cm radio continuum survey at 1420 MHz [18-19], and the all-sky 408 MHz survey [20].

In order to simulate the dust polarization, we assumed that the synchrotron signal is a good tracer of the galactic magnetic field and that the dust grains align very efficiently with this magnetic field. We used the synchrotron polarization angle to describe the dust polarization angle, consistent with the model presented by WMAP team [1]. For the intensity, we crudely assumed a constant overall polarization fraction of 5% relative to the total dust intensity at a given frequency. Using this fraction, we used the model 8 [2] of the maps [21] to simulate the polarized dust emission over EPIC's frequency range.

We ran our foreground removal algorithm on the 2 extreme EPIC configurations (NTD required and TES designed). The estimated CMB power spectra are presented in Fig. 2.2.2. We then estimated the lowest tensor achievable by these two setups at 99% confidence level. The required NTD configuration reaches r ~ 0.02 whereas the design TES configuration reaches r ~ 0.003.

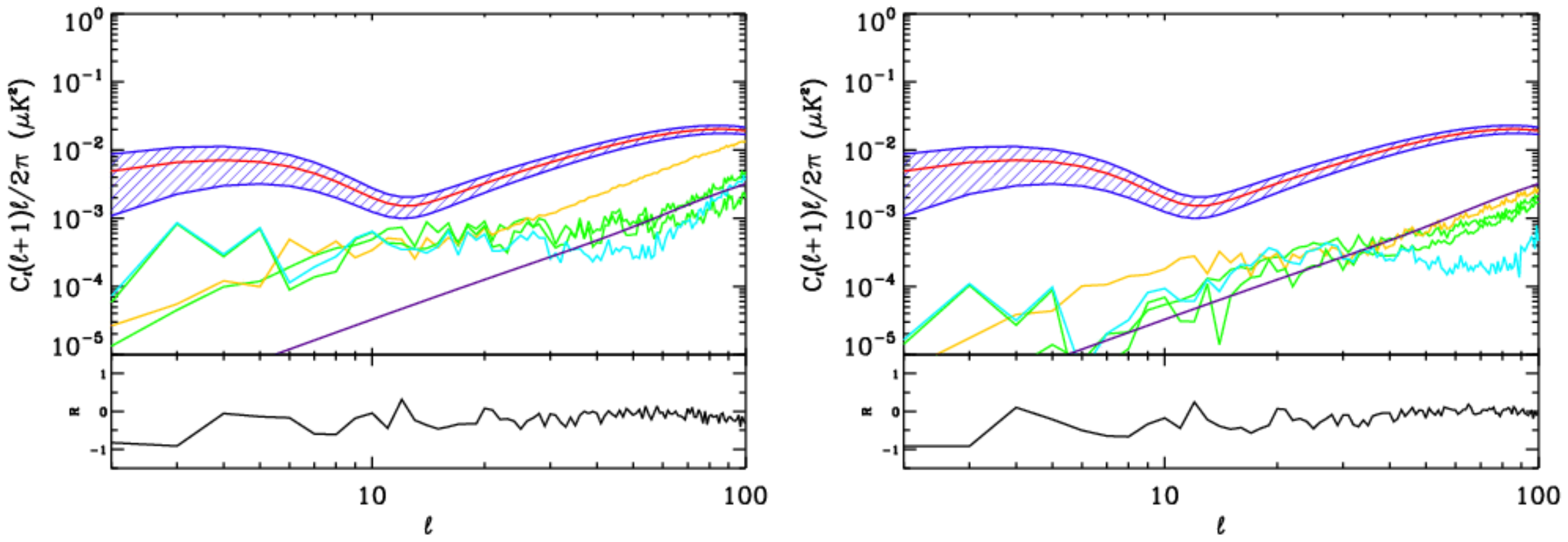

Fig 2.2.2 Estimated CMB power spectrum at r = 0.3 (red curve), dust (green) and synchrotron (cyan) residuals, noise level (orange) and CMB lensing (purple). The left plot represents the required NTD configuration, the right plot the TES design configuration.

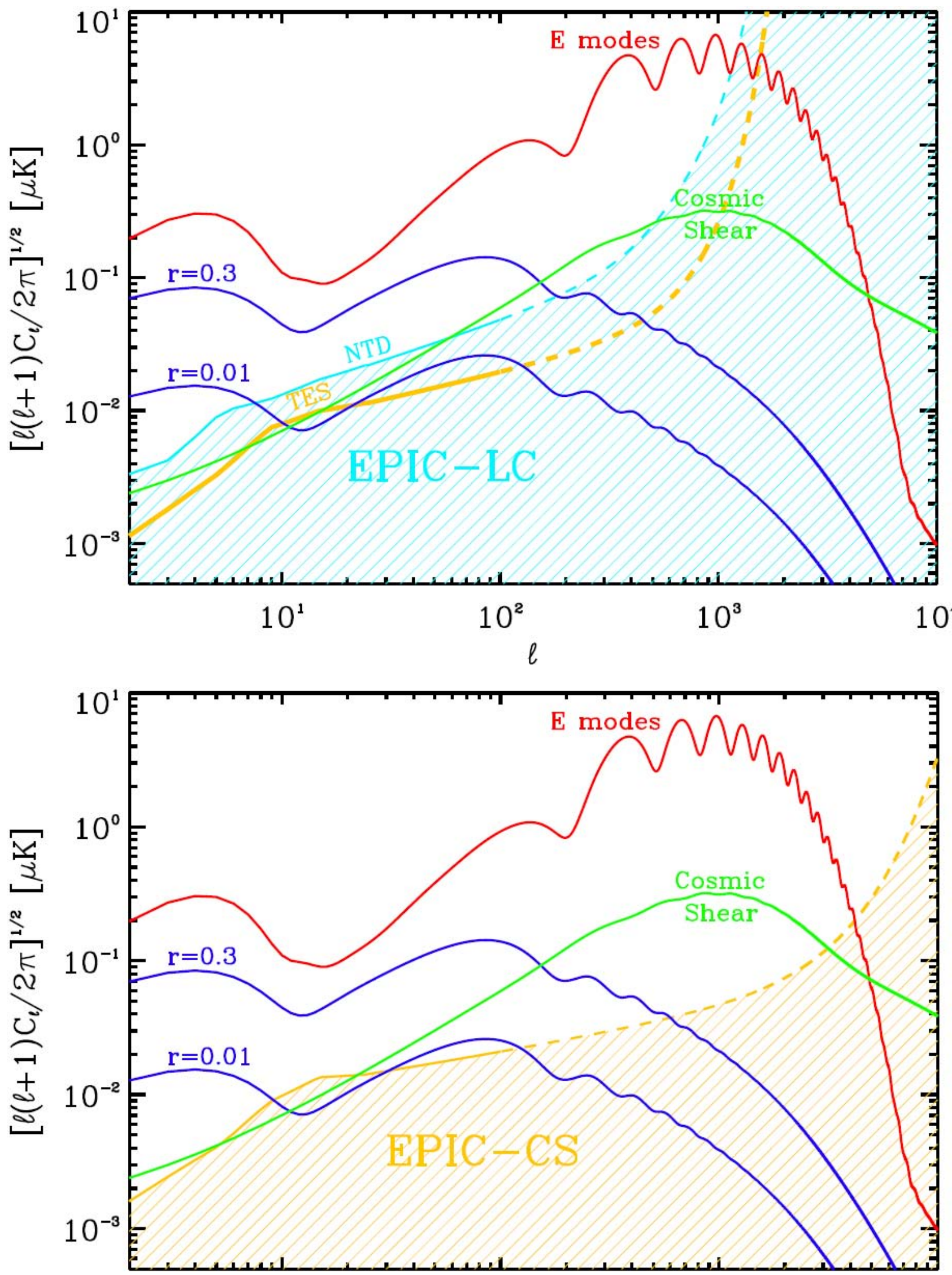

Fig 2.2.3 Estimated foreground residuals (shaded region) compared to primordial spectra of scalar polarization (red), CMB lensing (green), and B-modes with r=0.3 and r=0.01 (blue). The top plot represents the EPIC-LC option for both NTD and TES configurations as labeled while the bottom plot shows the EPIC-CS design configuration. The residuals are estimated using the Fourier-based cleaning techniques as described in Section 2.2.2. EPIC's goal of measuring to r = 0.01 across the spectrum is not satisfied for the EPIC-LC NTD case in this model, but is met for the TES options except a small region around ℓ ~ 15.

# 3. Systematic Error Control

Polarimetric fidelity must be integral to EPIC's design from the beginning in order to detect the nano-Kelvin level CMB signals imprinted by the inflationary gravitational wave background. All CMB observations confront systematic effects, many of which are traceable to optical imperfections and/or spurious couplings to radiative or thermal perturbations. Some of these effects are unavoidable, some may actually be magnified by poor design choices, and others can be eliminated by polarization modulation and judicious choice of scan strategy and optical design. We have modeled the impact of optical and thermal systematic effects. Many of these, such as thermal and electrical gain drifts, 1/f noise, far-sidelobes, and pointing errors, are already familiar from experiments designed for CMB temperature anisotropy. For polarimetry, a new class of error arises from the polarimetric fidelity of the optical system, which produces false B-mode polarization signals from much brighter temperature and E-mode polarization anisotropy.

Throughout this report we have defined our *requirement* on control of systematic errors such that the impact of the effect is below the science target of r = 0.01. As shown in Table 3.0.1, we require that the residual level of a systematic effect after correction be < 10% of the signal power expected for r = 0.01 at $\ell \leq 200$. For r = 0.01 the expected power is ~10 nK and thus we require that systematics be controlled to ~3 nK. All the systematic errors described in this report are correctable, given sufficient knowledge of the effect. Therefore our requirement is that systematic errors can be controlled post-correction to allow EPIC to just achieve its scientific goal of detecting an r = 0.01 gravitational-wave B-mode polarization signal. Our more ambitious *design goal* is to suppress the raw amplitude of systematic effects < 10% of binned statistical noise at $\ell \leq 200$ so that the effect is negligible without correction.

**Table 3.0.1** Systematic Error Requirements and Goals

| Instrument Criteria | Requirements* | Design Goals |
|---|---|---|
| Control systematic errors to negligible levels | Suppress systematic errors to < 10% of r = 0.01 signal, after correction to $\ell \leq 200$, in power | Suppress raw systematic effects to < 10% of statistical noise level to $\ell \leq 200$, in power |

*Taken from the Weiss Committee TFCR report

We then propagate these high-level goals and requirements to individual systematic effects, setting requirements on the degree of suppression and/or knowledge required for each effect. We first describe the challenges presented by systematics and then consider mission-specific methods for their mitigation in sections 5.2 and 6.2.

## 3.1 Description of Systematic Effects

Systematic errors in the measurement of polarization can be induced by imperfection in the optical beams, temperature drifts of the optics and detectors, scan synchronous signals from various sources including far-sidelobe response to local sources such as the sun, earth, moon and Galactic plane, 1/f noise in the detectors and readouts, and calibration errors. We pay particular attention to polarization and shape imperfections of the main telescope beams, since these are effects particular to polarimetry. Throughout this report we assume that the polarization is measured by the difference of matched detector pairs, where each bolometer is sensitive to linear vertical or horizontal polarization. Differencing the signals from the matched pair reduces common-mode signals from unpolarized radiation, as well as thermal drifts, pick-up, and stray magnetic fields. Furthermore we assume the signals are modulated by scanning the spacecraft at a relatively low spin rate ~1 rpm. Active polarization modulation (see section 6.4), which puts the signal band at higher frequencies and relaxes the requirements on control of 1/f noise, is considered an upscope of the baseline design and is not assumed in any of the calculations for the control of systematic errors.

It is common to distinguish between two broad classes of polarization systematics. Those that cause leakage of temperature anisotropy to polarization, and are called 'instrumental polarization', and those that cause leakage from E-mode to B-mode, and are called 'cross-polarization'. Of the two effects, instrumental polarization is generally more important since temperature anisotropy is brighter than E-mode polarization anisotropy. Nevertheless because B-mode polarization is still faint with respect to E-mode polarization, cross-polarization effects must also be considered carefully.

As listed in Table 3.1.1, there are numerous sources of systematic error that must be controlled. A short taxonomy of these effects is as follows:

***Main Beam Effects:*** The optical system can produce a variety of effects associated with polarization and shape deviations in the main beams. Instrumental polarization effects leak CMB temperature ($T$, $\nabla T$, $\nabla^2 T$, etc) into polarization, while cross-polarization effects leak CMB E-mode polarization into B-mode polarization. Because main beam effects are particularly important for polarimetry, they are described in more detail in section 3.2. For an instrument with single polarization analyzers, satellite pointing errors would enter as an instrumental polarization effect. However, for our system with dual analyzers, its signature is more complicated since we can extract one linear Stokes parameters without error due to pointing uncertainty. Pointing error only enters in when comparing Stokes Q and U, which are taken either by using different pixels in the focal plane or by comparing the same pixel after rotating the spacecraft or the wave plate. Thus pointing error is not easily categorized as just an instrumental polarization or a cross-polarization effect.

***Scan-Synchronous Effects:*** Any effect which does not average down over the scan pattern is a scan synchronous effect. The EPIC scan pattern (see section 5.2) covers each pixel on the sky with a wide range of view angles, or equivalently a large range in spin and precession angles. Therefore effects which depend only on spin or precession angle will tend to average down. Effects that are synchronous over the whole scan pattern are more pernicious because they do not average with the scan pattern. Scan-synchronous signals must typically arise from a geometry that is external to the spacecraft or optics.

Far sidelobe response to the sun, earth, moon and Galactic plane will produce a scan fixed pattern. Thus the optical system needs to have a very high degree of off-axis rejection to these sources of emission. Effects of solar heating can also give a scan-synchronous signal. Although EPIC always holds the sun at a fixed angle so that the average solar power is constant, shadows on the sun side of the spacecraft can produce a scan-synchronous signal by inducing a slight temperature variation associated with the ecliptic poles, which are observed in phase with the shadows. Pickup from magnetic fields at L2 will also produce a scan synchronous signal.

***Thermal Drifts:*** Temperature drifts of the optics produces time-varying optical signal on the detectors due to variations in thermal emission. To first order, this largely unpolarized signal is removed by the common-mode rejection of the detector pair difference. But since the common mode rejection is not perfect, the temperature of the emitting optic must be sufficiently stable, either through passive design or active control. Temperature fluctuations of the 100 mK focal plane also produce false signals on thermal bolometers, which mimic optical power but are due to variations in the thermal power flowing through the detector's isolating supports. These fluctuations are similarly removed by differencing detectors, to the extent that pairs of detectors are matched in their thermal conductivity. We assume that thermal drifts must be controlled on the time scale of a spin period of the spacecraft. In fact, this assumption is conservative since drifts are less serious for smaller angular scales. The requirement on the thermal stability of the optics depends both on the physical temperature and the source coupling to the focal plane. The stability of the focal plane is more demanding because here the signal couples through the temperature sensitivity of the bolometers.

***Other:*** In addition to the effects listed above, a wide variety of systematics can potentially result in spurious B-mode polarization signals. 1/f noise in the detectors and readouts can cause stripes in the map, resulting in a loss of sensitivity to particular CMB modes. This effect is at least partially mitigated by having a highly cross-linked scan strategy, to reduce the effect of stripes in the cross-scan direction. The focal plane can either be designed with sufficient intrinsic stability, as in the case of Planck, or the polarization signal can be actively modulated. Mismatched passbands between detector pairs will leak intensity from unpolarized foregrounds into polarization. Because the relative gain of detector pairs are calibrated on the CMB dipole, passband differences will cause differential gain to signals with a different spectral shape. An accurate knowledge of the passband from pre-launch measurements can be used to mitigate this effect. The relative gains of channels must be accurately characterized to prevent leakage of common-mode temperature anisotropy into polarization.

Table 3.1.1 summarizes the most challenging systematics, and how they imprint signals into the data stream. We have developed specific mitigation strategies to minimize these effects, as described in section 5.2.

| Systematic Error | Description | Azimuthal Symmetry | Potential Effect |
|---|---|---|---|
| *Main Beam Effects – Instrumental Polarization* | | | |
| **Δ Beam Size** | $FWHM_E \neq FWHM_H$ | Monopole: spin 0 | T → B |
| **Δ Gain** | Mismatched gains,<br>Mismatched coatings | Monopole: spin 0 | T → B |
| **Δ Beam Offset** | Pointing E ≠ Pointing H | Dipole: spin 1 | ∇T → B |
| **Δ Ellipticity** | $e_E \neq e_H$ | Quadrupole: spin 2 | $\nabla^2 T \rightarrow B$ |
| **Satellite Pointing** | Q and U beams offset | Complex | ∇T → B, E → B |
| *Main Beam Effects – Cross Polarization* | | | |
| **Δ Rotation** | E & H not orthogonal | Quadrupole: spin 2 | E → B |
| **Pixel Rotation** | E ⊥ H but rotated w.r.t. beam's major axis | Quadrupole: spin 2 | E → B |
| **Optical Cross-Pol** | Birefringence | Quadrupole: spin 2 | E → B |

| Systematic Error | Description | Azimuthal Symmetry | Potential Effect |
|---|---|---|---|
| *Scan Synchronous Signals* | | | |
| **Far Sidelobes** | Diffraction, scattering | - | T, E → B from sun, earth, moon and Galactic plane |
| **Thermal Variations** | Solar power variations | - | Temperature variation in optics, detectors |
| **Magnetic Pickup** | Susceptibility in readouts and detectors | - | Residual signal from ambient B field |
| *Thermal Stability* | | | |
| **Optics Temperature** | Varying optical power from thermal emission | - | Residual signals from temperature variations |
| **Focal Plane Temperature** | Thermal signal induced in detectors | - | |
| *Other* | | | |
| **1/f Noise** | Detector and readout drift | - | Striping in map |
| **Passband Mismatch** | Variation in filters | - | Differential response to foregrounds |
| **Gain Error** | Gain uncertainties between detectors | - | T → B |

## 3.2 Main-Beam Systematic Effects

Polarization experiments place demanding requirements on polarization effects created by the optics, unlike the temperature and scan-synchronous errors that are familiar from temperature anisotropy measurements. Main beam systematics can be classified according to instrumental polarization (IP) and cross-polarization (XP) effects, and their behavior under rotation of the beams about their symmetry axes for each pixel. This is the approach first taken by Hu, Hedman, & Zaldarriaga (2003) [1] who simulate systematic effects for coherent (RF amplifying) polarimeters. Unlike Hu *et al.*, our approach assesses the impact of systematic effects on the Stokes parameters directly, rather than on the electric fields. This complementary approach is specifically applicable for bolometric polarimeters such as EPIC. We performed calculations using two separate techniques; one assessing the systematics effects in the map domain and the other in the angular Fourier domain. Results from both approaches are in good agreement.

All effects discussed below have been included in a simulation tool described in section 5.2.3. This simulation allowed us to identify the most challenging effects and optimize our optical design, modulation strategy, and scan strategy accordingly. Because main beam effects are strongly dependent on the beam size, we summarize this calculation separately for the two mission configurations in section 5.2.3 and 6.2.

### *3.2.1 Instrumental Polarization Effects*

We parameterize optical systematic effects by their distortion of two nominally Gaussian beams associated with each of the two linearly polarized antenna planes that correspond to two matched bolometers. The antenna planes are referred to as ‘E’ and ‘H’ and thus each bolometer is sensitive to either an E or an H orientation of the polarization. Each antenna pattern is given by $G(\theta) = \exp(-\theta^2/2\sigma^2)$, where θ is the boresight angle and θ is the beamwidth.

Main beam effects we considered are shown graphically in Fig. 3.1.1. ‘Differential Gain’ occurs when the two detectors have unequal optical transmission/gain. Differencing the bolometer signals associated with each antenna leads to an apparently polarized signal. ‘Differential Beam Width’ occurs when the two beams are circularly-symmetric Gaussians, but have different beam widths $\sigma_E \neq \sigma_H$. Differencing the detector signals associated with each antenna-plane leads to instrumental polarization. Both differential gain and differential beam width possess monopole symmetry, i.e. the effect averages to zero by rotating the instrument through a full range of view angles on the sky. If each antenna in the pair produces an elliptically shaped beam then ‘Differential Ellipticity’ corresponds to the effect arising from the difference in ellipticities. Differential ellipticity has a quadrupolar symmetry and does not average down with instrument rotation. The effect of ‘Differential Beam Offset’ is caused when the direction of the centroid of the two beam patterns on the sky is not identical. The effect has dipolar symmetry and thus couples gradients in the CMB temperature anisotropy into polarization.

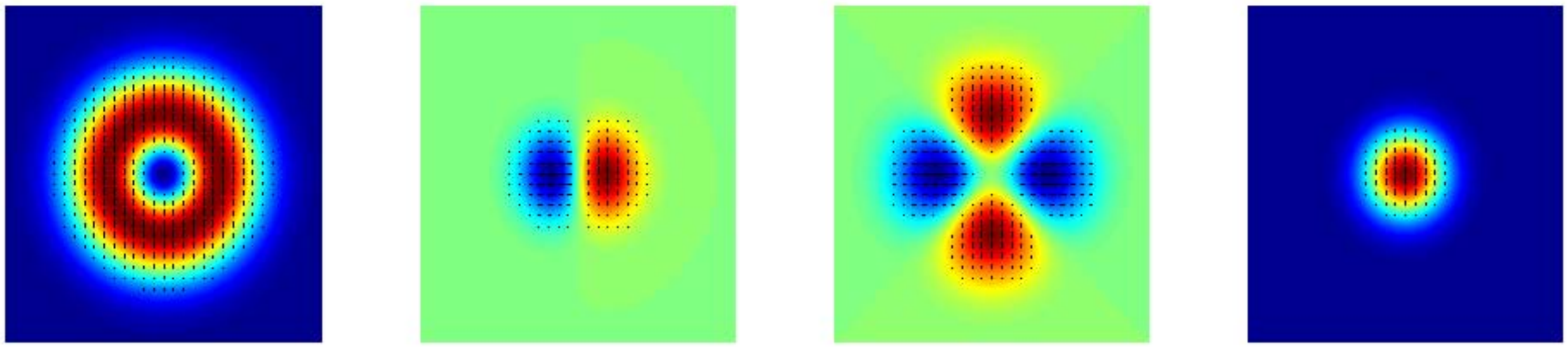

Figure 3.1.1. Systematic effects in real space. From left to right: differential FWHM (monopole effect), differential beam offset (dipole IP effect), differential ellipticity (quadrupole effect) and differential gain (monopole effect) .

### *3.2.2 Cross-polarization Effects*

Main beam cross polarization effects can be caused by the following sources. ‘Differential Rotation’ causes the two polarizations measured in a detector pair to be non-orthogonal. Since this effect is second-order, as it converts polarization, it is ignored by Hu *et al*. [1], but we include it. Differential rotation can be caused by misalignment between detector pairs. ‘Pixel Rotation’ corresponds to the axes of a pixel staying orthogonal but rotating on the sky. This effect may be caused by rotation of the optics with respect to the focal plane, or uncertainties in the satellite rotation angle. Both effects also enter simply due to uncertainty in measuring the polarization axes on the sky. ‘Optical Cross-Polarization’ produces both rotation effects due to imperfections in the optical system. Birefringence in the lenses can cause a differential rotation of the plane of polarization. This effect tends to be maximum on axis, where the lenses are thickest. Mismatch in the optical coatings of refractive optics can also produce cross-polarization. Finally cross-polarization enters in all optical systems due to distortion, which changes the magnification over the field of view but also rotates angles on the sky.

One way to reduce optical cross-polarization is to use optics with low intrinsic cross-polar coupling. Centered optical systems exhibit substantially less cross-polar response compared to de-centered designs. Although secondary effects having to do with beam shape are also possible in cross-polarization, we ignore them since they can generally be made small by appropriate design choices. For example, for a well-designed feedhorn or antenna, the maximum cross- polarization response enters in the sidelobes and is strongly suppressed relative to the main beam. Cross-polar effects can be corrected *post-facto* by measurement of known polarized sources. We note in particular that diffuse polarized emission in the Galactic plane is relatively bright will be precisely characterized by sub-orbital and ground-based measurements well in advance of EPIC.

A challenge with cross-polarization is that the effects have quadrupolar rotational properties, i.e. the same rotational symmetry as that of the true B-mode signal. Thus cross-polarization effects can not be distinguished through rotation of the instrument. Spurious second-order cross-polarization effects can also arise, which produce non-vanishing cross-correlations between E and B and their respective power spectra, effectively introducing "forbidden spectra" such as EB.

## 4. Mission Overview

EPIC will measure the polarization of the CMB with a significant advance in sensitivity and polarized systematic error control compared to WMAP and Planck. The most compelling goal of EPIC, searching for a signal from the gravitational wave background produced by inflation, can be accomplished with a low-cost experiment package with limited angular resolution. An expanded mission, with essentially the same focal plane and sensitivity, but with a larger aperture for higher angular resolution, can access the signals produced by gravitational lensing of the CMB by intervening matter, and extract the full cosmological information from the E-mode polarization signal. Both options will also be able to measure Galactic polarization, providing a new window on Galactic magnetic fields.

### 4.1 Scientific goals

*Primary Goal: B-mode Polarization from Inflationary Gravitational Waves*

The main scientific goal of EPIC is to definitely and comprehensively search for the B-mode polarization signal from inflationary gravitational waves. How deep a search is possible? While instrument sensitivity and control of systematic errors are demanding, these is a clear technological path forward to overcome these hurdles, guided by the rapid progress in bolometer array technology and the experience gained in implementing new capabilities in sub-orbital experiments. Instead, we expect the ultimate limit to a search for the IGW polarization signal may well be set by local foregrounds. It is worth noting that the CMB itself has a spectrum that is known to extraordinarily high precision, and therefore any emission spectrum that deviates from the CMB blackbody spectrum can be identified as a foreground. Unfortunately, polarized foreground signals are not well measured in the millimeter-wave band, so it is difficult to project how deeply they can be removed. We take our scientific goal from the Weiss Committee, to search down to $r = 0.01$. The instrument is designed to achieve this goal *after foreground subtraction* based on a simplified model. Because EPIC has higher sensitivity than needed to detect an uncontaminated polarization signal, it has the potential go deeper than $r = 0.01$. Probing an IGW B-mode signal at $r = 0.01$ will provide a strong test of models of Inflation at the GUT energy scale.

*Secondary Scientific Goals*

CMB temperature anisotropy has proven to be our best window on the physical state of the universe at recombination. The CMB provides a well-understood, linear physical probe that can be used to understand the primordial power spectrum, the geometry, and the composition of the early universe. The cosmology obtained from CMB temperature anisotropy, measured by WMAP and other experiments, is the central pillar of the standard model in cosmology. CMB E-mode polarization offers an independent measure of the cosmological information available from the CMB, and will be far from fully mined even after Planck. A secondary goal of EPIC is therefore to extract all of the cosmological information from E-mode polarization, measuring the E-mode signal to cosmic variance into the Silk damping tail, much as WMAP has done for extracting all of the cosmology from temperature anisotropy (which will be completed by Planck into the damping tail).

Reionization can be studied based on the amplitude and shape of the E-mode spectrum at low multipoles. The history of reionization, i.e. the evolution of the ionization fraction over time even if $\tau$ is held constant, produces smaller secondary features in the E-mode spectrum. Measuring the E-mode spectrum at low multipoles to cosmic variance allows all of the information on reionization available in the CMB to be recovered.

**Table 4.1.1 Mapping NASA Objectives to EPIC Science Goals to Instrumentation**

<table>
<tr><th>NASA Objective</th><th>EPIC Objective</th><th>Measurement Criteria</th><th>Instrument Criteria</th></tr>
<tr><td rowspan="5">Discover what powered the Big Bang… search for gravitational waves from the earliest moments of the Big Bang<br><br>Discover the origin, structure, evolution, and destiny of the universe<br><br>(NASA 2006 Strategic Plan)</td><td rowspan="5">Test Inflationary paradigm at GUT energy scales by probing Inflationary Gravitational Wave B-mode polarization signal to r = 0.01.</td><td rowspan="3">Detect BB signal at r = 0.01* after foreground removal</td><td>$w_p^{-1/2} < 6$ μK-arcmin†</td></tr>
<tr><td>30 – 300 GHz†</td></tr>
<tr><td>Control systematics to negligible levels</td></tr>
<tr><td rowspan="2">Positively detect both the ℓ = 5 and ℓ = 100 BB peaks</td><td>All-sky coverage</td></tr>
<tr><td>Low angular resolution (< 1°)†</td></tr>
<tr><td>Understand how the first stars and galaxies formed</td><td rowspan="2">Distinguish models of reionization history<br><br>Extract all available EE cosmology</td><td rowspan="2">Measure EE to cosmic variance</td><td rowspan="5">Parameters above</td></tr>
<tr><td>Determine the size, shape, and matter-energy content of the Universe</td></tr>
<tr><td rowspan="2">Measure the cosmic evolution of the dark energy, which controls the destiny of the universe</td><td>Measure lensing BB to determine neutrino mass and dark energy equation of state</td><td rowspan="2">Measure lensing BB to ~cosmic variance</td></tr>
<tr><td>Remove lensing BB using shear map</td></tr>
<tr><td>…Trace the flows of energy and magnetic fields… between stars, dust, and gas</td><td>Map Galactic magnetic fields</td><td>Measure synchrotron and dust polarization</td></tr>
</table>

Primary Objective

Secondary Objective

†Parameters recommended by Weiss Committee TFCR

Intervening matter slightly distorts the observed CMB by gravitational lensing. This lensing affects the temperature and E-mode polarization, and produces an apparent B-mode polarization. A precision measurement of these signals can be used to ascertain the distribution and evolution of the intervening dark matter at a later epoch, and can be used to assess the effects of neutrino mass and dark energy. The lensing polarization is also a possible foreground (albeit an extragalactic foreground) for the IGW B-mode search, assuming the search is not limited by Galactic foregrounds first. Although the lensing polarization signal cannot be distinguished on the basis of spectral color, a high resolution measurement of the lensing temperature and polarization signals can be used to partially remove this foreground, allowing a potentially deeper IGW B-mode search. A mission designed to extract the E-mode spectrum can also make a definitive measurement of the lensing signal.

Finally, EPIC offers a rich data set for Galactic physics. Measuring polarized Galactic emission in multiple bands from synchrotron and thermal and spinning dust grains provides a new view of the Galaxy, particularly the magnetic field structure in the Galaxy which governs these processes.

### 4.2 Two Mission Options

In this report we detail two possible mission concepts, a ***Low-Cost Option*** with multiple small apertures designed to deliver the primary science goal of the IGW search on large angular scales at minimum cost and risk, and a ***Comprehensive Science Option*** with essentially the same focal plane technology but with a 4-m ambient telescope for the higher angular resolution needed to deliver the secondary science goals. The low-cost option uses largely proven technologies, and could be flown as soon as the focal plane and waveplate technologies are demonstrated. The comprehensive mission is designed to address a wider scope of science goals and targets a longer mission life.

**Table 4.2 Parameters for the Two Mission Options**

| Instrument Criteria | Low Cost Mission | Comprehensive Mission |
|---|---|---|
| High sensitivity | System sensitivity < 2 $\mu K\sqrt{s}$ | |
| Subtract foreground signals to r < 0.01 | Remove foregrounds to r < 0.01 with optimized band coverage between 30 – 300 GHz | |
| Control systematic errors to negligible levels | Suppress systematic errors to < 10% of r = 0.01 signal, after correction | |
| Large angular scales | All-sky coverage | |
| Angular resolution | Low: 1° FWHM at 100 GHz | Moderate: 5′ FWHM at 100 GHz |
| Design Philosophy | Target IGW signal<br>Minimize cost<br>Minimize risk | Maximize science return<br>4-year mission life |

We note that although the low-cost scenario does not achieve all of the secondary science goals in full, it can still make important contributions to these themes. The low-cost option will measure E-mode polarization to cosmic variance for $\ell < 500 - 1000$, and either approaches or meets cosmic variance on lensing B-mode polarization for $\ell < 100 - 200$, depending on the final sensitivity. Thus the low-cost scenario provides a definitive measurement on the large angular scales where a space mission is necessary. Ground-based polarimeters will capable of measuring the signals at large ℓ, providing a nearly complete picture.

### 4.3 The Role of Space in CMB Measurements

Space provides the ideal environment to meet these science goals for the following reasons:

**All-sky coverage:** The IGW B-mode power spectrum has a distinctive shape, with broad peaks at $\ell \approx 5$ and $\ell \approx 100$. The two peaks provide a strong predictive test, and a thus a measurement objective of our IGW B-mode search. Measuring the power spectrum on these large angular scales requires an experiment that maps most or all of the sky with high fidelity, which is only possible from space. Measuring the entire B-mode spectrum also enables a consistency test of inflation through the relation between $n_s$ and $n_t$.

**Systematic error control:** Space enables the extreme control of systematic errors needed in these difficult measurements. The environment of an observatory at L2 allows for exquisite thermal stability. We employ a scanning pattern, only possible in space, that completely rotates the instrument with respect to the sky, while allowing continuous calibration on the CMB dipole. Space-borne measurements produce well-characterized and uniform data sets, with lasting legacy value. As a result, CMB satellite experiments (COBE/DMR, COBE/FIRAS, and WMAP) have all made watershed advances in our knowledge of the early universe.

**Frequency coverage:** EPIC requires multiple frequency bands to deeply remove Galactic foregrounds. While the range and band coverage needed is unclear due to our limited knowledge of foregrounds at present, the Weiss committee recommends covering the entire region from 30 – 300 GHz. Atmospheric absorption limits the full exploitation of this band from the ground, and even prevents observing in selected bands (60 GHz, 120 GHz) from balloon altitudes. Only space allows us to cover the entire band with uniform sensitivity.

**Sensitivity:** Space offers both high instantaneous sensitivity, due to the low background, and long integration times, with uninterrupted observations over several years, depending on the mission design. A space-borne instrument has a significant overall sensitivity advantage compared to what can be achieved on a balloon or from the ground.

## 5. Low-Cost Mission Architecture

### 5.1 Low-Cost Mission Overview

The EPIC low-cost (EPIC-LC) mission concept is designed to accomplish the primary science objective, searching for IGW polarization down to $r = 0.01$, with an emphasis of minimizing mission cost and using well developed technologies that minimize risk. The secondary science goals are not completely fulfilled due to the limited angular resolution, and instead we rely on a combination of space-borne and ground-based measurements to fulfill these objectives. A larger and more capable comprehensive science (EPIC-CS) mission architecture, described in section 6, can fulfill these objectives without secondary data sets. The high-level design requirements and goals are summarized in Table 5.1.1.

**Table 5.1.1 Low Cost Mission Option Design Requirements and Goals**

| Instrument Criteria | Requirement | Design Goal |
|---|---|---|
| **High sensitivity** | $w_p^{-1/2} < 6$ μK-arcmin | $w_p^{-1/2} < 2$ μK-arcmin |
| **Subtract foreground signals to negligible levels** | Remove foregrounds to below r = 0.01 science goal | Optimize bands for foreground removal based on best knowledge |
| **Control systematic errors to negligible levels** | Suppress systematic errors to < 10% of r = 0.01 signal, after correction | Suppress raw systematic effects to less than 10% of statistical noise level |
| **Maintain sensitivity on large angular scales** | All-sky coverage with redundant interleaved scan strategy | |
| **Angular resolution** | < 1° at 100 GHz | |

The low-cost mission architecture is summarized in the foldout section. The main features are described in detail in the following subsections:

**Systematic Error Mitigation (Section 5.2):** Control of systematic errors is integral to our designs from the beginning. EPIC uses a spinning and precessing scan strategy to uniformly cover the sky in both integration time and to rotate the crossing angle over a given region of the sky. This pattern is highly redundant, allowing for daily data comparisons (jack-knife tests) to identify systematic effects. EPIC covers half the sky in a single day, making the instrument immune to data interruptions, loss of pixels, and even loss of arrays. The optics, designed for low main-beam distortion and polarization, are placed behind a cooled waveplate that eliminates any polarization produced by the optics. The optics and absorbing baffles give extremely low off-axis response. The instrument is shielded to provide a highly stable thermal environment.

**Refracting Optics (Section 5.3):** EPIC-LC uses six 30 cm refracting telescopes (see section 5.3). The wide-field optics provide large AΩ throughput, needed for large focal plane arrays, and high sensitivity. The lenses are cooled to 2 K and define the telecentric focal planes appropriate for planar detector arrays. An absorbing cold stop at the primary lens provides beam apodization. Each telescope (except the highest frequency telescope) is monochromatic for optimized performance. This refracting design has low instrumental and cross-polarization, and excellent control of main beam asymmetry and polarization. The telescope and absorbing 40 K forebaffle provides extremely low off-axis response. A prototype of this system has been tested and fielded for the BICEP experiment.

**Rotating Half-Wave Plate (Section 5.4):** EPIC-LC uses a continuously rotating half-wave plate for polarization modulation. Because it rotates the polarization vector without changing the field distribution, the waveplate modulates the polarization direction without altering the beams. Because the waveplate operates at 2 K *in front of the optics*, it does not modulate any polarization or main beam asymmetries produced by the optics. Thus this approach is extremely immune to systematic errors introduced by main beam imperfections. For risk mitigation, it is possible to spin the waveplate continuously if the noise stability of the focal plane is sufficient, although we consider this an increase in mission scope.

**TES/SQUID Focal Plane (Section 5.5):** Large arrays of antenna-coupled TES bolometers lie at the heart of EPIC. This technology enables high sensitivity over the entire frequency band recommended by the Weiss Committee, 30 – 300 GHz. Because the antennas replace the optical functions provided by massive Cu scalar feedhorns, the focal plane mass is significantly reduced. SQUID multiplexing reduces the wire count and power dissipation compared to unmultiplexed NTD/Ge technology. We currently carry both options, and find that the NTD/Ge bolometer option with noise margin just meets the sensitivity requirements.

**Cooling to 100 mK (Section 5.6):** EPIC-LC uses a simple liquid-helium cryostat to deliver a 1-year required lifetime with ample margin. Cryostat technology is well developed, and eliminates potential interactions between the sensitive TES detectors and an active cooler, and simplifies integration and verification testing of the instrument on the ground. EPIC makes full use of radiative cooling, with the shell of the cryostat running at 40 K for minimum heat load. We describe a 4-stage adiabatic demagnetization refrigerator that provides continuous operation at 100 mK, although other options are possible. Developments in TES detectors could enable focal plane operating temperatures as high as 250 mK with little loss in total sensitivity.

**Deployed Sunshade (Section 5.7):** The scanning/precessing scan strategy requires a deployed multilayer sunshield to both defeat solar input power to the instrument and to give a stable thermal environment. We have developed a design that requires no technology development, and is based on metalized kapton sheets deployed on simple hinged booms.

**EL2 Halo Orbit (Section 5.8):** We have chosen a compact halo orbit at L2, in order to reduce variations in the earth and moon angles over the orbit, relaxing requirements on the sunshield size and the fixed downlink antenna. A full analysis of insertion and station-keeping requirements is presented.

**Spacecraft Requirements (Section 5.9):** EPIC-LC uses well-developed space-proven technologies to the maximum extent possible. The instrument is mounted to a commercial spacecraft close to the performance specifications delivered by catalog spacecraft provided by several industrial vendors. The pointing and power requirements are within the capabilities of standard hardware. EPIC-LC is designed to fit within the volume and mass capabilities of a Delta-II 2925 launch vehicle. Likely replacements for this vehicle (e.g. Atlas V 401, Delta IV 4040) provide more mass lift and larger payload volume.

**Telemetry (Section 5.10):** EPIC/LC uses a fixed antenna with a toroidal beam pattern for downlink, eliminating the complications of a gimballed and continuously rotating antenna. The data rate requirements can be satisfied with a moderate on-board transmitter and twice-daily links to either a 30-m DSN station, or a commercial 12-m station.

**Cost (Section 5.11):** We provide a cost estimate of this option.

**Table 5.1.2 Baseline Instrument Parameters Summary Table**

| | |
|---|---|
| **Instruments** | Six telescopes (30 cm diameter x 95 cm long) |
| **Bands** | 30, 40, 60, 90, 135, 200 & 300 GHz |
| **Detectors** | 830 (baseline NTD), 2366 (TES option) |
| **Sensitivity** | $w_p^{-1/2}$ = 5.9 μK-arcmin (NTD required), 3.0 μK-arcmin (NTD design)<br>$w_p^{-1/2}$ = 3.6 μK-arcmin (TES required), 1.8 μK-arcmin (TES design) |
| **Resolution** | 16 – 155 arcmin (FWHM), diffraction limited by band |
| **FOV** | 20 deg |
| **Pointing Knowledge** | 30" |
| **Focal Plane** | Antenna-coupled NTD bolometers (baseline)<br>Transition-Edge Superconducting (TES) bolometers (upscope) |
| **Read Out** | Si JFETs mounted at 40 K with warm AC bias and demodulation<br>Multiplexed SQUID current amplifiers (TES upscope) |
| **Pol. Modulation** | Half-wave plate before telescope |
| **Optics** | Six 30-cm wide-field refractors |
| **Cryogenics** | Passive to 40 K / LHe cryostat to 2 K / ADR to 0.1 K |
| **Payload Mass** | 898 kg including 43% contingency |
| **Payload Power** | 272 W including 43% contingency |
| **Average Data Rate** | 88 kbps including 100% contingency (NTD option) |

**Table 5.1.3a Detailed Baseline NTD Ge Bands and Sensitivities**

| Freq [GHz] | $\theta_{FWHM}$ [′] | $N_{bol}$[3] [#] | Required Sensitivity[1] | | | | Design Sensitivity[2] | | | |
|---|---|---|---|---|---|---|---|---|---|---|
| | | | NET[4] [μK√s] | | $w_p^{-1/2}$ [μK-′][5] | δTpix[6] [nK] | NET[4] [μK√s] | | $w_p^{-1/2}$ [μK-′][5] | δTpix[6] [nK] |
| | | | bolo | band | | | bolo | band | | |
| 30 | 155 | 8 | 83 | 29 | 90 | 530 | 59 | 21 | 45 | 270 |
| 40 | 116 | 54 | 73 | 9.9 | 30 | 180 | 51 | 7.0 | 15 | 90 |
| 60 | 77 | 128 | 61 | 5.4 | 17 | 100 | 43 | 3.8 | 8.2 | 49 |
| 90 | 52 | 256 | 53 | 3.3 | 10 | 60 | 37 | 2.3 | 5.0 | 30 |
| 135 | 34 | 256 | 49 | 3.1 | 10 | 56 | 35 | 2.2 | 4.7 | 28 |
| 200 | 23 | 64 | 59 | 7.3 | 22 | 130 | 41 | 5.2 | 11 | 67 |
| 300 | 16 | 64 | 120 | 15 | 44 | 260 | 82 | 10 | 22 | 130 |
| **Total[7]** | | **830** | | **1.9** | **5.9** | **35** | | **1.4** | **3.0** | **18** |

**Table 5.1.3b Detailed Bands and Sensitivities for TES Option**

| Freq [GHz] | $\theta_{FWHM}$ [′] | $N_{bol}$[3] [#] | Required Sensitivity[1] | | | | Design Sensitivity[2] | | | |
|---|---|---|---|---|---|---|---|---|---|---|
| | | | NET[4] [μK√s] | | $w_p^{-1/2}$ [μK-′][5] | δTpix[6] [nK] | NET[4] [μK√s] | | $w_p^{-1/2}$ [μK-′][5] | δTpix[6] [nK] |
| | | | bolo | band | | | bolo | band | | |
| 30 | 155 | 8 | 80 | 28 | 87 | 520 | 57 | 20 | 44 | 260 |
| 40 | 116 | 54 | 71 | 9.6 | 29 | 180 | 50 | 6.8 | 15 | 88 |
| 60 | 77 | 128 | 60 | 5.3 | 16 | 97 | 42 | 3.7 | 8.1 | 48 |
| 90 | 52 | 512 | 52 | 2.3 | 7.0 | 42 | 37 | 1.6 | 3.5 | 21 |
| 135 | 34 | 512 | 49 | 2.2 | 6.6 | 39 | 35 | 1.5 | 3.3 | 20 |
| 200 | 23 | 576 | 54 | 2.3 | 6.9 | 41 | 38 | 1.6 | 3.5 | 21 |
| 300 | 16 | 576 | 92 | 3.8 | 12 | 70 | 65 | 2.7 | 5.9 | 35 |
| **Total[7]** | | **2366** | | **1.2** | **3.6** | **22** | | **0.8** | **1.8** | **11** |

Notes:
[1]Sensitivity with √2 noise margin in a 1-year mission
[2]Calculated sensitivity with 2-year mission life
[3]Two bolometers per focal plane pixel
[4]Sensitivity of one bolometer in a focal plane pixel
[5]$[8\pi\, NET_{bolo}^2/(T_{mis}\, N_{bol})]^{1/2}(10800/\pi)$
[6]Sensitivity δT in a 120′ x 120′ pixel
[7]Combining all bands together

**Table 5.1.4 Sensitivity Model Input Assumptions**

| | | | | | |
|---|---|---|---|---|---|
| Optics temperature | $T_{opt}$ | 2 K | Focal plane temperature | $T_0$ | 100 mK |
| Optics coupling* | $\varepsilon_{opt}$ | 10% | Optical efficiency* | η | 40% |
| Wave plate temperature | $T_{wp}$ | 2 K | Fractional bandwidth* | Δν/ν | 30% |
| Wave plate coupling* | $\varepsilon_{wp}$ | 2% | NTD Ge heat capacity* | $C_0$ | 0.25 pW/K |
| Baffle temperature | $T_{baf}$ | 40 K | NTD time constant† | $\tau(d\theta/dt)/\theta_F$ | ≤ 1/2π |
| Baffle coupling* | $\varepsilon_{baf}$ | 0.3% | TES safety factor† | $P_{sat}/Q$ | 5 |

*Parameter based on experimental measurement
†Selectable design parameter, $\theta_F$ is FWHM

**Table 5.1.5 Comparison of EPIC and Planck Sensitivity $w_p^{-1/2}$**

| Freq [GHz] | EPIC Baseline | | EPIC TES Option | | Planck[1] |
|---|---|---|---|---|---|
| | Req'd | Design | Req'd | Design | Goal |
| 30 | 90 | 45 | 87 | 44 | 350 |
| 40 | 30 | 15 | 29 | 15 | 350 |
| 60 | 17 | 8.2 | 16 | 8.1 | 350 |
| 90 | 10 | 5.0 | 7.0 | 3.5 | 100 |
| 135 | 10 | 4.7 | 6.6 | 3.3 | 80 |
| 200 | 22 | 11 | 6.9 | 3.5 | 130 |
| 300 | 44 | 22 | 12 | 5.9 | 400 |
| **Total[2]** | **5.9** | **3.0** | **3.6** | **1.8** | **54** |

[1]Planck combined sensitivities in polarization for 1.2 year mission lifetime. Planck bands are shifted slightly to match the closest EPIC band.
[2]Total $w_p^{-1/2}$ is combined $w_p^{-1/2}$ from all bands in μK-arcmin

**Table 5.1.6 EPIC-LC Technology Readiness**

| Technology | TRL | Heritage |
|---|---|---|
| Focal Plane Arrays (NTD Ge bolometers) | | |
| NTD thermistors and JFET read outs | 8 | Planck & Herschel |
| Antennas | 4 | Demonstrated at 100 and 150 GHz |
| Wide-Field Refractor | 6 | BICEP |
| Wave plate (stepped every 24 hours) | | |
| Wave plate optics | 6 | SCUBA, HERTZ, MAXIPOL, etc. |
| Cryogenic stepper drive | 9 | Spitzer |
| LHe Cryostat | 9 | Spitzer, ISO, Herschel |
| Sub-K Cooler: Single-shot ADR | 9 | ASTRO-E2 |
| Deployable Sunshield | 4-5 | All components TRL = 9 |
| Toroidal-Beam Downlink Antenna | 4-5 | All components TRL = 9 |

## Table 5.1.7 Scientific Risk Assessment

| Instrument Criteria | Requirement | Impact of Not Meeting Requirement | Mitigations |
|---|---|---|---|
| **Sensitivity** | $w_p^{-1/2} < 6$ μK-arcmin | Sensitivity to r decreases | - NET has 1.4x margin<br>- Lifetime has 2x margin |
| **Foreground subtraction** | Remove foregrounds to below r = 0.01 science goal | | - Limited subtraction needed in clean regions<br>- Wide band coverage, flexible band weighting |
| **Systematic error control** | Suppress systematic error to < 10% of r = 0.01 signal, after correction | | - Multiple levels of polarization signal modulation<br>- Wave plate in front of telescope<br>- Temperature control<br>- High mapping redundancy to assess systematic error contributions |
| **Mapping large angular scales** | All-sky coverage with redundant interleaved scan strategy | Sensitivity at low ℓ reduced | - Similar scanning technique already demonstrated for WMAP |
| **Angular resolution** | < 1˚ at 100 GHz | Sensitivity at high ℓ reduced | - Chosen by design |

## Table 5.1.8 Detailed Risk Reduction Strategy

| Instrument Requirement Risk | Approach | Risk Mitigations |
|---|---|---|
| **Sensitivity** | NTD Ge detectors | Heritage from Planck & Herschel<br>Requirement Ies √2 noise margin<br>Up scope to TES bolometers when mature |
| **Subtract foregrounds to r < 0.01** | Antenna-coupled bolometers | Single technology covers 30 – 300 GHz |
| **Suppress systematic errors to < 10% of r = 0.01 signal, after correction** | Spinning/precessing scans | Uniform angular coverage on the entire sky |
| | Highly redundant scans | Daily maps cover > 50% of sky for jackknife tests.<br>Immunity to data interruptions, bad pixels, bad arrays<br>Two full maps in 1-year for systematic error testing |
| | Dual-polarization detector | Suppresses common-mode temperature signals, thermal drifts |
| | Wave plate modulator in front of telescope | Suppresses main beam systematics by modulating polarization without altering beam shapes<br>Suppresses 1/f noise, gain and temperature drifts by signal modulation if continuous |
| | Monochromatic refracting telescope | Low instrumental and cross-polarization<br>Low main beam asymmetries<br>Optimized low-reflection coatings<br>Low far-sidelobe response |
| **All-sky coverage** | 1-year required lifetime | Cryostat lifetime at L2 has > 100% margin |
| **Technical simplicity and cost** | 30 cm refracting telescope | Polyethylene lenses, simple AR coatings<br>Demonstrated technology in BICEP |
| | LHe cryostat | Low technology risk<br>Low integration risk: no microphonics, EMI or B-field disturbances<br>Readily allows systems-level testing |
| | Commercial spacecraft | Modest requirements on spacecraft |
| | Fixed downlink antenna | Eliminates risk of counter-rotating antenna |

## Table 5.1.9 Detailed Mass Summary

| Sub-Assembly | | Mass (CBE) [kg] | Contingency [%] | Allocated Mass [kg] |
|---|---|---|---|---|
| **Focal Planes** | Mass at 0.1 K per unit | 0.9 | 43 | 1.3 |
| | Mass at 0.4 K per unit | 1.0 | 43 | 1.4 |
| | Mass at 2 K per unit | 0.5 | 43 | 0.7 |
| | **Total Focal Plane Assemblies (6)** | **14.2** | **43** | **20.3** |
| **Tele-scopes** | Lenses at 2 K per unit | 2.1 | 43 | 3.0 |
| | Supports per unit | 1.6 | 43 | 2.3 |
| | Shields per unit | 0.9 | 43 | 1.3 |
| | **Total Telescope Assemblies (6)** | **27.8** | **43** | **39.8** |
| **Wave Plates** | Wave plate 3-stack ave per unit | 3.0 | 43 | 4.3 |
| | Suspended bearing/motor per unit | 1.5 | 43 | 2.1 |
| | Non-suspended mass | 2.5 | 43 | 3.6 |
| | **Total Wave plates (6)** | **41.7** | **43** | **59.6** |
| **Adiabatic Demagnetization Refrigerator** | | **5.7** | **43** | **8.2** |
| **Ejectable Telescope Covers (6)** | | **6.0** | **43** | **8.6** |
| **Cryostat and Shell** | Liquid Helium | 62.9 | 0 | 62.9 |
| | Helium Tank | 29.5 | 43 | 42.2 |
| | Vapor-Cooled Shields | 57.5 | 43 | 82.2 |
| | Vacuum shell | 185.0 | 43 | 264.6 |
| | MLI | 12.7 | 43 | 18.2 |
| | Fill/vent lines, valves, ports | 16.0 | 43 | 22.9 |
| | **Total Cryostat and Shell** | **363.6** | | **493.0** |
| Cabling | | 7.0 | 43 | 10.0 |
| Warm Electronics | | 40.0 | 43 | 57.2 |
| V-groove Radiators | | 51.3 | 43 | 73.4 |
| Deployed Sunshield | | 74.1 | 43 | 106.0 |
| Struts from S/C to Instrument | | 15.5 | 43 | 22.2 |
| **Subtotal for Wet Payload** | | **646.9** | | **898.3** |
| Attitude Control System | | 81.9 | 43 | 117.1 |
| C&DH | | 24.1 | 43 | 34.5 |
| Power | | 52.6 | 43 | 75.2 |
| Propulsion (dry) | | 22.1 | 43 | 31.6 |
| Structures and mechanisms | | 212.9 | 43 | 304.4 |
| Launch adapter | | 14.3 | 43 | 20.4 |
| Cabling | | 46.4 | 43 | 66.4 |
| Telecom + X-band Antenna | | 18.7 | 43 | 26.7 |
| Thermal | | 25.5 | 43 | 36.5 |
| Propellant [ΔV = 215 m/s] | | 172.0 | 0 | 172.0 |
| **Subtotal for Wet Spacecraft** | | **670.5** | | **884.8** |
| **Total Launch Mass** | | **1318** | | **1783** |
| **Launch Vehicle Maximum Payload Mass to L2 (C3 = -0.6)** | | | | |
| **Vehicle** | | **Pld Mass [kg]** | **Margin [%]** | **Margin [kg]** |
| **Atlas V 401** | | **3485** | **95** | **1702** |
| **Delta IV 4040** | | **2773** | **56** | **990** |

## 5.2 Systematic Errors for EPIC-LC

EPIC LC is designed to provide the highest polarization fidelity possible for a realistic CMB polarimeter capable of detecting B-mode polarization at the 10 nK level. We have calculated the performance necessary to suppress raw systematic errors to a negligible level, which is our goal for systematic error control as described in Table 3.0.1. As noted, this is a conservative goal since it is always possible to correct and remove systematic errors after the fact given sufficient knowledge of the effect. We have carried out detailed calculations for a variety of systematic errors, and have developed a new and precise formalism to estimate polarization effects induced by the optics. For all of the systematic errors we have considered, the level of performance required appears to be achievable by design. In several cases, namely thermal stability and far-sidelobe response, sufficient suppression has already been demonstrated in working pathfinder instrumentation systems.

Although we have gone to great lengths to obtain the highest possible polarimetric fidelity in EPIC's thermal, mechanical and optical design, it is possible that residual imperfections will persist. Therefore we have designed EPIC not only to suppress raw systematic effects, but to allow in-orbit jackknife tests in order to assess systematic errors in situ.

### *5.2.1 Goals and Requirements for EPIC-LC*

We have developed requirements for the precision to which each systematic effect must be suppressed or measured in EPIC-LC to meet our requirements (outlined in section 3.0), listed in Table 5.2.1. Calculations specific to the low angular resolution chosen for EPIC-LC relate to main beam effects and pointing requirements, and are calculated in section 5.2.3. The temperature stability requirements are taken for the more capable TES focal plane parameters from Table 5.1.3b.

For thermal fluctuations, we calculate instantaneous requirements by requiring that the leakage temperature noise is less than 10% of the detector NEP when all noise sources (detector, photon, and systematic) are added in quadrature. Following the conventions listed in table 3.0.1, "requirement" is to control the systematic effects below the required NEP level, and "goal" is similar but for systematic NEP compared to the goal NEPs. We also calculate a scan-synchronous temperature variation.

Table 5.2.1 lists the required and goal suppression factors for numerous potential systematic errors. For beam effects we take the goal to be $[\ell(\ell+1)C_\ell/2\pi]^{1/2} = 1$ nK at $\ell = 200$, the requirement to be 3 nK at $\ell = 200$. For scan-synchronous effects, we take the suppression to be simply flat at 3 $nK_{rms}$ (required) and 1 $nK_{rms}$ (goal), which is the approximate level required although without accounting for the spatial signature of the particular effect. We assume a common-mode rejection ratio in detector pairs to unpolarized sources of optical emission as 100, and to focal plane temperature variations as 20. Note again that scan-synchronous effects are conservative – effects associated with the instrument alone average down over the course of entire EPIC observing campaign as the satellite maps out a large range of spin and precession angle over the scan strategy (section 5.2.5). In particular, common mode temperature fluctuations, which may escape detection in individual pairs, will tend to strongly average down. Since different detectors view different parts of the sky, effects that are common to the entire focal plane will also benefit from this additional averaging, particular on small scales.

Recently a full time-domain simulation has been carried out for optimizing the observing strategy of the SPIDER polarization experiment [1]. This analysis, which illustrates the benefits of using a stepped waveplate, is generally consistent with the goals and requirements quoted for EPIC in Table 5.2.1. A clear future direction would be to carry out a similar time-domain error analysis for the EPIC observing strategy.

**Table 5.2.1.** Systematic Error Goals and Requirements for EPIC-LC

| Systematic Error | Description | Suppression to Meet Goal | Knowledge to Meet Requirement |
|---|---|---|---|
| *Main Beam Effects[1] – Instrumental Polarization* | | | |
| **Δ Beam Size** | $FWHM_E \neq FWHM_H$ | $(\sigma_1-\sigma_2)/\sigma < 4 \times 10^{-5}$ | $(\sigma_1-\sigma_2)/\sigma < 10^{-4}$ |
| **Δ Gain** | Mismatched gains | $(g_1-g_2)/g < 10^{-4}$ | $(g_1-g_2)/g < 3 \times 10^{-4}$ |
| | Mismatched AR coating | $\Delta n/n < 6 \times 10^{-4}$ | $\Delta n/n < 2 \times 10^{-3}$ |
| **Δ Beam Offset** | Pointing E ≠ Pointing H | Δθ < 0.14" raw scan<br>Δθ < 10" symm. scan | Δθ < 0.4" raw scan<br>Δθ < 30" symm. scan |
| **Δ Ellipticity** | $e_E \neq e_H$<br>$\Delta e = (e_1-e_2)/2$ | $\Delta e < 5 \times 10^{-4}$, $\psi = 0°$<br>$\Delta e < 6 \times 10^{-6}$, $\psi = 45°$ | $\Delta e < 1.5 \times 10^{-3}$, $\psi = 0°$<br>$\Delta e < 2 \times 10^{-5}$, $\psi = 45°$ |
| **Satellite Pointing** | Q and U beams offset | < 12" | < 36" |
| *Main Beam Effects – Cross Polarization* | | | |
| **Δ Rotation** | E & H not orthogonal | $\Theta_1-\theta_2 < 4'$ | $\theta_1-\theta_2 < 12'$ |
| **Pixel Rotation** | E ⊥ H but rotated w.r.t. beam's major axis | < 2.4′ | < 7.2′ |
| **Optical Cross-Pol** | Birefringence | $n_e-n_o < 4 \times 10^{-5}$ | $n_e-n_o < 10^{-4}$ |
| *Scan Synchronous Signals* | | | |
| **Far Sidelobes** | Diffraction, scattering | $< 1\ nK_{CMB}$ | $< 3\ nK_{CMB}$ |
| **Thermal Variations** | Solar power variations | | |
| **Magnetic Pickup** | Susceptibility in readouts and detectors | | |
| *Thermal Stability[2]* | | | |
| **40 K Baffle[3,5]** | Varying optical power from thermal emission | 5 mK/√Hz; 25 μK s/s | 15 mK/√Hz; 75 μK s/s |
| **2 K Optics[3,6]** | | 500 μK/√Hz; 1 μK s/s | 1.5 mK/√Hz; 3 μK s/s |
| **0.1 K Focal Plane[4,7]** | Thermal signal induced in detectors | 200 nK/√Hz; 0.5 nK s/s | 600 nK/√Hz; 1.5 nK s/s |
| *Other* | | | |
| **1/f Noise** | Detector and readout drift | 0.016 Hz (1 rpm) | 0.2 Hz (1 rpm) |
| **Passband Mismatch** | Variation in filters | $\Delta\nu_c/\nu_c < 1 \times 10^{-4}$ | $\Delta\nu_c/\nu_c < 1 \times 10^{-3}$ |
| **Gain Error** | Gain uncertainties between detectors | $< 10^{-4}$ | $< 3 \times 10^{-4}$ |

[1]Main beam effects calculated at 100 GHz, no averaging over the focal plane is assumed
[2]Calculated at 100 GHz, at signal modulation frequencies, expressed for instantaneous and scan-synchronous signals respectively.
[3]Assumes 1% matching to unpolarized optical power, calculated at 90 GHz to give $1\ nK_{CMB}$(rms).
[4]Assumes 5% matching to focal plane drifts, calculated at 90 GHz to give $1\ nK_{CMB}$(rms).
[5]Planck achieves < 30 μK/Hz at 4 K regulated on Sterling-cycle cooler stage
[6]Planck achieves < 5 μK/√Hz at 1.6 K regulated on open-cycle dilution refrigerator J-T stage
[7]Planck achieves < 40 nK/√Hz at 0.1 K regulated on focal plane with open-cycle dilution refrigerator

*5.2.2 Systematic Error Mitigation Strategy*

We have designed multiple levels of systematic error suppression into EPIC-LC, exploiting the natural advantages provided by a differential polarimeter. EPIC-LC's small aperture provides notable advantages. Aberrations and polarization associated with the on-axis refractive telescope can be controlled to extremely low levels based on our design analysis. By cooling the optics to 2 K, the effect of temperature drifts in the optics is reduced. Because the optics are compact, we can heavily baffle and shield the aperture for stray light. The illumination pattern on the primary aperture is controlled by the combination of antennas or feeds in the focal plane, and an absorbing tapered stop at 2 K surrounding the primary. Depending on the focal plane packing, we estimate the cold stop is illuminated at -10 to -20 dB. The cold stop thus terminates the sidelobe pattern of the focal plane antenna or feed on a black and temperature stable surface. Following the approach used on BICEP, we plan to use an absorbing baffle to control far-sidelobe response. The baffle, operating at ~40 K, introduces acceptably small optical power to the detectors, and significantly improves the far-sidelobe response. The EPIC-LC sun-shield prevents the sun from illuminating any 40 K surface, resulting in an extremely stable thermal environment. The sun shield also prevents the moon from shining on the inside surface of the baffle. The thermal impact of the moon shining on the outside of the baffle, wrapped in multi-layer thermal blanketing, is negligible.

While our design study shows that it is possible to control raw main beam effects to the appropriate level, we recognize that this has not yet been demonstrated to the required level in hardware. Therefore we plan to install a half wave plate in front of each telescope aperture, clearly a design choice not possible with a larger telescope. The wave plate serves to eliminate polarization downstream by the optics and detectors, and eliminates main beam effects. This can be understood by noting that the wave plate rotates the polarization vector on the sky but does not change the illumination pattern on the aperture, and therefore does not change the beam shape on the sky. In this way, main beam effects and polarization from the optics can be eliminated no matter how large these effects may be.

Of all the potential systematic effects, conversion of unpolarized intensity to polarized intensity is by far the most pernicious. While this conversion, or *leakage*, can potentially arise from numerous thermal sources, the most effective suppression technique is to exploit the differential nature of CMB polarization and difference two spatially co-located detectors. This differencing, or *polarization analysis,* is the decomposition of incoming polarization into its two constituent linear polarization states. For EPIC this is achieved using two bolometers for each pixel, with orthogonal axes of polarization sensitivity. This approach has significant technical heritage. It was first used on BOOMERANG, then later on QUAD and BICEP. Assuming that detector pairs can be matched in optical and thermal properties to allow the rejection of common-mode signals to a level of 1%, the resulting temperature control requirements are listed in Table 5.2.2. These requirements have largely been already met by the monitoring and control techniques developed for Planck. However we note that an additional level of control can be obtained by continuously rotating the waveplate, which modulates the polarization signal at a higher frequency than with scanning, reducing the effect of drifts.

Striping due to 1/f noise in the focal plane is mitigated by high intrinsic stability in the focal plane detectors and readout electronics. The level of stability required for EPIC-LC has already been demonstrated with NTD Ge detectors in ground-based tests of the Planck High Frequency Instrument. For TES detectors, the level of 1/f noise suppression has been shown in tests of detectors in the laboratory, but has not yet been demonstrated in a fielded experiment.

While we are optimistic that this level of stability can be achieved, 1/f noise can be virtually eliminated by continuously rotating the half wave plate.

Scan synchronous errors can be induced by off-axis response to the Galaxy and sun, as previously discussed, but also due to thermal effects induced by the sun, and magnetic field pickup in SQUID readouts. Shadowing on the sunshield from the spacecraft and solar panels can introduce a scan-synchronous thermal signal. An order of magnitude calculation of the impact of thermal excursions from shadowing on the sunshield indicates these fluctuations are probably completely negligible at 40 K. However, we plan to monitor and control the temperature of the 40 K forebaffle to suppress any such effect. The requirements for magnetic shielding of SQUID readouts, presented in Table 5.5.6, are significantly less severe than what is needed on terrestrial pathfinder experiments due to the low fields at L2.

EPIC's scan strategy greatly reduces residual main-beam effects. CMB polarimeters benefit from a scanning strategy that not only modulates the polarized intensity but also the polarization angle. By viewing each pixel at a variety of polarization angles, we make use of the spin-2 nature of polarization (360 degrees of signal modulation for each 180 degrees of physical rotation about the pixel center). For illustration of the scan strategy's importance we note that in the absence of detector noise, contamination by several main-beam systematic effects, such as the two rotationally-symmetric (monopole) effects (differential gain and differential FWHM) can be *completely eliminated* if sky pixels are scanned with more than one polarization angle. Uniform polarization angle coverage for each map pixel is considered ideal, however in practice this is obviously impossible to achieve. EPIC LC's scan strategy, however, is very close to ideal, with only slight ecliptic-latitude dependence.

The scan strategy is also highly redundant. We obtain fully sampled independent maps of more than half the sky after several precession cycles (a few hours to a day), and complete maps of the entire sky in six months. This redundancy allows for the application of multiple statistical jackknife tests. For example, by making maps in fixed scan angles, and by comparing maps before and after wave plate rotation, we can assess the amplitude of main beam polarization effects before they are removed by wave plate rotation and view angle rotation. We can construct difference maps on multiple time scales (hours, days, weeks, months, years) to accurately assess instrument noise. The absolute and relative gain of each detector is measured on the dipole on the same region of the sky on the time scale of several hours. Over the course of 6 months we can produce maps in fixed spin angle or fixed precession angle to assess spin synchronous signals. Finally the high-redundancy of the scan pattern mitigates against data interruptions, loss of pixels, and loss of arrays.

**Table 5.2.2** Systematic Error Mitigation Architecture in EPIC-LC

<table>
<tr><th>Systematic Error</th><th>Goal Suppression</th><th>Mitigation</th><th>Heritage</th></tr>
<tr><td colspan="4">Main Beam Effects – Instrumental Polarization</td></tr>
<tr><td>Δ Beam Size</td><td>$(\sigma_1-\sigma_2)/\sigma < 4 \times 10^{-5}$</td><td rowspan="4">Half wave plate in front of telescope<br>Refracting telescope<br>Scan crossings</td><td rowspan="4">SPIDER & SPUD‡<br>N/A<br>BICEP† & SPIDER‡</td></tr>
<tr><td>Δ Gain</td><td>$(g_1-g_2)/g < 10^{-4}$</td></tr>
<tr><td>Δ Beam Offset</td><td>$\Delta\theta < 0.14''$ raw scan<br>$\Delta\theta < 10''$ symm. scan</td></tr>
<tr><td>Δ Ellipticity</td><td>$\Delta e < 5 \times 10^{-4}$, $\psi = 0^\circ$<br>$\Delta e < 6 \times 10^{-6}$, $\psi = 45^\circ$</td></tr>
<tr><td>Satellite Pointing</td><td>< 12"</td><td>Dual analyzers</td><td>Planck</td></tr>
</table>

| Systematic Error | Goal Suppression | Mitigation | Heritage |
|---|---|---|---|
| | | Gryo + tracker system | Many |
| *Main Beam Effects – Cross Polarization* | | | |
| **Δ Rotation** | $\theta_1$-$\theta_2$ < 4′ | Half wave plate in front of telescope | SPIDER & SPUD‡ |
| **Pixel Rotation** | < 2.4′ | | |
| **Optical Cross-Pol** | $n_e - n_o < 10^{-4}$ | Measure and subtract | Planck |
| *Scan Synchronous Signals* | | | |
| **Far Sidelobes** | < 1 $nK_{CMB}$ | Refracting optics and absorbing baffle | BICEP† |
| **Thermal Variations** | | Passive thermal design | Planck† |
| **Magnetic Pickup** | | Focal plane shielding | SPIDER‡ |
| *Thermal Stability*[1] | | | |
| **40 K Baffle**[2,4] | 5 mK/√Hz; 25 μK s/s | Dual analyzers, Temperature monitoring & control | Planck* |
| **2 K Optics**[2,5] | 500 μK/√Hz; 1 μK s/s | | |
| **0.1 K Focal Plane**[3,6] | 200 nK/√Hz; 0.5 nK s/s | | |
| *Other* | | | |
| **1/f Noise** | 0.016 Hz (1 rpm) | NTD Ge detectors<br><br>Faster scan for TES; HWP modulation | EBEX & SPIDER‡<br><br>BOOM & MAXIMA†, SPIDER‡ |
| **Passband Mismatch** | $\Delta\nu_c/\nu_c < 10^{-4}$ | Match bands as closely as practical, measure to the required level | Planck† |
| **Gain Error** | $< 10^{-4}$ | Orbit-modulated dipole | WMAP† |

*Performance already demonstrated to level required for EPIC
†Proof of operation, but requires improvement for EPIC
‡Planned demonstration to level required for EPIC.

**Table 5.2.3** Systematic Error Checks in Flight

| Systematic Effect | In-Flight Checks |
|---|---|
| **Main Beam Effects** | Combine data in fixed view angles<br>Combine data in fixed HWP angles |
| **Instrument Noise Model** | Construct difference maps |
| **Spin Synchronous Signals** | Combine data in fixed spin and precession angles |
| **Relative Pair Gains** | Orbit-modulated CMB dipole using dipole as a transfer standard |
| **Instrument Gain Model** | |

While the hardware mitigation methods are powerful, and greatly reduce the raw systematics level, mitigation is not restricted to hardware solutions alone. We have modeled techniques to reduce or remove residual systematic effects *post-observation*. Some strategies can only be implemented once the data has been acquired. For example, with knowledge of the main-beam and CMB temperature anisotropy, it is possible, in principle, to subtract the spurious systematic polarization resulting from the beams, if any artifact remains. We note that in cases where the beam parameters are not measured to the required precision it is still possible to estimate the degradation of the IGB signal by marginalizing over the unknown beam parameters.

Additionally, "forbidden spectra" such as the cross-correlation between temperature anisotropy and B-modes can be used to diagnose systematic effects. These spectra should not arise in the standard cosmological model, and their presence is indicative of leakage from temperature anisotropy to polarization. Similar techniques have been proposed to be used to extract gravitational lensing information, and in this application second-order, non-Gaussian artifacts can result.

### *5.2.3 Modeling and Analysis of Main Beam Effects*

To assess the tolerable level of optical systematic effects for which our scientific goal of detecting the IGB signal is not compromised we performed simulations of the main beam effects shown in Fig. 3.1.1. Two independent simulations pipelines were developed to appraise the nominal level of systematic effects associated with deviations of the main-beam from ideal.

Previous studies to quantify systematic susceptibility [2] used Jones matricies, an excellent choice for coherent amplification polarimeters such as WMAP. However, the Jones matrix formalism is not appropriate for bolometric polarimeters such as EPIC. Our formulations are instead based on Stokes matrices. The two methods we developed carried out calculations in map space and Fourier space. A detailed explanation of the map-based calculation is described in appendix A.

To quantify the impact of the main-beam systematics, we treated each focal-plane pixel as a separate polarimeter, with two linearly polarized detectors. For each pair of detectors in a single spatial pixel, one is aligned with the E-plane of the antenna, and the other with the H-plane. Any deviation in the shape, pointing or gain of the E- and H-plane beams from a nominally symmetric Gaussian leads a spurious polarization. The linear polarization is described by the Q and U Stokes parameters which are deduced by subtracting the intensity measured with two beams whose polarization sensitive axes are (ideally) orthogonal. Rotating about the boresight by 45 degrees and repeating the procedure yields the U parameter.

The main-beam optical systematic effects we simulated (see Fig. 3.1.1.) were 1) differential beamwidths for the E and H planes, 2) differential ellipticity of the two beams, 3) differential optical or detector gain between the two polarization planes, and 4) differential rotation or misalignment between the E and H plane beams. These effects can be further classified according to their behavior under rotation of the beams for each pixel about their symmetry axes. This is the approach first taken by [2] who simulated systematic effects in coherent (RF amplifying) polarimeters such as DASI and WMAP.

Once the beam imperfections are specified we investigate the dependence of the spurious polarization on the level of mismatch between the two beams. We first generate a realization of the CMB temperature anisotropy using Healpix [3]. Realizations including the E-mode polarization are generated when we investigate XP effects, but the B-mode power is always set to zero so that any B-mode power which appears is spurious. We then convolve the simulated maps with the beam, including each of the main-beam systematic effects. Next, the sky is scanned using EPIC's scan strategy and a coverage map is constructed. Then we reconstruct the observed temperature and polarization maps and finally, we extract temperature and polarization power spectra from the simulated maps using SpicePol [4]. The systematic parameters are varied over a wide range of values and from the dependence of the B-mode power spectrum we can compute EPIC's susceptibility.

***Main-Beam Effects***

The main-beam effects introduced in Section 3 mix CMB TT and EE signals into BB, giving a systematic BB polarization signal on all angular scales irrespective of the beam profile. *Differential gain* directly leaks CMB temperature to polarization. This effect can be removed by accurately calibrating the relative gains of detector pairs on the CMB dipole. *Differential rotation* effect cases E-B mixing. A second class of beam systematic is related to the matching of the beam shapes, more specifically, to the difference of the components of the gradient of the beam profile which couples to the underlying T, Q and U on the sky; these are *differential beamwidth* (monopole symmetry), *differential pointing* (dipole symmetry), and *differential ellipticity* (quadrupole symmetry). These effects mix gradients in TT and EE into BB polarization. The differential beam width effect is described as different FWHMs of the two Gaussian beams used to construct a Stokes Q or U parameter. *Differential pointing* is the effect induced by identical beams but with differential pointing centroids. *Differential ellipticity* is due to mismatched ellipticity of the beam profiles. This effect possesses the *same* quadrupolar symmetry as does the polarization of the CMB, and is thus not reduced by rotating the telescope. In general, all these systematics are coupled and can occur within the same system. Fig. 3.1.1 shows the polarization patterns and profiles of the various main beam systematics.

***Map-Based Calculation***

Rosset *et al*. (2007) [5] used simulations of the Planck beams to account for realistic beam systematics. For EPIC we modified this methodology and incorporated EPIC's scanning strategy into the formalism, and calculated the gradients of the underlying temperature and polarization fields on the spherical sky in map space with Healpix and assessed the level of spurious B-mode polarization. The relevance of scanning strategy to the calculation of the systematics is especially transparent in the case of the monopole effects. For example, for the *differential gain* effect, since the polarized beam patterns are perfectly symmetric, an ideal scan strategy with uniform polarization angle coverage will yield no spurious polarization of this type. However, real experiments never have this property of perfect sky coverage, and additionally, the coupling between the beam shape and imperfection-parameters as well as unavoidable inhomogeneity in the systematic parameters induces non-vanishing monopole and differential-gain signals. There will also be similar, albeit second order, contributions from scanning strategy to the other systematics discussed here.

Once the BB residuals were calculated, we investigated their dependence on the level of mismatch between the two beams. We first generated a realization of the CMB TT and EE anisotropy using Healpix [3]. The B-mode power is always set to zero so that any B-mode power which appears is spurious. We then convolved the simulated maps with the beam, including each of the four systematic effects. Then we reconstructed the observed temperature and polarization maps using the scan strategy map. Finally, we extracted temperature and polarization power spectra. The systematic parameters were varied over a wide range of values and from the dependence of the B-mode power spectrum so we can compute EPIC's susceptibility to the four main-beam systematic effects.

***Multipole-Space Calculation***

In addition to the time-ordered data approach discussed above, we appraised the impact of systematics in Fourier space [6] from the outset. Ignoring the important issue of scanning strategy, it is natural to work in Fourier space because the final product of our calculation is the power spectra of the contamination due to the beam systematics. These power spectra will eventually be subtracted from the *raw* power spectra to recover the various cosmological power spectra from which the cosmological parameters are deduced. In carrying out calculations in multipole space, we assume a perfectly uniform scan strategy.

It is computationally faster to work in Fourier space since beam imperfections contaminate the data when the beams are convolved with the sky in real space; in Fourier space this operation involves only calculating the *product* of two Fourier transforms: that of the fields T,E or B, and that of the beam itself. We then calculate the spurious polarization effects associated with the mismatch between the two beams to form the spurious polarization fields. The power spectra are obtained by simply taking the modulus squared of these pseudo-polarization fields and averaging over all directions in the 2-D Fourier space. It is straightforward to carry this calculation out by invoking the statistical isotropy of the underlying sky. The output is the spurious power spectra in terms of the real power spectra and a `mixing matrix' which depends on the beam mismatch parameters and the angular scale. Our results can be expressed in terms of elementary functions, and therefore may be useful during the map deconvolution and data analysis steps of CMB experiments.

The main objective of our preliminary calculations [6] was to use the six power spectra of the underlying sky (including the non-vanishing, or forbidden spectra such as $C_\ell^{TB}$ and $C_\ell^{EB}$, which are used as monitors) and the beam imperfection-parameters. The output was six power spectra which include the effects of all five systematics (i.e., four IP effects, and one XP effect) and the various couplings between them – coupling occurs if nonlinear, higher order corrections are important. While, in practice, this effect may be small, in the analogous case for gravitational lens cleaning it is important. In any case, it was straightforward to incorporate higher-order effects such as these in the multipole-domain pipeline.

We sort the power spectra in a 6-D vector (representing the power spectra for TT, TE, EE, BB, TB and EB). Couplings between the underlying spectra are encoded by a *mixing matrix,* which depends on the angular scale, and all the main beam systematic parameters. The mixing matrix encapsulates all the leakage/mixing processes we studied and is expressed in terms of analytic functions, which are extremely efficient to process. For simplicity, we work in the flat-sky approximation, although for some purposes the full sky calculation was simulated. The calculations were further simplified by working in multipole space from the outset, which resulted in exact analytic expressions. The results of these two calculations are in nearly perfect agreement.

***<u>Simulation Results</u>***

In the following figures we illustrate the results of our map-based calculations [7]. However, as pointed out above, the multipole space pipeline was used to appraise second-order beam effects, which were neglected by the time-domain pipeline. Together the two pipelines represent the most comprehensive study of main beam systematics to date. In addition to the beam effects described above, we simulated the effect of satellite pointing errors after reconstruction.

It is important to note that all of these effects are calculated with a single focal plane pixel at a single frequency (100 GHz). To the extent that parameters vary over the focal plane, these effects will partially average down to give a smaller residual signal, and this estimate will be conservative. Beam effects have various dependences on the beam width $\sigma$. In power spectrum units ($\mu K^2$), differential gain and differential rotation are independent of beam size, but differential beam width and differential ellipticity scales as $\sigma^4$. Differential pointing scales in a complicated manner, but for our uniform scan strategy we found it scales as $\sigma^2$.

Satellite pointing errors produce a systematic effect in a complicated manner. With dual analyzers, we instantaneously extract a single linear Stokes parameter (Q or U) in each beam that is not susceptible to pointing error. To extract the second parameter (U or Q), we must wait for the beams to rotate on the sky. To the extent that the satellite pointing is off, Q and U will be obtained from displaced beams on the sky. Thus the effect is not simply described as a dipole, and has both smaller amplitude and a different shape from differential beam offset, as evident from Fig. 5.2.1.

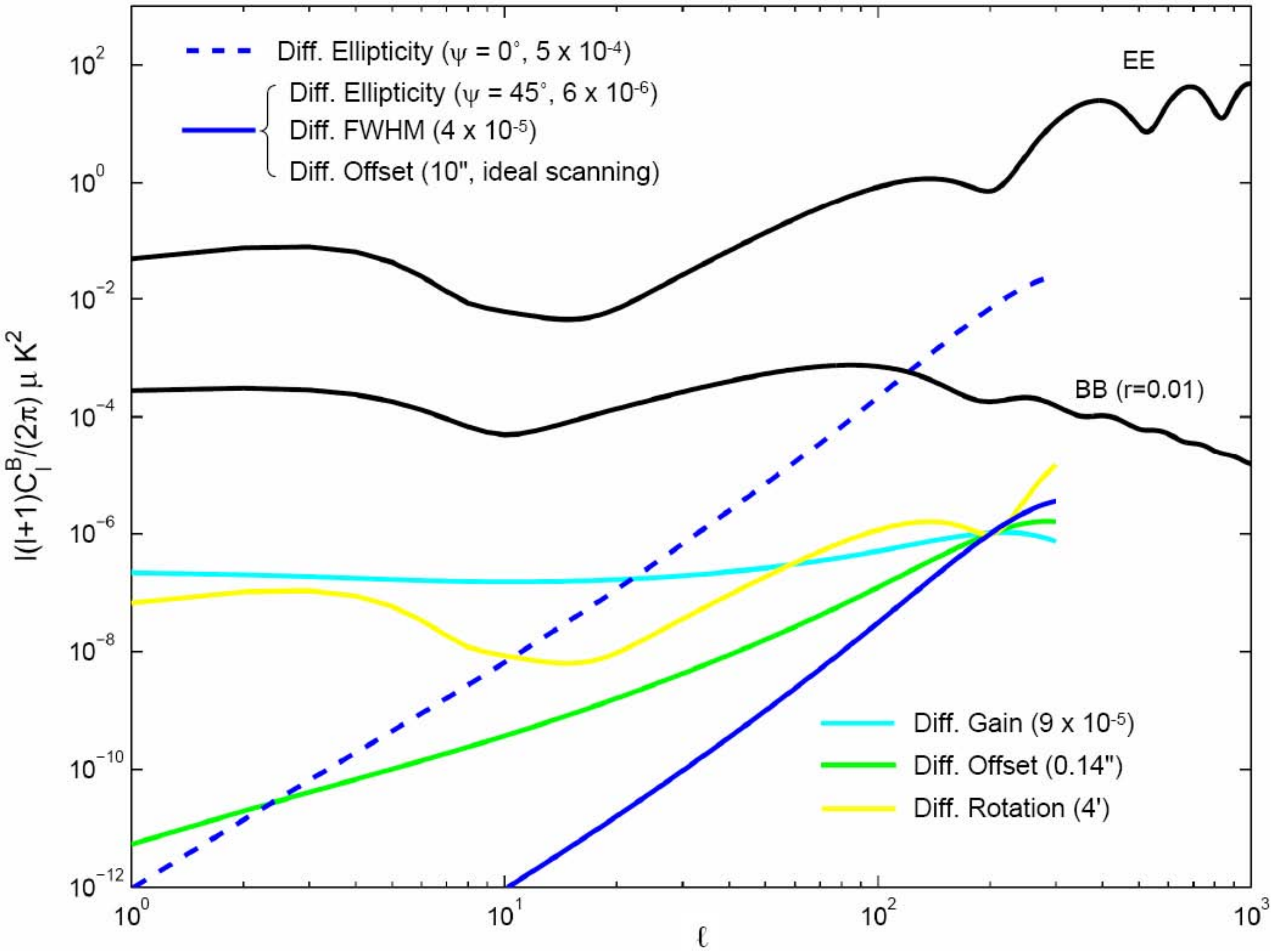

Figure 5.2.1. Spurious B-mode IP power spectra for refractor optics with 60′ beams at 100 GHz. The amplitudes of the effects are all chosen to be equivalent at $\ell$ = 200. The legend for each trace indicates the level of systematic of each type which produces spurious polarization signals. Note that the solid blue curve corresponds to three separate effects which have the same power spectrum. Differential ellipticity is shown for $\psi$ = 0˚, which only produces E-mode polarization, and for $\psi$ = 45˚, which only produces B-mode polarization. Shifts in the beam centroids, differential beam offset, is shown for two cases. One case is for the EPIC scan strategy, the other case is for an idealized scan pattern covering all scan angles uniformly over the entire sky. With the present scan pattern it may be possible to approximate the ideal scan pattern by mathematically weighting scans. These spectra indicate the level of the raw effect, and further reduction is possible given prior knowledge of the beam effects.

We note that the requirement on differential beam offset is very stringent. As evident from Fig. 5.2.1, this effect is greatly reduced by an ideal scanning pattern, which covers all scan angles uniformly on each piece of sky, compared to the present EPIC scan pattern. We think it is possible that the data in the present scan pattern can be manipulated to significantly reduce the effects of differential beam offset by taking advantage of the wide range of scan angles available on any pixel on the sky.

For example, we note that angular dependence of the measured signal at each pixel is a superposition of the lowest few multipoles of the polarization angle, alpha, viz:

$$I(p) = T(p) + R\cos\alpha + S\sin\alpha + Q\cos 2\alpha + U\sin 2\alpha + \text{higher order multipoles},$$

at a pixel "p". The second and third terms in this expression arise from the dipole, or first order differential pointing contribution to the instrumental polarization. “R” and “S” play a similar role to the true polarization terms Q and U. It is clear that the $R\cos\alpha + S\sin\alpha$ cannot represent true cosmological polarization, which is quadrupolar in nature, being modulated twice for each single physical rotation in $\alpha$. The R and S terms result from convolution from the underlying sky with the beam (in this case - the differential pointing error). Since the polarization angle $\alpha$ is recorded for each pixel, one can remove all data taken at pixel p with $\alpha$ values that *fail* to measure at both the polarization angle $\alpha$ *and* $\alpha + 180$. Discarding all measurements which don't have their mirror-counterpart does not mitigate higher order spurious modes such as the quadrupole or octopole (which are very small in any case) but at least removes the most pernicious (in practice) main-beam systematic - the 'dipole'. For a given pixel, the polarization angle coverage should resemble a bow-tie pattern.

A more refined strategy might be to weight scans mathematically to recover ideal scanning. This may be possible because the scan strategy has good coverage of $\alpha$ and $\alpha + 180$ degrees, for a large number of $\alpha$ values. Further analysis is required to evaluate these strategies. Of course removing some data comes with a noise penalty. In practice, however, this is a negligible, percent-level effect. This is confirmed in both our frequency domain and time domain studies using the EPIC scan strategy.

*5.2.4 Scanning Strategy*

Scan strategy is a central consideration for removing systematic polarization errors. Rotating the view angle on the sky allows us to separate systematics associated with a preferred direction in the focal plane. Rotating the view angle also allows us to remove or mitigate many of the polarization artifacts associated with main beam effects. Furthermore, scan redundancy provides an important check on many systematic effects, by allowing us to compare maps on identical regions of sky over multiple time scales. These multiple maps can be compared (‘jackknifed’) to evaluate systematics that vary over time, vary over the orbit, vary with respect to the angle from the sky region to the earth or sun, or are associated with spacecraft view angles. Therefore we have designed a scan strategy with rapid modulation of the view angle and high redundancy on a short time scales.

EPIC’s scan strategy consists of spinning the payload about the boresight axis, and precessing the boresight axis about the anti-solar direction (see Fig. 5.1.3). The observation direction makes an angle of 55 degrees about the boresight axis and the payload spins at ~1 rpm. The boresight axis is at 45 degrees of the antisolar direction and precesses with a period of ~3.2 hours (we have varied the precession period to give avoid scan overlaps so as to provide a fully sampled map in the shortest time). To illustrate this, Fig. 5.2.1 shows the fraction of the sky covered by one detector located at the center of the focal plane, in one spin period (1 minute), 3 spin periods (3 minutes) and one full precession period (3.2 hours).

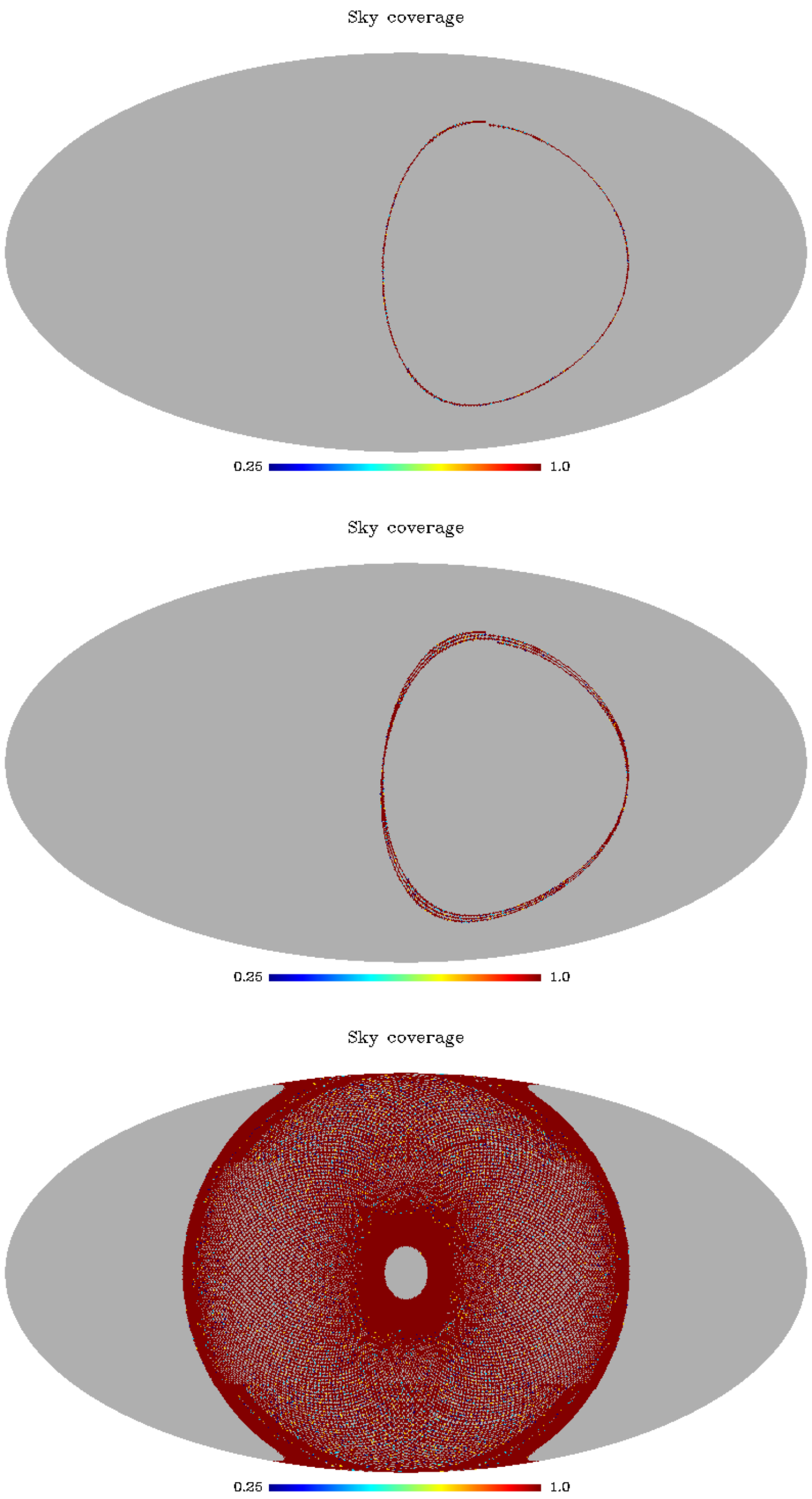

Figure 5.2.2: Fractions of the sky observed by a detector located at the center of the focal plane of EPIC for one spin period, three spin periods and one full precession period.

With this choice of parameters, EPIC covers 55% of the sky each day. Thus each day of observation provides the required data for a complete analysis of all angular scales. This will be invaluable for jackknife tests during the analysis of the data. It also connects all angular scales with a broad range of time frequencies which again will be an advantage in the rejection of time domain systematic effects.

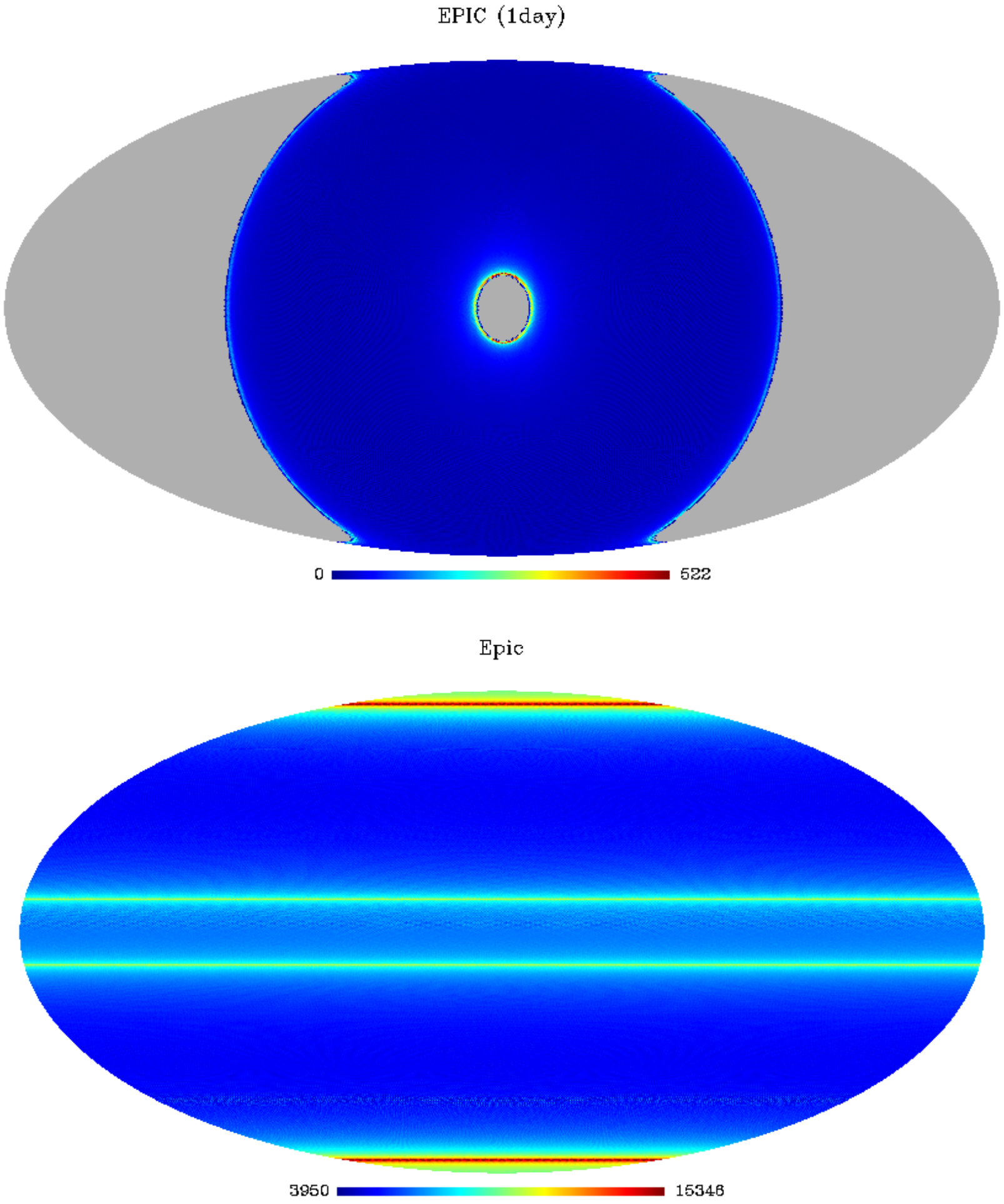

Figure 5.2.3. Number of hits per pixel after 1 day and 1 year of observation by EPIC in ecliptic coordinates. In 1 day, EPIC covers 55% of the sky.

This scanning strategy also optimizes the measure of view angle rotation, by maximizing the angular coverage of each point of the sky. The accuracy on polarization depends both on the observation time and the range of view angles (i.e. suppose that the same point of the sky is observed an infinite number of times with only one orientation of a polarized sensitive bolometer, then one of the polarization parameters, say Q, is determined with an infinite signal to noise, but no information is obtained on U and therefore the actual polarization state is undetermined). To obtain a precise measure of polarization, both Q and U must be determined with the same signal to noise, and this requires a uniform distribution of the view angle. A good estimator of this uniformity is the quantity $\langle \cos 2\alpha \rangle^2 + \langle \sin 2\alpha \rangle^2$ (see Appendix A for details) where the brackets denote average over all the samples of a given point of sky. The lower this figure of merit, the more uniform the angular coverage. Fig. 5.2.3 shows how well EPIC does compared to WMAP and Planck for a year of observation. Fig. 5.2.4 shows how this quantity evolves together with the observation redundancy for 1 day of observation, 1 month and 6 months.

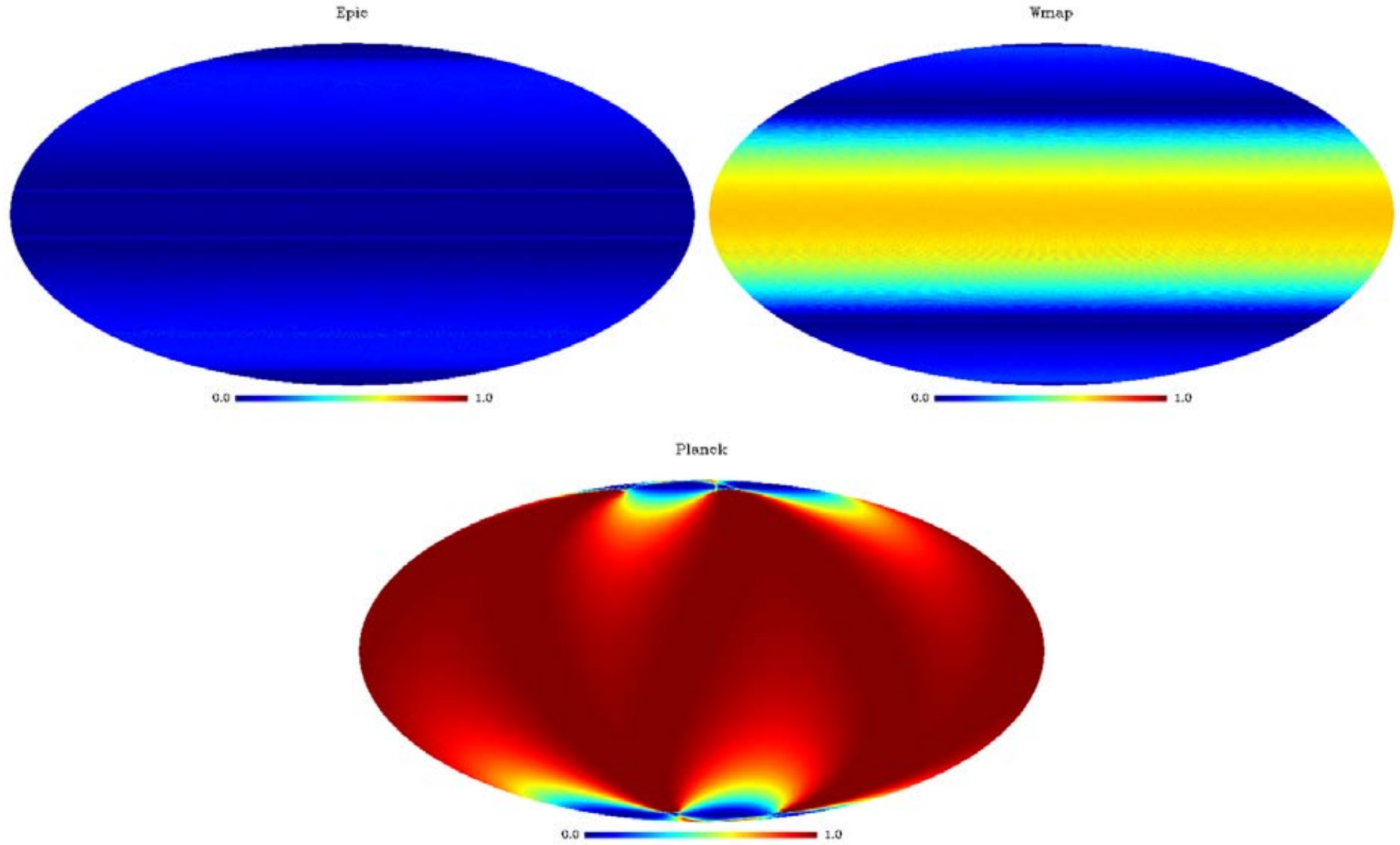

Figure 5.2.4. Estimator of the quality of the angular coverage of EPIC, WMAP and Planck for polarization. Epic has superior angular coverage uniformity over the sky.

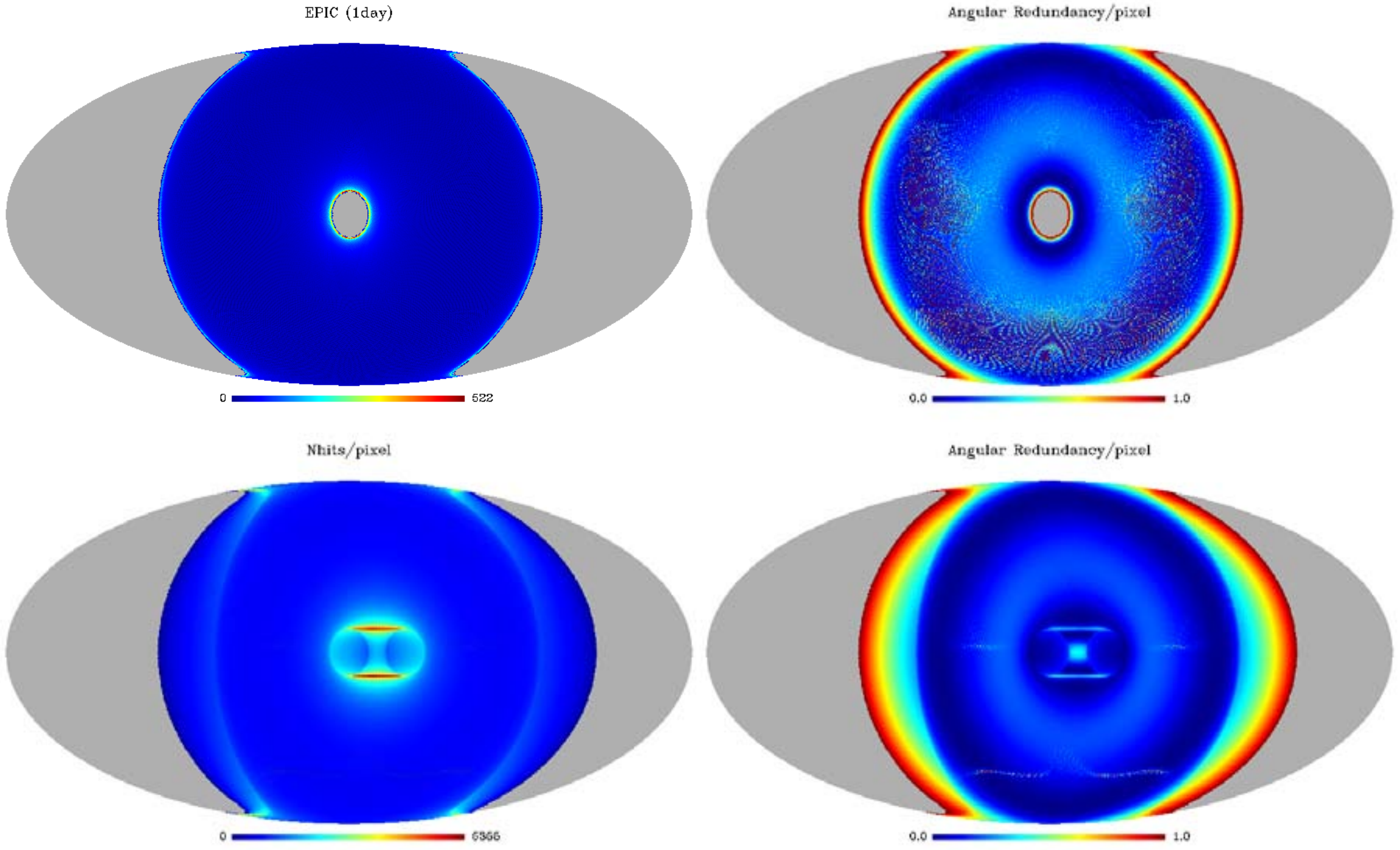

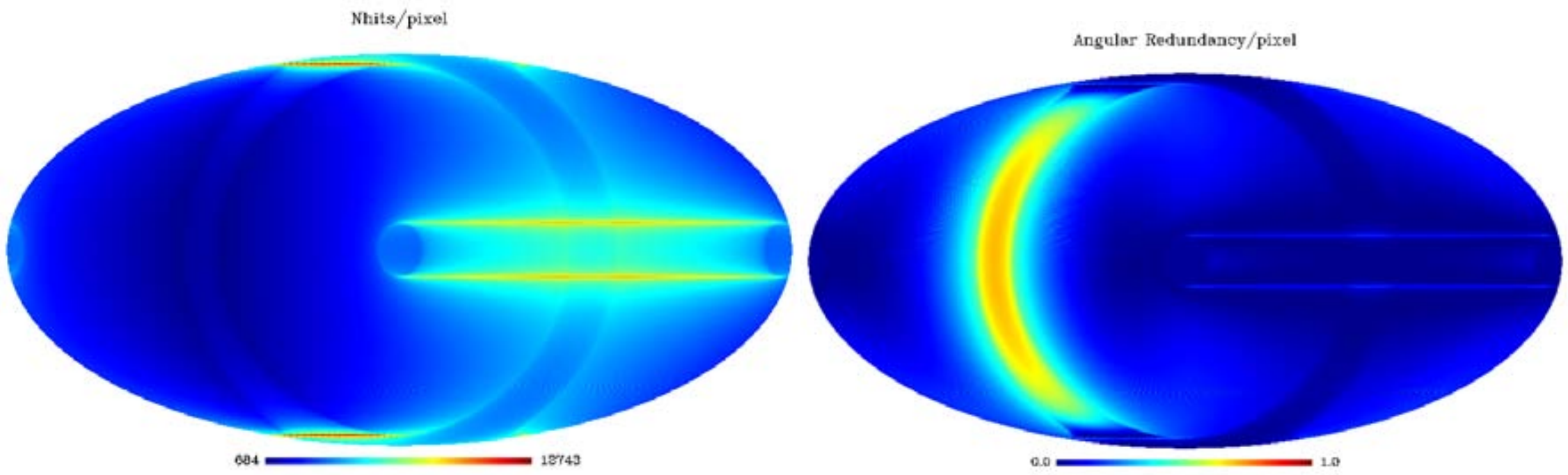

Figure 5.2.5: Number of observation per sky pixel and angular redundancy for (top to bottom) 1 day, 1 month and 6 months of observation by EPIC. Each day covers 55% of the sky with already excellent angular redundancy. The left hand column of figures shows uniformity in observation time per pixel; the right hand column shows angular uniformity.

## 5.3 Refracting Optics

The optical system for EPIC is designed to provide a large throughput or A , the product of the aperture area times the field of view solid angle, in order to incorporate a sufficiently large focal plane to reach the sensitivity requirements. Furthermore, the optics must meet the demanding requirements of low polarization with minimum beam imperfections, such as differential ellipticity, differential beam width, and differential gain, over this large field of view. Our solution for the EPIC-LC mission is to use a compact wide-field refractor, with each telescope operating in a narrow wavelength band. Some of the attractive features of this solution have already been demonstrated with the BICEP experiment, which was the first to implement it for studies of CMB polarization. The optical system provides low aberrations and polarization, and allows the refracting optics to be optimized over the limited spectral band. In front of the telescope, we place a half wave plate that is stepped during observations. Because the half wave plate rotates the polarization direction without appreciably changing the field distribution on the aperture, rotating the wave plate does not vary the beam shape on the sky. Thus we can measure polarization signals by differencing signals where the polarization on the sky has been rotated by 90˚ by rotating the wave plate. Signals from non-ideal beam shape are virtually the same between the two measurements, and their effect is subtracted and systematic errors associated with main beam effects are thus eliminated.

### *5.3.1 Design*

The throughput requirements for EPIC-LC are given in Table 5.3.1. For the optical design we assume the throughput requirements set by the option with TES detectors, which has a larger number of elements in the focal plane than the NTD option (see Table 5.1.3). The throughput requirements are satisfied by implementing 6 independent refracting telescopes of identical optical design. The design is shown in Fig. 5.3.1. It is an f/1.7, 300 mm entrance aperture system comprised of two polyethylene lenses. A half wave plate (HWP) is placed at the entrance aperture of the system and is the first element in the optical train. The lens surfaces are all conics of revolution. The 6-receiver LC design includes single frequency receivers at each of 60 and 135 GHz, two single frequency receivers at 90 GHz, a combined 200/300 GHz receiver, and a combined 30/40 GHz receiver. The dual-frequency receivers have the low frequency detectors arranged in a ring around the central high-frequency section of the array.

*5.3.2 Diffraction and Polarization Properties*

The optical properties of the system were analyzed with ray tracing using ZEMAX and CODE V. The polarization properties were quantified in terms of the Mueller matrices of the system, as calculated from CODE V, and using ZEMAX's physical optics package to include the effects of diffraction. The HWP has been included in the analysis as a disc with a single index of refraction that was the average of its ordinary and extraordinary indices (changing the index between the ordinary and extraordinary produces negligible change in the performance parameters that we report below).

The telescopes provide Strehl ratios that are much higher than 0.8 over the entire FOV of each of the telescopes, see Table 5.3.1. An optical system that has a Strehl ratio of 0.8 is considered diffraction limited. The mixing between Q and U Stokes parameters that is encoded by the QU terms of the Mueller matrix of the telescope is negligible. As an example we give the values for the 135 GHz in Table 5.3.2. Instrumental polarization, which is a leakage of the intensity term I into either the Q or U terms, is of less importance because rotation of the HWP modulates the polarization with negligible effect on the instrumental polarization. Thus this systematic effect can be canceled by appropriately differencing the signals measured at the two HWP rotation states. For completeness however we give the instrumental polarization terms (IQ, IU) of a detector at the edge of the field of view of the 135 GHz telescope in Table 5.3.2. The IQ/IU terms arise from differences in reflection of the two polarization states at the surface of the lens. The magnitude of the terms is dominated by potential non-idealities in the antireflection coatings. Table 5.3.3 gives the target performance of the telescope in terms of suppression of the systematic effects discussed in Section 5.2, and its calculated performance for an edge field detector at a frequency of 135 GHz. In this analysis we propagate two orthogonally polarized beams through the system using physical optics and then calculate the performance parameters described in the table. Note that calculated performance is taken as the worst performance over the focal plane, generally at the edge of the field of view, whereas the requirement is an average value. The performance is calculated based on the optical performance alone and does not take into account the additional mitigating effect of the HWP. For most of the anticipated instrumental effects the predicted performance is better than the target performance.

**Table 5.3.1** Parameters for EPIC-LC Optics

| Number of Receivers | Frequency [GHz] | Throughput[1] [$cm^2$ sr] | FOV[2] (deg) | Strehl Ratio[4] |
|---|---|---|---|---|
| 1 | 30 | 13 | 24.4[3] | 0.99 |
| | 40 | 48 | 16.8[3] | 0.99 |
| 1 | 60 | 50 | 17.2 | 0.91 |
| 2 | 90 | 89 | 16.3 | 0.96 |
| 1 | 135 | 40 | 15.3 | 0.99 |
| 1 | 200 | 20 | 13.2[3] | 0.98 |
| | 300 | 9 | 7.3[3] | 0.99 |

[1] The product of throughput per pixel and the total number of pixels at a given frequency. A pixel on the focal plane contains two polarization sensitive detectors.
[2] Pixels are arranged on a square grid with a circular boundary. We give the diameter of the FOV.
[3] The low frequency pixels are arranged in an annulus around the higher frequency ones.
[4] Ratio given at the outermost diameter of the frequency band.

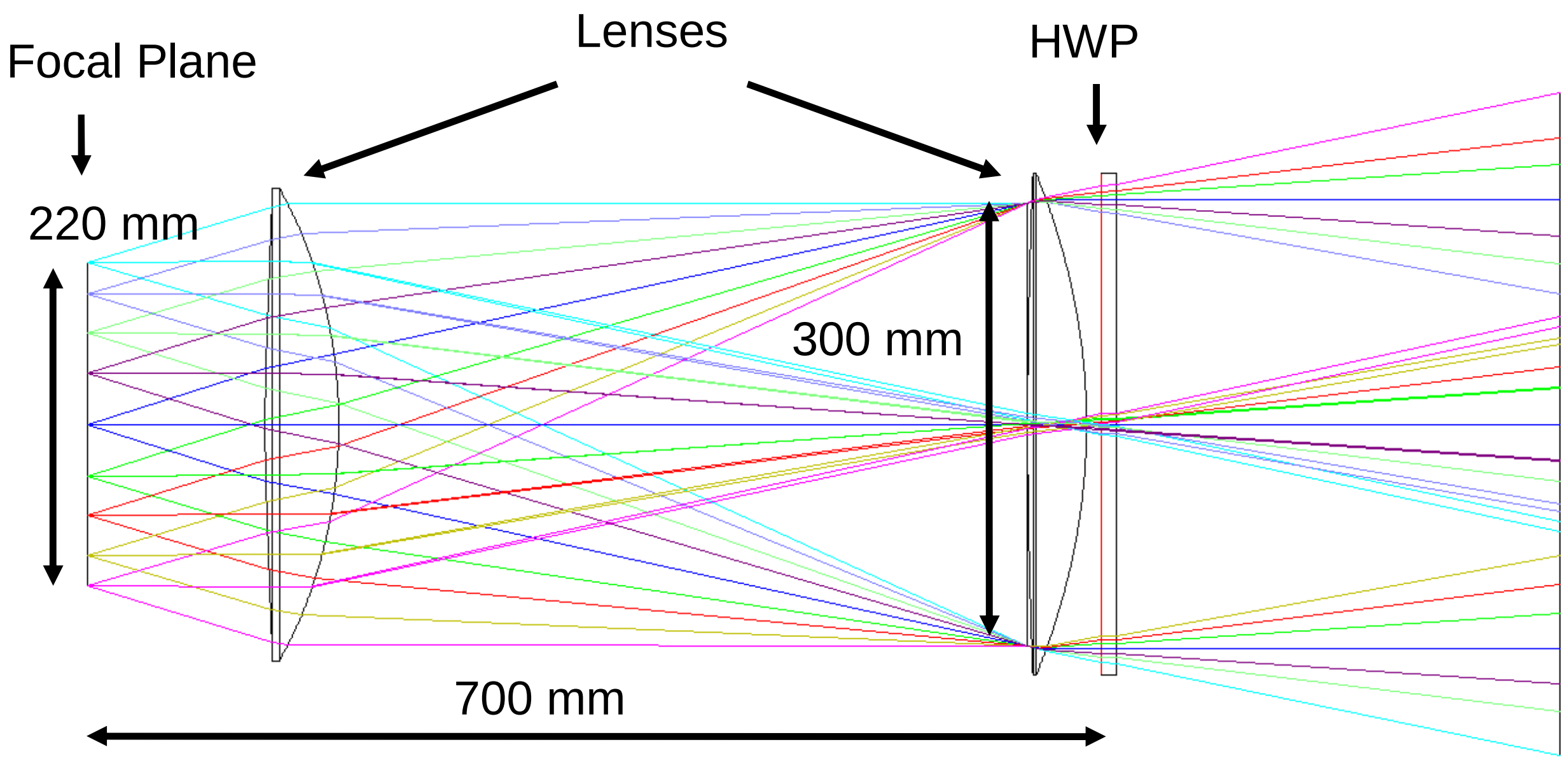

Figure 5.3.1. EPIC-LC employs 6 refracting telescopes, each of them identical to the design shown in this figure except for the HWP, which is the first element in the optical train. The specifications of the each of the HWPs are given in Table 5.4.1.

A telescope with a design similar to that of EPIC-LC was built for the BICEP CMB polarization experiment. BICEP is designed for simultaneous ground-based observations at 100 at 150 GHz and does not presently use a HWP in its optical path. Because BICEP has non-optimized anti-reflection (AR) coatings, with each surface giving ~1% reflection, a 100 mm thick zotefoam window, and two AR-coated Teflon thermal blockers that would not be used in EPIC, its performance should be considered an upper limit. Table 5.3.3 gives the measured beam performance of the BICEP optical system in terms of main beam systematic errors. The most pernicious main beam effect, differential ellipticity, is at $8{\cdot}10^{-4}$ and very close to the EPIC goal. However, our experience with BICEP has also uncovered some non-ideal effects. We measured polarized ghost reflections at the level of ~1% of the main beam, attributed to the blocker anti-reflection coatings that had to be optimized for both 100 and 150 GHz simultaneously. We measured a 0.7% differential pointing shift between the beam centers of matched polarized beams, possibly due to the lens material. Finally the pixels show larger rotation than expected from the optics, although this measurement includes the rotation of the hand-assembled polarization-sensitive bolometers which were aligned to a less demanding specification. We are investigating the physical origin of these effects so they can be reduced. We emphasize that the wave plate in front of the EPIC telescope should cancel main beam effects altogether. Our analysis shows there is no fundamental constraint to meeting the EPIC goals, and if sufficient improvement can be realized in a fully tested system, we may be able to descope the wave plate at a later date.

**Table 5.3.2. Polarization Properties of the EPIC-LC 135 GHz Telescope**

| Element | Ideal Coatings | index 10% high | Coating 10% thin |
|---|---|---|---|
| IQ | $4 \times 10^{-5}$ | 0.0034 | 0.00074 |
| IU | $< 1 \times 10^{-5}$ | $< 1 \times 10^{-5}$ | $< 1 \times 10^{-5}$ |
| QU | $< 1 \times 10^{-5}$ | $< 1 \times 10^{-5}$ | $< 1 \times 10^{-5}$ |

Table 5.3.2: Mueller matrix elements for the edge of the 135 GHz band of the EPIC-LC telescope. The ideal ARC is a $\sqrt{n}$, $\lambda/4$ anti-reflection coating for 135 GHz. The column labeled 'index 10% high' assumes an index that is 10% higher compared to the ideal index and the 'coating 10% thin' assumes a thickness that is 10% thinner than the ideal thickness. The values are given at the focal plane, after propagation through all the lenses.

**Table 5.3.3 Polarization Requirements and Performance for EPIC-LC Telescope**

| Effect | Quantity | Goal at 100 GHz | Predicted[1] | Measured[1,2] |
|---|---|---|---|---|
| Differential Beam Size | $(\sigma_1-\sigma_2)/\sigma$ | $4\cdot10^{-5}$ | $1\cdot10^{-4}$ | $< 2\cdot10^{-3}$ |
| Differential Beam Offset | $\Delta\theta$ | 0.14" raw scan<br>10" symm. scan | 0.007" | 10" |
| Differential Ellipticity | $(e_1-e_2)/2$ | $5\cdot10^{-4}$ ($\psi = 0^\circ$)<br>$6\cdot10^{-6}$ ($\psi = 45^\circ$) | $1\cdot10^{-4}$<br>- | $< 1\cdot10^{-3}$<br>- |
| Differential Gain | $(g_1-g_2)/g$ | $1\cdot10^{-4}$ | $2\cdot10^{-4}$ | $< 5\cdot10^{-3}$ |
| Polarization Rotation | $\Delta\theta/2\pi$ | $2\cdot10^{-4}$ | $5\cdot10^{-6}$ | $5\cdot10^{-3}$ |

[1]Calculated for a 135 GHz telescope.
[2]Median value over the BICEP focal plane in an end-to-end optical test, combining 100 and 150 GHz pixels.

Table 5.3.3: Target, and predicted beam effects for the EPIC-LC optics. We define the different beam effects in section 3.2. The target column is reproduced from the column labeled 'suppression to meet goals' of Table 5.2.2. The prediction is calculated at the edge of the FOV for a frequency of 135 GHz, assumes ideal anti-reflection coating and do not include the mitigating effects of the HWP. End-to-end measurements were carried out on a similar telescope designed for the BICEP CMB polarization experiment. Note that differential ellipticity with $\psi = 45^\circ$ was not calculated or measured, and would be a subject of future work.

### *5.3.3 Sidelobe Performance*

The off-axis response of the BICEP telescope was measured with a Gunn oscillator at 100 GHz, placed in the mid-field ~10 m from the aperture. The intensity response, shown in Fig. 5.3.2, drops to -40 dBi 40 degrees off axis, and is roughly azimuthally symmetric. The sidelobes, where measurable, show an almost featureless polarization of ~20% (not shown in the figure). The co-moving absorbing baffle provides an additional attenuation of at least 15 - 20 dB at large angles, where the measurement hits the noise floor. The coupling to the baffle was measured as 0.3% by measuring the total loading looking a zenith with and without the forebaffle in place. While we do not have an accurate model for the off-axis behavior, it appears that radiation is reflected and/or scattered to large angles by the lenses, blockers, and zotefoam window. The forebaffle serves to attenuate response at large angles, where it prevents a direct view angle of the window. A calculation of the sidelobe response to the Galactic plane, shown in Fig. 5.3.3, indicates that the level of far-sidelobe response is already at the level of control needed for EPIC.

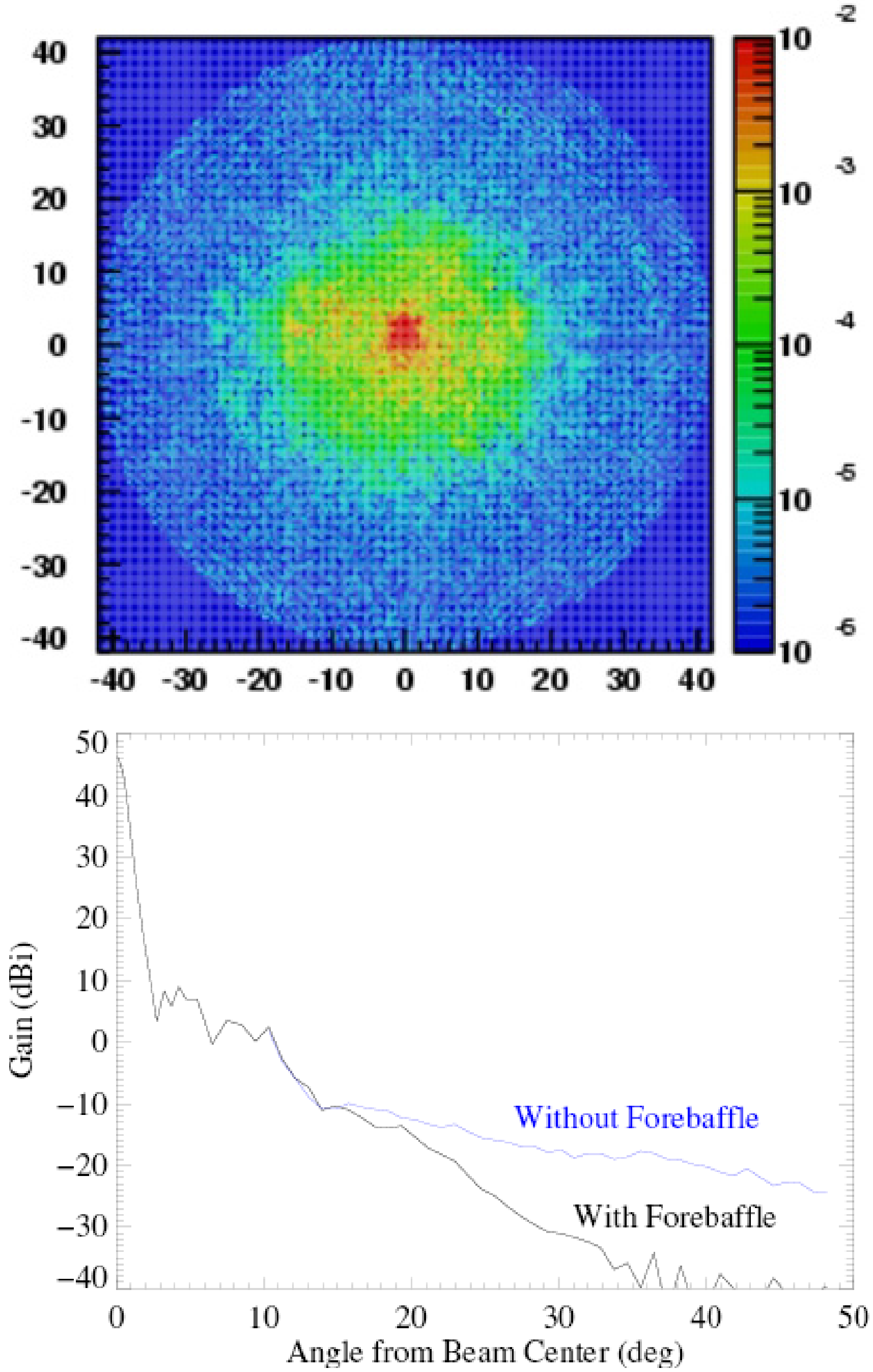

Fig. 5.3.2. Top: Measured off-axis intensity response of the BICEP telescope. Plot is g(θ), the response to a point source normalized to unity on axis, corrected for the non-linear response of the detectors when viewing the source at high gain. The far-sidelobe response is nearly featureless and drops below the per pixel noise level ~20˚ off axis. Bottom: A comparison of the telescope response with and without a black fore-baffle. We show the azimuthally averaged response, in units of antenna gain, $G(\theta) = (4\pi/\Omega)\, g(\theta)$. The response beyond 40 degrees falls to the noise level of the measurement.

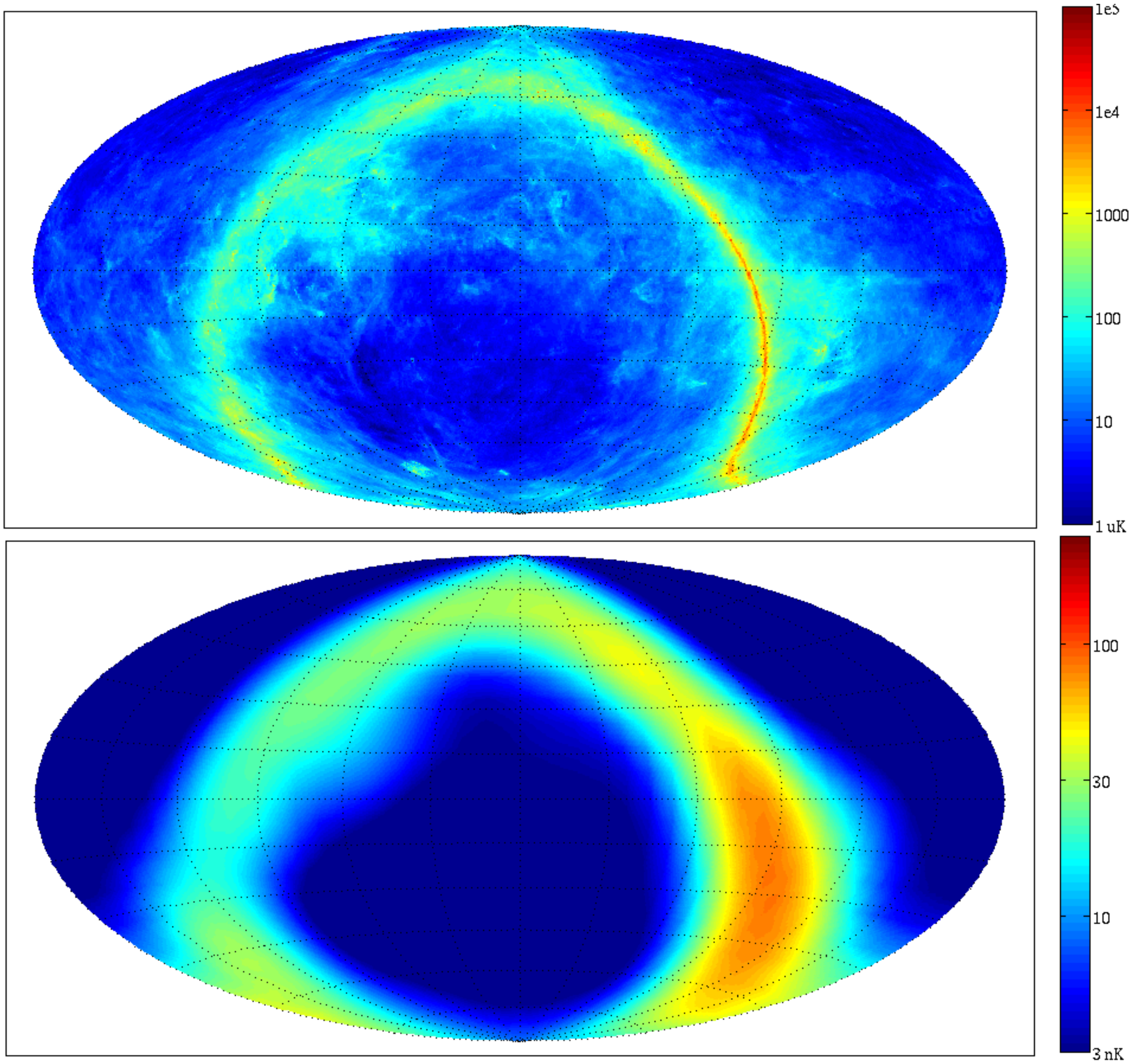

Figure 5.3.3. Emission from the Galaxy (top panel, FDS model 8 at 150 GHz in μK) is convolved with the measured sidelobe response of the BICEP telescope (bottom panel, polarized response in nK). To perform this calculation, we fit the measured response to a Gaussian profile, assumed a 20% polarization, and excluded the main beam where the response is > 10 dBi. Some caution must be noted because the off-axis response could not be measured for angles larger than 40 degrees, and the sidelobes are assumed to have a constant 20% polarization. The inferred level of response from the Galaxy falls below 10 nK about 20˚ off the Galactic plane.

### *5.3.4 Anti-Reflection Coating*

Polyethylene lenses that have a cryogenically robust anti-reflection coating have already been used successfully with the QUAD and BICEP CMB polarization instruments (see Fig. 5.3.4). These lenses are coated with a single layer of expanded polyethylene to achieve an average ~1% reflection over the frequency range used in these experiments, 75 – 175 GHz. In a limited $\Delta\nu/\nu$ = 30% band, the same coating technology gives ~0.2% reflection. Broader band coatings using multiple layers, such as those developed by the Cardiff group, promise improved performance.

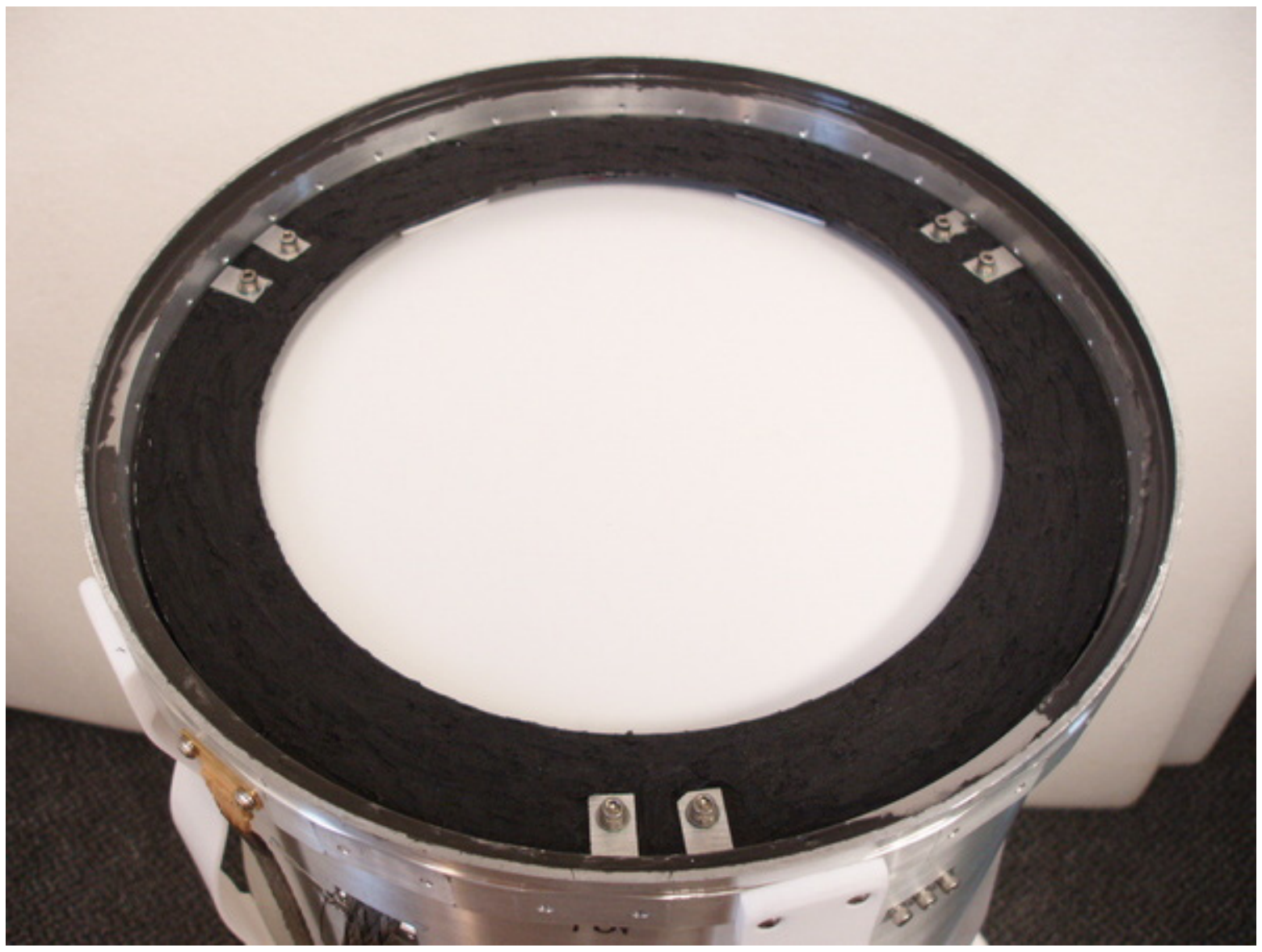

Fig. 5.3.4. 30-cm polyethylene lens, the primary optics for the BICEP telescope. This lens is anti-reflection coated with an expanded polyethylene sheet bonded to the polyethylene surface. More highly curved lenses were coated for the QUAD experiment using the same technique.

## 5.4 Half-Wave Plate Polarization Modulator

EPIC-LC employs a half wave plate in front of each telescope, see Fig. 5.3.1. Six HWPs are located at the entrance aperture to each of the telescopes, each optimized for a specific band. In this arrangement, only the sky polarization is modulated providing a strong discriminant against spurious polarization arising from the instrument. The principal function of the HWP here is to remove polarization produced by the optics or effects from main beam imperfections. Rotating the half wave plate causes the polarization angle to rotate, but does not change the illumination pattern on the entrance of the wave plate, and thus does not change the beam shapes on the sky. By taking the difference between measurements at two orientations of the wave plate, one extracts the polarization on the sky. Polarization artifacts associated with the optics or the beam shape, no matter how imperfect they may be, do not change and are subtracted and removed. Polarization analysis with EPIC is otherwise accomplished by a combination of dual analyzers in the focal plane. These are polarization sensitive bolometer pairs which share a common optical path and extract either Stokes U or Q in each pixel. The rotation of the spacecraft rotates the view angle of the telescope on the sky, so that both Stokes parameters can be measured using a single detector pair.

The baseline strategy is thus to step the wave plate by 45 degrees every ~24 hours, when independent maps are produced with a high degree of scan crossings (see Fig. 5.2.2). Signal modulation is accomplished by the scanning motion. Continuous rotation of the wave plate would relax requirements on instrument stability, but requires a low-power dissipation mechanism, potentially a magnetic bearing [1] and is considered an upscope.

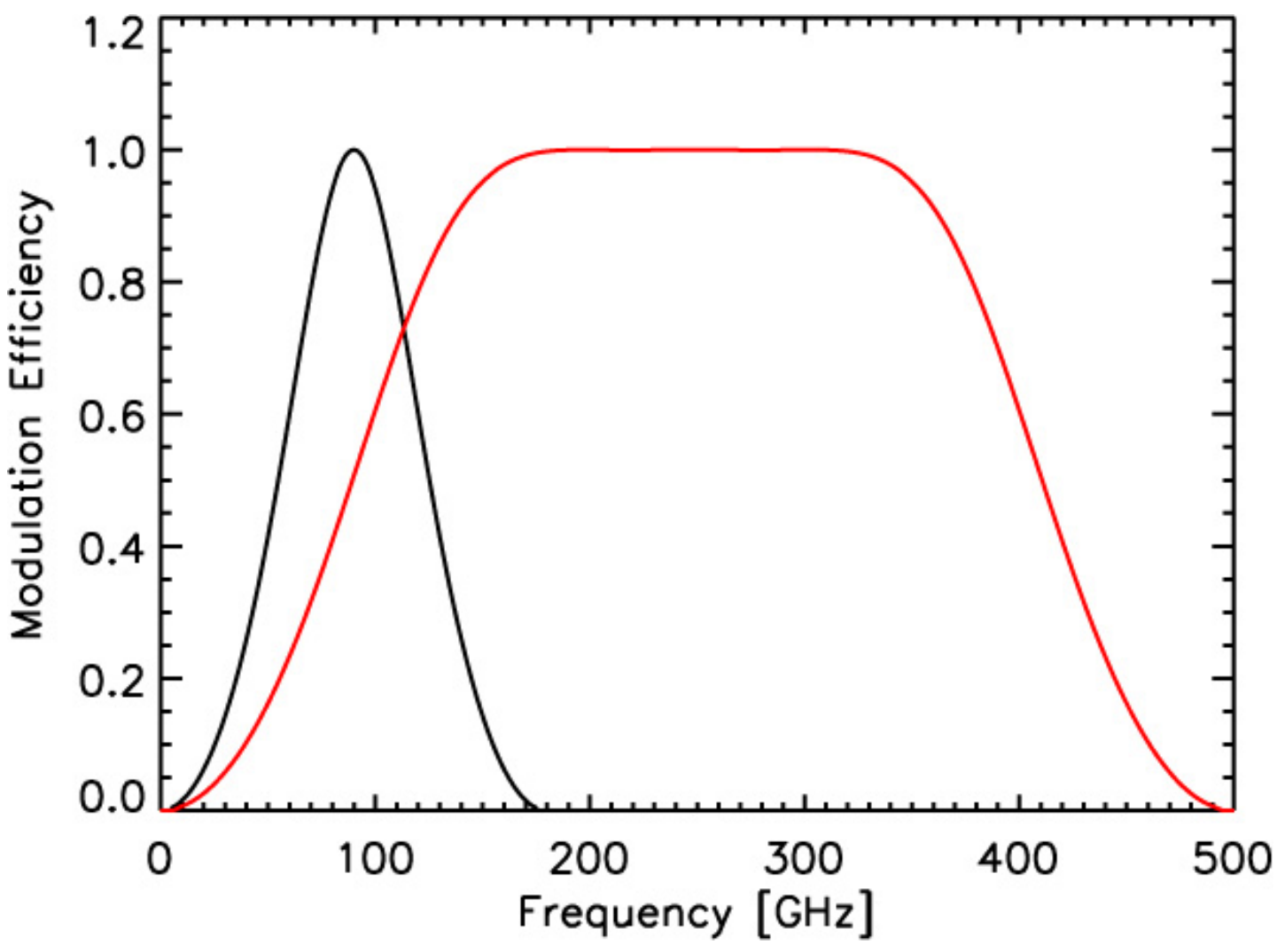

Figure 5.4.1. The polarization modulation efficiency of the HWP for the 90 GHz telescope (black) and for the 200/300 GHz telescope (red). For frequencies of 60 and 90 GHz the HWP is made of a single disc of sapphire. It provides 20% fractional bandwidth with modulation efficiencies larger than 95%. For other frequencies we use an achromatic HWP made of a stack of three plates. The calculation assumes an ideal anti-reflection coating and does not include reflections between the plates for the achromatic HWP.

### *5.4.1 Wave Plate Optical Design*

The HWPs are made of 33 cm diameter discs of birefringent sapphire crystal. The discs are oversized relative to the 30 cm entrance aperture of the telescopes. Discs of sapphire with this diameter are available commercially. For the 60, 90, and 135 GHz bands the HWP is made of a single disc of an appropriate thickness. The thickness, weight and bandwidth of the HWP for each of the frequency bands is given in Table 5.4.1. Such a HWP will provide a modulation efficiency of 95% for fractional bandwidth of nearly 30% (see Figure 5.4.1). For the other frequency bands we will use an achromatic HWPs (AHWP). The AHWPs are made by stacking three plates each of the same thickness. The second plate is rotated by 58 degrees relative to the first, which is co-aligned with the last. With this construction the AHWP has a modulation efficiency that is larger than 95% over a much broader range of frequencies in comparison with the single HWP.

**Table 5.4.1 Waveplate Specifications**

| Receiver [GHz] | Type of HWP | Center Frequency [GHz] | Bandwidth [GHz] | Thickness [cm] | Mass [kg] |
|---|---|---|---|---|---|
| 30/40 | 3-stack | 35 | 32 | 3.99 | 13.6 |
| 60 | Single | 60 | 17 | 0.79 | 2.7 |
| 90 | Single | 90 | 26 | 0.53 | 1.8 |
| 135 | Single | 135 | 39 | 0.35 | 1.2 |
| 200/300 | 3-stack | 250 | 206 | 0.56 | 1.9 |

Table 5.4.1: Specifications of each of the HWPs in the EPIC-LC design. The bandwidth is for modulation efficiency larger than 95%. The center frequency gives the frequency at which a single plate is an ideal HWP. The mass assumes an aperture of 33 cm. The two lowest and two highest frequencies each share an achromatic HWP to ensure sufficient bandwidth. We note that if the 30 GHz band is descoped then the mass of a single plate at 40 GHz would be only 4 kg. The mass and thickness do not include a layer of anti-reflection coating on the front and back of each of the HWP.

*5.4.2 Anti-Reflection Coating*

Sapphire has an index of refraction that is larger than 3 at the mm-wave band and therefore reflections from the HWP can exceed 50% if they are not minimized. A broad-band, cryogenic, anti-reflection coating (ARC) has been developed by Ade's group at Cardiff. Special materials are glued on both sides of the HWP to produce a multi-layer ARC. A 2 inch diameter sample has already been tested cold and tests on larger diameter achromatic waveplates are now ongoing as part of the development of the CLOVER and EBEX instruments.

*5.4.3 HWP and Systematic Errors*

The balloon borne MAXIPOL has already demonstrated a successful use of a continuously rotating HWP polarimetry with a CMB polarization experiment [1]. The use of a HWP as a polarization modulator provides the following important advantages in discrimination against systematic errors.

- Instrumental polarization coming from sources that are on the detector side of the halfwave plate is not modulated and therefore does not affect the signals from the sky. Specifically, the rotation of the HWP provides a strong discrimination against systematic errors that arise when the antenna pattern of the main beam is different for two orthogonal polarization states in terms of their angular size or ellipticity, when there is a differential pointing between the two antenna patterns, or when there is differential gain. We referred to these effects as differential beam size, differential ellipticity, differential beam pointing offset, and differential gain, respectively (see section 3.2). The rotation of the HWP only rotates the incident polarization vector without affecting any of these sources of systematic errors.
- Any reflection or differential absorption from the wave plate itself is modulated at the rotation rate and twice the rotation rate, respectively. Polarization however is rotated at four times the rotation rate. Thus a series of measurements on the sky in steps of 45 degrees allows us to also remove these effects.

The most important advantage of a stepped HWP comes from modulating the polarization on the sky side of the instrument without substantially affecting the beam shapes. However, because of the birefringent nature of the HWP, the beam pattern does shift slightly as a consequence of the rotation of the HWP. This effect does not affect the beam shapes as long as the wave plate is sufficiently oversized. Calculations made for the EBEX optical system, which take these effects into account, show that the magnitude of the effects is expected to be negligible.

There are additional advantages for a HWP when it is turned in a continuous rotation.

- Sky signals are constrained to a band of frequencies that are high compared to typical 1/f noise sources, thereby, reducing the 1/f noise requirements on the detection system. This relaxes requirements not only on the focal plane and readout electronics, but on sources of unpolarized 1/f noise such as thermal drifts.
- Since the signal is modulated more rapidly, one can use an individual detector to make independent measurements of all Stokes parameters for each pixel on the sky. No detector differencing is required. This eliminates errors produced with detector differencing arising from (a) uncertainty in the difference in gain between the detectors, (b) differences in beam pattern, and (c) differences in noise levels between the detectors.

Thus a continuous wave plate relaxes requirements on system stability and noise uniformity. If the receiver is stable, gain and beam pattern differences between detector pairs are stable and can be removed by stepping the waveplate every ~24 hours. With a highly stable receiver, the approach used in Planck, BICEP and QUAD, the wave plate can be stepped. With a continuous wave plate, such as used in MAXIPOL and EBEX, stability requirements on the focal plane are greatly simplified.

The data collected by the MAXIPOL experiment, which used a continuously rotating HWP for its CMB polarization measurements, demonstrate the features we listed, see Figure 5.4.2. The power spectrum of the Q and U Stokes parameters is flat to frequencies as low as 1 mHz after demodulation, the data show a white spectrum consistent with detector noise and there are no detectable systematic errors.

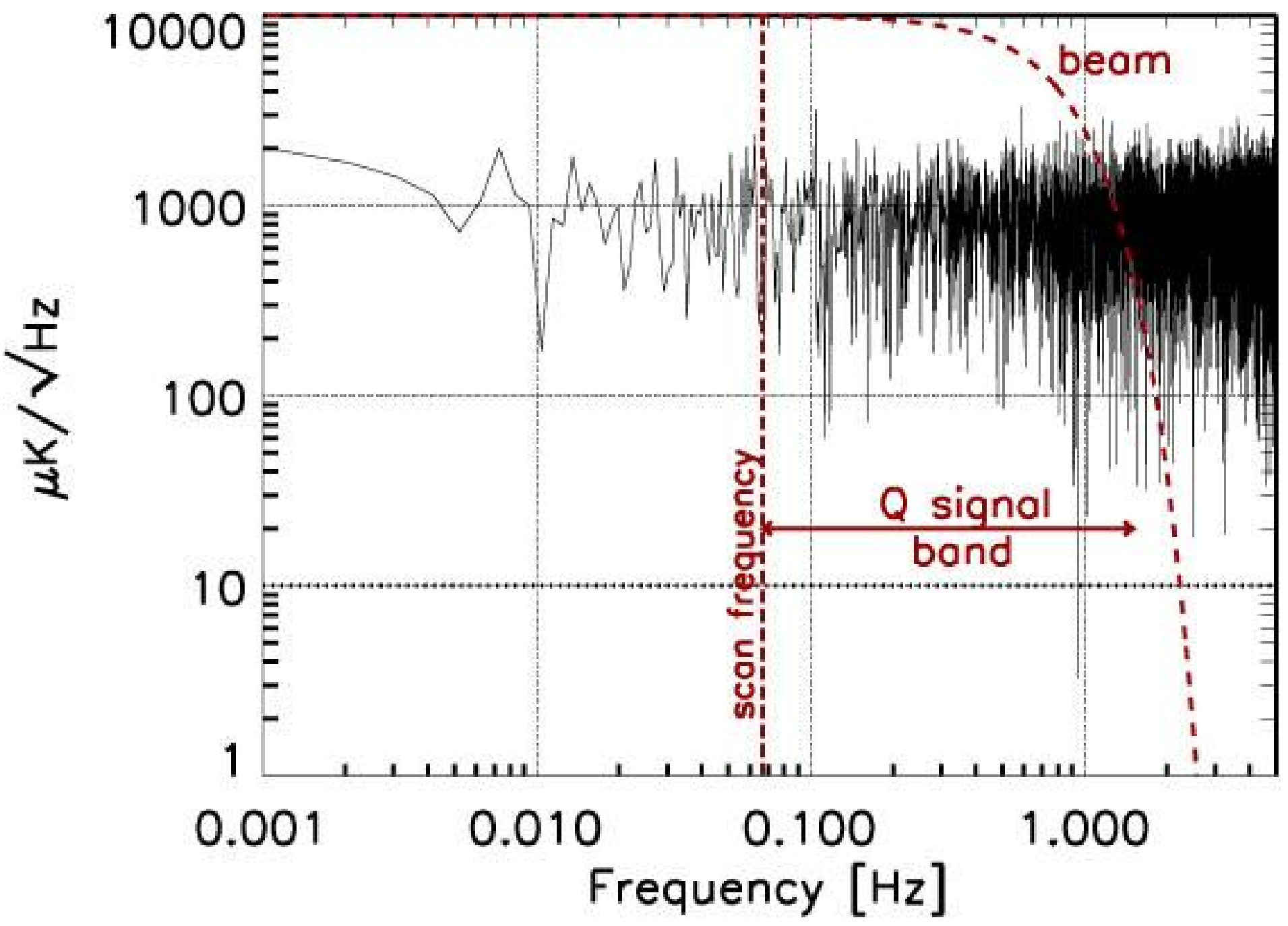

Figure 5.4.2. Power spectrum of time domain data of the Q Stokes parameter from the data of MAXIPOL. The data show white noise spectrum at frequencies as low as 1 mHz. The white noise level is consistent with detector noise. A lock-in has been applied to the raw time domain data, which encodes all Stokes parameters. For subsequent data analysis, the data is band-passed to reject frequency bands that have no sky signals. The bandwidth of the sky signal (red dash), is determined by the scan frequency and the beam size.

### *5.4.4 HWP Drive and Rotational Encoding*

In the baseline design the HWP is stepped every 24 hours by 45 degrees. The simplest implementation includes a cryogenic stepper motor that drives the HWP with a gear. The HWP is mounted with mechanical bearings operating at 2 K.

Due to the advantages for turning the HWP continuously, an upgrade for EPIC-LC will include a mechanism for continuous rotation. A generic rotation mechanism includes a rotor, a stator, bearings between the rotor and the stator, a driver for the rotor, and an encoder for the rotational position. Several candidate technologies can be implemented for each of these elements. We will concentrate on a subset of the available options.

In the purely mechanical option the rotor, which carries the HWP, is connected to the stator with cryogenic ball bearings. Tests conducted for current ground and balloon-based experiments suggest that the microphonic noise arising from the rotations of these bearings is small compared to detector noise in low resistance transition edge sensor (TES) bolometers. However, experience from the MAXIPOL balloon-borne instrument suggests that this technology will not be suitable for B-mode polarization measurements using high resistance NTD-based bolometers. An additional concern with ball bearings is the amount of energy dissipated due to friction in the bearings. Initial laboratory tests suggest that frictional heat dissipation may be a limiting factor in the implementation of ball bearings for a satellite mission.

Hanany et al. (2003) [1] proposed to use a superconducting magnetic bearing as a replacement for the mechanical bearings. The rotor, which holds the HWP, is made of a strongly magnetized NdFeB magnet and the stator is made of YBCO high temperature superconducting (HTS) material. The construction is traditionally called a superconducting magnetic bearing (SMB) [2]. In an SMB the HTS is cooled below the transition temperature and the magnet is held mechanically in place. Below the transition temperature the mechanical constraints on the magnet are released and it levitates in position relative to the HTS. It maintains its position in all directions as if constrained by stiff springs. Typical spring constants are few thousand N/m [1, 3]. The rotor of the SMB can be driven mechanically, or by means of motors. Various techniques are available for rotational encoding. They include using a cryogenic laser and a chopper wheel as an optical encoder, placing hall sensor below the rotor to encode the small rotational inhomogeneity of the rotating magnet, or using mechanical encoding in the case of a mechanical drive for the rotor.

## 5.5 Focal Plane Detectors

### *5.5.1 Focal Plane Parameters*

EPIC is designed to be an order of magnitude more sensitive than ESA's Planck spacecraft -- comparable to the jumps from COBE to WMAP and from WMAP to Planck. This large step in capability will produce an exceedingly rich set of cosmological measurements, including a possible detection of the gravitational-wave signal from inflation. EPIC will have a large focal-plane array of bolometric detectors cooled to 100 mK that can cover the required frequency range of 30-300 GHz. The choice of frequency bands was discussed in section 2 on foregrounds. In our study, we have baselined a focal plane with Neutron-Transmutation-Doped Ge (TES) bolometers with unmultiplexed JFET amplifiers, the same technology developed for the Planck HFI. These detectors are antenna-coupled to allow operation to the lowest frequency of 30 GHz. As an upscope, we have studied the use of SQUID-multiplexed TES bolometers. This technology offers higher sensitivity, through larger detector formats, and the possibility for higher operating temperature due to intrinsically faster speed of response.

Bolometers can be antenna coupled through a planar phased-array antenna, with lens-coupled antennas, or as line probes coupled to scalar horn antennas. In contrast to mesh-absorber bolometers, the active volume of planar-coupled bolometers does not increase at lower frequencies. With this assumption, bolometers achieve comparable sensitivity to HEMTs at frequencies below 100 GHz, while maintaining a major advantage at frequencies above 100

GHz. Antenna-coupled bolometers thus provide a single technology capable of spanning the entire frequency range of interest.

We calculate the sensitivity of the focal plane by first computing the optical power from the sky, optics, and forebaffle. The waveplate and lenses are all housed at 2 K. The forebaffle is assumed to be at 40 K, and couples 0.3% to the detectors, as measured in the BICEP receiver. This coupling is probably pessimistic, since EPIC will reduce the coupling paths in BICEP by using monochromatic anti-reflection coatings, and eliminating the window.

| **Table 5.5.1. Low-Cost Option Instrument Parameters** | | | |
|---|---|---|---|
| Aperture | D | 30 | cm |
| Optics Temperature | $T_{opt}$ | 2 | K |
| Coupling to 2 K Stop | $\varepsilon_{opt}$ | 10 | % |
| Waveplate Temperature | $T_{wp}$ | 2 | K |
| Waveplate Absorption | $\varepsilon_{wp}$ | 2 | % |
| Baffle Temperature | $T_{baf}$ | 40 | K |
| Coupling to Baffle | $\eta_{baf}$ | 0.3 | % |
| Pixel size | $d/f\lambda$ | 1.7 - 2.1 | |
| Fractional bandwidth | $\Delta\nu/\nu$ | 30 | % |
| Optical efficiency | $\eta_{opt}$ | 40 | % |
| Base temperature | $T_0$ | 100 | mK |
| **NTD Bolometer Parameters** | | | |
| Heat capacity at 100 mK | $C_0$ | 0.25 | pJ/K |
| Detector time constant | $\tau(d\theta/dt)/\theta_F$ | ≥ 1/2π | |
| Thermal conductance | $G_0T_0/Q$ | ≥ 3 | |
| Amplifier noise | $V_n$ | 10 | nV/√Hz |
| **TES Bolometer Parameters** | | | |
| Transition temperature | $T_c$ | 215 | mK |
| Alpha | dln(R)/dln(T) | 100 | |
| Transition temperature | $T_c$ | 215 | mK |
| Heat capacity at 100 mK | $C_0$ | 0.15 | pJ/K |
| Bolometer saturation margin | $P_{sat}/Q$ | 5 | |
| Multiplexer noise | $I_n$ | negl. | pA/√Hz |

By using cold optics, instrument emission is eliminated (see Fig. 5.5.1), and the optical power on the detector is dominated by the CMB itself. In fact, only in the highest band does the instrument emission from the baffle exceed the CMB. This allows the highest possible sensitivity per detector (see Fig. 5.5.2), roughly a factor of two better per detector than in the TFCR report, which assumed warm optics and included the appropriate noise margin.

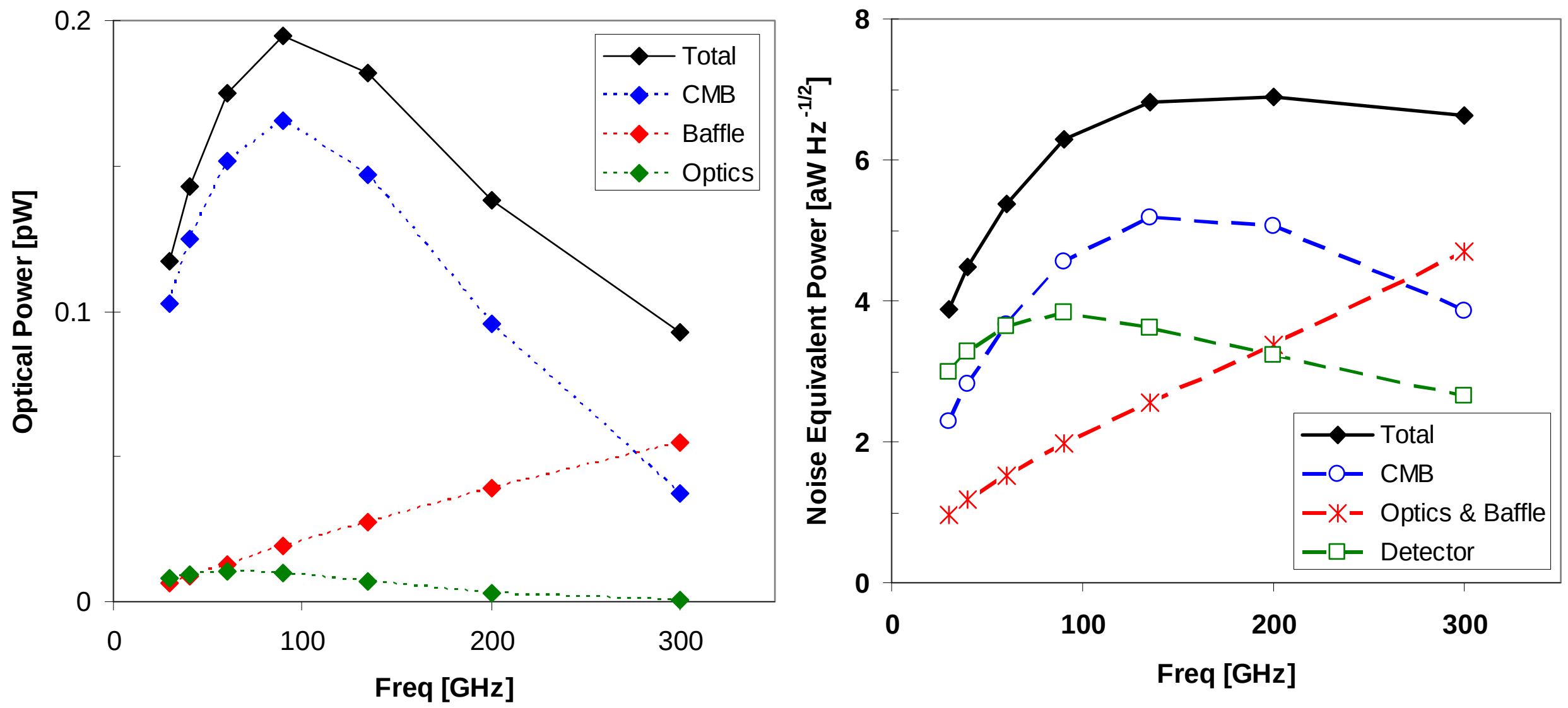

Fig. 5.5.1. (Left) Contributions to the total optical power on the detectors from CMB, optics and waveplate, and the baffle. (Right) Contributions to the total noise from the CMB, optics, and detector and readout. The total noise is dominated by the CMB for 70 < ν < 200 GHz. The noise calculation was carried out for a TES bolometer, but the result is similar for the NTD bolometer case.

**Table 5.5.2 Summary of NTD Ge Bolometer Parameters**

| Freq [GHz] | Optical Loading | | | $G_0$ [pW/K] | $G_0T_0$/ Q | NEP | |
|---|---|---|---|---|---|---|---|
| | CMB [fW] | Baffle[1] [fW] | Optics[2,3] [fW] | | | photon [aW/√Hz] | bolo [aW/√Hz] |
| 30 | 100 | 6 | 8 | 3.3 | 3 | 2.4 | 3.2 |
| 40 | 125 | 9 | 9 | 4.1 | 3 | 2.9 | 3.5 |
| 60 | 150 | 13 | 11 | 5.0 | 3 | 3.9 | 3.9 |
| 90 | 165 | 19 | 10 | 5.5 | 3 | 4.8 | 4.1 |
| 135 | 145 | 28 | 7 | 8.6 | 5 | 5.6 | 4.2 |
| 200 | 95 | 39 | 3 | 15 | 11 | 5.9 | 4.6 |
| 300 | 35 | 55 | 1 | 25 | 27 | 6.1 | 5.5 |

[1]Baffle at 40 K with 0.3% coupling, coefficient measured with BICEP
[2]Waveplate at 20 K with 2% coupling
[3]Optics at 2 K with 10% coupling

**Table 5.5.3 NTD Ge Focal Plane Design Sensitivity**

| Freq [GHz] | $\theta_{FWHM}$ [arcmin] | $N_{bol}$[1] [#] | τ [ms] | $NET_{bol}$[2] [μK√s] | $NET_{band}$ [μK√s] | $w_p^{-1/2}$[3] [μK arcmin] | δTpix[4] [nKrms] |
|---|---|---|---|---|---|---|---|
| 30 | 155 | 8 | 53 | 59 | 21 | 45 | 270 |
| 40 | 116 | 54 | 44 | 51 | 7.0 | 15 | 90 |
| 60 | 77 | 128 | 35 | 43 | 3.8 | 8.2 | 49 |
| 90 | 52 | 256 | 32 | 37 | 2.3 | 5.0 | 30 |
| 135 | 34 | 256 | 22 | 35 | 2.2 | 4.7 | 28 |
| 200 | 23 | 64 | 15 | 41 | 5.2 | 11 | 67 |
| 300 | 16 | 64 | 10 | 82 | 10 | 22 | 130 |
| Total[5] | | 830 | | | 1.4 | 3.0 | 18 |

[1]Two bolometers per focal plane pixel
[2]Sensitivity of one bolometer in one polarization, $NET_{pix} = NET_{bol}/\sqrt{2}$
[3]Sensitiivity δT in a pixel $\theta_{FWHM}$ x $\theta_{FWHM}$ times $\theta_{FWHM}$, assuming 2-year design life
[4]Sensitivity δT in a pixel 120′ x 120′, with a 2-year design life
[5]Combining all bands together

Table 5.5.3 shows the system parameters for the NTD-based mission, namely the frequency bands, the number of feeds per band, beamsizes, Noise-Equivalent Powers (NEPs), and Noise-Equivalent Temperatures (NETs). There are a total of 830 bolometers, two per pixel, distributed over six frequency bands. The total number of bolometers is set by practical limits on the number of wires (2 per bolometer) amplifiers (1 per bolometer), and Si JFET power dissipation to the 40 K passively-cooled shell. The optical loading, as shown in Table 5.5.2, is dominated by the CMB except at the highest frequencies. This is even true if the waveplate has to operate at a somewhat higher temperature (~20 K) for reasons of power dissipation. The thermal conductivity is chosen for sensitivity, with $G_0T_0/Q \sim 3$ in the lowest frequency bands. In the highest frequency bands, the detector time constants become an issue, set by the beam size and the 1 rpm spin rate. For these channels, we increase the thermal conductivity to meet the scanning time constant specification, at some loss in sensitivity. The detectors operate close to the background limit, i.e. the ratio of the total NEP to the photon noise NEP varies from 1.6 at 30 GHz to 1.3 at 135 GHz. The combined sensitivity of the entire focal plane is 1.4 μK√s which gives an average noise per 2° pixel of 18 nK in a two-year mission.

In addition to the high technology readiness NTD-based system, we have also studied a more capable TES bolometer instrument. The NETs for the TES detectors are similar to the NTD devices, however the multiplexing allows for larger focal plane arrays. Furthermore, the TES detectors have significantly faster speed of response, allowing a greater range of possible scan speeds. This faster speed of response means that the thermal conductivities of the higher frequency channels can be optimized for sensitivity instead of speed, resulting in improved NETs. The combined sensitivity of the entire TES focal plane is 0.8 μK√s, which is a factor of 1.8 better than the NTD focal plane.

**Table 5.5.4 Low-Cost Option TES Focal Plane Design Sensitivity**

| Freq [GHz] | $\theta_{FWHM}$ [arcmin] | $N_{bol}$[1] [#] | $NET_{bol}$[2] [μKcmb√s] | $NET_{band}$ [μKcmb√s] | $w_p^{-1/2}$[3] [μK arcmin] | $\delta T_{pix}$ (2°x2°)[4] [nKrms] |
|---|---|---|---|---|---|---|
| 30 | 155 | 8 | 57 | 20 | 44 | 260 |
| 40 | 116 | 54 | 50 | 6.8 | 15 | 88 |
| 60 | 77 | 128 | 42 | 3.7 | 8.1 | 48 |
| 90 | 52 | 512 | 37 | 1.6 | 3.5 | 21 |
| 135 | 34 | 512 | 35 | 1.5 | 3.3 | 20 |
| 200 | 23 | 576 | 38 | 1.6 | 3.5 | 21 |
| 300 | 16 | 576 | 65 | 2.7 | 5.9 | 35 |
| Total[5] | | 2366 | | 0.8 | 1.8 | 11 |

[1]Two bolometers per focal plane pixel
[2]Sensitivity of one bolometer in one polarization, $NET_{pix} = NET_{bol}/\sqrt{2}$
[3]Sensitiivity δT in a pixel $\theta_{FWHM}$ x $\theta_{FWHM}$ times $\theta_{FWHM}$, assuming 2-year design life
[4]Sensitivity δT in a pixel 120′ x 120′, with a 2-year design life
[5]Combining all bands together

In order to calculate the required sensitivity, we apply an overall sensitivity margin of √2 in sensitivity (similar to Planck HFI), and assume a one-year lifetime to derive the required parameters in Table 5.1.3. Figures 5.5.1 and 5.5.2 show the NET per feed as a function of frequency for the TES and the semiconductor based focal planes. Figs. 5.5.3 and 5.5.4 show the sensitivity to CMB polarization anisotropy.

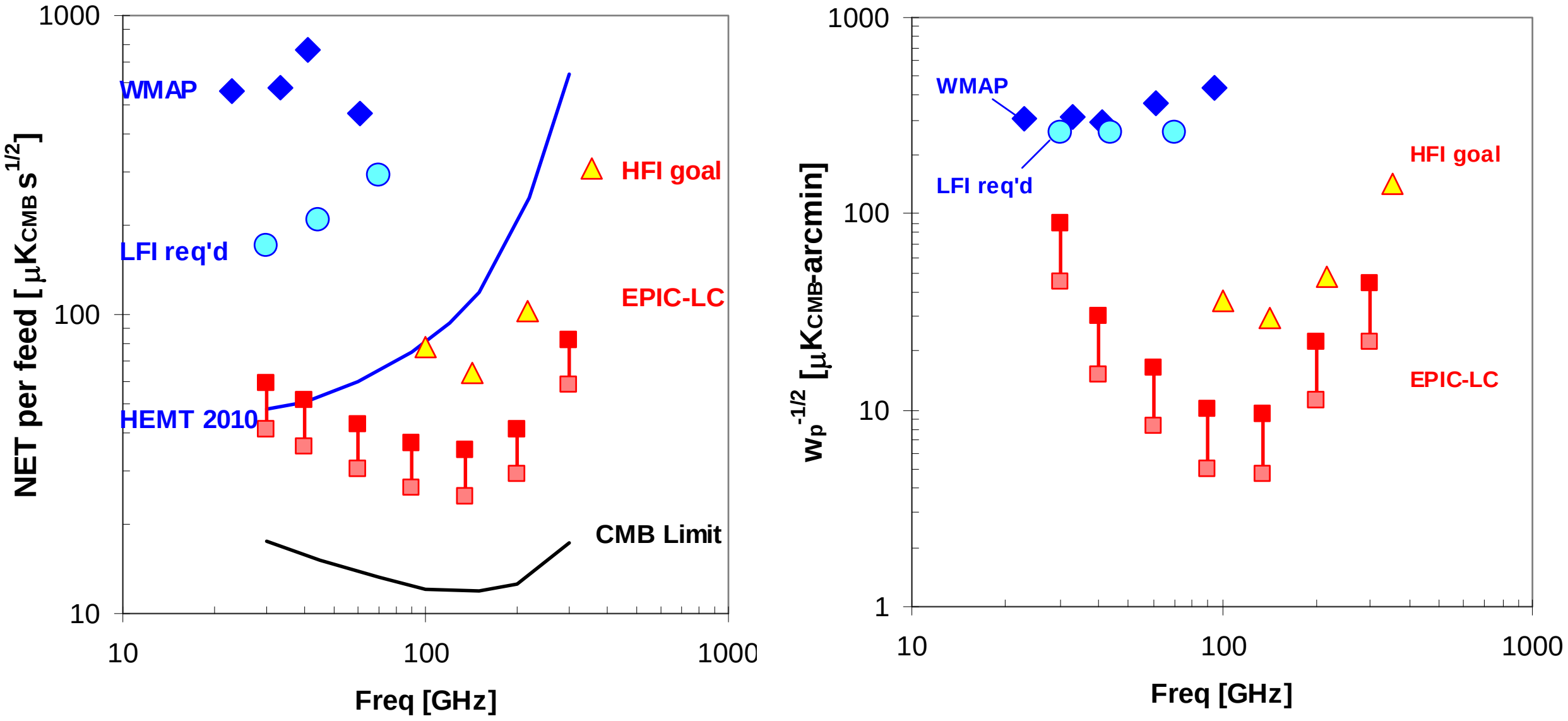

Figure 5.5.2. (Left) Noise equivalent temperature (NET) per feed for the EPIC NTD-based focal plane, showing a range from design sensitivities (shaded points) to requirements (solid points), where the required sensitivity is √2 times larger than the design value. For comparison, we plot the in-flight sensitivity of WMAP, and the projected sensitivities of the polarized channels in Planck-LFI and -HFI. We show requirements for LFI, and goals for HFI, to correspond to the sensitivities demonstrated in pre-flight testing. The projected sensitivity of future HEMT amplifiers obtaining Stokes' Q and U simultaneously, taken from the Weiss Committee Report, shown as a solid blue line is comparable to the EPIC sensitivities for $\nu < 100$ GHz, but rapidly degrades for $\nu > 100$ GHz. The design goals of EPIC approach the ultimate sensitivity floor from CMB photon noise with 100% optical efficiency, shown by the black line. (Right) Comparison of the figure of merit $w_p^{-1/2}$, defined as $[8\pi\, NET_{bolo}^2/(T_{mis}\, N_{bol})]^{1/2}(10800/\pi)$ in $\mu K_{CMB}$ - arcmin. We compare the sensitivity of WMAP after 8 years of observations, Planck in 1 year of observations, and EPIC. For EPIC we assume required sensitivities and the 1-year required lifetime (solid points), and the design sensitivities and the design 2-year lifetime (shaded points).

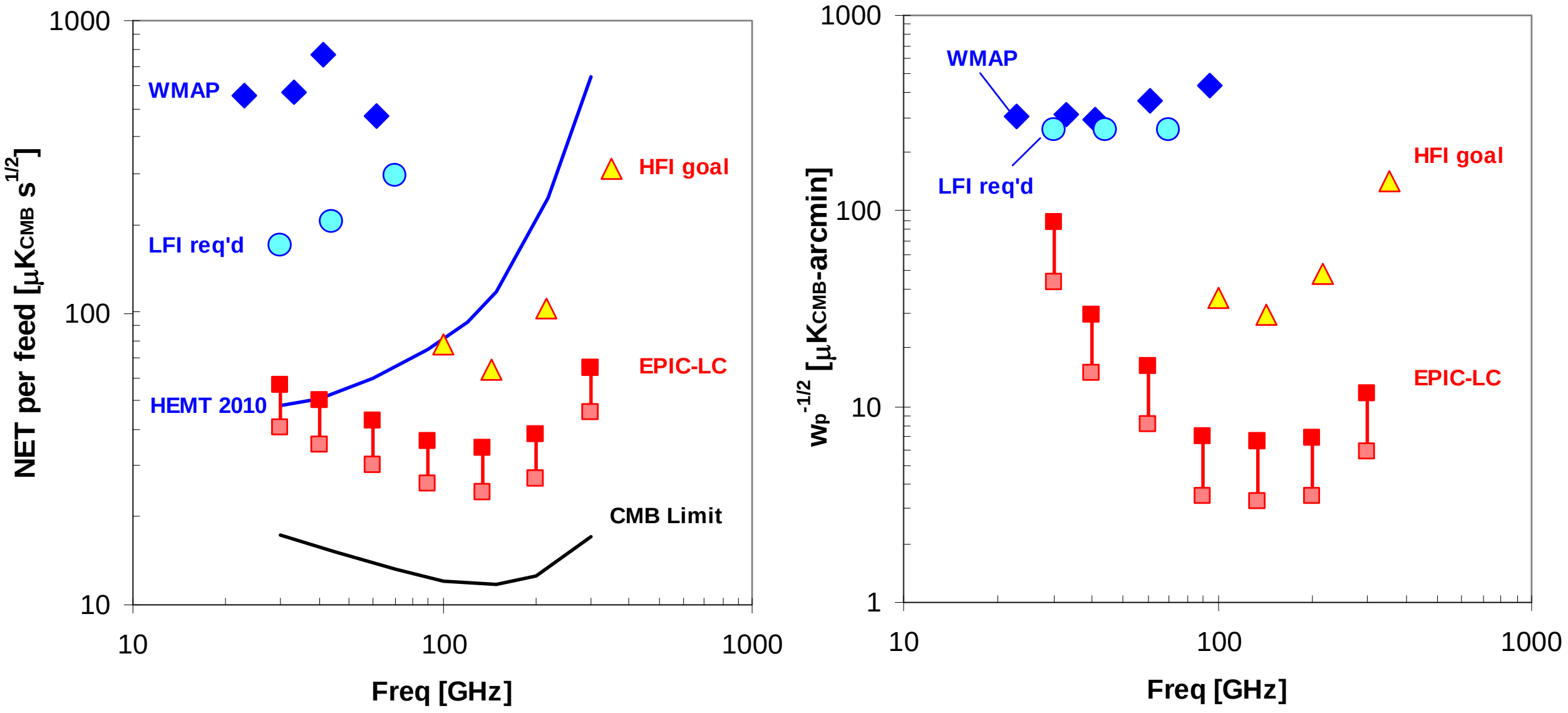

Figure 5.5.2. Same as Fig. 5.5.1 except for the TES option. Note the sensitivity per feed is similar, except at the highest frequencies where time constants are not an issue for TES bolometers and the sensitivity improves. The overall sensitivity improves due to larger detector formats.

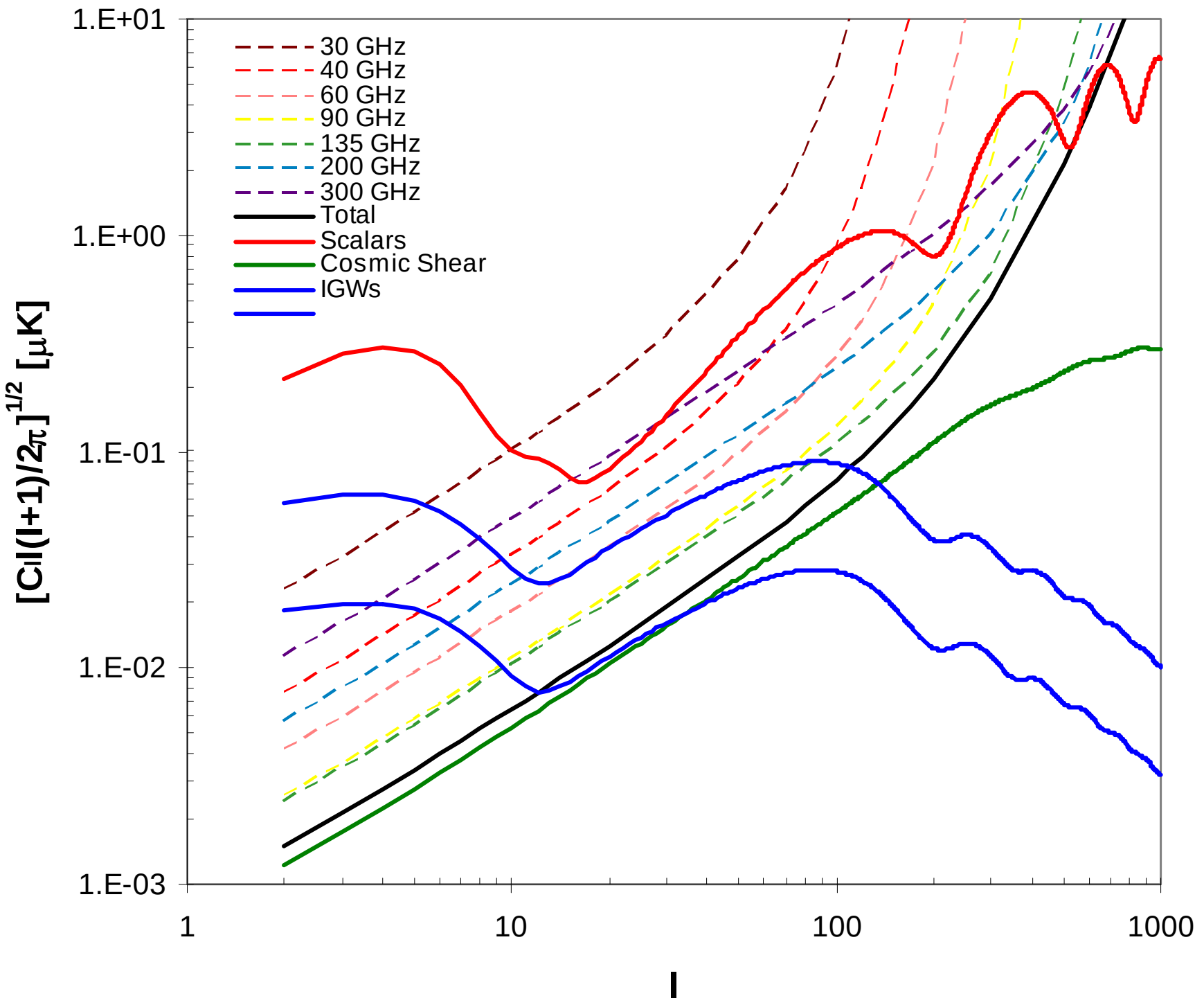

Fig. 5.5.3. Noise Cls calculated for the required NTD-Ge bolometer case (√2 sensitivity margin, 1 year mission life) for all of the bands (dashed), and combined (solid black) compared to scalar EE (red), cosmic shear (green), and IGW BB (blue at r = 0.1 and r = 0.01).

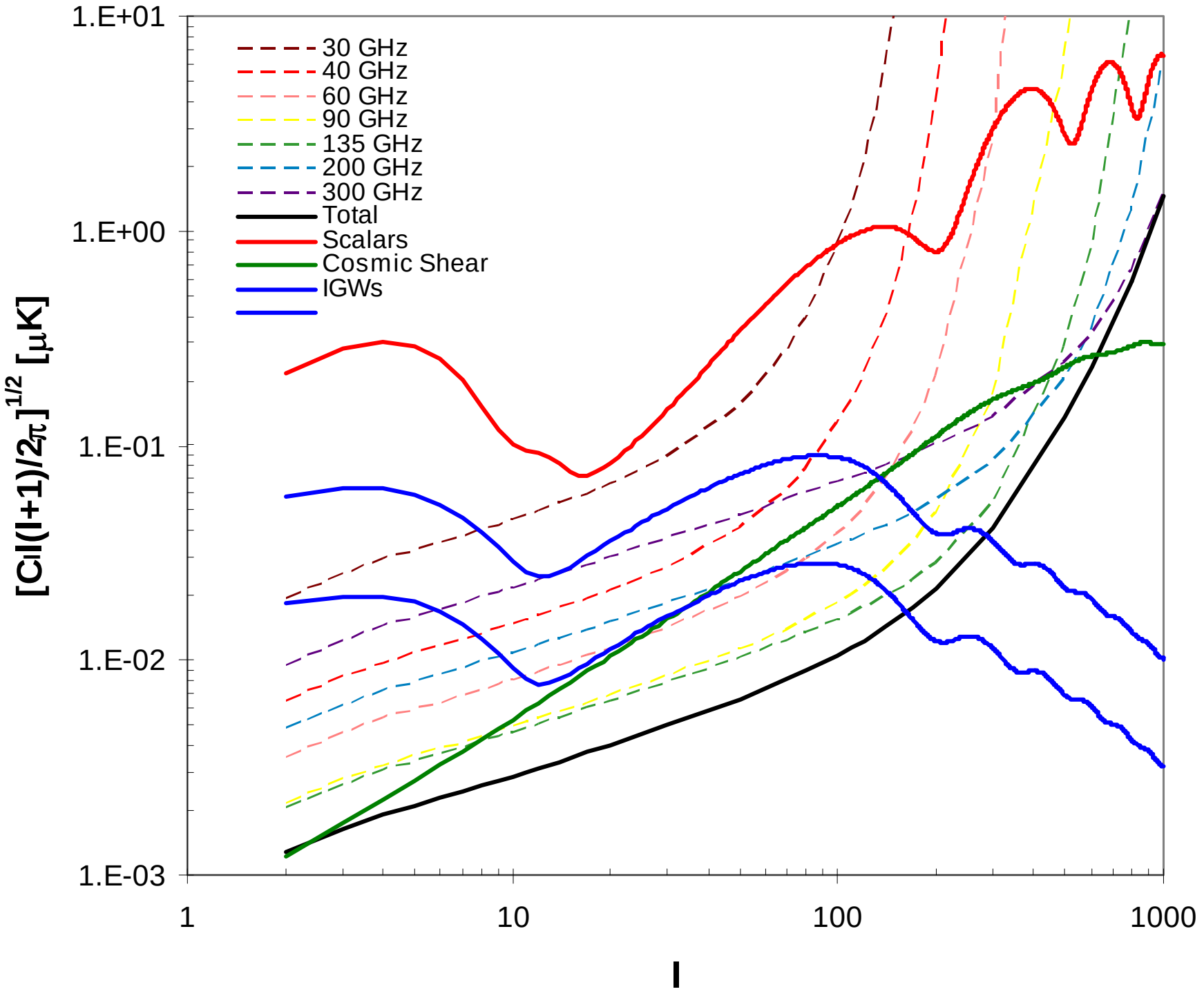

Fig. 5.5.4. Errors on Cls calculated for the required NTD-Ge bolometer case (√2 sensitivity margin, 1 year mission life) for all of the bands (dashed), and combined (solid black) compared to scalar EE (red), cosmic shear (green), and IGW BB (blue at r = 0.1 and r = 0.01). The calculation assumes fsky = 0.8, $\Delta\ell/\ell = 0.3$ binning, and ignores sample variance.

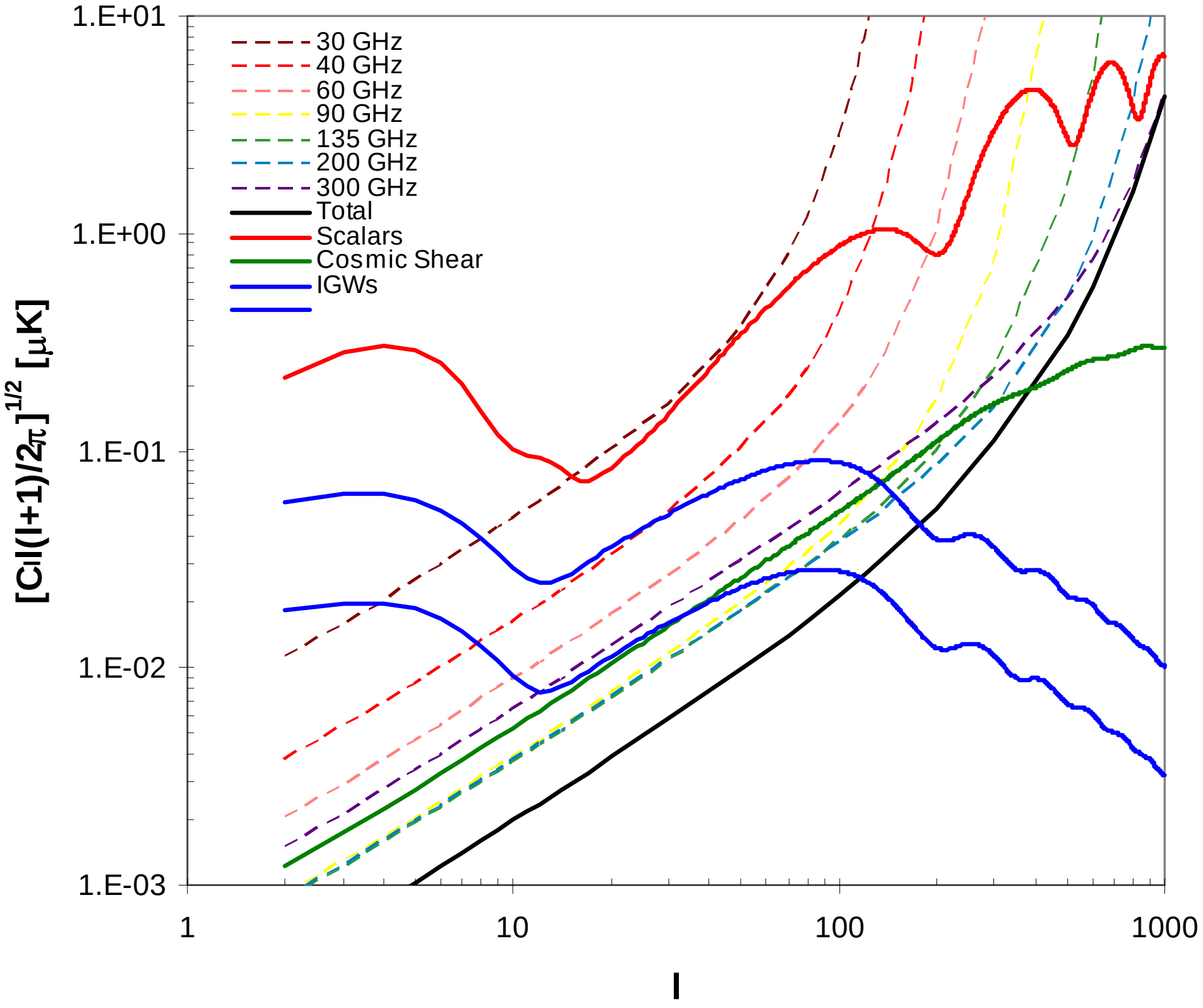

Fig. 5.5.5. Noise Cls calculated for the design TES bolometer case (no sensitivity margin, 2 years mission life) for all of the bands (dashed), and combined (solid black) compared to scalar EE (red), cosmic shear (green), and IGW BB (blue at r = 0.1 and r = 0.01). Note that this option measures scalar EE to sample variance out to ℓ ~ 1000, and can measure lensing BB to "sample variance" (a convenient misnomer since the statistics are non-Gaussian) out to ℓ ~ 300.

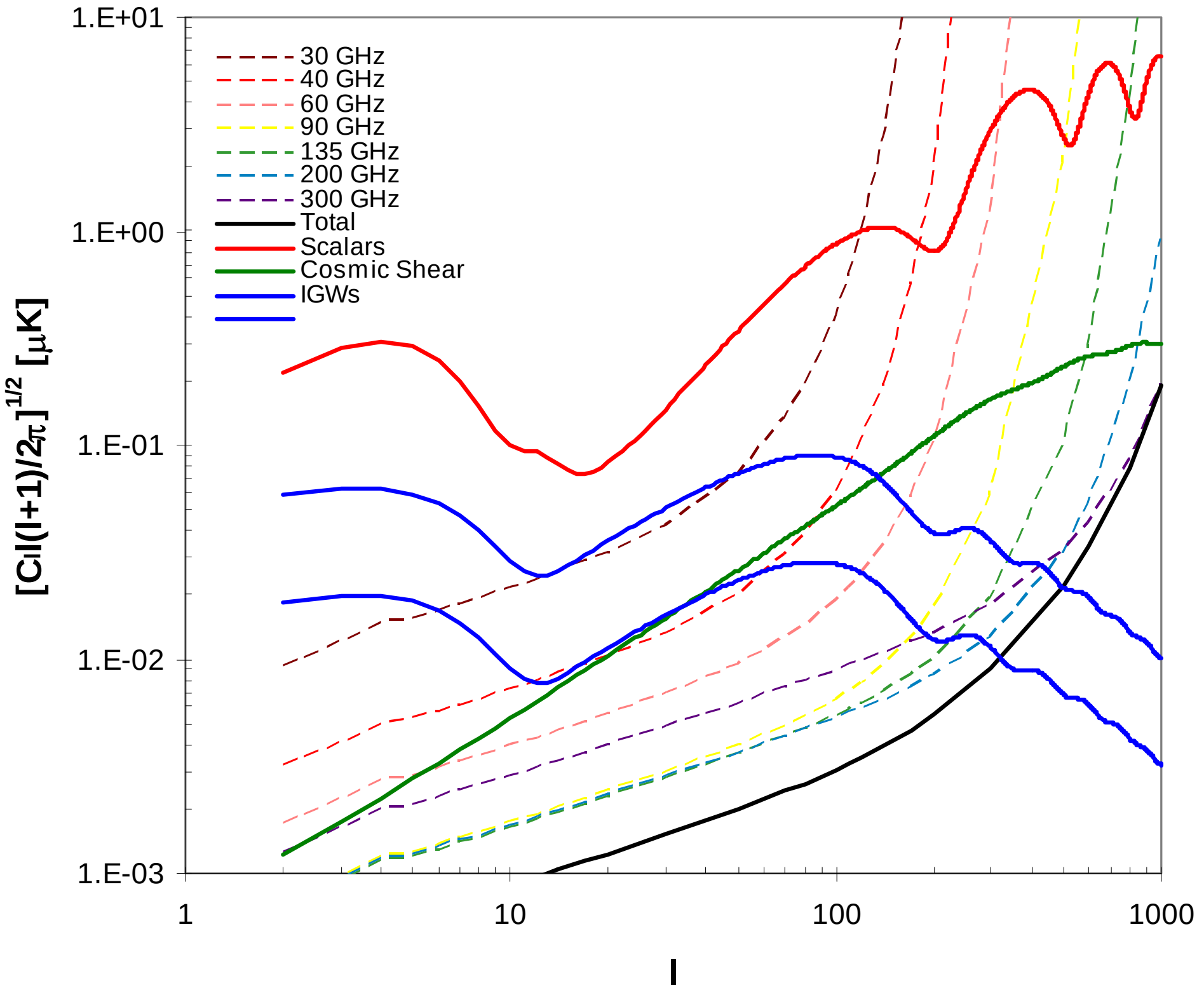

Fig. 5.5.6. Errors on Cls calculated for the design TES bolometer case (no sensitivity margin, 2 year mission life) for all of the bands (dashed), and combined (solid black) compared to scalar EE (red), cosmic shear (green), and IGW BB (blue at r = 0.1 and r = 0.01). The calculation assumes fsky = 0.8, Δℓ/ℓ = 0.3 binning, and ignores sample variance.

*5.5.2 TES Detector System Implications*

The TES-based focal plane allows larger focal plane arrays, and places different requirements on instrument resources, many of which are beneficial. We note that planar antennas are key to eliminating the otherwise large mass of sub-K Cu scalar feedhorns. We estimate the masses of the 0.1 K, 0.4 K, and 2.0 K stages as 5.2 kg, 6.0 kg, and 3.0 kg for a planar-antenna coupled focal-plane as compared to 24.8 kg, 11.0 kg, and 31.3 kg for a scalar-horn-coupled focal-plane. TES arrays reduce the wire count and eliminate the large power dissipation from the JFETs. We assume the JFET power can but thermally sunk to 40 K, but significant thermal engineering will be needed to place the JFETs as close as possible to the 100 mK focal plane.

**Table 5.5.5 Resource Tradeoffs with TES and NTD Focal Planes**

| **Resource** | **TES / SQUID TDM** | **TES / SQUID FDM** | **NTD / JFET** |
|---|---|---|---|
| Detectors | 2366 | | 830 |
| System NET (w/o margin) | 0.8 | | 1.4 |
| Time Constant | 0.8 – 1.7 ms | | 10 – 53 ms |
| Focal plane mass at 0.1 K | 5.2 kg | | |
| Focal plane mass at 0.4 K | 6.0 kg | | |
| Focal plane mass at 2 K | 3.2 kg | | |
| Power at 40 K | N/A | | 148 mW* |
| Power at 0.1 K | 1.0 μW[1] / 0.1 μW[2] | Zero | N/A |
| Wires to 0.1 K | 520 | 280 | 1660 |
| Warm electronics power | 340 W[1] / 170 W[2] | 570 W[1] / 190 W[2] | 95 W[1] |

[1]Currently demonstrated
[2]Expected with optimization

TES detectors also offer the possibility for operation at higher focal plane temperatures, greatly simplifying the requirements on the sub-K cooling system. While NTD-Ge detectors must be cooled to 100 mK in order to have sufficient speed of response, TES detectors have strong electro-thermal feedback which provides a significant speed up. Thus the thermal conductivity can be optimized for sensitivity at a higher operating temperature. A comparison of the resulting sensitivities are shown in Fig. 5.5.7 and Table 5.5.6. Operating at a higher temperature put more severe requirements on the focal plane: we have a smaller safety factor for the 250 mK TES focal plane, and because of the slower response times, the focal plane must have better stability. Nevertheless, except in the highest frequency bands where the G must be chosen for speed of response, the reduction in sensitivity is only ~15% at 250 mK compared to 100 mK.

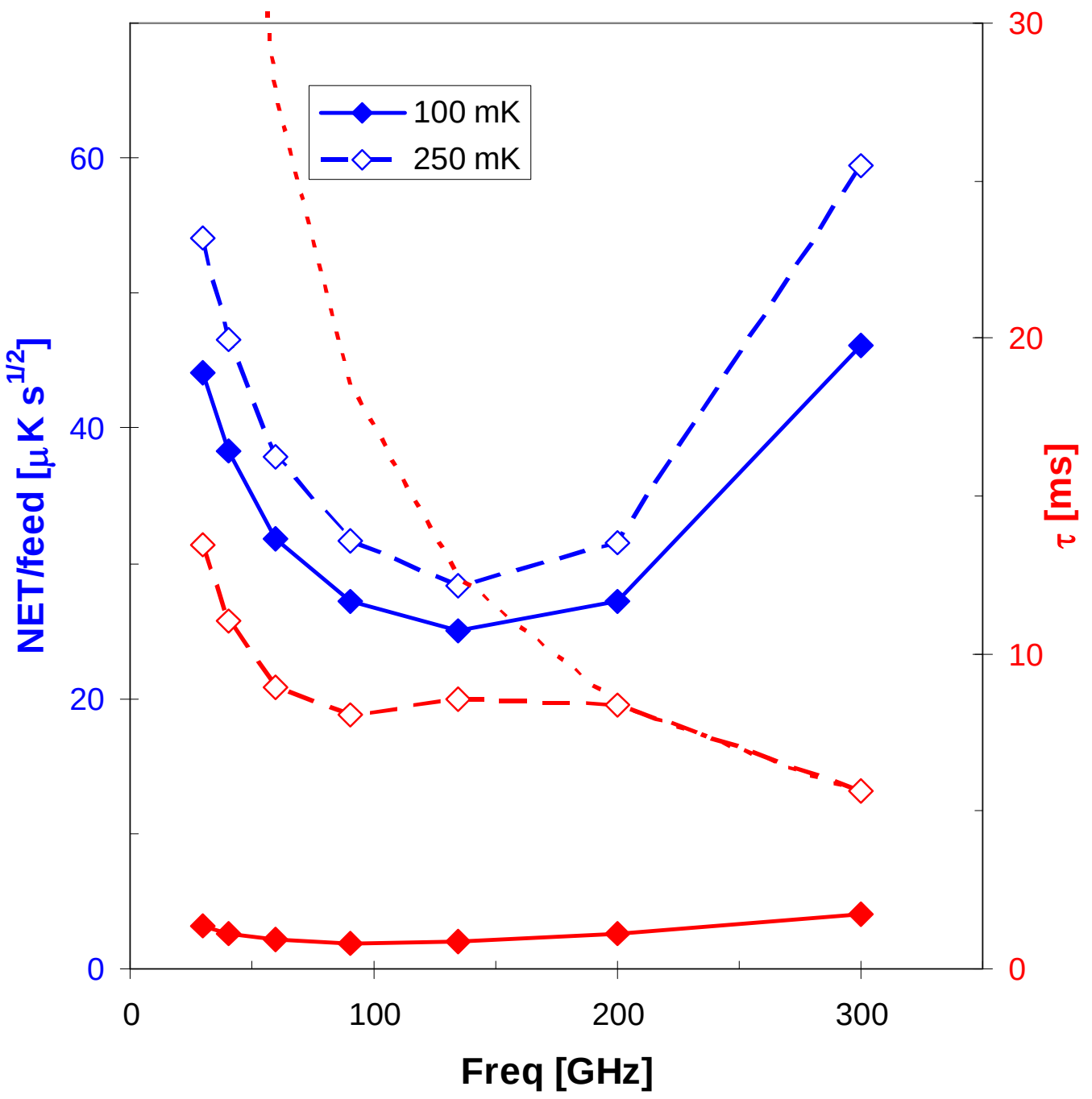

Fig. 5.5.7. A comparison of the sensitivity and time constant for TES focal planes operating at 100 and 250 mK. Calculated NETs for 250 mK (open blue) are only 15 - 20% larger than the NETs calculated at 100 mK (solid blue), except at 300 GHz. 100 mK devices (solid red) easily meet the time constant requirement (dashed red) for a spin rate of 1.5 rpm, whereas 250 mK devices (open red) must be tailored to meet the speed requirements in the 200 and 300 GHz bands.

**Table 5.5.5. TES Bolometers for Two Operating Temperatures**

| Quantity | Value | | Units |
|---|---|---|---|
| $T_0$ | 100 | 250 | mK |
| Psat/Q | 5 | 3* | |
| $G_0$ (min) | 0.7 | 3.2 | pW/K |
| β | 1.5 | 2.5 | |
| $C_0$ | 0.2 | 0.4 | pJ/K |
| α | 100 | 100 | |
| Rop | 10 | 30 | mΩ |
| NEP $\sqrt{\tau}$ | 1 | 4 | e-19 J |
| Spin rate | 3 | 1.5 | rpm |
| 1/f knee | 50 | 25 | mHz |

Note: All other detector parameters from Table 5.5.1
*200 and 300 GHz bands have higher $P_{sat}/Q$ to meet required τ.

TES detectors are close to the requirements for noise stability, although this has not been demonstrated in a full working instrument to date. If sufficient stability can be realized (15 – 50 mHz depending on the scan speed), EPIC can avoid the use of a polarization modulator for reasons of noise stability. Fig. 5.5.8 shows an example of a detector with a 1/f knee of 40 mHz.

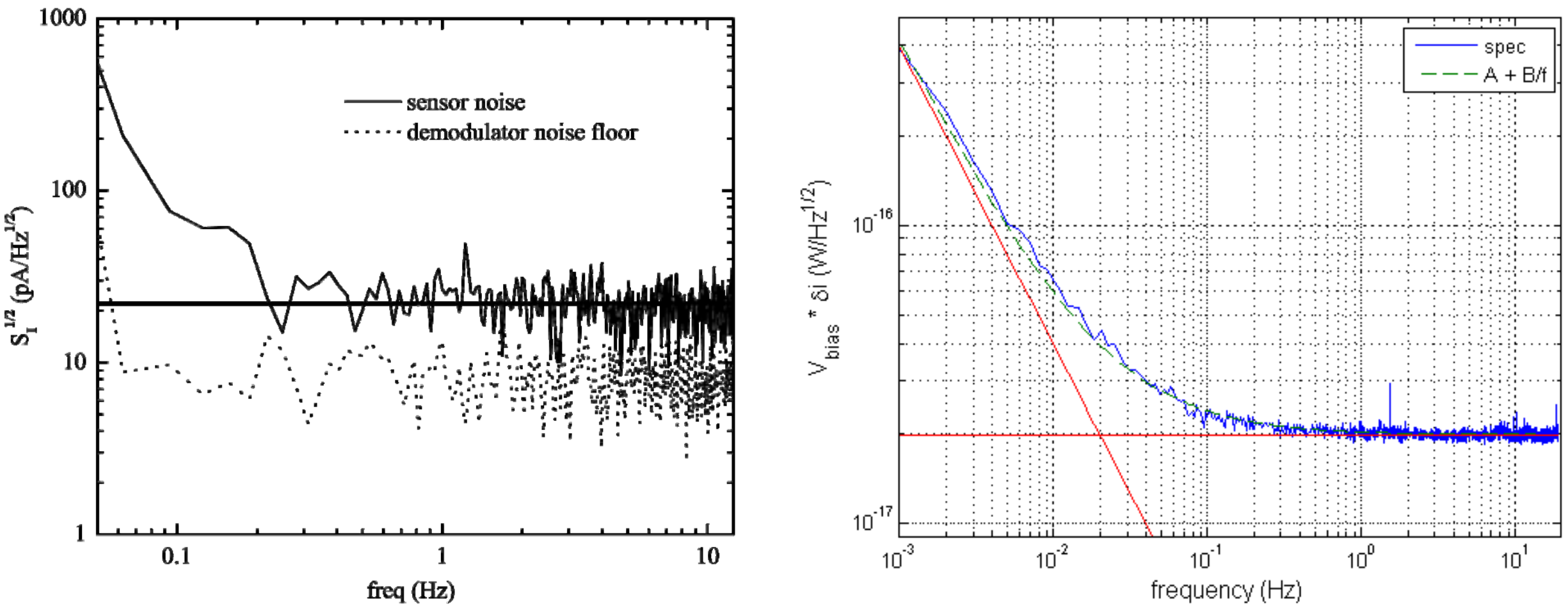

Figure 5.5.8. (Left) Noise as a function of frequency for a TES sensor readout using a frequency-domain multiplexer. The noise level agrees with theoretical expectations indicated by the solid horizontal line. (Right) Measured noise on a Ti TES sensor with a single SQUID readout. The noise spectrum, taken over a period of 14 hours without deglitching, shows excellent stability with a 1/f knee frequency of 40 mHz.

Shielding is an important consideration for a space-borne system, due to the associated mass at low temperature. We concentrate on time-domain multiplexing, because the shielding requirements appear to be more restrictive than the frequency-domain multiplexing. A SQUID, of course, measures current via the magnetic field in its input inductor coil. TES bolometers are sensitive to magnetic field through the dependence of Tc on applied field. Fortunately, these susceptibilities are well characterized in the lab so one can specify the magnetic shielding and then develop an optimal design.

Our very conservative but achievable criterion is to require that any spurious field signal be made less than the expected map rms on 1o pixels, which ensures that not even a scan-synchronous or sky-synchronous field signal could be a significant map contaminant. These requirements are set for shielding ambient DC fields, 50 µT for earth’s field and 5 nT at L2. Fields can also be generated locally by motors and electronics, but there can be shielded at the source and fall off quickly with distance. Table 5.5.6 presents the ambient fields, field sensitivity, and resulting shielding requirements for a balloon experiment (SPIDER) and EPIC. It is clear that operating in earth's field is more of a shielding challenge than L2.

The most magnetically sensitive focal plane components are the 1$^{st}$–stage SQUIDs, which are about 200 times more sensitive that the TESs (we have measured ~1 mK/50 µT for Ti). The key experimental parameter is SQUID "effective area", which converts from applied field to magnetic flux. The most recent design (mux06a) has an effective area of (12 µm)$^2$. We show in Table 5.5.6 how this susceptibility feeds through the system. However, because this field signal is additive, not multiplicative, it should be possible to move it outside the signal frequency band by using an AC modulation/lockin technique for the TES bias. Such a technique would no doubt mitigate this susceptibility by the factor of $10^3$ needed to render it smaller than the intrinsic TES susceptibility.

**Table 5.5.6. Magnetic Shielding Requirements**

| Quantity | SPIDER | EPIC | Units |
|---|---|---|---|
| CMB Depth | 0.5 | 0.05 | $\mu K_{CMB}$ |
| Ambient Field | 50 | 5e-3 | μT |
| **TES B-Field Sensitivities** | | | |
| Bolometer G | 20 | 5 | pW/K |
| Operating Temp | 300 | 100 | mK |
| TES Responsivity | 7 | 30 | $mK/K_{CMB}$ |
| Field Sensitivity | 2.7 | 0.7 | $\mu K_{CMB}/nT$ |
| Residual Field Req't | 180 | 70 | pT |
| Attenuation Req't | 3e5 | 70 | |
| **SQUID B-Field Sensitivities** | | | |
| B-Field to TES current conversion | 0.5 | 0.5 | A/T |
| TES current to CMB responsivity | 0.7 | 1.5 | $pA/\mu K_{CMB}$ |
| Field Sensitivity | 700 | 400 | $\mu K_{CMB}/nT$ |
| Residual Field Req't | 0.7 | 0.14 | pT |
| Attenuation | 7e7 | 4e4 | |

Notes: This calculation is carried out for 150 GHz, given the parameters in Table 2, assuming a TES field sensitivity of 1 mK/50 μT, and an input coil to SQUID mutual inductance of 275 pH.

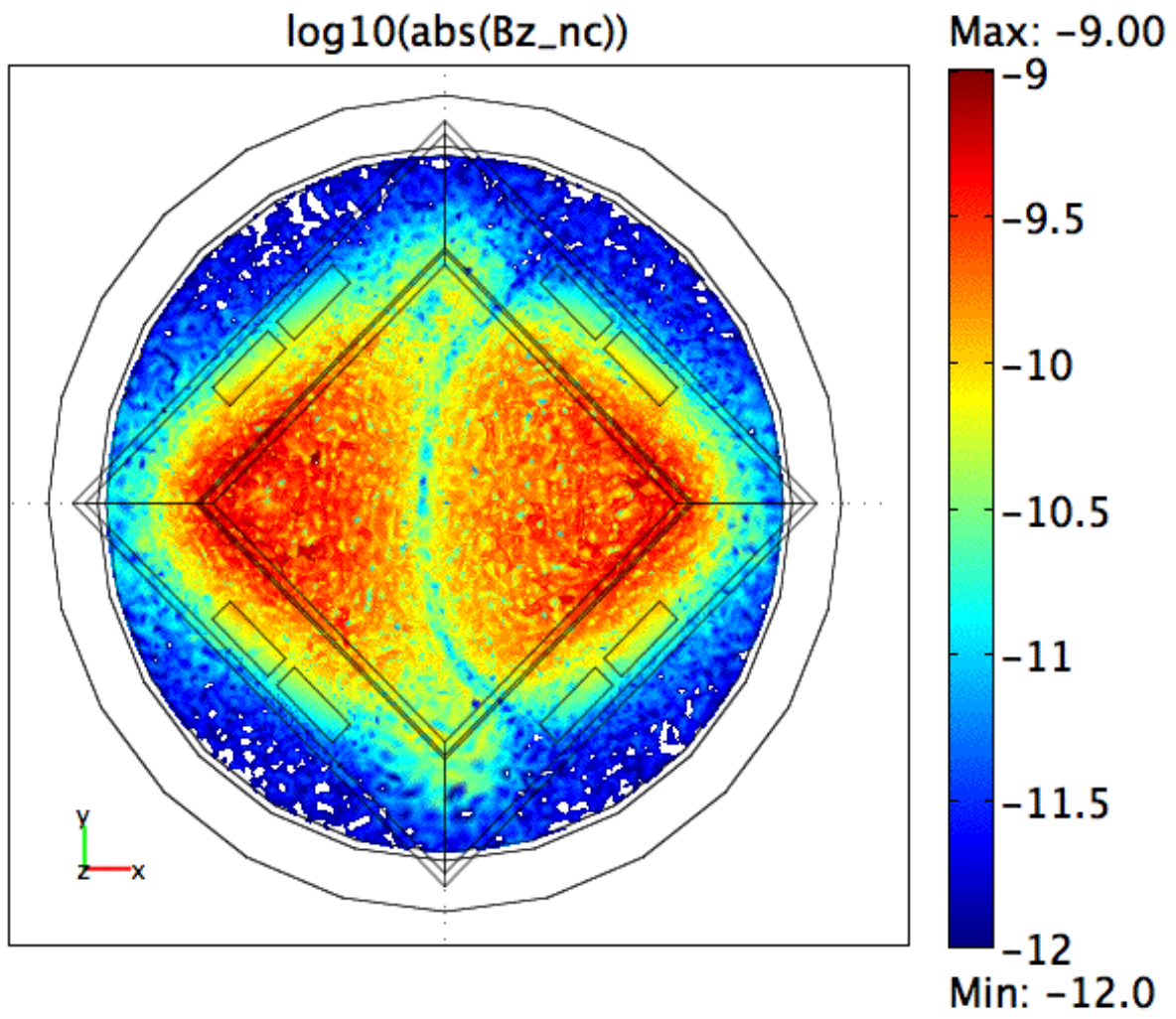

Fig. 5.5.9. Calculated fields in the focal plane for the shielding design developed for SPIDER, incorporating a single layer of cryoperm at 4 K and 250 mK Nb box with a flared opening through which light enters, and a Nb plane just under the array. A 50 μT field has been applied at a 45° angle to the optical axis and yields a residual field of < 300 pT over most of the focal plane. The inner diamond is the bolometer array, while the nearby rectangles are the SQUID chips. While this shielding arrangement provides adequate performance, its mass is a strong design consideration for a space mission.

We have done exploratory work on magnetic shielding design using COMSOL that indicates we can meet the TES susceptibility goals. Fig. 5.5.9 displays field profiles in the focal plane of SPIDER, which approximately meets the requirements in Table 5.5.6. EPIC has less stringent needs than SPIDER, so its field goal is certainly achievable. Shielding requirements can be dramatically reduced by building immunity into the focal plane, by optimized SQUID design, the use of monitor channels such as dark SQUIDs, and common-mode rejection in differencing matched pixels. These mitigations are not assumed in our shielding calculations, and we expect they will provide a large reduction in shielding requirements.

*5.5.3 Bolometer Technologies*

TES is a maturing detector technology that will be incorporated in upcoming ground-based and sub-orbital experiments. TES bolometers cooled to temperatures of 100 mK can have a sensitivity that is nearly limited by photon arrival statistics over much of the frequency range of interest. They have two properties that are essential for building large focal-plane arrays (i) They are simple to fabricate using optical photolithography, and (ii) their readout can be "multiplexed" so that a row of detectors can be readout using a single amplifier – this greatly reduces the complexity of the cryogenic wiring.

The TES is a superconducting film biased in the middle of its transition. It is voltage biased, and in this mode it has high stability and linearity due to negative feedback that occurs between the thermal and electrical "circuits" of the bolometer. The signal from a TES is measured using a Superconducting QUantum Interference Device (SQUID) ammeter, which can operate at cryogenic temperature. Our team has experience building TES detectors over the last decade, including Al/Ti, Cu/Mo and Au/Mo proximity sandwiches, and elemental Ti. Figure 5.5.8 shows noise data for a Ti TES sensor.

There are two categories of polarimeters. A differencing polarimeter measures polarized radiation by subtracting the signals from two detectors that are sensitive to orthogonal polarizations. A modulated polarimeter periodically changes the polarization transmitted to a single detector. Most current bolometric CMB polarization experiments, such as BOOMERanG, BICEP, QUAD, and Planck HFI use differencing polarimeters. A modulated polarimeter has advantages in minimizing systematic errors since it does not require matching the parameters of two detectors. Systematic error mitigation is discussed in Sections 3 and 5.2. Current and future funded CMB polarization experiments with modulated bolometric polarimeters include MAXIPOL, EBEX, and PAPPA. SPIDER will use only scan modulation and waveplate stepped on a timescale of many hours. In the case of SPIDER, the waveplate is used to mitigate main beam effects, not for signal stability. Our baseline design for the refractor version of EPIC is based on scan modulation with stable NTD detectors and a stepped waveplate. It is not yet clear if active modulation will be required for EPIC.

Presently all antenna designs brings out vertical and horizontal polarizations into two detectors, and the difference is used to extract a single Stokes parameter. However, one can extract both Q and U in a single pixel by splitting half the power in each polarization into a 180˚ degree hybrid, followed by a pair of detectors. This scheme requires 4 detectors per pixel, with half the power in each detector. TES detectors can be designed to this lower background with negligible overall loss in sensitivity. Alternatively, the focal plane can be alternated between Q and U by using +/- 90˚ hybrids and two detectors per pixel. This arrangement gives better instantaneous Q/U coverage in a single scan. A detailed systematics simulation must be carried out to determine the best approach.

We have studied three methods for optical coupling of the focal plane: (i) phased-array planar antennas, (ii) lens-coupled planar antennas, and (iii) planar-probe-coupled scalar horns. All three are viable, and further study will be required before a single technology can be chosen for EPIC.

**Phased-array planar antennas:** The angular size of an antenna's beam becomes smaller due to diffraction as the effective area of the antenna grows. Most planar antennas have a size that is comparable to the wavelength of the radiation and a correspondingly broad antenna pattern. A phased-array antenna combines a large number of small antenna elements to form a larger antenna. Phased-array antennas are common at radio wavelengths for e.g. radar and communications.

The millimeter-wavelength monolithic phased-array antenna has been developed recently by the JPL/CIT group. A photo of an array coupled to a circuit board with SQUID multiplexer readout chips is shown in Fig. 5.5.9. The antenna is made from a large number of slot dipoles, and the RF signals are added coherently by a network of microstrip transmission lines. After addition, the signals are bandpass filtered, and finally the signals are detected by TES bolometers.

Advantages of the phased-array antenna over other candidates include minimal focal-plane mass, efficient use of the focal-plane area, and a completely monolithic fabrication process which can be critical for making large arrays. The current development status is that single pixels with a complete antenna/filter/bolometer have been measured and show symmetric beam patterns closely matching theoretical predictions, low cross-polarization, a spectral bandpass with the expected width, and high optical efficiency. Figure 5.5.10 shows a measured antenna pattern and frequency band shapes. The engineering parameters for making a large array (TES uniformity and reproducibility, stripline index and reproducibility, yield) have been studied in detail and appear to all be sufficiently controllable. An array has been developed and is now being integrated into a multiplexed focal plane.

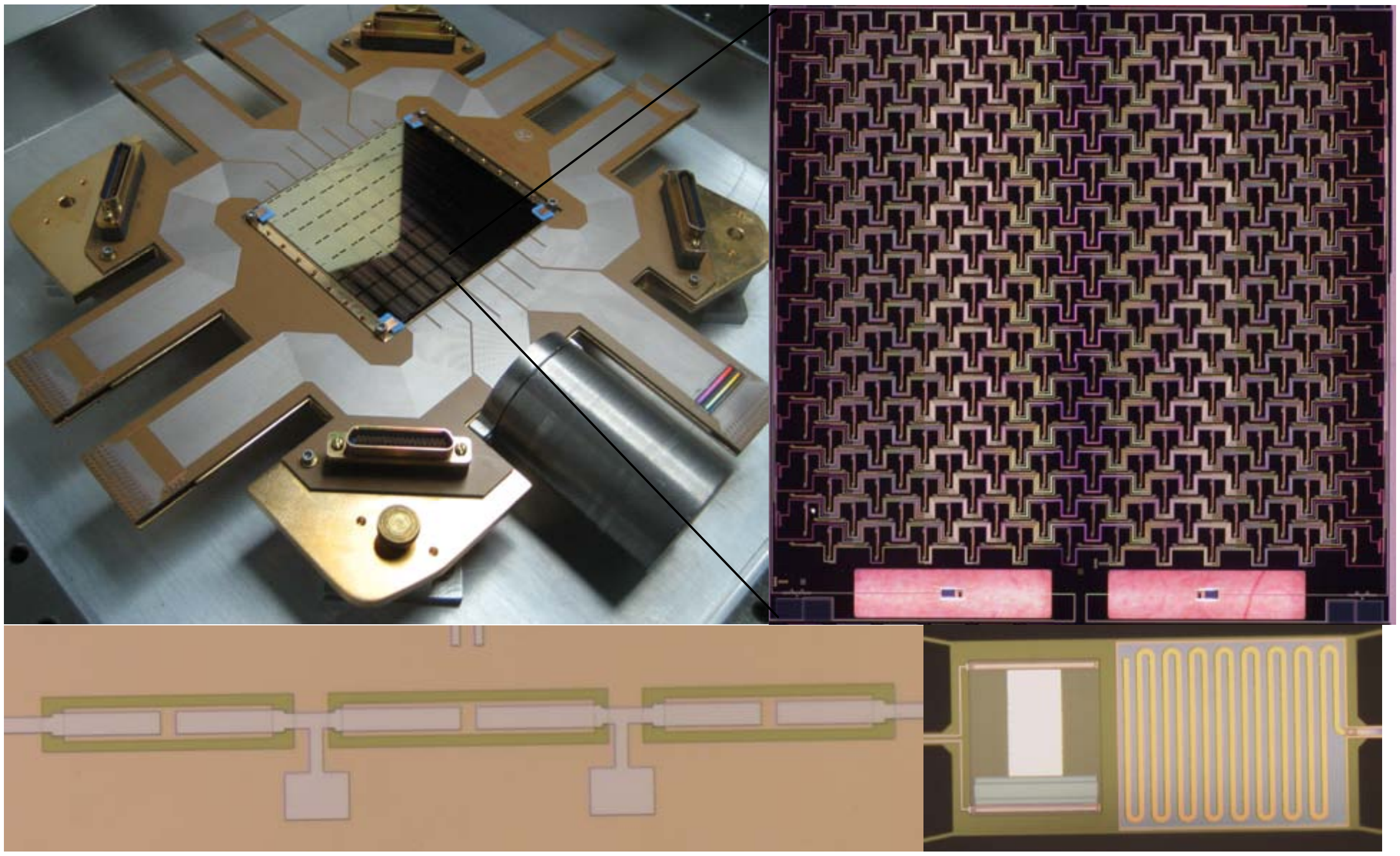

Figure 5.5.9. 8x8 TES bolometer array designed for operation at 150 GHz (upper left). Optical coupling is achieved with planar phased-array antennas as shown in the closeup (upper right). This monolithic array includes antennas, band-defining filters (lower left), and TES bolometers (lower right). The band-defining filter uses lumped component built in coplanar waveguide. This layout illustrates the JPL planar antenna concept.

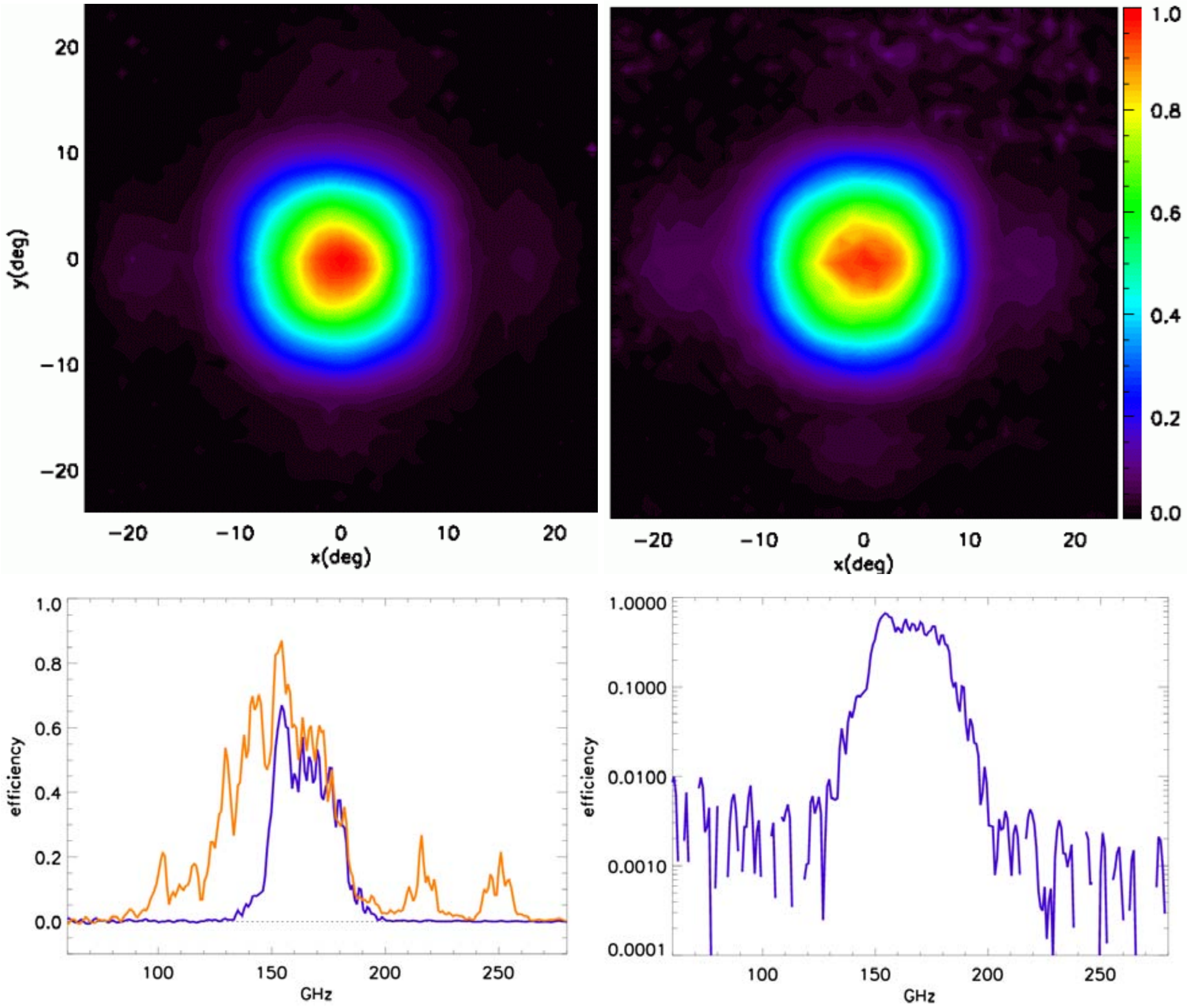

Fig. 5.5.10. Measured beam maps of the antenna-coupled bolometer shown in Fig. 5.5.9 for the vertically polarized bolometer (top left) and the horizontally polarized bolometer (top right). The beam shapes are in excellent agreement with theoretical predictions. Spectral response (bottom left) of the antenna manufactured with (blue) and without (orange) the transmission-line filter shown in Fig. 5.5.9. The response is plotted in units of optical efficiency, determined by measuring the response to a cryogenic blackbody source. This device was tested without a backshort, which is expected to improve the efficiency by ~15%. No leaks (bottom right) are evident in the spectral response down to the measurement noise floor of ~1e-3.

**Lens-coupled planar antennas:** A small antenna that is comparable to a wavelength in size can be attached to a small contacting lens to give a suitable beam for coupling to a telescope. This approach has been well studied in the engineering and sub-mm mixer community. Much of the area under the lens is available for components such as filters, switches, and readout components. Although, the refractor version of EPIC uses several single-color focal planes, a long term advantage of lens-coupled planar antennas is that they can be built to sense multiple frequency bands in a single pixel. The Berkeley/LBNL group has been developing detectors using the lens-coupled planar antenna. Figure 5.5.11 shows an array and closeup of a single pixel. This array has single-color pixels for 90, 150, and 220 GHz. The current status is that single pixels including lens, antenna, band-defining filters, and TES detectors have been tested and prototype arrays are currently under test.

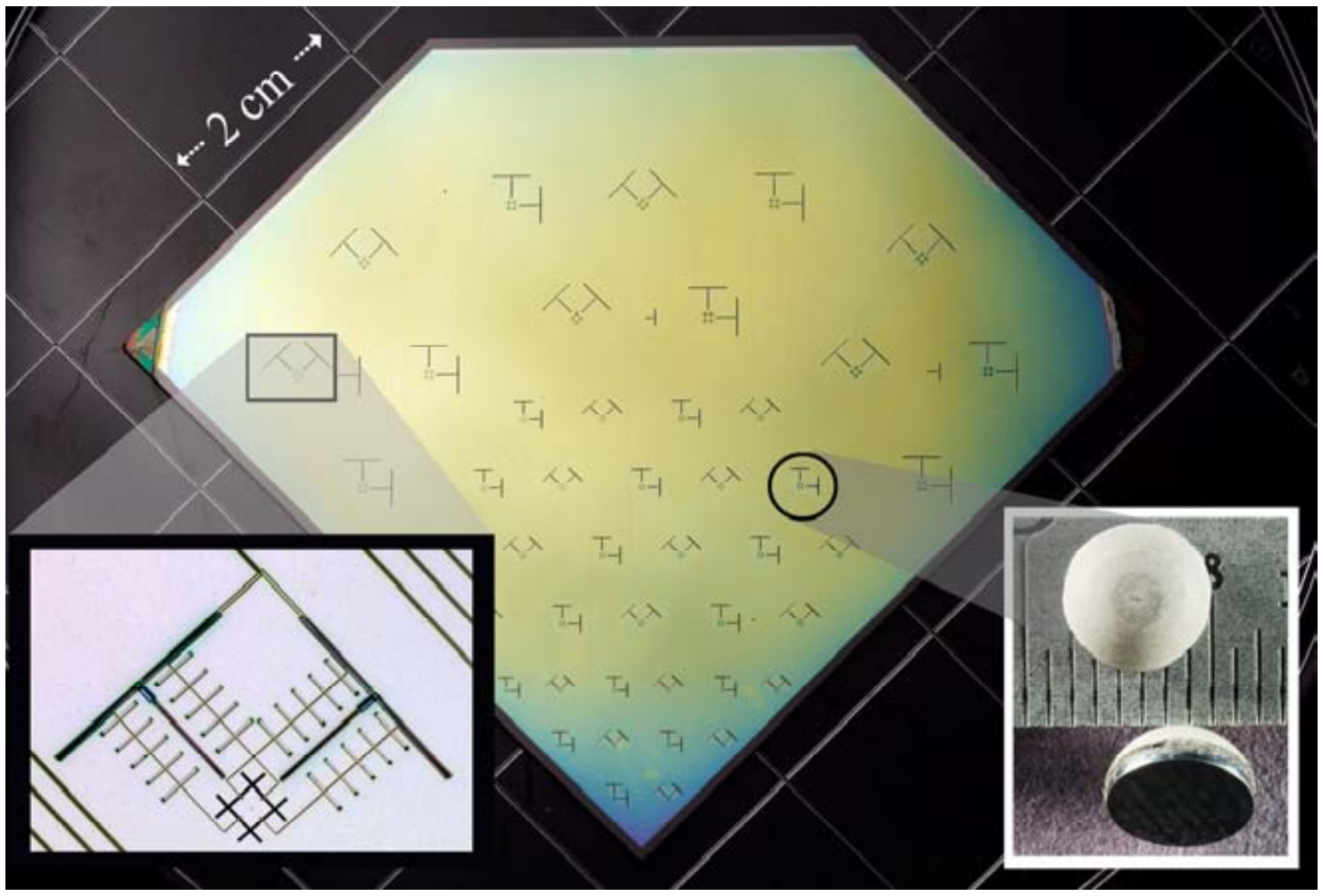

Figure 5.5.11. Array of TES bolometers optically coupled by a combination of planar antenna and contacting hemispherical lens. Left inset shows a close of of a single pixel where the antenna is at bottom and the RF filters connect the antenna to the “T” shaped TES bolometer. Right inset shows and hemispherical lens with antireflection coating. This array has 90 bolometers distributed between 90, 150 and 220 GHz. The bandpass filters use a distributed design with ¼ wavelength stubs. The anti-reflection coating is made from stycast and have been demonstrated to work optically and to withstand thermal cycling. Optical testing of single pixels from this wafer has been done.

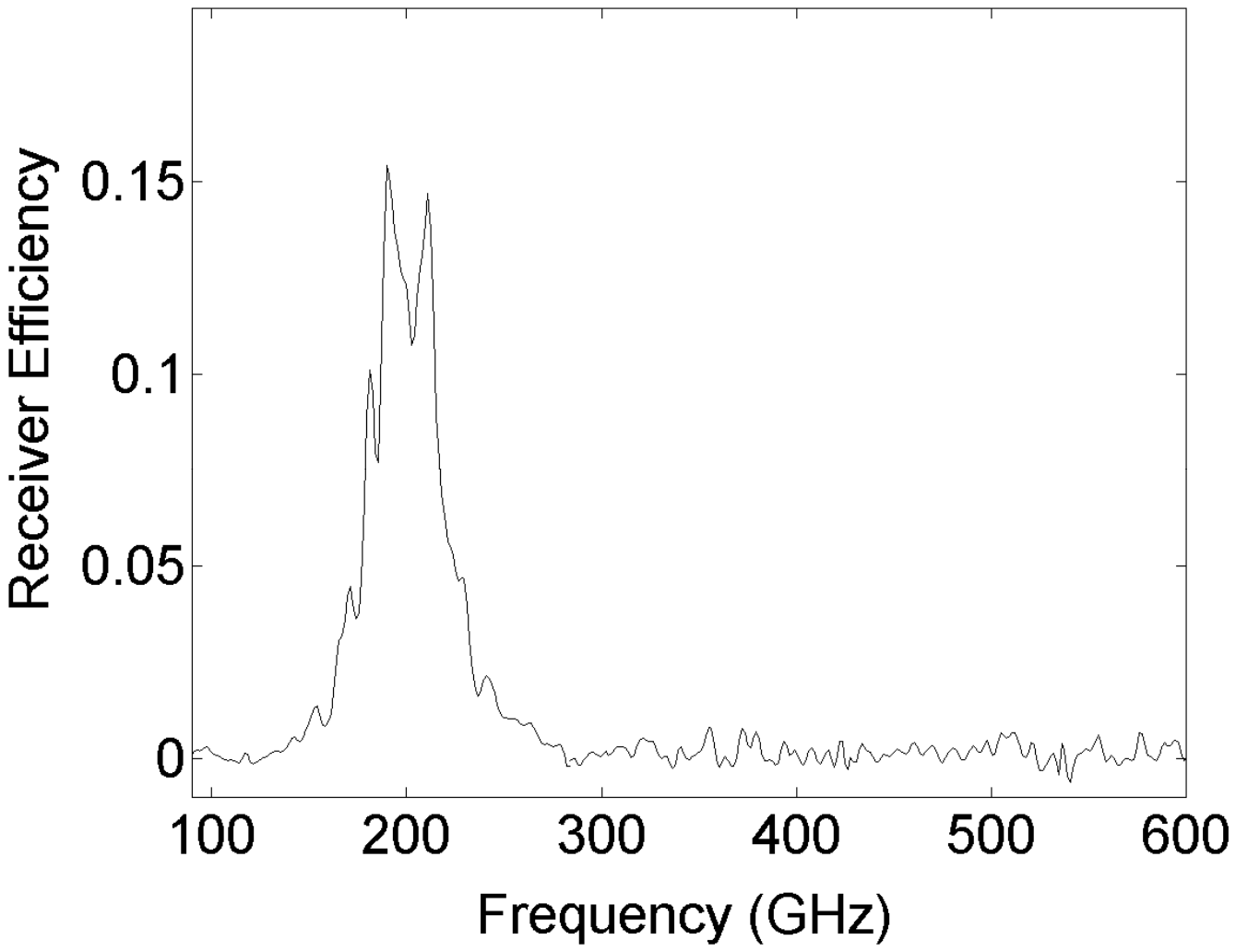

Figure 5.5.12. Measured spectral response of a planar-antenna coupled pixel similar to that shown in Fig. 5.5.11. The efficiency is end-to-end for the entire receiver and would be improved by a factor two by adding antireflection coatings to the optics.

**Planar-probe-coupled scalar horns:** Scalar-horn antennas have a strong heritage in CMB experiments. For CMB polarization experiments, scalar horns have advantages of highly symmetric beams, low cross-polarization, and low-sidelobes. Arrays of scalar horns can be coupled to a monolithic array of bolometers by use of planar, lithographed OMTs as are being developed by the GSFC group. The scalar-horn is most advantageous for systems with minimal optics, e.g. those with no cold aperture stop, where the horn defines the optical performance rather than the rest of the optics. Such designs have usually have tradeoffs in feedhorn packing density, related to the available field-of-view, size and weight of the focal plane, and level of sidelobe control that is required.

All three of these focal-plane technologies are viable for EPIC. As described in the technology development plan section 7, near term sub-orbital experiments using these technologies will clarify the tradeoffs and give a basis for down selection for EPIC.

**SQUID multiplexing:** There are several readout multiplexer technologies that are reaching maturity, and they can be broadly divided into techniques that divide signals in either time or frequency domains. A time-domain readout multiplexer that uses SQUID switches to sequentially choose the detector that is read with the single output amplifier has been developed at NIST. The time-domain multiplexer can read 32 detectors with a single readout amplifier with no loss in bolometer noise performance or bandwidth. It has been used in an 8-channel system at the Caltech Submillimeter Observatory, and it will be used with arrays of several thousand pixels in several upcoming experiments including SCUBA2 and ACT.

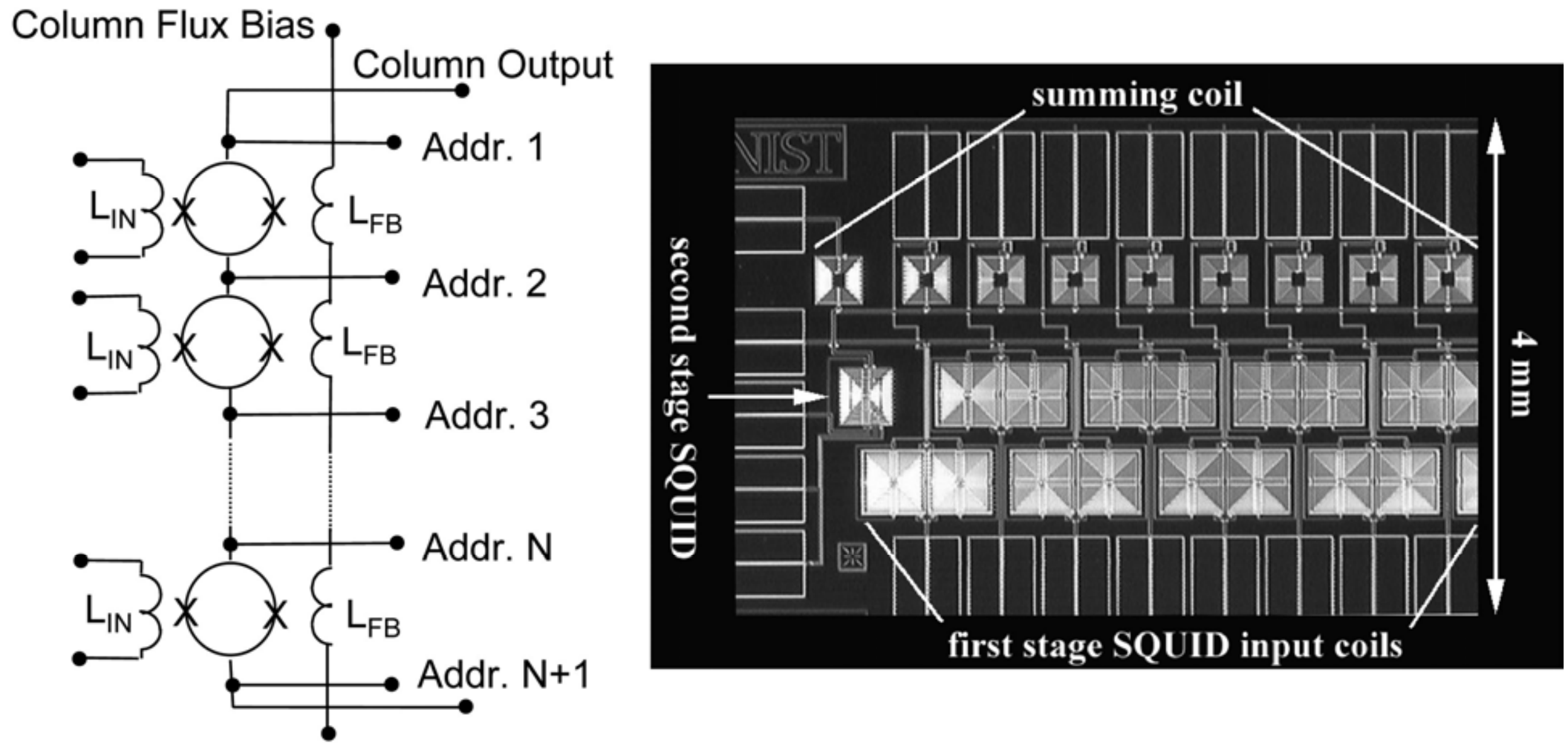

Figure 5.5.13. Left: Circuit schematic for an early version of a NIST time-domain SQUID multiplexer, showing its basic functions. Each input inductor $L_{IN}$ is connected in series with a TES detector that is always biased and a bias resistor (neither shown here). The N SQUID loops are connected in series. One SQUID is turned on sequentially using the address lines. Since the other SQUIDs are not biased, they remain in the zero voltage state. The column output is that of the one SQUID that is biased. The feedback flux for each SQUID is stored digitally between cycles and applied during the on state. Right: Photograph of the kilopixel SQUID multiplexer under development for SCUBA2. Figure and photograph courtesy of Kent Irwin.

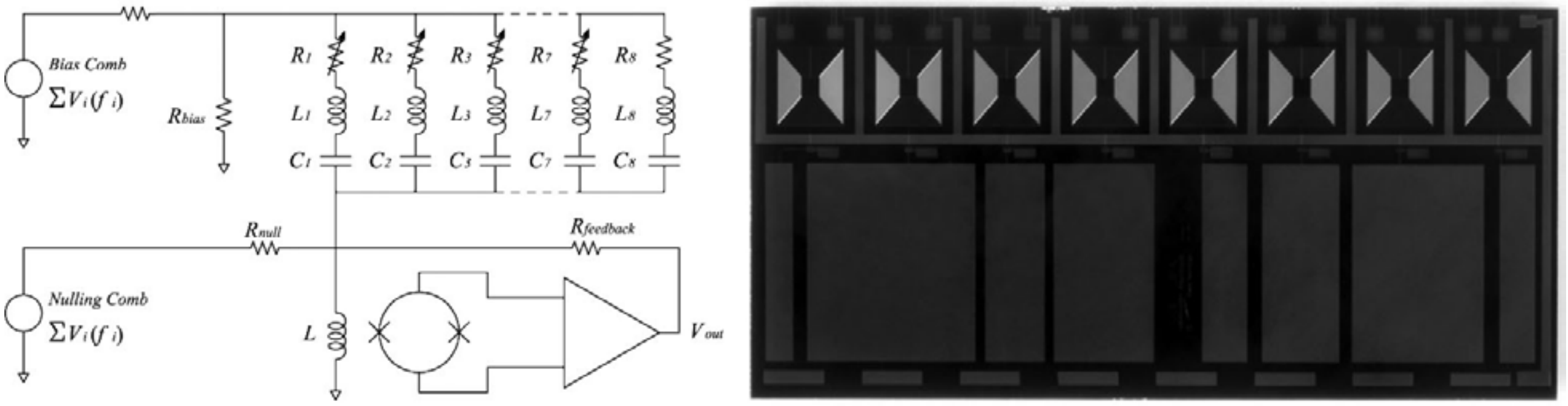

Figure 5.5.14. Left: Schematic of a current-summing frequency-multiplexer circuit. The TES devices, represented as variable resistors, are sinusoidally biased each with a different frequency. An LC resonator in series with each TES filters out wideband Johnson noise that would contribute noise to the other channels. The currents from all muxed TESs are added together at the input inductor of the SQUID. Shunt feedback is used to reduce the input impedance at that point creating a virtual ground. A 180° phased shifted bias signal can be added to the SQUID input to null each of the carriers from the TES partially. The amplitude of this nulling current can be adjusted on longer time scales than those characteristic of astronomical observations. Right: Photograph of a niobium LC filter chip fabricated by TRW (now Northrup-Grumman). The center frequencies vary from 380 kHz to 1 MHz, with 80 kHz channel spacing. A 32 channel multiplexer can be implemented by using four chips with interdigitated frequencies with a resulting spacing of 20 kHz.

Several groups are independently working on frequency-domain readout multiplexing. In this scheme, each detector is biased using a sine wave with a unique frequency, the bias signals are amplitude-modulated by the bolometers, and the sum of all the currents is measured using a single SQUID ammeter. This type of multiplexer will be used for 1000 pixel arrays in several upcoming funded experiments such as SPT and EBEX.

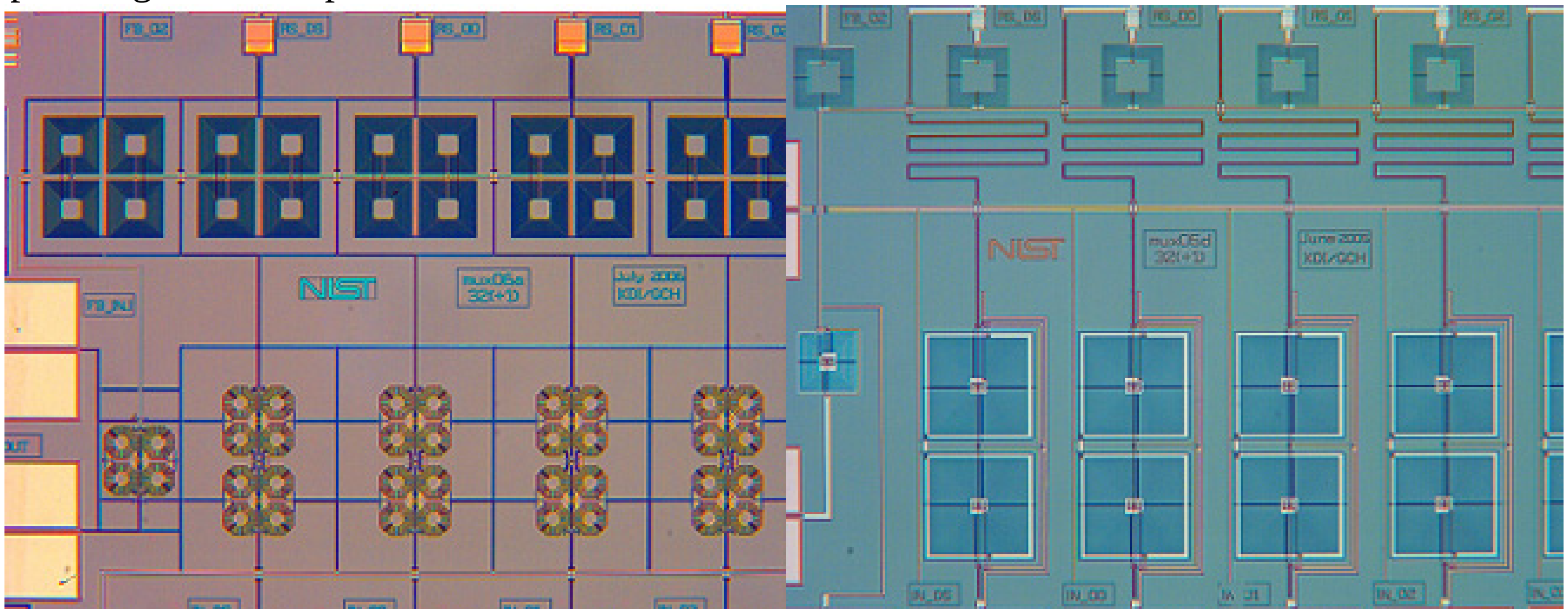

Fig. 5.5.15. Photomicrographs of the left four pixels on an older mux05d version time-domain SQUID multiplexer (left), and the newer fourth-generation gradiometric mux06a SQUID multiplexer (right). Mux06a uses 2$^{nd}$-order gradiometer SQUIDs (along the bottom and left) and 2$^{nd}$-order gradiometer flux-summing coils (along the top) to achieve 100 times improved field immunity.

Important tradeoffs between these two types of multiplexing include magnetic shielding, wiring constraints, power dissipation, and heat leak to 100 mK from wires to the higher temperature stages. In the current state of development, the time-domain multiplexer will dissipate 1.0 µW at 100 mK for the 2366 bolometers in the EPIC-LC option. The use of the time-domain multiplexer would benefit from the development of the continuous ADR which has large cooling power. The frequency-domain multiplexer is completely passive at 100 mK with zero power dissipation. It could be used with the Planck dilution refrigerator without further development. However, the frequency-domain multiplexer requires the development of an additional cooled amplifier stage to drive the long cables from 4 K to ambient temperature. This development is already underway.

## 5.6 Cooling System

The EPIC cryogenic system is based on the design used successfully in *Spitzer*, a superfluid liquid helium dewar. By passive cooling the shell of the liquid helium dewar to ~40 K, parasitic heat loads onto the liquid helium are minimized. In this scenario, instrument heat loads, summarized in Table 5.6.1 for the NTD Ge and TES focal plane options, become dominant. In the design of the EPIC cryostat, we conservatively assume the larger of the two heat loads in Table 5.6.1, whichever is worse. The EPIC cryostat is designed to provide at least 2 years of observational lifetime at L2 with suitable design margin. The requirement is 1 year.

The cryogenic design is designed for operational simplicity. There is no thermal piping or straps to interface to the spacecraft – the interface is basically the mechanical mount to the bipods. Because the cryostat is enclosed in a vacuum shell, we can operate and test the full instrument in a laboratory environment. This is a major simplification from *Planck*, which requires a large cooled vacuum chamber to in order to operate the instruments on the ground, which makes tests of the integrated system extremely costly and time-consuming. Following the approach used on *Spitzer*, the passive cooling system is verified by a program of materials testing, design margin, and testing on a scale model. We do not plan to carry out an integrated system-level test of passive cooling. The sunshield largely does not participate in passive cooling, contributing only ~1% of the heat load on the 40 K cryostat shell stage, and the passive cooling performance is thus insensitive to the parameters of the sunshield.

**Table 5.6.1 Instrument Heat Loads for Focal Plane Options**

| Heat Load | Temp [K] | NTD Ge Option | TES Option |
|---|---|---|---|
| Focal Plane Dissipation | 0.1 | - | 1 μW |
| ADR Heat Load | 2 | 3 mW | 3 mW |
| JFET Dissipation | 44 | 128 mW | - |
| Focal Plane Wiring | 0.1 - 300 | 830 pairs* | 300 pairs* |

*40 awg manganin in stainless shields

**Table 5.6.2 Temperature Stability Requirements and Goals**

| Stage | T | Requirement | | Design Goal | |
|---|---|---|---|---|---|
| | [K] | Instantaneous[3] | Scan Synch | Instantaneous[3] | Scan Synch |
| Focal Plane[1,4] | 0.1 | 600 nK/√Hz | 1.5 $nK_{rms}$ | 200 nK/√Hz | 0.5 $nK_{rms}$ |
| Optics[2,5] | 2 | 1.5 mK/√Hz | 3 $\mu K_{rms}$ | 0.5 mK/√Hz | 1 $\mu K_{rms}$ |
| Baffle[2,6] | 40 | 15 mK/√Hz | 75 $\mu K_{rms}$ | 5 mK/√Hz | 25 $\mu K_{rms}$ |

[1]Assumes 5% matching to focal plane drifts
[2]Assumes 1% matching to unpolarized optical power
[3]At signal modulation frequencies
[4]Planck achieves < 40 nK/√Hz at 0.1 K regulated on focal plane with open-cycle dilution refrigerator
[5]Planck achieves < 5 μK/√Hz at 1.6 K regulated on open-cycle dilution refrigerator J-T stage
[6]Planck achieves < 30 μK/Hz at 4 K regulated on Sterling-cycle cooler stage

*5.6.1 Superfluid Liquid Helium Cryostat*

EPIC-LC uses a conventional cold-launched superfluid helium-cooled cryogenic design to minimize cost and to allow simplified ground testing. The telescopes, focal planes, and superfluid helium tank are surrounded by vapor-cooled shields inside of a vacuum vessel / shell. The radiation shields and vacuum vessel provide thermal insulation for the superfluid helium before launch. A fixed 3-stage V-groove radiator cools EPIC's IR-black painted vacuum shell to < 44 K. The V-groove radiator couples to a deployed sunshield described in section 5.7, although the sunshield merely reflects the thermal radiation from the V-groove and does not itself provide radiative cooling. The superfluid helium absorbs heat from the ADR, focal planes, and parasitic heat leaks from the wires, struts and radiation.

The EPIC-LC superfluid helium cryostat is based on the proven thermal and mechanical performance of the *Spitzer* cryostat. The EPIC cryostat is based on mature technology from many past missions including *Spitzer* and *IRAS*. Specific design features include wide flow range porous plugs, low torque vents, shielded twisted-pair low thermal conductivity wires, alumina/epoxy bipods, thermal coatings, contamination control methods, and the dust cover and mechanisms. Heritage for the deployed vacuum-tight telescope covers is from *IRAS*. Superfluid helium dewars have been used reliably for several space missions and are considered a mature technology.

EPIC's low-cost mission concept is designed for a 24-month cryogen lifetime with 20% margin using a model which includes preliminary sizing of the support tubes, configuration-dependent multilayer insulation (MLI) surface areas, plumbing, instrumentation, aperture loads, and instrument heat loads. The heat rates used for this preliminary sizing of the EPIC cryostat are conservative.

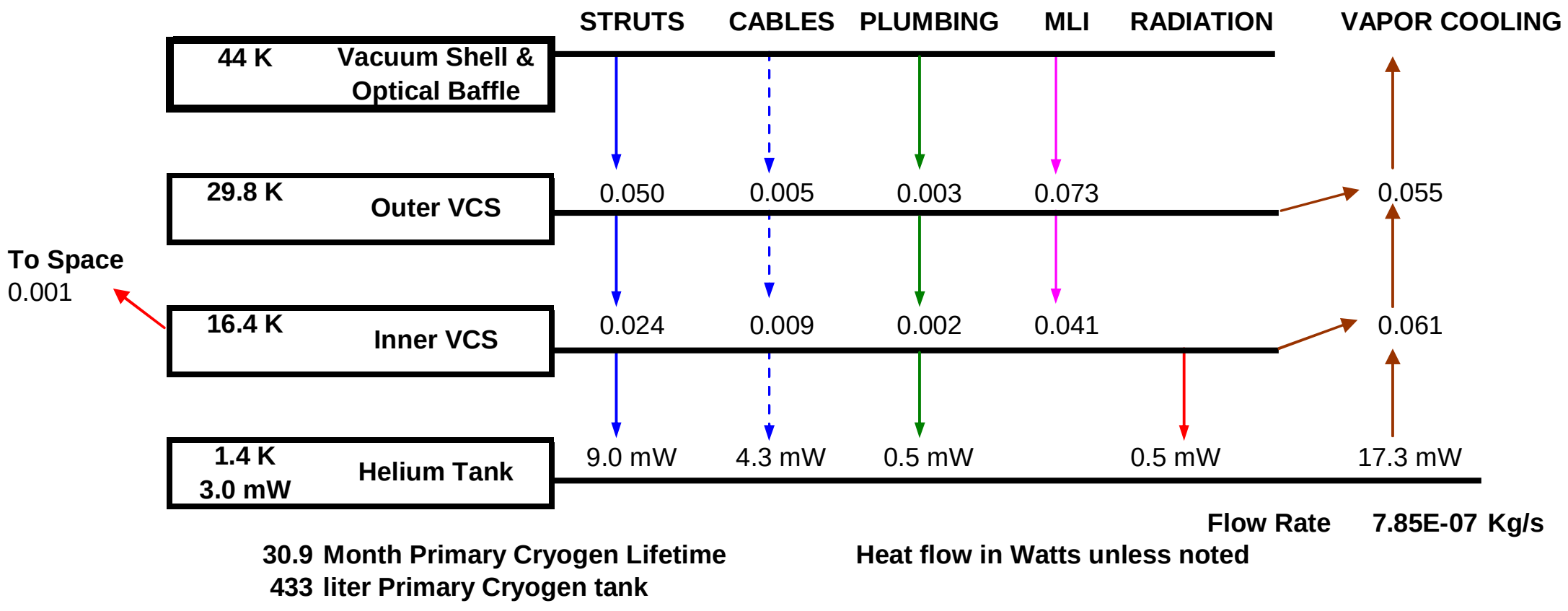

Figure 5.6.1 EPIC Cryostat steady state heat map

Figure 5.6.1 shows the steady state heat flow diagram for the cryostat. The ADR and focal planes are thermally connected to the helium tank through high conductivity flexible thermal straps. The ADR, focal planes and helium tank are supported by a set of six low thermal conductivity gamma-alumina support struts. Two vapor-cooled shields are mounted on the struts and intercept conductive heat loads from the wires and struts and thermal radiation from the 44 K vacuum shell. Electrical signals from the detectors are carried by 830 wire pairs of low thermal conductivity shielded twisted ribbon cables. High temperature superconducting current leads minimize the heat leak from leads to the ADR magnets. A parallel set of normal conducting leads between the vacuum shell and inner VCS facilitate ground testing of the ADR. Separate deployed aperture covers on each of the 6 telescopes close out openings in the outer VCS and vacuum shell. These covers are opened in space once the outer shell cools to about 150 K and contamination around the spacecraft dissipates. Vacuum jacketed bayonet fittings on the vacuum shell will be located where fairing access is provided and where there is no interference from the stowed sunshield. MLI blankets are used on the outside of the outer and inner VCS to improve the ground hold performance.

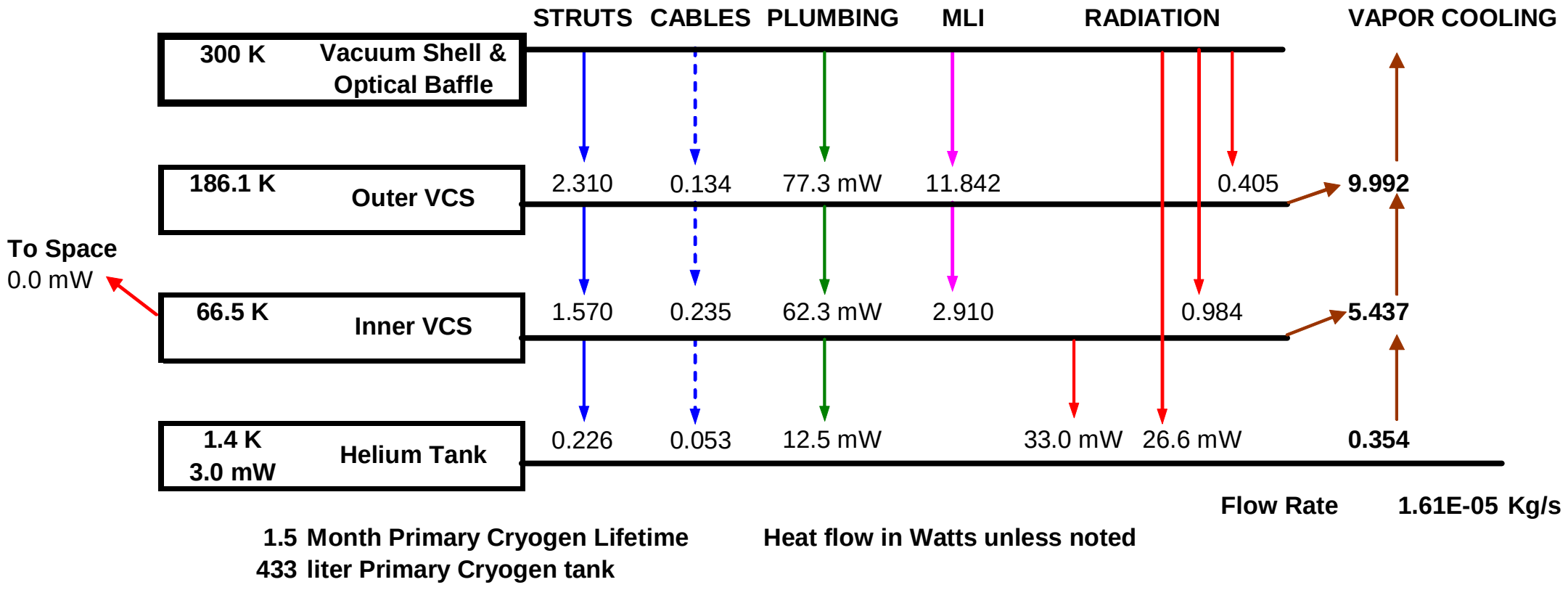

Figure 5.6.2 EPIC Cryostat ground hold heat map

The parameters and lifetime of the EPIC-LC cryostat are summarized in Table 5.6.3. This performance can be compared with the Spitzer system which will achieve a 66 month lifetime at L2 for a 360 liter dewar, an instrument heat load of 3 mW, and an achieved shell temperature of 34 K. Note that the lifetime of the cryogen depends on the heat loads and the overall thermal design of the system, which determines the temperature of the outer shell of the cryostat, much more than on the details of the cryostat design. Our current shell hold time estimate may be conservative because the 3 mW dissipation specification for the ADR is conservative, and the 44 K shell temperature assumes JFET dissipation on the shell, doubles the estimated ADR wiring load, and discounts any cooling contribution from the deployed sunshields.

**Table 5.6.3 Summary of Cryostat Parameters**

| | |
|---|---|
| Design Lifetime at L2 | > 2 years |
| Required Lifetime at L2 | 1 year |
| Liquid Helium Volume | 450 liters |
| Cryostat Mass (CBE Dry) | 301 kg |
| Cryostat Mass (CBE Wet) | 364 kg |
| Instrument load on 2 K | 3.0 mW |
| Cryostat shell temperature | 44 K |

Before launch and during ground testing, the dewar vacuum shell is at room temperature. The EPIC Cryostat ground hold heat map in Figure 5.6.2 shows that there is ample dewar lifetime for ground testing and to minimize boil off during ground hold and while the vacuum shell is cooling at the beginning of the mission. A dynamic model of the sunshield cool down indicates that the outer shell reaches operating temperature about 4 days after launch. The higher boiloff rate during ground hold and the cool down results in only 1 month reduction in mission duration.

Any periodic variation in the stray thermal radiation that is synchronous with the observatory rotation rate and seen by the detectors leave a possible scan-synchronous systematic signature. A first order dynamic thermal model of the sunshield shows that there should be no measurable periodic variation in the outershell temperature caused by the rotation of the sun's shadow on the sunshield. The high thermal mass of the vacuum shell damps all but low frequency temperature variations. However, the orientation of the moon relative to EPIC's apertures may cause a periodic variation in stray radiation scattered into the instrument. Therefore, a moon baffle is included to prevent the moon from shining on each telescope's 40 K black baffle. Although we expect their temperature to be intrinsically stable, the instrument will be highly sensitive to any temperature changes of the black baffles. We plan to install an active PID closed loop temperature control and low noise temperature readouts on each baffle.

The EPIC cryostat integration and test is typically performed with thermal/mass models of the telescope and focal planes in order to allow for parallel development of the cryostat/ADR system. The ADR and telescope/focal plane assemblies would be developed in separate cryogenic test facilities. The test sequence of the cryostat includes the following steps:

1. Integrate cryostat components and thermal/mass models
2. Aperture cover release test
3. Cold vibration test with helium (warm shell)
4. Health check
5. Cryostat thermal performance with cooled outershell
6. De-integrate thermal/mass models
7. Instrument integration
8. Instrument performance tests

*5.6.2 Passive Thermal Cooling System*

We carried out a passive cooling design study to control the thermal background and determine boundary temperatures to aid in defining active refrigeration requirements. The thermal shielding concept for both mimics the 3-stage shielding adopted by ESA and Thales Alenia Space for Planck, which has undergone a complete thermal/mechanical design and analytical process, as well as some thermal testing. TC Technology (which did the work reported here under contract to JPL) was intimately involved in the early thermal design and analysis of the Planck passive cooling system.

Figure 5.6.3 shows the basic elements of the EPIC-LC configuration. The launch shroud limits the rigid portions of the system to 3 m maximum diameter. In the Planck system, all parts of the thermal control system are rigid. For EPIC, a deployable shield is necessary to keep sunshine off of the optics due to the more demanding 45˚ solar angle set by the scan strategy. To defeat thermal radiation from the warm sunshields, we require 3 shields. The shields are arranged such that they are extensions of each of the 3 V-groove radiators. Thermally, the sunshield does no 'work', as the thermal conductance of each shield is assumed to be zero. The sunshield enters the cooling performance mostly in letting radiated thermal power from the V-groove escape to cold space.

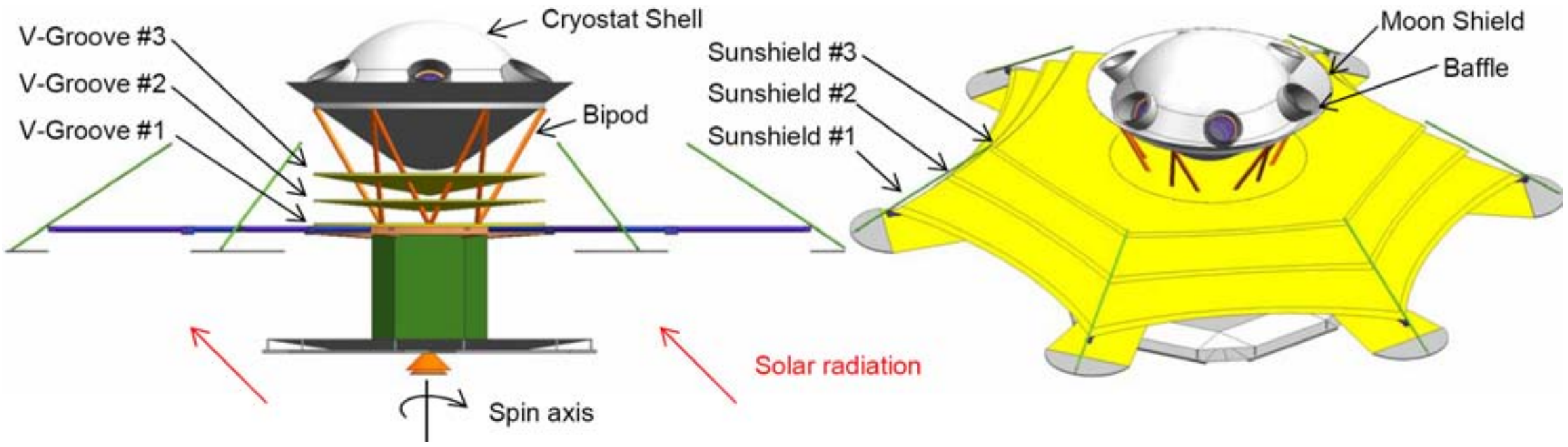

Figure 5.6.3. Basic architecture of the EPIC-LC passive cooling configuration. The left side shows the a sideview with the sunshields removed, to better display the 3-stage V-groove radiators which provide the bulk of the passive cooling, and intercept conducted heat flow from the bipod support of the cryostat. On the right the deployed sunshields are shown.

The first and largest shield is the flat Sun shield. The rigid portion resides on the anti-Sun side of the spacecraft bus, and is assumed to operate at room temperature, 293 K. A flat deployable shield section is attached to this first rigid section. We decided that the deployed section should be flat in order to keep the solar power input constant over the spin/precession motion. In practice, shadowing from the spacecraft causes the solar power to be constant averaged over the shield, but to vary spatially over the shield during a spin. We expect that this variation is highly damped so that spatial/temporal variations on the innermost shield are negligible; however a dynamic analysis of the shields exceeds the scope of this study. The first shield is coated on its Sun-facing side with silver-Teflon second-surface mirror that passively results in the shield operating at approximately 213 K when illuminated at 45 degrees by the Sun. The anti-Sun side of the first shield, as well as all surfaces of successively colder shields and the exterior of the telescope enclosure, is coated with 'specular' aluminum to radiatively decouple the shields.

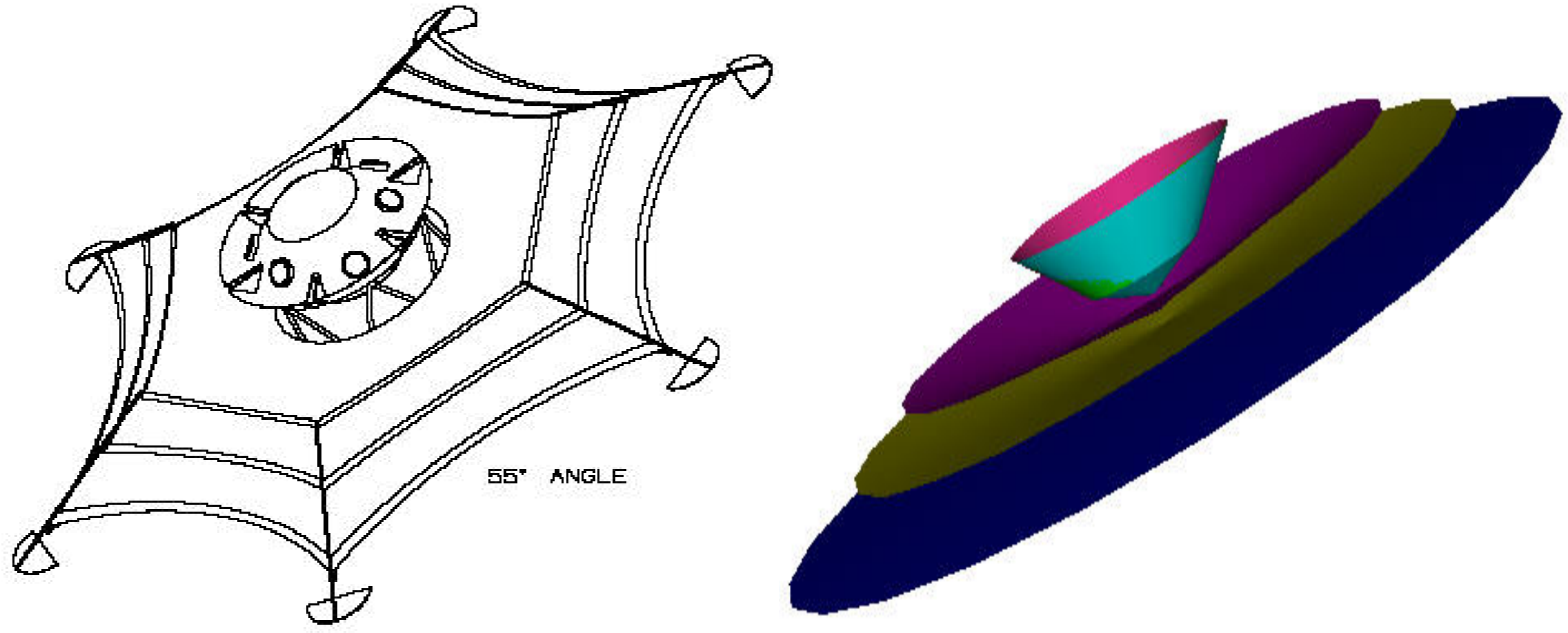

Figure 5.6.4. CAD and NEVADA (thermal radiative) model representations of EPIC-LC. The major dimension of the largest deployed shield is 7.1 m (tip to tip).

The separation of adjacent shields at the cone apexes is 25 cm. The angular separation is 5 degrees. The central, rigid portions of the two conical shields are assumed to be aluminum honeycomb, to support negligible internal thermal gradients. The single-node deployable portions of all 3 shields are also assumed to be isothermal, but are also (for simplification) conductively isolated from their corresponding rigid shield sections. This modeling infidelity does not significantly alter the conclusions of this study.

The aluminized surfaces of the shields are assumed to be 98% specular, conservatively low for the long thermal wavelengths involved. The shields vary widely in temperature, so the broadband IR emissivity of the shields is varied accordingly. The support structure has negligible radiative impact on the overall performance, and is therefore ignored in the radiative models.

A preliminary structural analysis resulted in the selection of gamma alumina struts; corresponding temperature-dependent linear conductances were input to the SINDA/G thermal model. Harness wires are heat sunk along the support struts; they are allocated to be the equivalent of 830 #36 AWG manganin twisted pairs, modeled as 16 Herschel/SPIRE cables, and high current leads for the ADR, all shielded and insulated. The conductive contributions associated with these wires are small. Shield deployment hardware is also modeled, and has insignificant impact on performance.

The adiabatic demagnetization refrigerator (ADR) wires were modeled, and the temperature effects are not severe, a ~2 K increase in the temperature of the cryostat shell. We additionally added the heat load for the JFETs, assumed to dissipate power to the cryostat shell, the result being a 4 K rise in the shell temperature.

The simplified heat flow map shown in Figure 5.6.5 summarizes the thermal performance predictions. To minimize confusion, the heat rejection rates to space from the deployable shield sections are not shown, since they are indicative of radiative parasitics only, and since these shields do no useful thermal 'work'.

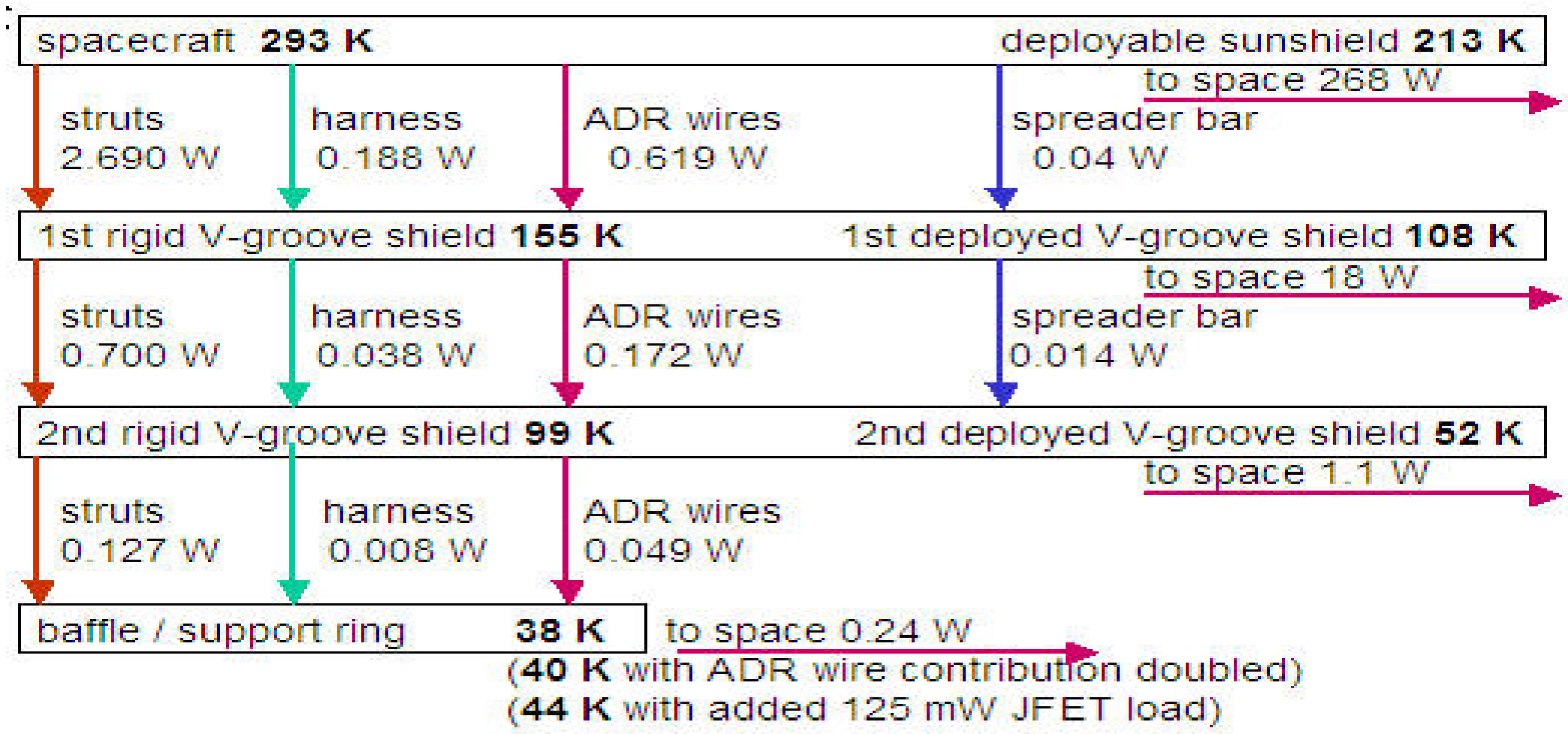

Figure 5.6.5. EPIC-LC simplified heat flow and temperature map.

For EPIC-LC, 3 cases were addressed. The 1st case includes only the passive ADR wire loading due to conduction. The 2nd case doubles the ADR wire passive conduction from case 1, conservatively including thereby the effects of resistive wire heating. The 3rd case adds 125 mW of estimated JFET loading to the coldest passively cooled stage. The heat flow map shown in Fig. 5.6.6 is for the first case only. The temperature changes for the three cases are shown only for the coldest stage for simplification; temperatures at warmer levels are relatively constant. As noted, the cryostat design is predicated on the worst-case shell temperature.

*5.6.3 Adiabatic Demagnetization Refrigerator*

The sub-Kelvin architecture shown Fig. 5.6.6, used to cool the bolometric detectors for EPIC, consists of a detector stage, a thermal intercept stage, and a base attached to a superfluid $^4$He bath. The thermal intercept stage serves to intercept heat which reduces the load on the detector stage cooler and to buffer variations in the thermal environment. We considered 3 high technology readiness level (TRL) cooler types to provide cooling at the detector and intermediate stages, an adiabatic demagnetization refrigerator [1] (ADR), a pumped $^3$He evaporative cooler [2] and an open cycle $^3$He/$^4$He dilution cooler [3]. Single shot designs of the ADR and pumped $^3$He coolers have flown in space. Pumped $^3$He and the open cycle dilution cooler have been built and will fly in space on *Herschel* and *Planck*. The baseline design for EPIC are two detector-intercept stage pairs of single shot ADRs with a hold time of 48 hours. The intercept stage is held at ~500mK to minimize the total mass of the ADR cooler system. Each pair cools half of the detectors to enhance the reliability and to enable an upscope option to provide continuous cooling at 100mK using parallel [4] or serial [5, 6] cycling of the ADRs.

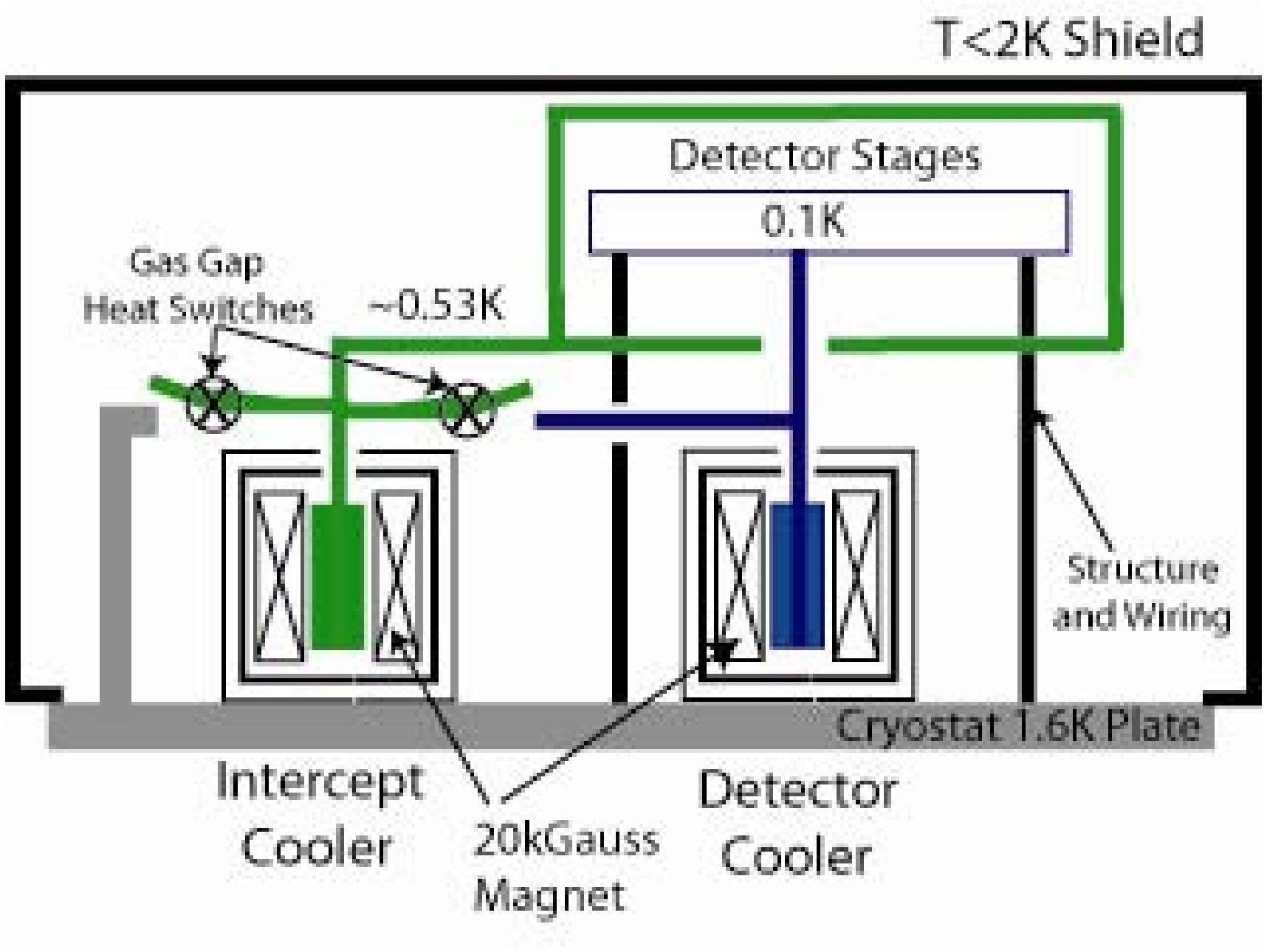

Figure 5.6.6 Graphical layout of the baseline cooler for EPIC.

The architecture for EPIC consists of 6 telescopes each with a single color focal plane of antenna-coupled voltage-biased superconducting Transition Edge Sensors (TES) cooled to 100mK. We base the cooler design on the time-domain SQUID multiplexer, since it places more demanding resource requirements on the cooler than the frequency-domain SQUID multiplexer. For each telescope, the detectors are designed as described in section 5.5 and voltage biased using a load resistance ~ 4% of the bolometer operating resistance mounted on the detector stage. The detectors are readout with an N x M array of multiplexed first stage DC SQUID ammeters mounted on the detector cold stage. Each set of N first stage SQUIDs are readout, via stripline cables, to M second stage high bandwidth series array SQUIDs cooled to 1.6 K by a pumped superfluid $^4$He bath described in section 5.6.1. The detector stage is supported with low thermal conductance supports from an intercept stage. A radiation baffle with infrared blocking filters mounted from intercept stage surrounds the entire detector stage assembly. The intercept and detector stage assembly is supported from the superfluid $^4$He tank. The supports between each stage are sized to support launch load with standard factors of safety to yield (1.2), ultimate (1.4), and buckling (2.0). The baseline design is made from high strength titanium which has very high yield and ultimate strength, 115 and 140 ksi respectively and low thermal conductivity, 150 $T^{2.7}$ μW/cm K. Supports made from tensioned Kevlar would have about a factor ~2 lower heat load on the stages but with additional design elements including the support posts, alignment fixtures and preloading springs to compensate for long term creep. The detector readout cables are striplines similar to those used for SPIRE and heat sunk at the intercept stage and superfluid $^4$He bath.

When the intercept stage is allowed to cool passively to steady state between the detector stage and superfluid $^4$He bath temperatures, the heat load to the detector stage, shown for the system in Table 5.6.4, is dominated by the parasitic heat. The cooling power required to lift this heat at the detector stage is $P_d = \Delta S(T_h, T_d)\, T_d/t_c$. Here $\Delta S(T_d,T_h)$ is the entropy lifted to maintain the detectors at $T_d$ and dissipated at $T_h$ by the liquid helium bath every cycle period $t_c$. For an adiabatic cooler, $\Delta S$ is constant and the heat load (or waste heat) from the cooler on the liquid helium is given by

$$P_h = \Delta S(T_h, T_d)\, T_h/t_c = P_d\, \Delta S(T_h, T_d)\, T_h/T_d.$$

The hold time or cycle period $t_c$, can be increased or the waste heat $P_h$ at $T_h$ can be reduced, both significantly, by using a part $\Delta S_i$ of the available entropy $\Delta S$, to cool the intercept stage temperature $T_i$ below the passive steady state temperature as shown in Fig. 5.6.7. The reduced heat load on the detector stage is lifted by the remaining entropy $\Delta S_d = \Delta S - \Delta S_i$. As shown in Fig. 5.6.7, the *two stage* ADR achieves a useful hold time of 48 hours when operated in single shot fashion. By comparison, the mass of *single stage* ADR, scaled from the Astro-E II ADR, is not feasible for EPIC as shown in Table 5.6.4. For the superfluid cryostat proposed for EPIC, the reduction in $P_h$ for the two stage configuration is favorable, but not critical to the cryostat performance but is significant if a cryocooler were used since flight qualified 2K coolers have a total capacity at 1.6 K of < 5 mW.

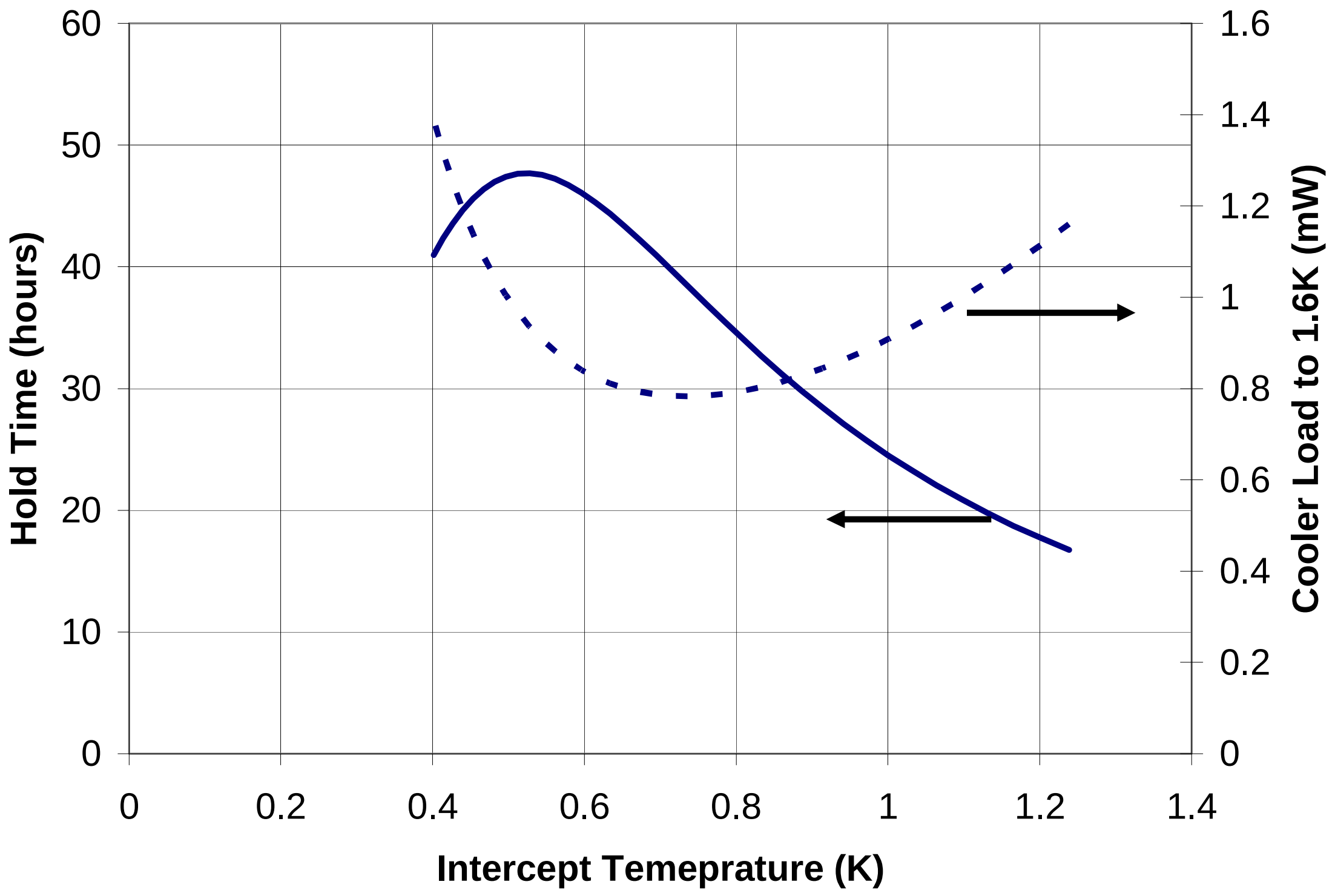

Fig. 5.6.7. Single shot hold time or cycle period and heat load into the 1.6K heat sink as a function of intercept stage temperature for an adiabatic cooler.

**Table 5.6.4 ADR Parameters with and without Intercept Stage Cooling**

| | Units | Intercept Cooling | No Intercept Cooling |
|---|---|---|---|
| Detector System Power | μW | 1 | 1 |
| Parasitic Power | μW | 2.7 | 25 |
| Intercept Temperature | K | 0.53 | 1.3 |
| Heat Strap Mass | kg | 3.5 | 15.3 |
| ADR Mass for 48 hour hold time | kg | 9.7 | 53.4 |
| ADR Cooler Load to $^4$He | mW | 0.9 | 9.4 |
| ADR *System* Load to $^4$He | mW | 2 | 10.5 |
| Total $^4$He Consumption | ℓ/yr | 20 | 111 |

The steady state temperature of the intercept stage with no active cooling is ~1.3K. The mass includes (2) two stage ADRs. The ADR system load includes parasitic heat leak through the high Tc superconducting leads from the inner VCS at 16.4K.

### *5.6.3.1 Heat Straps*

The detector focal planes in EPIC are distributed over nearly ~70 cm circumference around the cryostat at the center. Large thermal straps are necessary to conduct heat during operation and to expedite the initial cool down. The gradient along a metallic strap of width w, thickness t, and conductivity $\kappa_0 T$, where $\kappa_0$ is constant, $dT = P_j l_j / wt$, should be kept small $dT/T \ll 1$, where $P_j$ is the power dissipated by the detector and SQUID amplifier for telescope j located $l_j$ from the cooling source. This criterion fixes the values of w and t. By substituting the $wt = m\rho l_j$, where m is the mass of the strap and ρ is the density of the strap material, into the expression for the gradient and solving for m, to obtain an expression in terms of fixed system parameters, $m = P_j \rho\ l_j^2/\kappa_0 TdT$. We size copper thermal straps with $\kappa_0$ = 1 W/cm K for a 1% gradient to connect the detector stage and intercept stage of each of 6 telescopes to the centrally located cooler. Three telescopes require long (40 cm) straps and the other three require shorter (30 cm) straps. The mass of all 6 detector stage and all 6 intermediate stage straps is 2.5 kg and 1.0 kg respectively. Without the intercept stage, the mass of all 6 detector stage straps is nearly 15 kg.

*5.6.3.2 Paramagnetic Salts*

Adiabatic demagnetization of a paramagnetic salt was the first method used to achieve sub-Kelvin temperatures [7]. Adiabatic demagnetization refrigerators (ADRs) with superconducting solenoid magnets are commercially available for laboratory instruments and have been flown on balloons, rockets [8] and spacecraft [1]. In the cooling cycle shown in Fig. 5.6.8, the paramagnet is magnetized isothermally at $T_{DA}$ through path DA, where the heat of magnetization is conducted to a heat sink using a heat switch. Once at peak field, the heat switch is opened and the paramagnet is demagnetized to temperature $T_{AB}$ through path AB. At temperature $T_{BC}$, the stage absorbs heat isothermally at a much slower demagnetization rate through path BC. The steady state temperature of each stage is chosen by the magnetic field at which isothermal demagnetization begins. The ability to easily choose stage temperature with an ADR suits the two stage design for EPIC.

Paramagnets used for space flight ADRs that cool detectors to 100 mK are hydrated salts containing paramagnetic ions, either Chrome Potassium Alum (CPA) or Chrome Cesium Alum (CCA) grown onto gold or copper skeletons [1,9]. The gold or copper skeletons are connected to a gold plated copper bolt cold finger and the salt is encapsulated in welded stainless steel container to prevent long term dehydration. This assembly is commonly called a salt pill. The salt pill is supported within a 2-4 Tesla superconducting magnet using low thermal conductivity support such as tensioned kevlar. For the intercept stage at ~500 mK, Gadolinium Sulfate Octohydrate [4] (GdS) is attractive since it has a larger ion density and spin quantum number J than the chrome alums and does not order magnetically until ~180 mK.

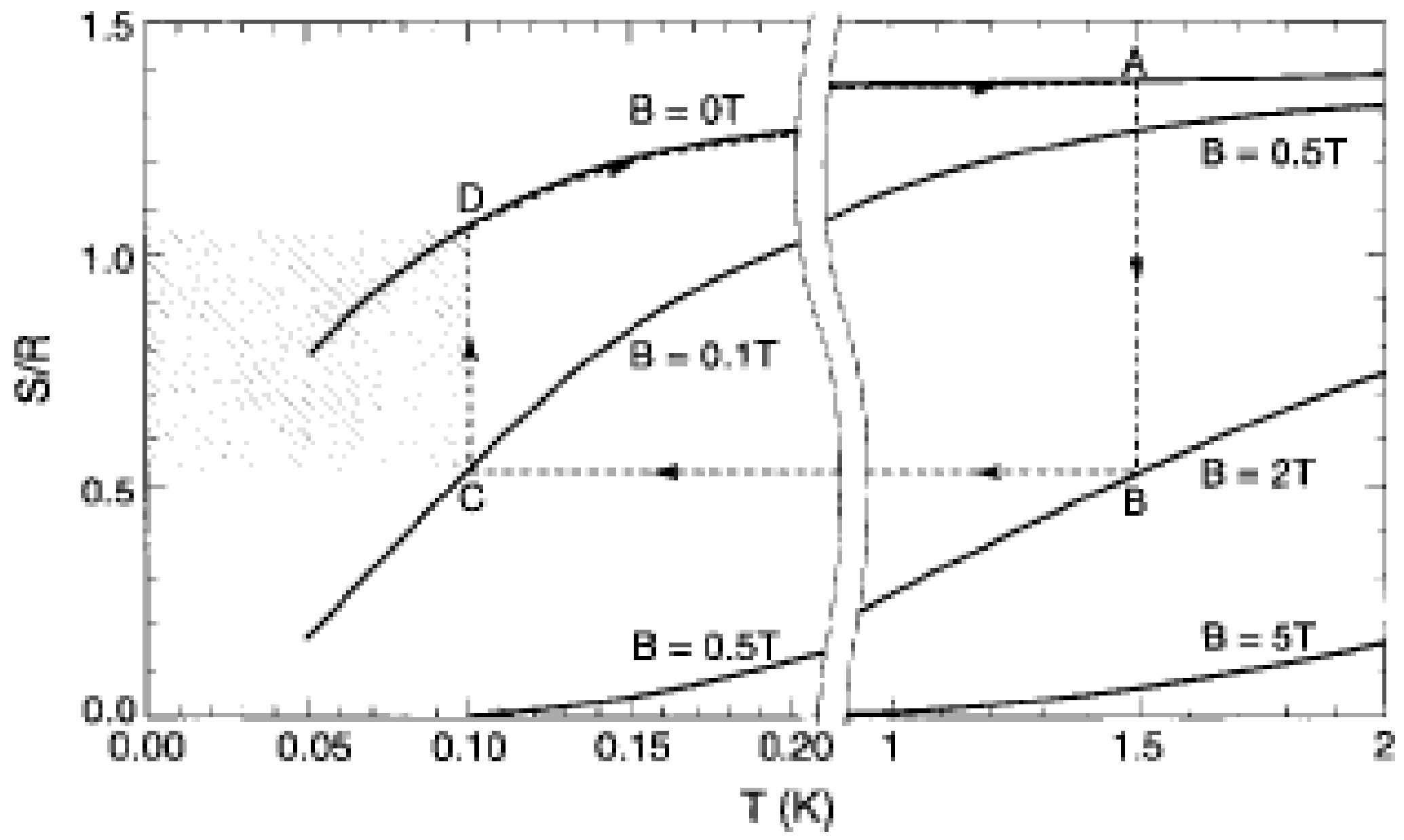

Figure 5.6.8 Generic thermodynamic cycle for an ADR taken from Hagmann [10].

The most common method of operating an ADR is 'single shot' mode. In this mode, the detector stage warms and cools with the paramagnet through the cycle A-B-C-D-A with a quick recycle time at $T_h$ such that the duty cycle at $T_d$ is high > 90%. To date, all commercial systems, such as the Janis ADR, ground based ADRs on telescopes, (PYTHON and ZSPEC), balloon borne ADRs, MAX and MAXIMA, rockets, and space have been single shot. A central single shot ADR would be prohibitively massive for a usable hold time and not feasible for EPIC. The baseline cooler for EPIC is a pair of two stage ADRs. Each ADR cools 3 of the 6 focal planes. The salt pill for each ADR is sized to maintain the focal planes at 100 mK for 48 hours.

*5.6.3.3 Heat Switches*

For the baseline cooler, two gas gap heat switches [2,12] are required for each two stage ADR, for a total of 4 heat switches for the entire cooler system. The detector stage heat switch is thermally anchored to the intercept stage to reduce the heat load to the detector stage when the heat switch is in the off state. The intercept stage heat switch is anchored to the liquid helium bath. The heat switches are nominally identical. The thermal design is driven such that the on state conductance of the intercept heat switch is high enough to conduct the heat of magnetization of both the detector and the intercept stage salt pills to the liquid helium bath at 1.6 K in <2.5 hours so that the duty cycle is > 95%. This sets a typical on state conductance at 1.6 K of 5 mW/K. The main body of the heat switch is a pair of gold plated copper plugs each with a set of copper fins that, when assembled, are interleaved with a fin separation of 100 – 150 μm. The interleaved section of copper fins is encased in low thermal conductance thin walled stainless steel or titanium alloy tubing to form an assembly about the size of a roll of coins. At operating temperature the switch is normally off. It is activated by heating a small charcoal capsule connected to the main body of the switch with a thin walled capillary. When heated, the charcoal desorbs $^3$He gas which thermally shorts the copper fins.

*5.6.3.4 Magnets and Magnet Leads*

For the design study we considered a solenoid magnet wound on an aluminum spool with ferromagnetic shield to complete the magnetic flux circuit. The mass of each magnet and shield assembly depends primarily on the spacing of the wires and the desired field to current ratio. Stock commercial superconducting magnets are made with Nb-Ti wire spaced ~250 μm diameter Nb-Ti wire. Reducing the wire spacing increases the field to current ratio but increases the cost and risk of magnet failure to thermal and field cycling. With this wire winding, the magnet and shield assembly mass increases dramatically for field to current ratios in the range 0.3 - 0.5 T/A. For our mass estimate we use a current to field ratio of 0.4. The magnets are all mounted to the cryostat cold stage at < 2 K and shielded with ferromagnetic material sized to return all magnetic flux at peak field [10].

The magnet leads on the < 2 K cold stage are all superconducting Nb-Ti wired in parallel with high purity copper wire. AC losses in the magnet system are expected to be negligibly small [10]. High temperature superconductor (hiTc) wire such as the HTS Cryoblock wire available from American Superconductor [11] are used between the cold stage, vapor cooled shields, cryostat shell and to the second (coldest) rigid V-groove. The second rigid V-groove is always below 100 K where the HiTc wire is well within the current rating rated at 10A with no resistive losses so there is little risk of the wire becoming resistively unstable. We use a power law fit to the thermal conductivity data of the HiTc wire, $\kappa = \kappa_0 T^\beta$ where $\kappa_0$ = 34 mW/cm $K^\beta$ and $\beta$ = 0.62, a wire cross section of $A = 0.0044\ cm^2$, and wire length $\ell \sim 100$ cm along support struts in the thermal models of the cryostat and radiators. For ground test, the maximum temperature of the inner vapor cooled shield (VCS), shown in Figure 5.6.2, is expected to cool to 66 K which is just safely below the resistive transition temperature ~116 K of the hiTc wire. To facilitate ground testing with a 300 K vacuum shell and warm V-groove radiators, each HiTc wire is paired with a normal resistance wire, such as brass, that is sized to have the same conductive heat leak. The contribution of the ADR wires to the parasitic heat leak in the cryostat is shown in Table 5.6.5.

**Table 5.6.5 ADR Wiring Parasitic Heat Leak.**

| From | To | Steady State (mW) | Ground Hold (mW) |
|---|---|---|---|
| Vacuum Shell | Outer VCS | 5.1 | 133 |
| Outer VCS | Inner VCS | 3.6 | 92 |
| Inner VCS | Liquid Helium (1.4K) | 1.1 | 10 |

The contribution of the ADR wires to the total parasitic heat leak in the cryostat in the steady state, Figures 5.6.1, in space and on the ground with a warm cryostat shell Figure 5.6.2.

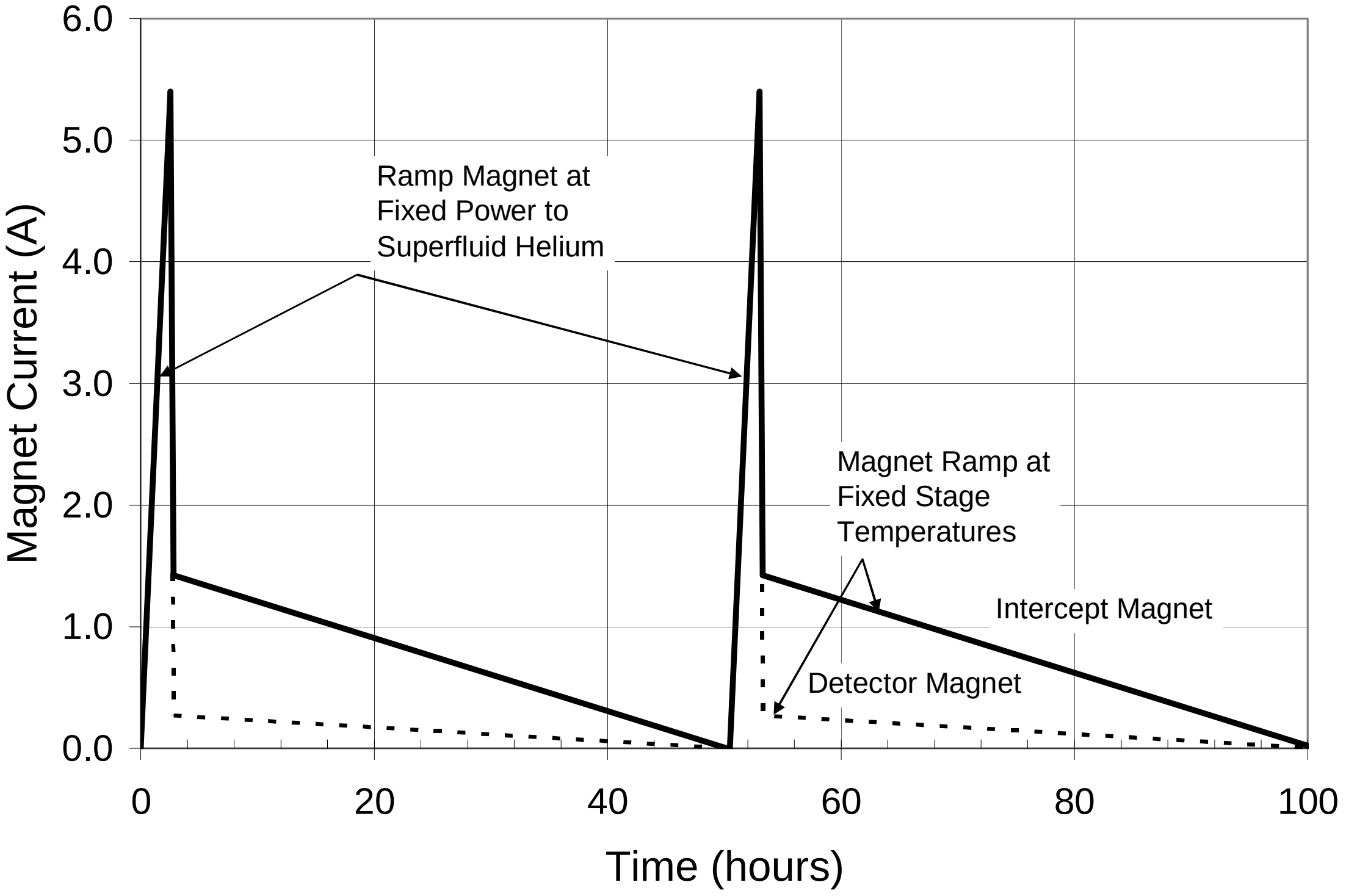

Fig. 5.6.9 Magnet current as a function of time for the detector and intercept stage magnets.

Normal resistive wiring for the magnet current leads is required from second V-groove up to the spacecraft bus at ~300 K. The normal resistance wires are heat sunk at each thermal shield. The length of the wires is fixed by the spacecraft layout. The diameter of these wires is computed so that the average Joule heating per magnet cycle is equal to the parasitic heat leak. The Joule heating per cycle in each wire is, $P = \langle I^2 \rangle_c R$, where $\langle I^2 \rangle_c$ is the square of the magnet current averaged over one cycle, and R is the resistance of one of the two current leads driving a magnet. For typical low thermal conductance wires such as constantan or brass which have a resistivity nearly constant with temperature, the thermal conductance, $G = \kappa A/\ell$ can be approximated using the Wiedemann-Franz law $G = LT/R$, where $L = 24.5$ nW Ohm/$K^2$ is the Lorentz number. The total power into thermal stage j for N wires from thermal stage j+1 is $P_j = (NL/R)(T_{j+1}^2 - T_j^2)/2 + 1/2\, N\langle I^2\rangle_c R$, where 1/2 of the Joule heating power is conducted to each stage. The minimum $P_{ij} = 2\,(L\langle I^2\rangle_c(T_{j+1}^2 - T_j^2))^{1/2}$ for a pair of wires N = 2 to a single magnet is achieved for the optimum resistance $R = (L(T_{j+1}^2 - T_j^2)/\langle I^2\rangle_c)^{1/2}$ for each wire.

*5.6.3.5 Continuous Cooling*

Many systematic effects from the cooling system on EPIC can be eliminated by continuous cooling at 100 mK over the whole mission. Continuous cycle ADRs have been proposed [4] and demonstrated in the laboratory [5,6]. The baseline design can be operated to provide continuous cooling be modifying the heat strapping and cycling pattern of the 4 salt pills by adding heat switches. First, all 6 intercept stages would be connected with additional heat straps. Second, two gas gap heat switches would be added to the intercept stages so that the two intercept salt pills could be alternately cycled to maintain the intercept stage at fixed temperature. The detector stages would be modified and cycled similarly. This design is a parallel mode continuous ADR. However, a superconducting heat switch would be used to alternatively link the 100 mK salt pills to the detector stage [13,14]. This type of switch consists of foils of modest quality 1100 series aluminum, with gold plated contacts on each end clamped to copper rings. At 100 mK, the aluminum is in the superconducting state so the heat switch is normally off. The heat switch is turned on by driving the aluminum into its normal state with a magnetic field perpendicular to the conducting axis. For the on-state conductance, we assume a with a conductance of 50 μW/mK cm at 100 mK to give an on state conductance of ~4 μW/mK for 1 $cm^2$ square 100 μm thick foils. The off state heat leak to the intercept stage is negligible compared to other parasitic heat sources.

For this study we consider a fixed cooler mass. The added mass of the 4 heat switches and additional heat straps is compensated for by reducing the mass of the two stage ADRs. This yields a hold time (or cycle period) for each two stage ADR of ~20 hours. This design is robust to the failure of any one heat switch at the cost of the reduced single shot hold time.

A different type of continuous ADR operated in serial mode [5] has been demonstrated. The serial mode continuous ADR requires only 4 heat switches total, not 8. In this design, the detector stage temperature is controlled by isothermally magnetizing and demagnetizing one of the detector stage salt pills. This design is elegant and has shown cold point stability of a few microK. However, there are several disadvantages compared to the parallel operation proposed here. First, unlike single shot ADRs, the current in the magnet leads is flowing most of the time for a cycling ADR which makes the hiTc leads critical. Second, the ADR fails to cool if one heat switch fails. This single fault can mitigated by adding an additional set of 4 ADR units for a total of 8 with a modest increase in mass. However, this doubles the required number of hiTc leads and hence doubles the heat load into the liquid helium and requires additional current drive electronics. A parallel mode continuous ADR with only 6 ADR units (not 8) would be at least single fault tolerant for continuous operation and at least double fault tolerant with a usable single shot hold time. A summary of the continuous ADR design study is available in Holmes *et al*. 2007 [15].

*5.6.3.6 Other Coolers*

**Pumped $^3$He:** Pumping pure liquid $^3$He is a very common method to cool detectors to < 300 mK and is used in many ground based experiments (ACBAR, ZSPEC, PYTHON, Bolocam, QUaD, BICEP) balloon borne (Boomerang, MAXIMA, TopHat) and space borne instruments (IRTS [2] and Herschel). These self contained single shot instrument coolers consist of a still or evaporator containing the liquid $^3$He, a condensation point heat sunk to a < 2 K cold plate, and a charcoal sorption pump all connected with thin walled stainless steel or titanium alloy tubing. The cooler is cycled by heating the pump to ~40 K to desorb the $^3$He which condenses and collects in the still. Once all of the $^3$He is condensed, the pump cools to ~2 K and begins re-adsorbing the $^3$He which cools the liquid to < 300 mK. For a typical single shot $^3$He sorption cooler [2], 99 J is dissipated into the 2 K heat sink for each 1 J of cooling at 300 mK. The Carnot efficiency of the single shot cooler referred to a 2 K heat sink is < 7%. Of the 99 J, 50 J is the heat of adsorption to pump the $^3$He in the charcoal. During the recycling phase, the remaining 49 J of the 99 J consists of ~25 J to cool the $^3$He gas from the desorption temperature ~40 K and condense it at 2 K and 24 J "ballast heat" to cool the charcoal pump back to 2 K. A system of 2 or more sorption units phased to provide continuous cooling at 300 mK and cycled slowly at near 50% duty cycle enables a lower desorption temperature. This reduces the peak power into the cryogen and the integrated heat required to cool the $^3$He gas. Also, cross linking the pumps with heat switches to use the ballast heat to raise the temperature of one pump as another cools improves the power balance and reduces further the heat load during recycle. A reduction of the recycling power by nearly 50% is possible. The simplicity and flight heritage of the $^3$He system is attractive. Even with the proposed improvements, the heat load into the cryogenics, >10mW, has a significant impact on the cryogen hold time for the cryostat or requires ~factor 2 improvement in the heat lift capability of existing <2K cryocoolers.

**Open Cycle Dilution:** At temperatures below the tri-critical mixing temperature < 0.86 K, mixtures of $^3$He and $^4$He phase separate into $^3$He rich and $^4$He rich ($^3$He dilute) phases. At T = 0 K, the $^4$He rich phase has a $^3$He content of ~6.5%. Driving flow of $^3$He, $dn_3/dt$, from the $^3$He rich phase to the dilute phase gives $dQ/dt = 84\ T^2\ dn_3/dt$ of cooling power dQ/dt at temperature T. The open cycle dilution refrigerator cools by pumping precooled streams of pure $^3$He and pure $^4$He into a mixing chamber. On combination, the $^4$He fraction forms bubbles saturated with $^3$He at the combination temperature. The $dn_3/dt$ flow across the interface between $^3$He and $^4$He bubbles works without gravity [3] and therefore is suitable for space borne cooling. Above ~350 mK, cooling by evaporation of $^3$He dominates [16]. Below ~350 mK, flow of $^3$He into the pure $^4$He the cooling power given is larger than evaporative cooling of pure $^3$He. The primary dilution cooling lifts $Q_d$ at the temperature $T_d$ of the detector stage. Additional cooling $dQ_i$ at the intermediate stage at temperature $T_i$ is obtained by flowing more $^3$He than required to provide $Q_d$ at $T_d$.

We compute the helium flow rates, $dn_3/dt$ and $dn_4/dt$, to cool the EPIC detector arrays to $T_d$ = 100 mK. The heat lift required at the detector stage $Q_d$ uniquely determines the flow rate $dn_4/dt$ since excess $^3$He will form phase separated pure $^3$He bubbles. The temperature of the intermediate stage $T_i$ is uniquely determined by $dQ_i(T_i)$ at the intercept stage and additional $^3$He which can be diluted, $x(T_i) - x(T_d)$, in the fixed flow of $^4$He $dn_4/dt$. Operational parameters for a single system that cools all detector are given in Table 5.6.6. Cooling power is provided by a Joule Thompson expansion valve (JT valve) on the input stream of the Planck 100 mK cooler. Estimated additional cooling at the 1.6 K is scaled from the results for the Planck cooler and assumes a 4.7 K precooling stage on the input helium flow at the same input pressure (295 bar) and flow impedance of the JT valve in the Planck 100 mK cooler. We note that the open-cycle dilution cooler option becomes much more attractive if the focal plane detectors can operate at somewhat higher temperature (see section 5.5.2). At the higher detector temperature ~0.25 K, a volume of $^3$He comparable to that used for Planck would allow a 4 year EPIC mission. The dilution cooler operating at 100 mK may also be attractive if the SQUID readout dissipation can be reduced, for example going to the frequency-domain multiplexer.

**Table 5.6.6 Continuous Flow Dilution Cooler Parameters**

| | Units | EPIC | Planck |
|---|---|---|---|
| Intercept Temperature | K | 0.387 | ? |
| Detector Stage Dissipation | nW | 1000 | < 100 |
| $dn_3/dt$ | μmole/s | 13.6 | 6.7 |
| $dn_4/dt$ | μmole/s | 51 | 20 |
| Cooling at 1.6 K | μW | 275 | 100 |
| $^3$He per year | ℓ(STP) | 9600 | 4730 |

Comparison of a continuous flow dilution cooler estimated for EPIC at 100 mK to the baseline operation of the flight cooler installed in Planck.

## 5.7 Sunshade

We have developed a deployable multi-layer sunshield to meet the thermal requirements of the EPIC-LC mission option. The sunshade fits stowed for launch in a 3-m fairing and deploys once in space. The sunshield's primary function, keeping sunlight off the progressively colder parts of the instrument, is accomplished with 3 shades, each an extension of the 3 stages of the rigid V-groove cooler. The layers of the sunshade are arranged so the V-grooves have a large view angle out to space for maximum radiative cooling. The thin aluminized kapton layers of the sunshade have low thermal conductivity and thus do not contribute to the effective area for passive cooling. The deployment design is based on a simple hinged scheme using high-TRL components.

### *5.7.1 Design Requirements*

The refractor telescope is sized for launch in a three stage Delta II 2925 rocket with a 3-m stretched composite fairing. While in a halo orbit at L2, the spacecraft will rotate at 1 RPM about its central axis for the telescope. To prevent mechanical disturbances resulting from this rotation, the lowest natural frequency of the sunshade in the plane of rotation should be at least 10 times the rotation rate. Thus, the first mode should occur above 0.167 Hz. In addition to this frequency requirement, the load carrying members of the sunshade must have an acceptable factor of safety against buckling and material strength failure. Because the three layers of the sunshade membranes are tensioned to minimize film wrinkling, the support struts are loaded in compression. Furthermore, because the top two sunshield layers are additionally tensioned by the spreader bars to maintain their separation and support, the spreader bars themselves are subjected to bending stresses. Based on this mission description, a summary of the requirements for the deployable sunshade is given in Table 5.7.1.

**Table 5.7.1 Sunshade Parameters**

| Requirement | Low-cost Telescope |
|---|---|
| Maximum shade diameter (tip to tip) | 7.1 m |
| Minimum shade diameter (scallop to scallop) | 5.5 m |
| Stowabe inside rocket fairing | Delta II 2925 |
| Fairing diameter | 3-m Stretched Composite |
| Fundamental frequency (in the plane of sunshade) | >0.167 Hz |
| Factor of safety on buckling of struts | >6.0 |
| Factor of safety on bending strength of spreader bars | >3.0 |
| Viewable area of spreader bar cross section | 1 $cm^2$ |
| Number of reinforced aluminized shade layers | 6 |
| Maximum mass for sunshade and deployment hardware | 100 kg |

### *5.7.2 Technical Approach*

The EPIC-LC refractor design uses a relatively small diameter sunshade that can be folded and stowed vertically, as shown in Fig. 5.7.1. The design of the sunshield is summarized in Williams *et al*. 2007 [1], and references found therein and in section 6.7 of this report. Thin-walled round struts are used, since they carry only compressive loads, which are hinged in the middle, folded in half, and stowed in a vertical position in order to properly fit inside the launch shroud. Spreader bars are latched along the outboard portion of the struts during launch. The tube struts are also hinged at their base to a ring that supports the sunshield and is attached to the spacecraft. To establish stability during launch, the struts and layers of membrane films are secured in place during with cords that are cut prior to deployment in space.

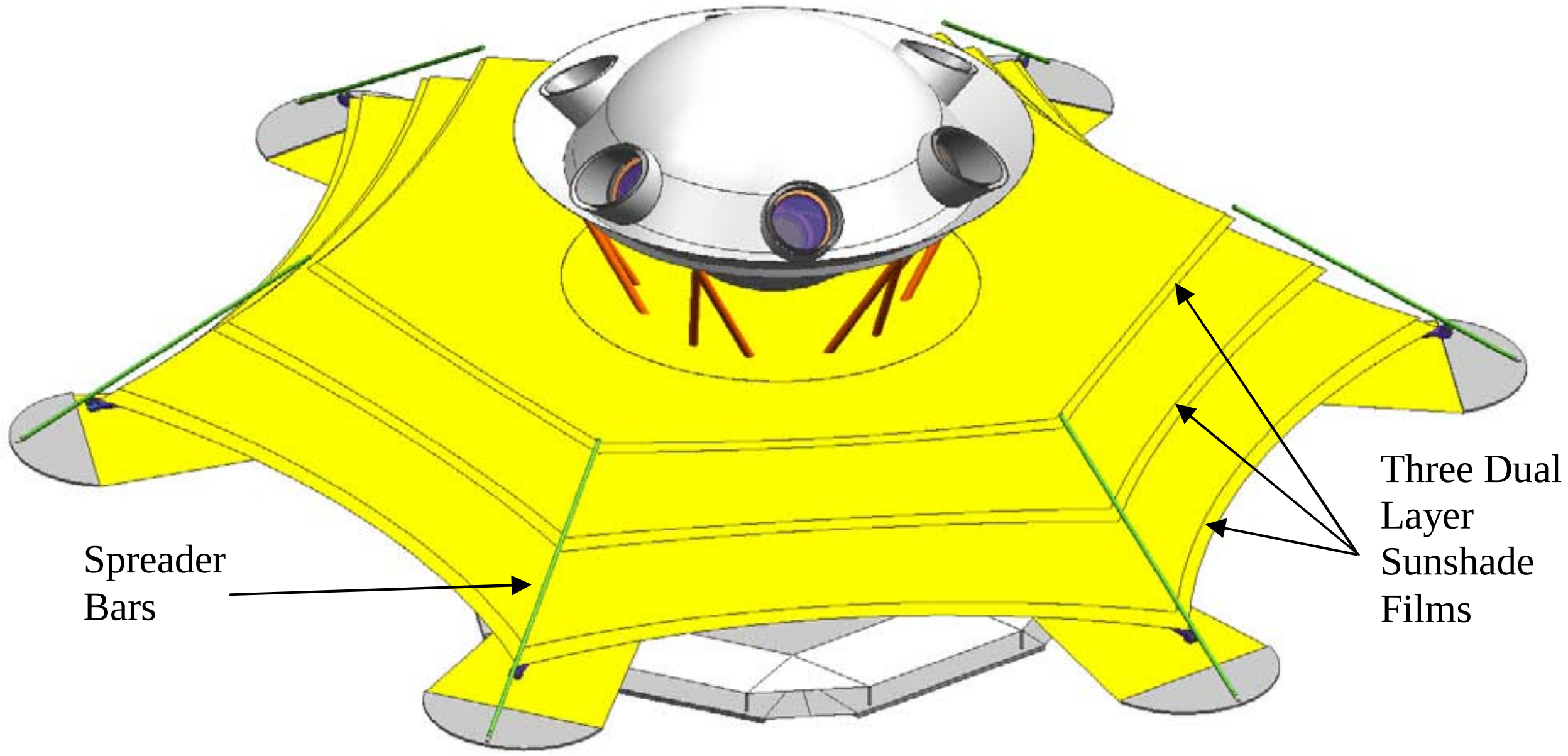

Figure 5.7.1: The Deployed EPIC Sunshade

We have carried out a mechanical analysis and have determined the design meets the dynamic and strength requirements in the deployed configuration, and provides the thermal performance necessary for the telescope cooling system to maintain the appropriate temperatures for the various optical components. The estimated mass of the sunshield and deployment hardware is 74 kg, plus 51 kg for the V-groove radiators.

*5.7.3 Stowage and Deployment*

The refractor design requires a sunshade whose diameter is small enough to be folded once and then stowed vertically inside the launch vehicle fairing. This section describes the on-orbit, petal-like deployment of the EPIC sunshade shown in Figure 5.7.1. The stowed configuration of the round, thin-walled struts is presented in Figure 5.7.2, where the folded membranes are omitted for clarity.

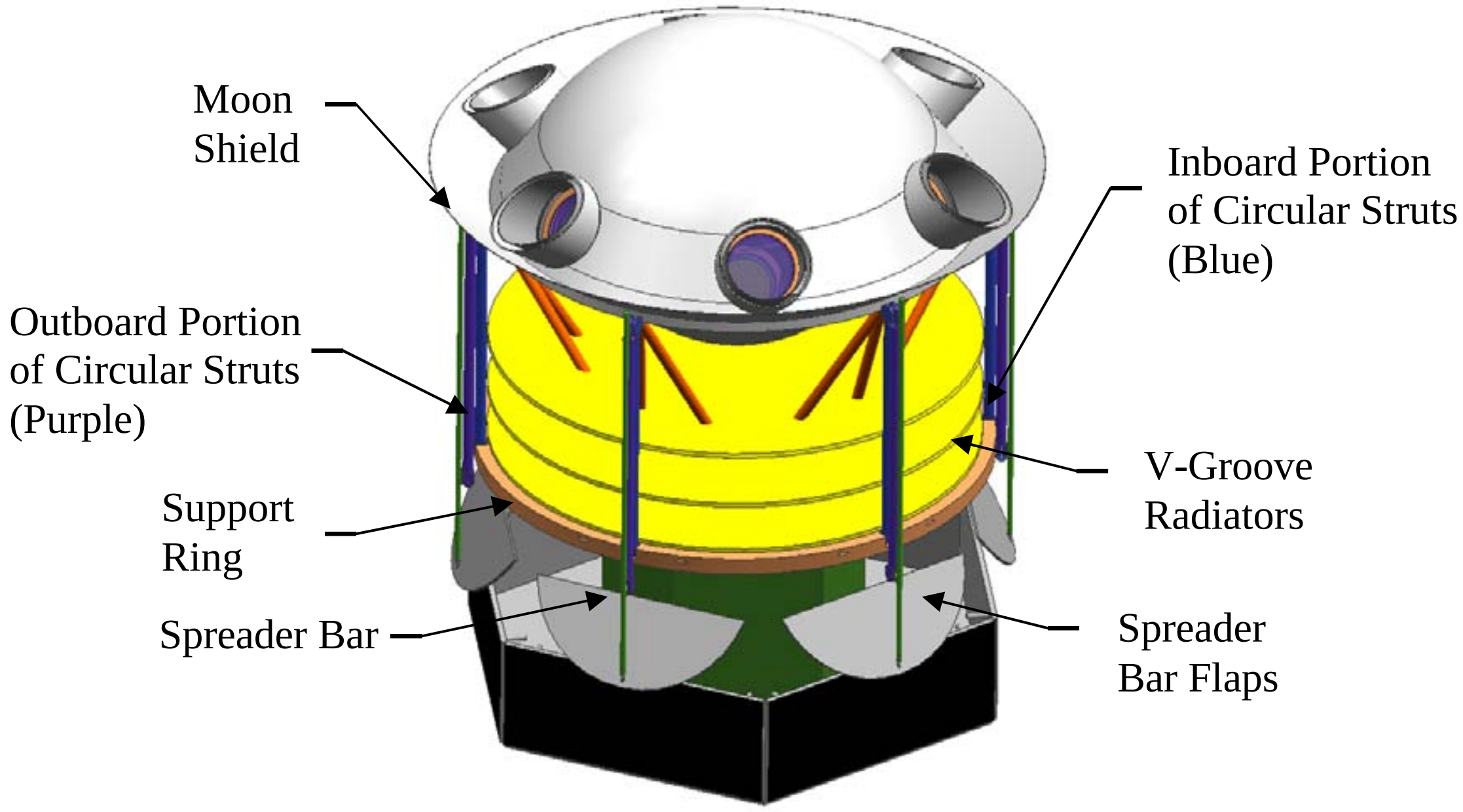

Figure 5.7.2: Stowed Configuration of Sunshade Struts (Membranes Omitted for Clarity).

For this refractor design, round, thin-walled struts are used to carry only compressive loads (a different approach for the larger EPIC-CS sunshield using wrapped ribs is described in section 6.7). The struts are hinged in the middle, folded in half, and stowed in a vertical position in order to properly fit inside the launch shroud. The spreader bars are latched along side the outboard portion of the struts during launch. The tube struts are also hinged at their base to a ring that supports the sunshield and is attached to the spacecraft, as shown in Figure 5.7.3.

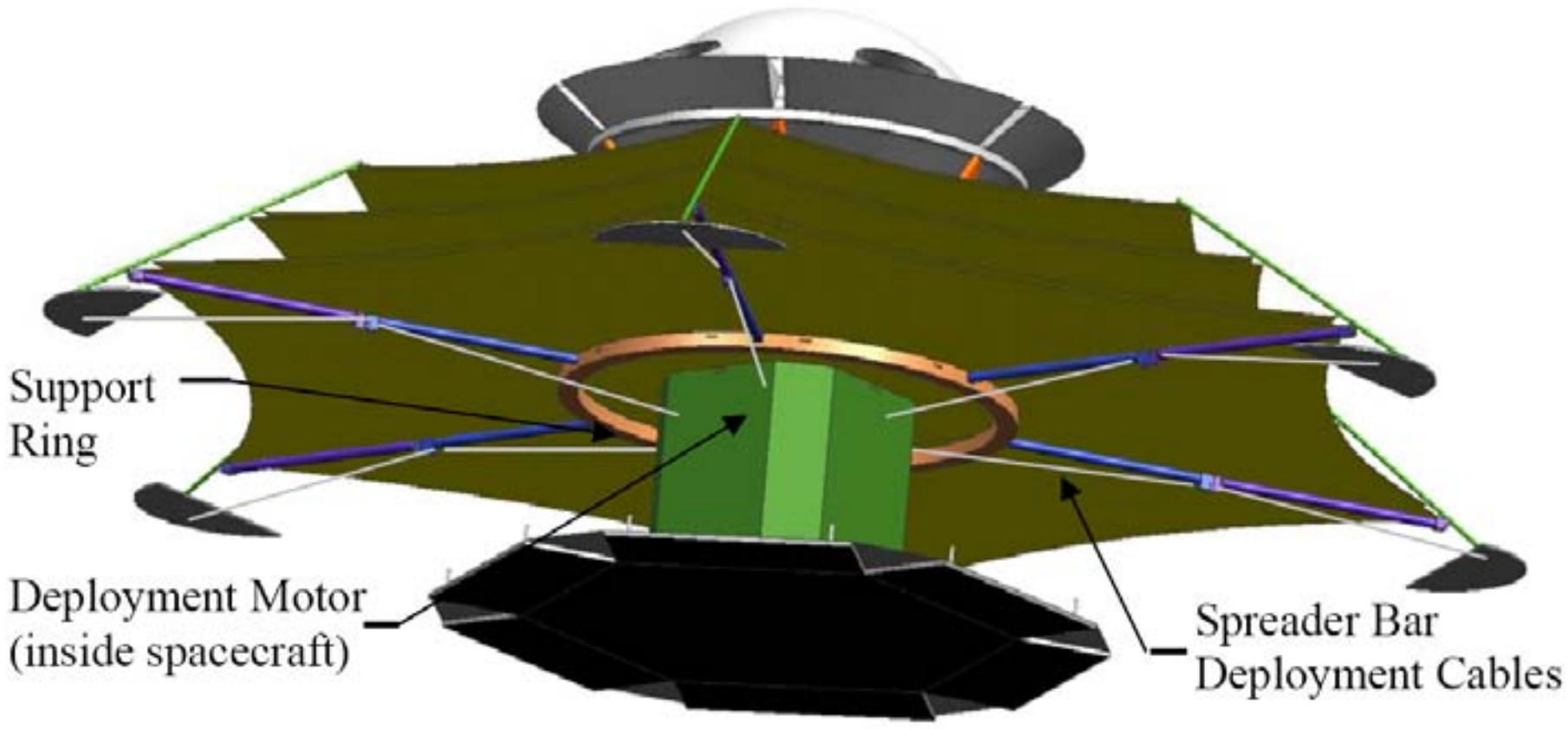

Figure 5.7.3: Bottom View of Deployed Sunshade

During assembly of the telescope, the ring will be attached to the spacecraft, and then the layers of the sunshade attached to their respective V-groove. Next, the reflective film is loosely folded between the stowed struts as depicted in Figure 5.7.4.

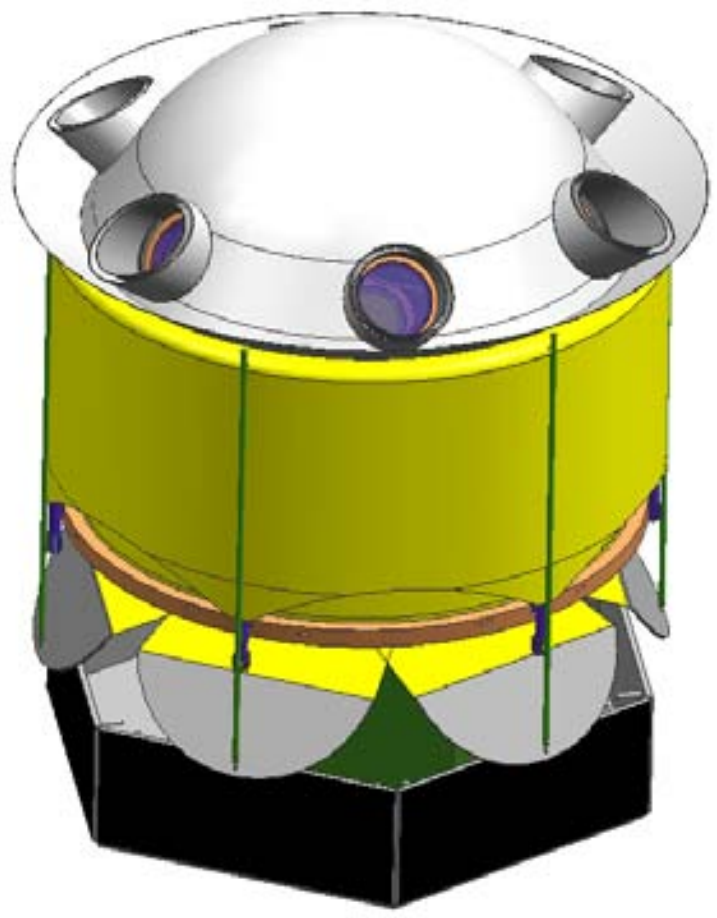

Figure 5.7.4: Stowed EPIC Sunshade Film.

To establish stability during launch, the struts and layers of membrane films are secured in place during launch with a series of cords that can be automatically cut after launch when the sunshade is ready to deploy. In order to not damage the membrane film, the struts must deploy in a slow and controlled fashion. Such a deployment is actuated by the spreader bar deployment cables shown in Figure 5.7.3. Each of the six struts is controlled by its own cable, and all 6 cables are attached to a single take-up mandrel that is controlled by a single deployment motor. This motor and mandrel are located near the center of the bottom sunshade for simplicity and to prevent unwanted unbalance vibration disturbances during the spinning operational phase of the telescope. The spreader bar cables are attached near the end of the spreader bars, which remain locked down to the outboard portion of the struts during launch and are not released until the very end of the sunshade deployment. Restraining the spreader bars in this manner allows one cable (per strut) to pull at the bottom of the spreader bar to deploy each strut. To best illustrate the deployment process of the sunshade, first consider the behavior of the hinged struts (blue and purple) shown in Figure 5.7.5.

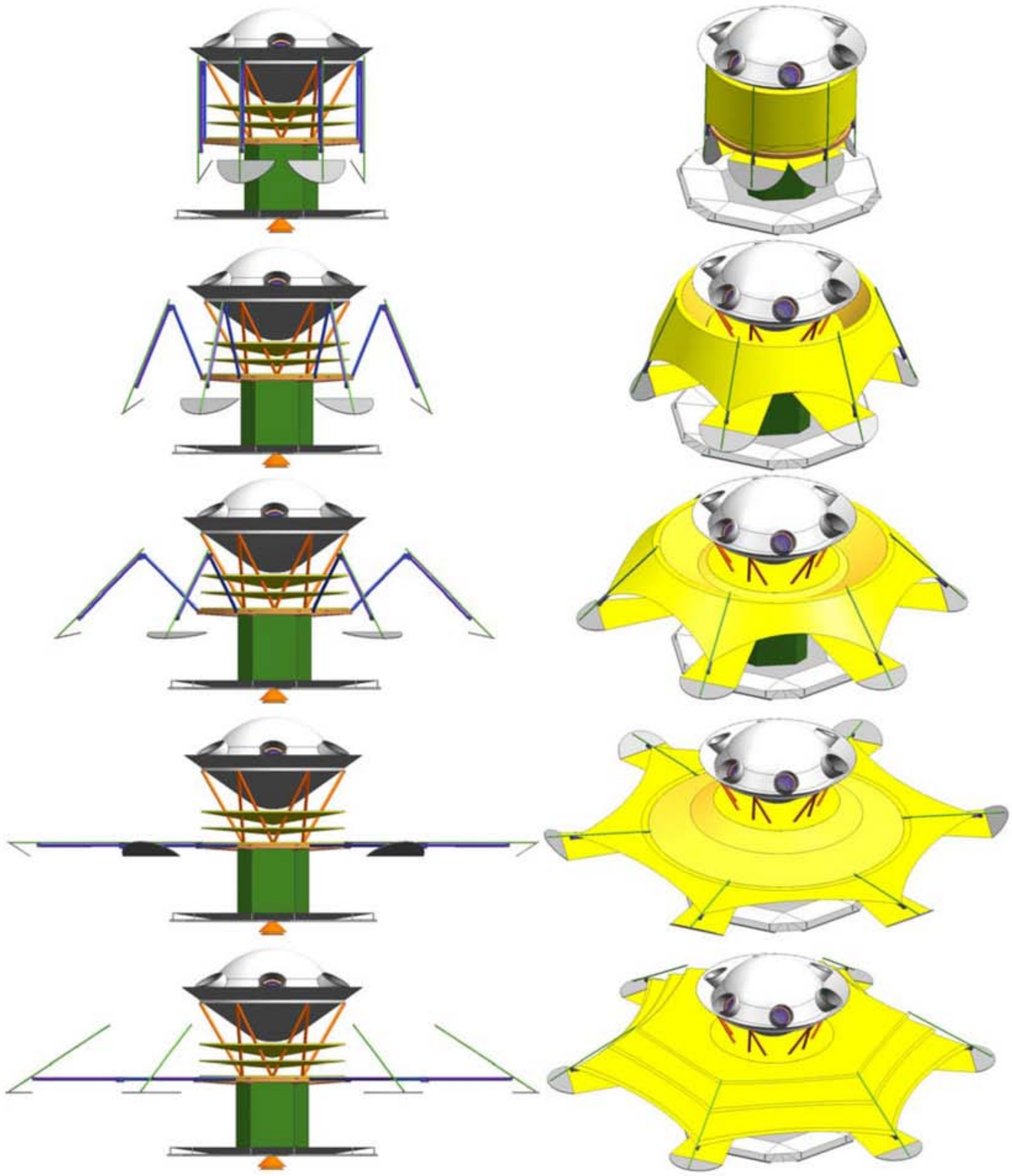

Figure 5.7.5. Sequence of sunshield deployment. The reinforced aluminized kapton sunshield material is omitted on the left side to show the struts. After the restraining cords are jettisoned from around the sunshade in (a), its deployment begins in (b). As the cables, which run from the deployment motor over a pulley system at the mid-span hinge of the struts, are taken-up, they pull on the bottom of the restrained spreader bars creating a moment about the hinge. This moment rotates the outboard section of the struts from its vertical, inverted position outward as shown in (b). A small rod-linkage parallel to the inboard half of the struts synchronizes the two halves of the strut, causing the inboard sections to rotate from the vertical position downwards at the same angular rate. In (c) the struts continue to unfold, with both hinge joints opening at the same angular rate. Such a motion ensures that the ends of the struts do not impact the deployed solar panels and the tops of the spreader bars do not impact the moonshield. Furthermore, deployment in this manner makes the membrane and stowage less complicated. In (d), the struts are fully deployed in their straight configuration, and the mid-span hinges locks into place. Finally, in (e), the spreader bars are released and erected by the spreader bar cables, which are synchronized by being wrapped around the same spool attached to a single spreader bar deployment motor. Deployment of the spreader bars separates the three layers of the sunshade and adjusts them to the proper tension.

*5.7.3 Deployed Configuration*

Given the architecture for stowing and deploying the sunshade, the details of the deployed configuration are now discussed. The fully deployed sunshade is shown in Figure 5.7.2, and Figure 5.7.6 shows a detailed view of a spreader bar region.

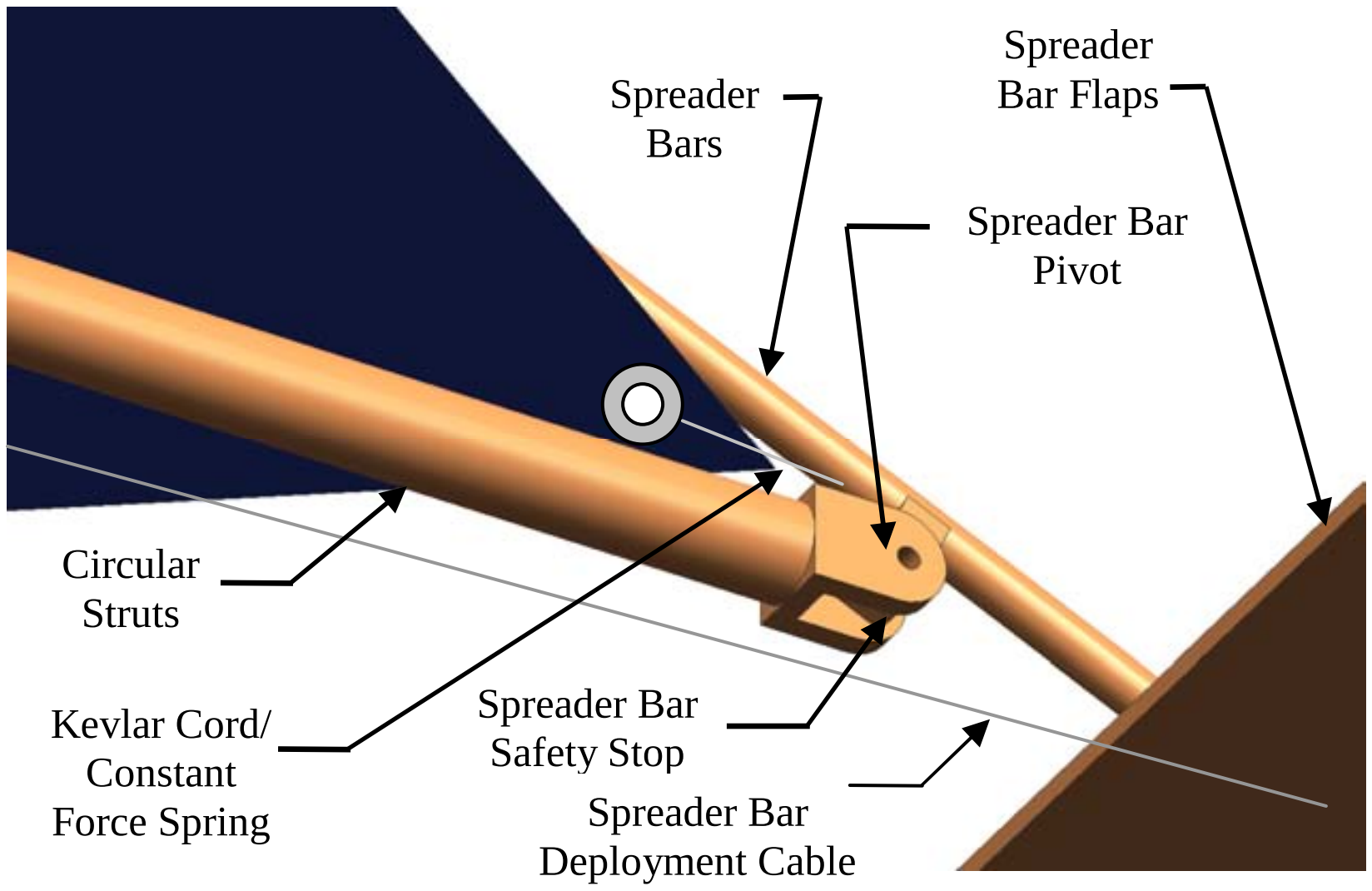

Figure 5.7.6: Detailed View of Spreader Bar Region.

The spreader bar is attached to the end of the strut using a hinge mechanism with a built-in hard-stop that allows the spreader bar to be stowed alongside the circular strut during launch and unfolding of the struts, and then be released to rotate as necessary during spreader bar deployment. When the spreader bars reach their fully deployed position, the spreader bar deployment cable tension is adjusted using the spreader bar deployment motor to properly position the three dual-layers of reflective film. A safety stop will also be used to prevent the spreader bar from over-rotating during deployment and damaging the shade, the constant force springs, or the Kevlar cords that attach the springs to the films. The exact position of the films is controlled by adjusting their tension using constant force springs and Kevlar cables that are contained inside and exit the strut or spreader bar pivot. These constant force springs will be designed to maintain the proper tension in all of the layers at all times, even in the presence of thermally-induced deformations, such as the Kevlar cords expanding or the Kapton films contracting when they are cooled. The spreader bar flap attaches to the bottom of the bottom film to keep the spreader bar in the shade at all times to eliminate radiation heat transfer into the telescope. As described earlier, the telescope is spinning at 1 RPM. This rotation causes a small amount of tensile force in the struts, which would tend to slightly increase the fundamental natural frequency. Therefore, the present analysis, which does not include this rotation, presents a conservative design. During the life of the mission, various components of the sunshade will see different temperatures. The lower temperatures will slightly decrease the damping in the composite struts. However, high damping is not required as the system is designed to have its resonance much higher than the 1 RPM excitation frequency. Also, since the circular struts are always in the sun and the sunshade is deployed while warm (early in the mission), no major effects are expected due to temperature dependent changes in material properties.

*5.7.5 Materials*

With the deployed configuration given above, the materials for the key components are selected. Table 5.7.2 lists the key components and the corresponding type of material. The circular struts are made from graphite epoxy composite to minimize weight, while the V-grooves are aluminum-honeycomb to conduct heat on their faces while being as light as possible. The inter-hub struts (bipods) and spreader bars are gamma alumina and S-glass epoxy composite, respectively, to reduce conduction and radiation heat transfer from the warmer shields into the telescope. The pulleys, motors, motorized hub, constant force springs, and spreader bar deployment cables can be metal as they are on the warm side of the sunshade and are not coupled thermally to the colder telescope components.

Table 5.7.2: Key Structural Components and Selected Material

| Item | Material |
|---|---|
| Circular Struts | Graphite/Epoxy composite |
| V-Grooves | Aluminum-faced honeycomb |
| Hinges | Aluminum |
| Ring support struts | Gamma alumina |
| Sunshade membrane | Aluminum coated, reinforced Kapton |
| Spreader bars | S-glass/epoxy composite |
| Spreader bar pivot | Aluminum |
| Constant force springs | Spring steel |
| Membrane attachements | Kevlar cord |
| Spreader bar deployment cables | Steel |
| Pulleys and motors | Aluminum |

*5.7.6 Specifications*

With the deployed configuration given above, a structural analysis was performed in order to design the structure to meet the requirements given in Table 5.7.1. While the details of this analysis are omitted here for simplicity and presented fully in the Appendix, they allow the geometry of the sunshade to be completely designed and the total system mass estimated. Figure 5.7.7 shows the circular strut root cross-section design for the sunshade as well as the first in-plane mode shape of the sunshade. Table 5.7.3 presents the required circular strut geometry for the desired sunshade configuration, while sunshade area is calculated in Table 5.7.4. Table 5.7.5 presents the geometric and material properties used in the present analysis.

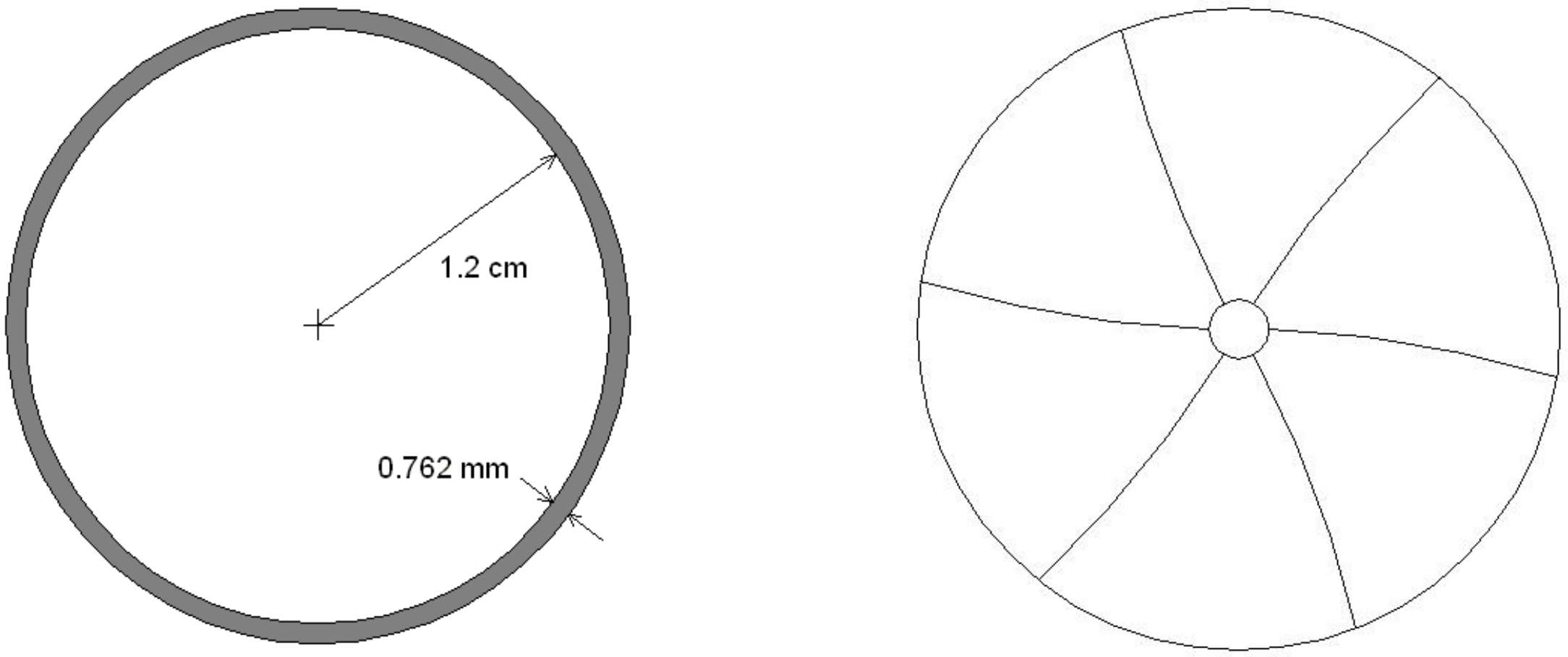

Circular Strut Root Dimensions — First in-plane mode: ~2.79 Hz

Figure 5.7.7: a) Circular cross section for sunshade and b) first In-plane mode frequency.

Table 5.7.3: Required Circular Strut Cross Section Geometry.

| Parameter | Value |
|---|---|
| Root radius (cm) | 1.2 |
| Tip radius (cm) | 1.1 |
| Length (m) | 2.45 |
| t (cm) | 0.0762 |
| Frequency (Hz) | 2.79 |

Table 5.7.4: Sunshade Geometry

| Bottom shield (i=1) | | | Middle Shield (i=2) | | | Top Shield (i=3) | | |
|---|---|---|---|---|---|---|---|---|
| Leff= | 3.55 | m | L= | 3.24 | m | L= | 2.99 | m |
| R1= | 2.75 | m | R= | 2.48 | m | R= | 2.27 | m |
| h= | 0.325 | m | h= | 0.325 | m | h= | 0.32 | m |
| R2= | 5.01 | m | R2= | 4.20 | m | R2= | 3.60 | m |
| theta= | 0.72 | radians | theta= | 0.79 | radians | theta= | 0.86 | radians |
| theta(degree) | 41.50 | degrees | theta(degree) | 45.38 | degrees | theta(degree) | 49.06 | degrees |
| shade area | 28.1 | $m^2$ | shade area | 23.0 | $m^2$ | shade area | 19.4 | $m^2$ |
| Total Area (6 shades) | | 141.0 | $m^2$ | | | | | |

Table 5.7.5: Strut, Film, and Material Properties.

| Item | Symbol | Value | Units |
|---|---|---|---|
| V-groove spacing (center) | $d$ | 0.135 | m |
| V-groove spacing (edge) | $d_{edge}$ | 0.25 | m |
| Film stress | $\sigma$ | 20,684 | Pa |
| Film thickness | $t$ | 2.54E-05 | m |
| Film density | $\rho_{film}$ | 0.09 | $kg/m^2$ |
| Spreader bar thickness | $t_{SB}$ | 7.62E-04 | m |
| Strut modulus | $E_{strut}$ | 72.8 | GPa |
| Strut density | $\rho_{strut}$ | 1522 | $kg/m^3$ |
| Strut thickness (5.5-m) | $t_{lenticular}$ | 7.62E-04 | m |

Based on the designed deployed geometry and the properties of the selected materials, the sunshade mass estimate is given in Table 5.7.6. These mass estimates include only the sunshade film and the structural support and deployment hardware. The masses of the rigid, central aluminum-honeycomb V-groove radiators are not included. It should be noted that the mass of the three aluminum-honeycomb V-groove radiators is not included in this estimate, but together add another 51.3 kg to the sunshield as shown in Table 5.1.9.

Table 5.7.6: Mass Estimate for Sunshade

| **Low Cost Telescope, Earth Protected, 5.5 m Diameter Folding Sunshade** | | | | |
|---|---|---|---|---|
| **Item** | **Mass Per (kg)** | **Qty** | **Substructure mass (kg)** | **Comments** |
| Circular Struts | 0.21 | 6 | 1.23 | Give f=2.79 Hz, FS>6 |
| Spreader bar pivot | 0.10 | 6 | 0.62 | 50% of strut mass |
| Spreader bars | 0.10 | 6 | 0.60 | FS=30 |
| Pulleys | 0.01 | 30 | 0.42 | |
| Constant Force Spring | 0.05 | 18 | 0.81 | |
| Connectors | 0.005 | 18 | 0.09 | |
| Kevlar cord | 0.03 | 1 | 0.03 | |
| Aluminized-Kapton Film (6 shades) | 16.49 | 1 | 16.49 | 90 $g/m^2$, 0.001" thick, 30% for seams |
| Spreader bar flaps | 0.04 | 6 | 0.21 | 1.63m dia, semicircle |
| Support rod | 0.08 | 6 | 0.49 | 0.5 cm dia |
| Ring support strut to Vgroove attachments | 1.00 | 36 | 36.00 | wag for a bracket |
| Deployment system | | | | |
| Spring-loaded hinges at hub | 2.00 | 6 | 12.00 | wag |
| Spreader bar deployment cable | 0.11 | 1 | 0.11 | wag |
| Spreader bar deployment motor | 5.00 | 1 | 5.00 | wag |
| | | | 74.11 | **Total Mass, kg** |

*5.7.7 Future Work*

While this effort has outlined a preliminary design for a deployable, lightweight sunshade that meets or exceeds the requirements for the EPIC telescope, further work is needed as this project moves forward. Previous experience with motorized hubs controlling the deployment of rigidizable, inflatable struts indicates that precautions must be taken in order to successfully deploy all types of struts. A more detailed analysis of the power required to safely deploy the system and of the required strengths of the hub, bottom plate, and mounting hinges for the struts must be performed in order to reduce the overall system mass. The next step in a more detailed, preliminary design would also focus on the spreader bar pivots, ring support strut-to-V-groove connectors, as well investigate the best set of material properties (tailorable for composite struts). Testing will be needed in the folding of such large, sectioned membranes and the effects that creases will have on the thermal performance of the sunshade system. It may be possible to use a tensioned cord around the perimeter of the sunshade, which would increase the natural frequency by inducing clamped-pinned-type mode shapes. Likewise, the structural model ignores the small shear stiffness contribution of the membrane film. Either modification to the model would result in lighter struts. Thus far, no analysis has been performed to determine if the proposed design would survive the mechanical and acoustical conditions imposed on the stowed sunshade during launch. A small, proof-of-deployment-concept study for the membrane folding, storage, and wrap-rips could be performed in the near-term.

This report has outlined the concepts and analytical tools required to develop a large, deployable sunshade using circular, thin-walled struts. The proposed sunshade is a viable design that is capable of meeting the requirements for this current refractor telescope design. The design is stowable for launch and readily deployed on-orbit. The dynamic and strength requirements are met while maintaining a reasonable and workable configuration.

### 5.8 L2 Orbit Analysis

The EPIC-LC orbit at L2 has been designed to minimize the halo diameter in order to reduce the size of the shields needed to keep the sun, moon, and earth off the cryostat and optics, and to allow for the use of a high-gain fixed antenna with a toroidal beam pattern. Our analysis shows this orbit can be accomplished, all the while avoiding partial eclipses of the sun, without a significant increase in the propellant requirements. The propellant requirements are dominated by statistical uncertainties in the trajectory provided by the launch vehicle and upper stage, followed by maneuvers required for station-keeping.

We used LTOOL (Winter06) to generate several transfer trajectories from an Earth parking orbit to a Lissajous orbit about EL2. For this analysis, the gravitational forces from three bodies in the solar system were used when integrating the trajectory of the EPIC spacecraft. The gravitational force of the Earth was used as the primary force, while the gravitational force from the Moon and the Sun were modeled as secondary perturbing forces. Solar radiation pressure, spherical gravity harmonics, and any other perturbing forces were not incorporated in this study.

All values calculated in this investigation are given in the Sun-Earth Rotating Frame. According to the LTOOL documentation, this rotating coordinate frame is characterized by the X-axis pointing from the Sun to the Earth, the Z-axis is perpendicular to the instantaneous Earth orbital plane formed by the position and inertial velocity vectors, and the Y-axis completes the right handed coordinate system. All trajectories pictured in this document are shown in this rotating frame with either the Earth or EL2 as the center.

Although the Earth parking orbit used in this analysis is not consistent with the launch vehicle being considered for the EPIC mission (Delta 2925H), it still provides suitable values for this study. A discrepancy of this nature is usually optimized to null at a later stage. The Earth parking orbit used was a 250 km altitude circular orbit with a 28.5° inclination. Furthermore, this simulation does not model the exact time for launching from a KSC launch pad. It merely uses the time of injection into the transfer trajectory for the launch time.

### *5.8.1 Input Constraints on Orbit*

Several constraints were imposed on this study. First, the total spacecraft mass is limited due to the performance of the launch vehicle. Hence, the required deterministic and statistical ΔV must be minimized to maximize payload mass. The transfer trajectories investigated in this analysis include a lunar fly-by to drastically reduce the post-launch deterministic ΔV. Furthermore, it would be beneficial to also minimize the C3 to be able to insert more mass into orbit.

Second, the preferred Lissajous orbit is one where the Earth-view angle is as small as possible. The Earth-view angle is defined as the angle between the Earth-EL2 position vector and the Earth-EPIC position vector at maximum radius. The Earth-view angle (α) is shown in Figure 5.8.1. This analysis will investigate whether stable Lissajous orbits with an Earth-view angle of less than 9° are possible.

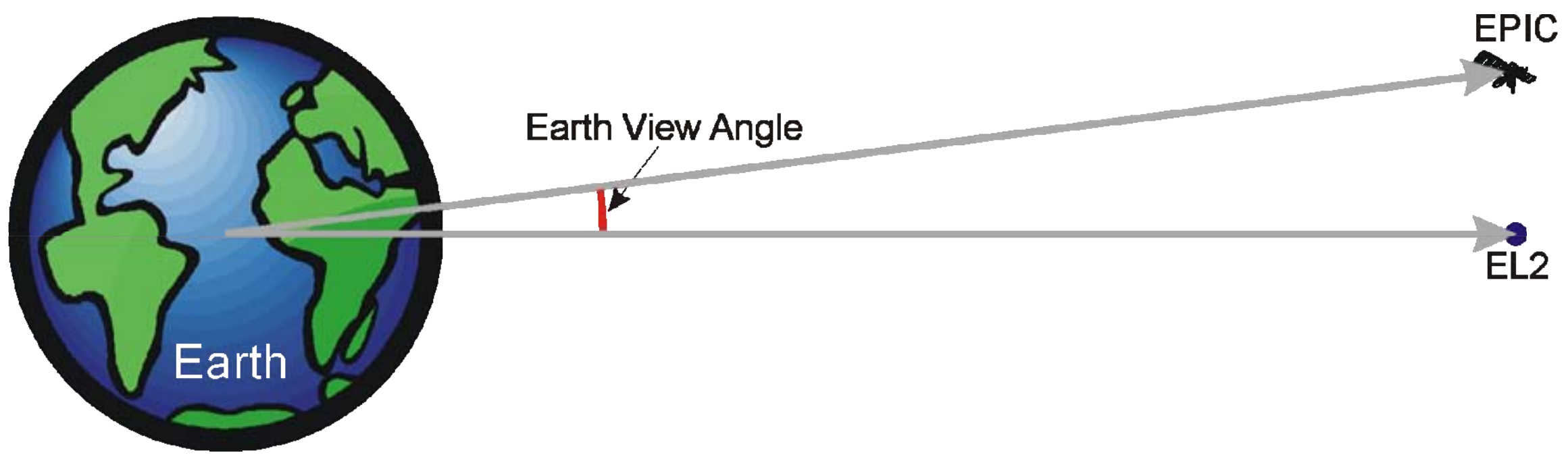

Figure 5.8.1. The EPIC spacecraft is operated such that the precession cone is centered on the Sun-spacecraft vector so as to keep solar input power constant over the orbit. The finite size of the halo orbit at EL2 thus causes the earth-view angle to vary, as shown.

While the desire is to have the smallest value of α as possible, the minimum α value is restricted by a science phase requirement. During the science phase of the mission, the spacecraft must never enter any type of eclipse due to thermal reasons. Therefore, the minimum Earth-view angle is forced to be large enough to avoid entering the Earth's penumbra during the mission. Using average values for the distance between the Sun, Earth, and EL2, the average minimum α value to avoid entering into the Earth's penumbra is 0.53°. However, as the distance between the Sun and Earth changes during the mission, the minimum angle needed to avoid entering the penumbra also deviates from this average value. Therefore, the minimum Earth-view angle used shall always be larger than this value.

Fourth, the transfer time between launch and Lissajous orbit insertion would, ideally, be less than 180 days. This constraint is due to the need to carry cryogens for the scientific instruments. The cryogens boil away during the entire transfer, consequently, the shorter the transfer, the less cryogen mass the needs to be launched for the science phase. The last constraint concerns the duration of the science-phase. The science-phase begins just prior to the insertion into a Lissajous orbit about EL2 and continues for 4 years.

*5.8.2 Analysis Results*

First we confirmed the results of a previous study that showed by using a lunar fly-by, several trajectories exist that require 0m/s deterministic ΔV. These trajectories require no deterministic maneuvers after injection into the transfer orbit and include injection into a Lissajous orbit at EL2. Furthermore, it was found that there are Lissajous orbits about EL2 that have an Earth-view angle is less than 9°.

Using LTOOL, 13 transfer trajectories were found from the previously defined Earth parking orbit to various-sized Lissajous orbits about EL2. These 13 transfer trajectories are captured in Table 5.8.1, which displays the resulting C3, deterministic ΔV, Lissajous orbit parameters, and significant event times for the transfer trajectories. Before further discussion of the results, the variables and term definitions used in Table 5.8.1 are described below:

**C3** - Launch Energy (inversely proportional to mass injected into orbit)
**X max, Y max, Z max** - Maximum distance between the EPIC spacecraft and EL2 along the x-axis, y-axis, and z-axis of the Sun-Earth Rotating Coordinate Frame, respectively.
**A max, A avg** - The maximum and average values of α throughout Lissajous orbit, respectively.
**Earth to Science** - The time from Earth parking orbit to a point where the science may begin. The science may begin before injection into the Lissajous orbit due to the slow orbital speed at this point in the trajectory.
**Earth to Insertion** - The time from Earth parking orbit to Lissajous injection.
**After Insertion @ EL2** - The time that the EPIC spacecraft is in the Lissajous orbit.

Table 5.8.1 shows that, indeed, there are stable Lissajous orbits about EL2 that have values of α smaller than 9° and require no deterministic ΔV for transfer or orbit maintenance. By plotting the value of maximum α versus C3 as shown in Figure 5.8.2, there is an observable almost-linear correlation between the size of the Lissajous orbit and the value of launch energy for Lissajous orbits smaller than 9°. However, there are a few cases that do not conform to this trend (i.e. Cases 4, 11, 12, and 13.) More investigation is needed to discern the reasons for these anomalous cases.

The data illustrated in Figure 5.8.2 reveals that as the size of the Lissajous orbit decreases, the value of C3 increases; this means that as the orbit size is reduced, so does the amount of mass that can be delivered to that orbit by the launch vehicle. However, the variation in launch mass over these 13 cases is only 1.3 kg[2], which poses minimal impact to the design of the spacecraft.

These 13 transfer trajectories take into account only three of the four constraints outlined previously; the one constraint that was not enforced was the eclipse constraint. The Lissajous orbits in Table 5.8.1 were allowed to enter into eclipse during the science-mission phase to allow the smallest Lissajous orbits possible. The next step in the analysis is to determine the consequences of enforcing the no-eclipse constraint on the required deterministic ΔV.

---

[2] This calculation made using the LV tool, which utilizes the KSC data for the Delta-II (2925H-9.5). Difference in launch mass for C3 values ranging from -0.48 to -0.44 $km^2/s^2$.

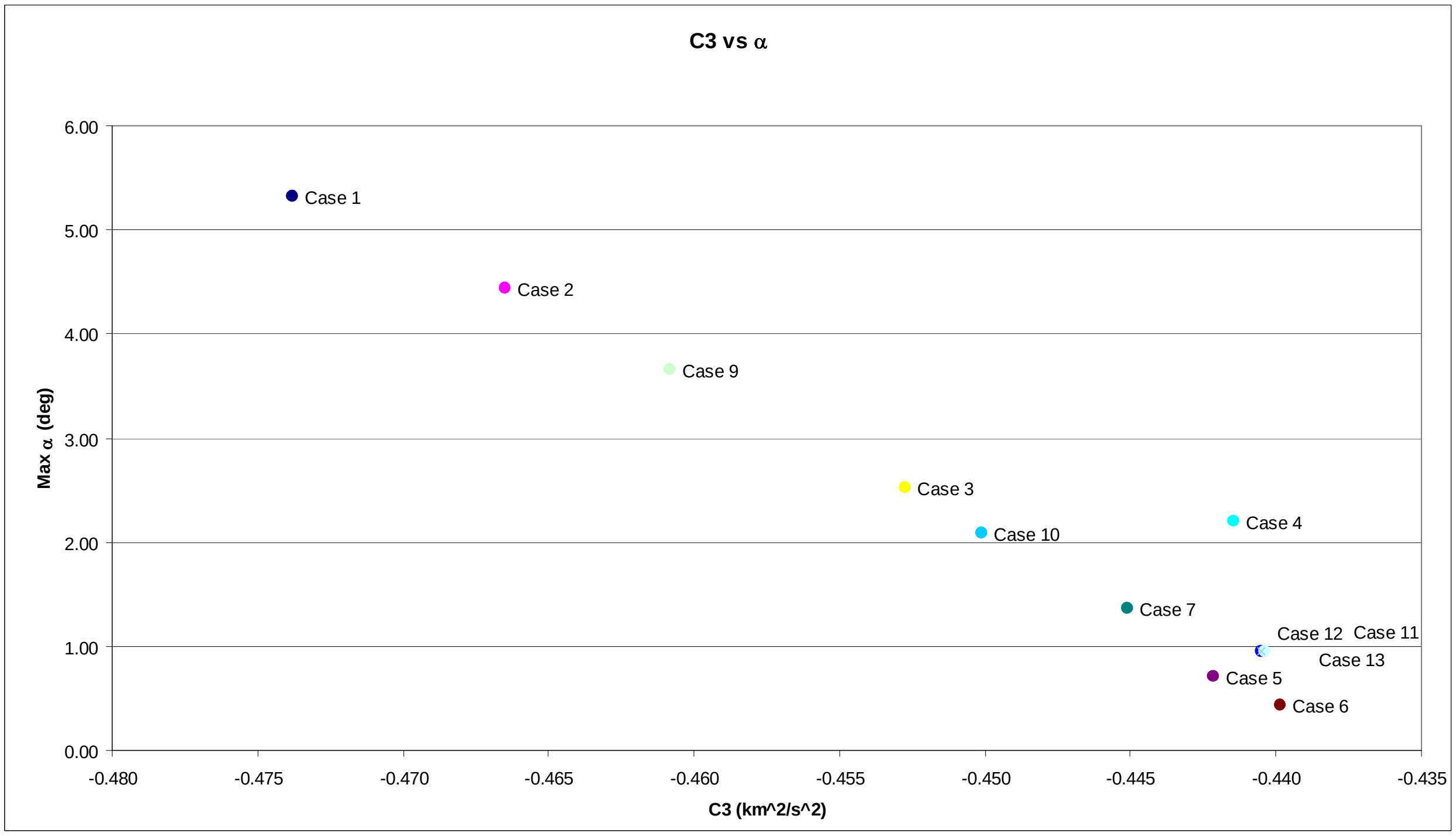

Figure 5.8.2. Comparison of C3 and Maximum Earth-View Angle for a selection of trajectories.

### *5.8.3 Enforcing the No-Eclipse Constraint*

In order to be consistent with the given constraints, limits were placed on the minimum and maximum allowable Earth-view angle while traversing the Lissajous orbit over the science mission duration. These limits ensure that the EPIC spacecraft stays out of any kind of eclipsing. As stated previously, the minimum Earth-view angle limit is 0.53° while the maximum limit is confined to values less than 2° to keep the Lissajous orbit as small as possible. Table 5.8.2 details the four trajectories found that satisfy all of the given constraints.

The Eclipse Case 4 trajectory has the largest average Earth-view angle, which provides the largest margin to remain outside of the eclipse during the science-phase. Therefore, this trajectory is recommended for the EPIC mission's nominal trajectory and is used as the basis for the launch period and statistical ΔV analysis. The entire nominal trajectory is shown in Fig. 5.8.3 while the Lissajous orbit is illustrated in finer detail in Figs. 5.8.4 through 5.8.6.

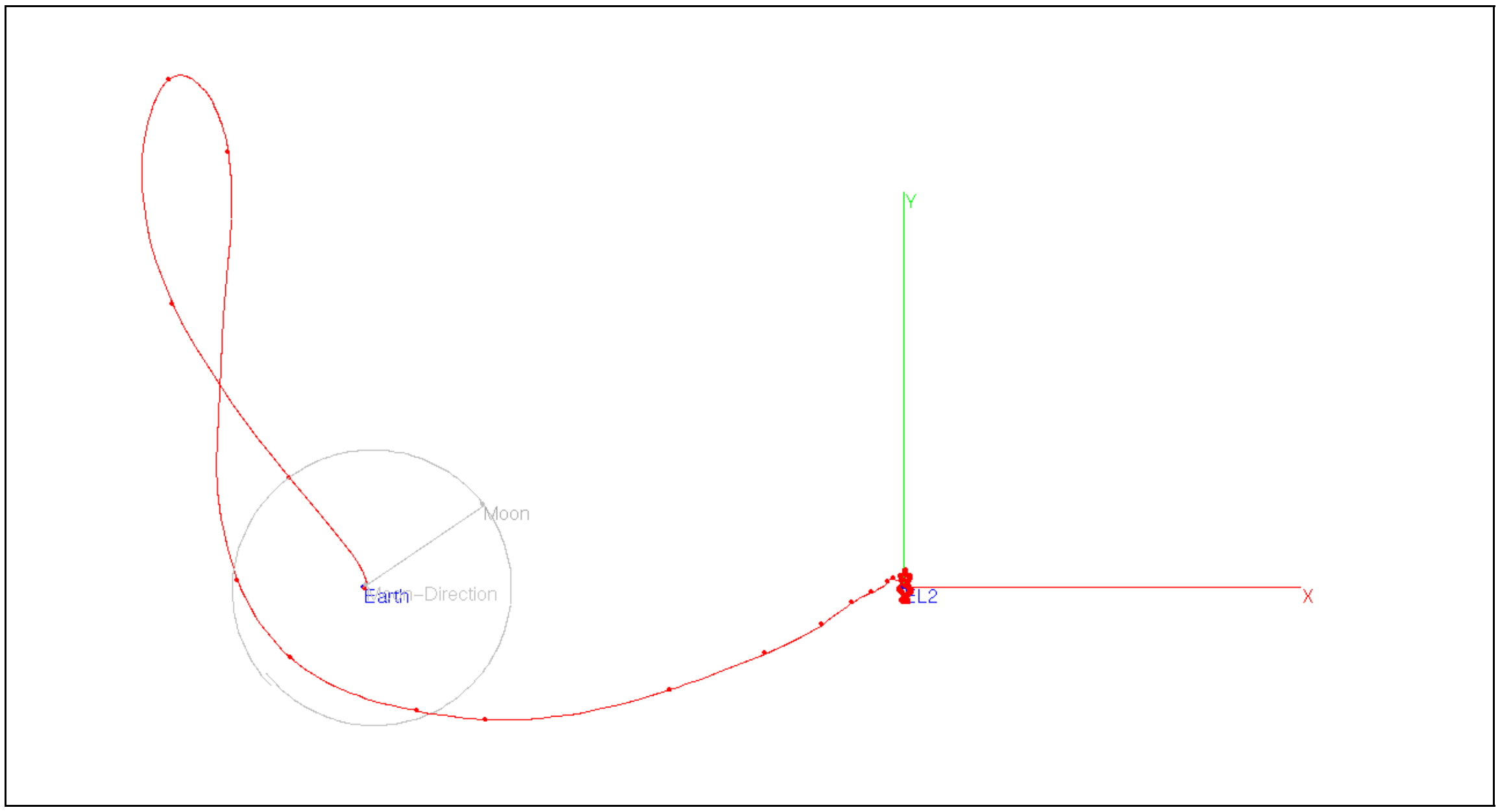

Figure 5.8.3. Nominal trajectory to EL2.

**Table 5.8.1. Transfer Trajectory Summary**

| | Deterministic ΔV | | | Lissajous Orbit Parameters | | | | | Event Times | | |
|---|---|---|---|---|---|---|---|---|---|---|---|
| | C3 | Mid-Course | Insertion | X max | Y max | Z max | A max | A avg | Earth to Science | Earth to Insertion | After Insertion |
| | $km^2/s^2$ | m/s | m/s | 1e6 m | 1e6 m | 1e6 m | deg | deg | Days | Days | Days |
| Case 1 | -0.47 | 0 | 0 | 48.5 | 127.5 | 59.5 | 5.3 | 3.40 | 78 | 159 | 294 |
| Case 2 | -0.47 | 0 | 0 | 41.2 | 102.9 | 58.1 | 4.4 | 2.86 | 78 | 160 | 294 |
| Case 3 | -0.45 | 0 | 0 | 26.3 | 53.2 | 40.6 | 2.5 | 1.57 | 80 | 161 | 295 |
| Case 4 | -0.44 | 0 | 0 | 19.6 | 32.8 | 51.5 | 2.2 | 1.48 | 80 | 161 | 293 |
| Case 5 | -0.44 | 0 | 0 | 14.9 | 16.0 | 11.2 | 0.7 | 0.41 | 81 | 162 | 295 |
| Case 6 | -0.44 | 0 | 0 | 12.0 | 6.3 | 9.6 | 0.4 | 0.26 | 82 | 163 | 296 |
| Case 7 | -0.45 | 0 | 0 | 17.2 | 23.5 | 28.2 | 1.4 | 0.82 | 81 | 162 | 295 |
| Case 8 | -0.66 | 0 | 0 | 93.0 | 262.4 | 255.8 | 13.3 | 8.94 | 72 | 154 | 287 |
| Case 9 | -0.46 | 0 | 0 | 35.3 | 83.1 | 49.8 | 3.7 | 2.32 | 79 | 160 | 294 |
| Case 10 | -0.45 | 0 | 0 | 23.3 | 43.3 | 34.0 | 2.1 | 1.26 | 80 | 161 | 295 |
| Case 11 | -0.44 | 0 | 0 | 12.5 | 8.7 | 24.9 | 1.0 | 0.62 | 81 | 162 | 295 |
| Case 12 | -0.44 | 0 | 0 | 12.6 | 8.8 | 24.9 | 1.0 | 0.62 | | 162 | 295 |
| Case 13 | -0.44 | 0 | 0 | 12.5 | 8.7 | 24.9 | 1.0 | 0.62 | 81 | 162 | 295 |

**Table 5.8.2. Transfer Trajectories w/Eclipse Constraint**

| | Constraints | | Deterministic ΔV Information | | | Lissajous Orbit | | |
|---|---|---|---|---|---|---|---|---|
| | Min. α | Max α | C3 | TCM ΔV | LOI ΔV | A max | A min | A avg |
| | deg | deg | $km^2/s^2$ | m/s | m/s | deg | Deg | deg |
| Eclipse Case 1 | 0.53 | 2 | -0.451 | 0 | 0 | 1.61 | 0.52 | 1.12 |
| Eclipse Case 2 | 0.55 | 2 | -0.451 | 0 | 0 | 1.64 | 0.54 | 1.13 |
| Eclipse Case 3 | 0.55 | 2 | -0.453 | 0 | 0 | 1.99 | 0.53 | 1.37 |
| Eclipse Case 4 | 0.60 | 1.95 | -0.453 | 0 | 0 | 1.95 | 0.60 | 1.36 |

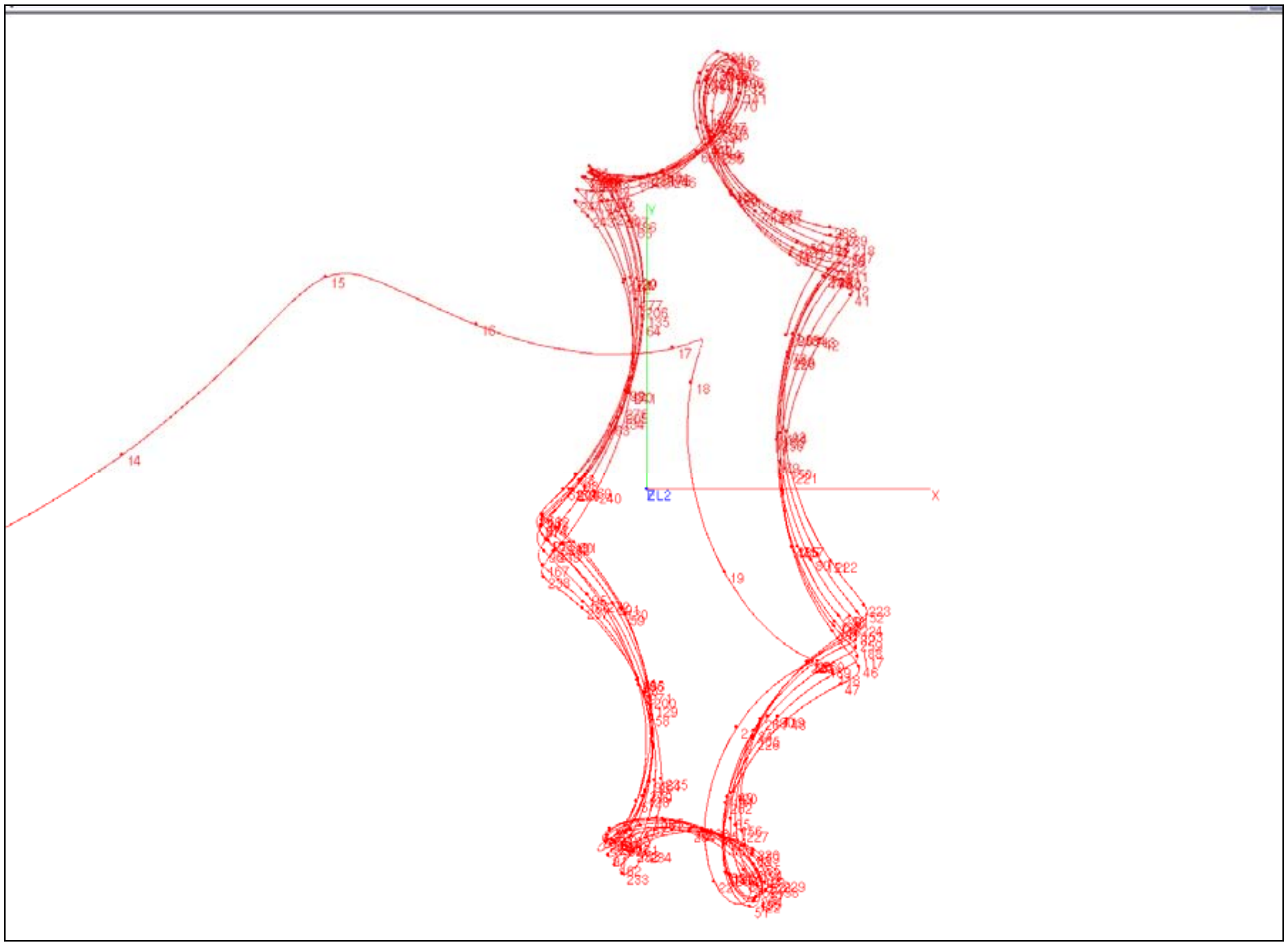

Figure 5.8.4. Nominal Lissajous Orbit - XY Plane View.

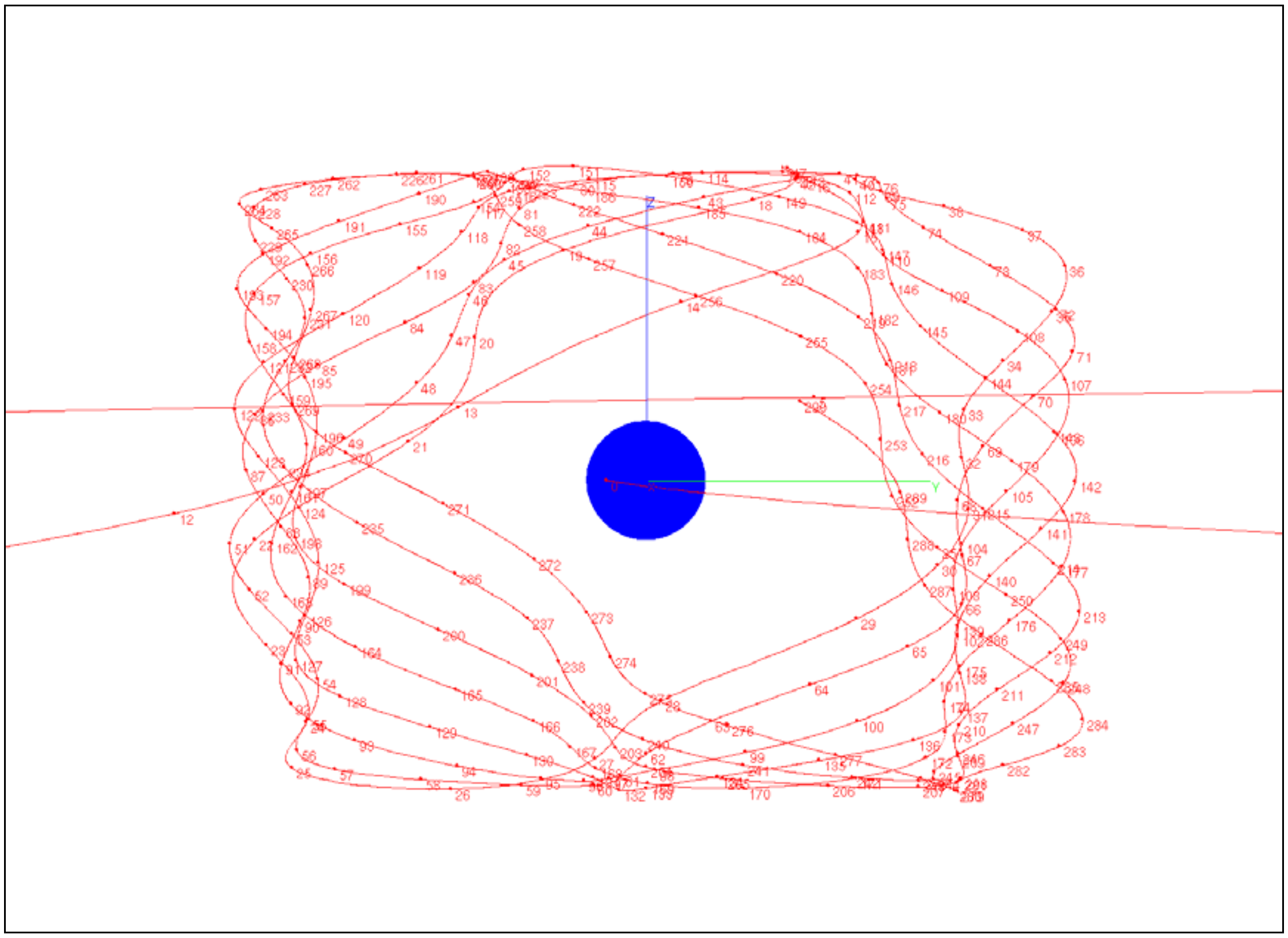

Figure 5.8.5. Nominal Transfer and Lissajous Orbit - YZ Plane View.

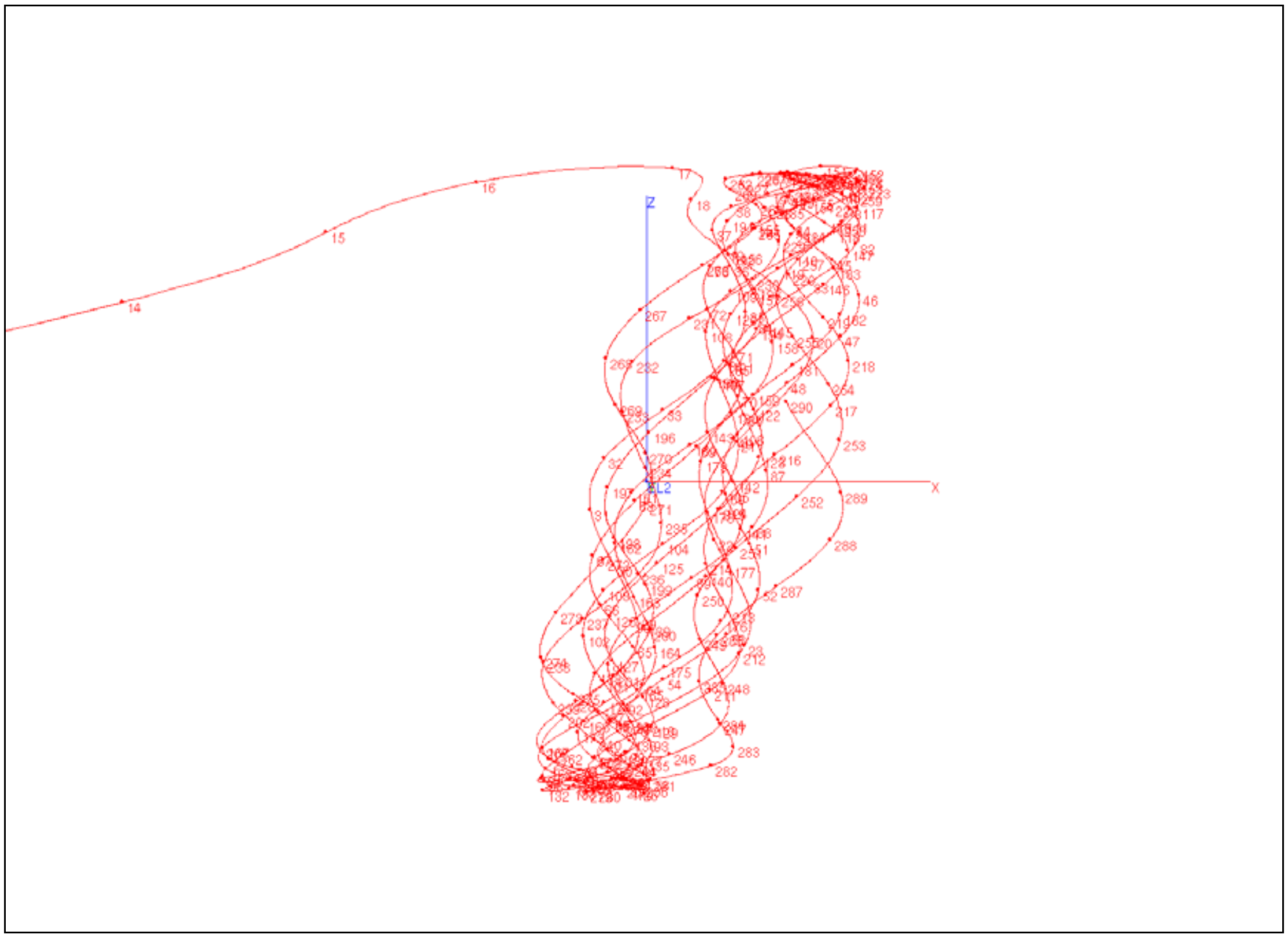

Figure 5.8.6. Nominal Lissajous Orbit - XZ Plane View.

**Table 5.8.3. Additional Nominal Trajectory Parameters**

| | |
|---|---|
| Lissajous Orbit Start | 2015 AUG 25 22:06:03 |
| Lissajous Orbit End | 2019 OCT 25 10:43:02 |
| Years in Lissajous Orbit | 4.166 years |
| Lissajous YZ max | 52318 km |
| Lissajous YZ min | 15662 km |
| Lissajous YZ ave | 35909km |
| Lissajous Max | 23747 x 45956 x 32961 km |
| Lissajous A max | 1.95° |
| Lissajous A min | 0.60° |
| Lissajous A ave | 1.36° |

**Table 5.8.4. Nominal Orbital Parameters**

| | |
|---|---|
| Transfer Time | ~170 Days |
| Halo Dimensions | 24,000 km×46,000km×33,000km |
| Max Earth-view Angle | 2 ° |
| Lunar Fly-by Distance | 64,000 km |
| C3 | -0.45 $km^2/s^2$ |
| Max Launch Mass | 1437.8 kg (LV: Delta II 2925H-9.5) |
| Deterministic ΔV | 0 m/s |

*5.8.4. Launch Period Analysis*

Starting with the nominal trajectory discussed previously, a launch period analysis was performed to encapsulate the deterministic ΔV needed to extend the launch period for ten days both before and after the nominal launch date. By changing the launch date, a mid-course maneuver must be implemented in order to arrive at the nominal Lissajous trajectory. This mid-course maneuver is optimized to occur at roughly Launch + 50 days. The maneuver is placed at this time because this is when the spacecraft is moving at its slowest velocity therefore reducing the required amount of deterministic ΔV to change its trajectory.

Table 5.8.5 catalogs several parameters calculated during the launch period analysis while Table 5.8.6 provides a detailed description of the labels used in Table 5.8.5. The most notable of the given parameters is the C3 and the required amount of mid-course ΔV (DV), which is calculated with respect to the nominal trajectory launch date. The data shows that as the launch date increases from the nominal launch date, in either direction, the amount of deterministic ΔV linearly increases, with one exception (highlighted). If the launch is delayed by five days, the ΔV cost significantly increases compared to delaying the launch for either four or six days. The dramatic increase in ΔV stems from the interference of the Moon at launch.

Figs. 5.8.7 and 5.8.9 show the Moon geometry at launch time for the nominal launch date trajectory (red) and the 5-day delayed launch date trajectory (yellow), respectively. Starting from Earth, the EPIC spacecraft moves along the red or yellow path while the Moon continues in its counter-clockwise orbit denoted by the gray line. Approximately two days after the launch, the EPIC spacecraft crosses the orbit of the Moon for the first time (the spacecraft will cross the Moon's orbit three times total.) When the spacecraft first crosses the Moon's orbit in the nominal launch date case (Fig. 5.8.8), the geometry is such that the Moon itself is not in a position to significantly perturb the spacecraft's trajectory.

**Table 5.8.5. Launch Period Deterministic ΔV**

| NITR | LAUNCH | DLY | NODE | MEAN | DV0 | C3 | FT | DR | DV | ST |
|---|---|---|---|---|---|---|---|---|---|---|
| | | (day) | (deg) | (deg) | (m/s) | ($km^2/s^2$) | (day) | (km) | (m/s) | |
| 16 | 26-Feb-15 | -10 | 345 | 129 | 3207 | -0.473 | 71 | 0.00 | 86 | o |
| 16 | 27-Feb-15 | -9 | 345 | 130 | 3207 | -0.473 | 70 | 0.00 | 78 | o |
| 17 | 28-Feb-15 | -8 | 345 | 131 | 3207 | -0.472 | 69 | 0.00 | 70 | o |
| 16 | 1-Mar-15 | -7 | 345 | 132 | 3207 | -0.471 | 68 | 0.00 | 62 | o |
| 17 | 2-Mar-15 | -6 | 346 | 132 | 3207 | -0.469 | 67 | 0.00 | 53 | o |
| 16 | 3-Mar-15 | -5 | 346 | 133 | 3207 | -0.466 | 66 | 0.01 | 45 | o |
| 16 | 4-Mar-15 | -4 | 347 | 134 | 3207 | -0.464 | 65 | 0.00 | 36 | o |
| 19 | 5-Mar-15 | -3 | 347 | 135 | 3207 | -0.461 | 64 | 0.01 | 27 | o |
| 22 | 6-Mar-15 | -2 | 348 | 136 | 3207 | -0.458 | 63 | 0.00 | 18 | o |
| 55 | 7-Mar-15 | -1 | 348 | 137 | 3207 | -0.456 | 62 | 0.01 | 9 | o |
| 158 | 8-Mar-15 | 0 | 349 | 137 | 3207 | -0.453 | 45 | 0.00 | 0 | o |
| 52 | 9-Mar-15 | 1 | 349 | 138 | 3208 | -0.450 | 61 | 0.01 | 9 | o |
| 18 | 10-Mar-15 | 2 | 351 | 139 | 3208 | -0.446 | 60 | 0.14 | 19 | o |
| 60 | 11-Mar-15 | 3 | 352 | 138 | 3208 | -0.439 | 58 | 0.00 | 29 | o |
| 45 | 12-Mar-15 | 4 | 357 | 136 | 3209 | -0.430 | 57 | 0.00 | 39 | o |
| 132 | 13-Mar-15 | 5 | 7 | 136 | 3206 | -0.492 | 74 | 0.00 | 145 | o |
| 85 | 14-Mar-15 | 6 | 360 | 129 | 3201 | -0.592 | 57 | 0.00 | 46 | o |
| 21 | 15-Mar-15 | 7 | 352 | 138 | 3202 | -0.575 | 57 | 0.00 | 55 | o |
| 17 | 16-Mar-15 | 8 | 351 | 141 | 3202 | -0.562 | 56 | 0.08 | 65 | o |
| 63 | 17-Mar-15 | 9 | 351 | 143 | 3203 | -0.562 | 55 | 0.02 | 75 | o |

**Table 5.8.6. Legend for Table 5.8.5**

| |
|---|
| NITR = Number of NPOPT iteration |
| LAUNCH = launch date |
| DLY = delay offset with respect to the nominal launch date |
| NODE = ascending node |
| MEAN = mean anomaly |
| DV0 = injection delta-v |
| C3 = injection energy |
| FT = flight time until insertion delta-v |
| DR = position uncertainty at insertion |
| DV = insertion delta-v |
| ST = NPOPT state |
| o = optimum solution found |

Conversely, when the launch is delayed for five days from the nominal launch date, the geometry allows the Moon to get much closer to the spacecraft at the first orbit-crossing point as seen in Fig. 5.8.10. The result of this close fly-by is that the spacecraft is pulled far off the nominal trajectory by the Moon and a large amount of ΔV is needed to correct back to the nominal path. Therefore, launching five days after the nominal launch date is not recommended.

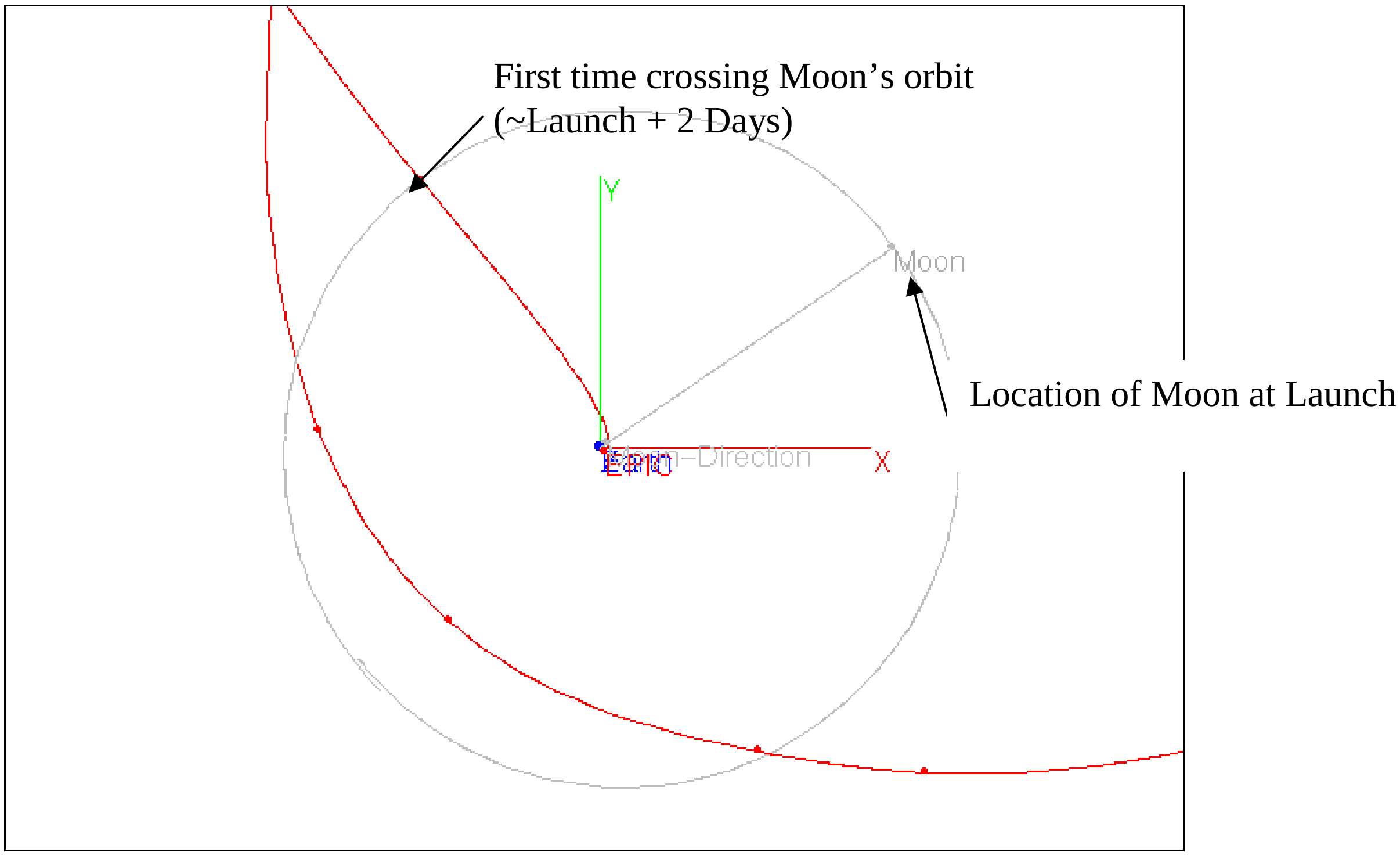

Figure 5.8.7. Moon Location at Launch (Nominal Launch Date).

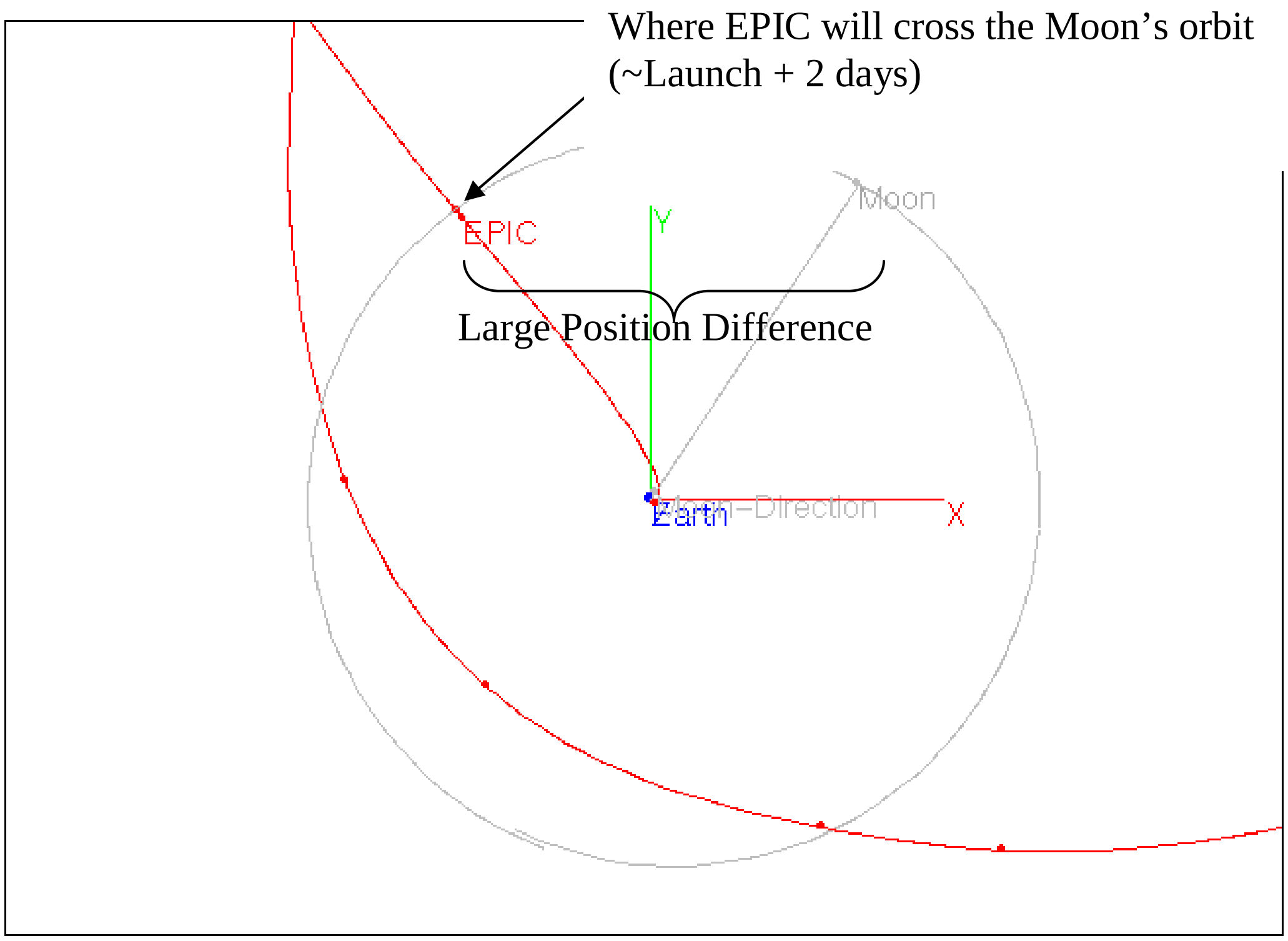

Figure 5.8.8. Moon Close-Approach (Nominal Launch Date).

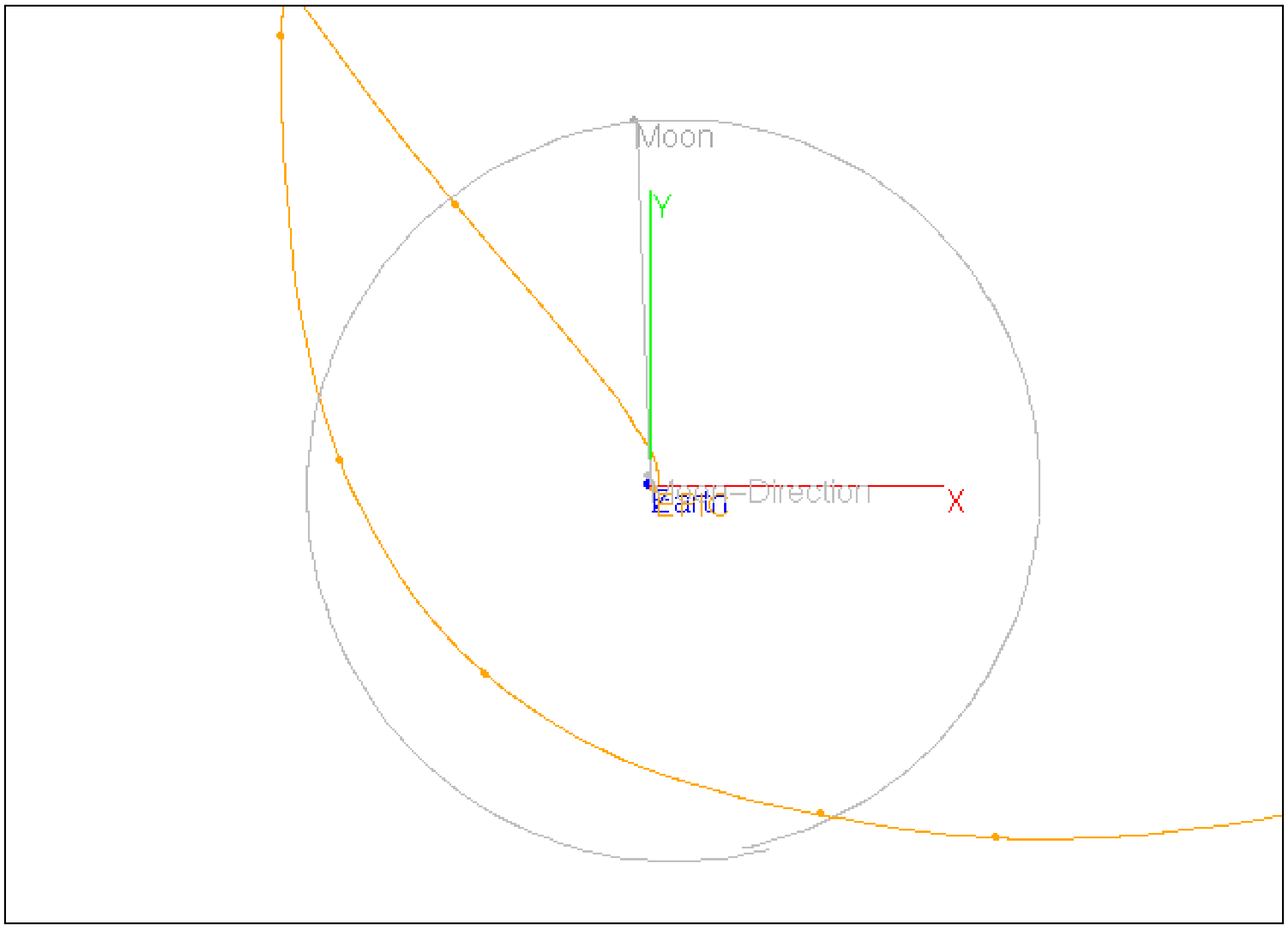

Figure 5.8.9. Moon Location at Launch (Nominal Launch + 5 Days).

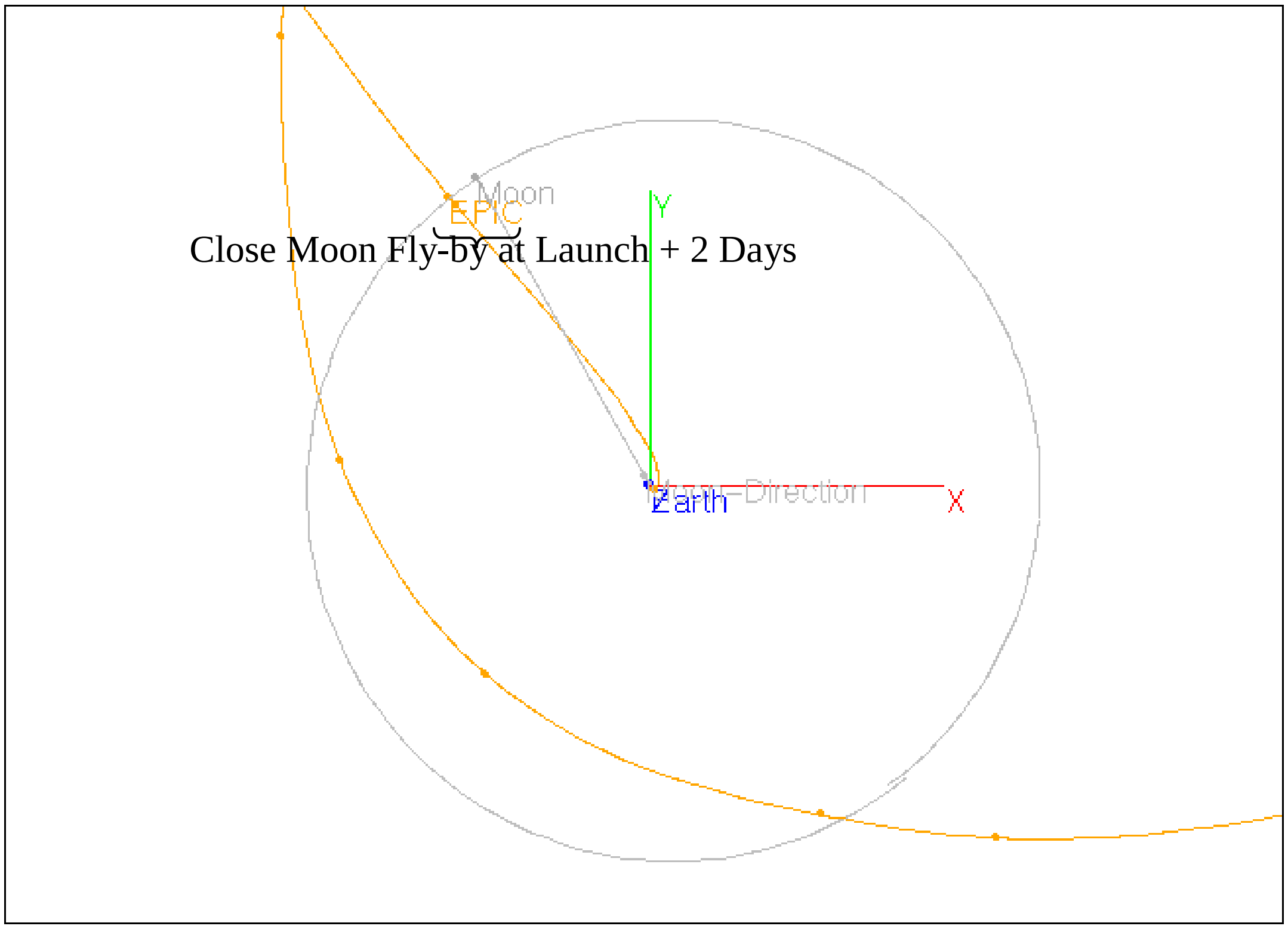

Figure 5.8.10. Moon Close-Approach (Nominal Launch + 5 Days).

The trajectories produced by Table 5.8.5 can be further optimized by re-optimizing C3 and the Lissajous portion of the trajectory. The resulting deterministic ΔV can, therefore, be slightly reduced. Once the Lissajous portion of the orbit is re-optimized, it is no longer identical to the nominal Lissajous orbit. Therefore, the final science orbit becomes dependant on the actual launch date. Nevertheless, the Earth-view angle is still constrained between 0.06 and 1.95° for these re-optimized orbits, which corresponds to staying 2,000 km above the penumbra. Table 5.8.7 shows the comparison between the C3 and ΔV values of the original launch period and the re-optimized trajectories. This table also includes the distance between the spacecraft and the moon at Launch + 2 Days, which is given in units of Lunar Radii.

**Table 5.8.7. Comparison of C3 and Mid-Course ΔV for Re-Optimized Lissajous**

| | | Original Values | | Re-Optimized | | |
|---|---|---|---|---|---|---|
| **LAUNCH** | **DLY** | **C3** | **DV** | **C3** | **DV** | **Dist** |
| | (day) | ($km^2/s^2$) | (m/s) | ($km^2/s^2$) | (m/s) | Moon Radii |
| 21-Feb-15 | -15 | - | - | -0.43 | 89 | - |
| 22-Feb-15 | -14 | - | - | -0.43 | 80 | - |
| 23-Feb-15 | -13 | - | - | - | - | - |
| 24-Feb-15 | -12 | - | - | -0.44 | 68 | - |
| 25-Feb-15 | -11 | - | - | - | - | - |
| 26-Feb-15 | -10 | -0.47 | 86 | -0.45 | 59 | - |
| 27-Feb-15 | -9 | -0.47 | 78 | -0.45 | 54 | - |
| 28-Feb-15 | -8 | -0.47 | 70 | -0.45 | 48 | - |
| 1-Mar-15 | -7 | -0.47 | 62 | -0.45 | 43 | - |
| 2-Mar-15 | -6 | -0.47 | 53 | -0.45 | 37 | - |
| 3-Mar-15 | -5 | -0.47 | 45 | -0.45 | 32 | - |
| 4-Mar-15 | -4 | -0.46 | 36 | -0.45 | 26 | - |
| 5-Mar-15 | -3 | -0.46 | 27 | -0.45 | 20 | - |
| 6-Mar-15 | -2 | -0.46 | 18 | -0.45 | 14 | - |
| 7-Mar-15 | -1 | -0.46 | 9 | -0.45 | 8 | - |
| 8-Mar-15 | 0 | -0.45 | 0 | -0.45 | 0 | > 200 |
| 9-Mar-15 | 1 | -0.45 | 9 | -0.45 | 9 | 200 |
| 10-Mar-15 | 2 | -0.45 | 19 | -0.45 | 16 | 160 |
| 11-Mar-15 | 3 | -0.44 | 29 | -0.45 | 23 | 121 |
| 12-Mar-15 | 4 | -0.43 | 39 | -0.44 | 35 | 82 |
| 13-Mar-15 | 5 | -0.49 | 145 | -0.47 | 48 | 47 |
| 14-Mar-15 | 6 | -0.59 | 46 | -0.6 | 43 | 56 |
| 15-Mar-15 | 7 | -0.58 | 55 | -0.59 | 48 | 86 |
| 16-Mar-15 | 8 | -0.56 | 65 | -0.58 | 60 | 129 |
| 17-Mar-15 | 9 | -0.56 | 75 | -0.57 | 70 | 174 |
| 18-Mar-15 | 10 | -0.57 | 84 | -0.58 | 80 | 217 |

In addition to avoiding launch 5 days after the nominal launch date, it is probably not wise to launch on the dates when the lunar distance at Launch + 2 Days is below 100 lunar radii due to the large injection errors associated with the solid third stage of the Delta II class. Therefore, launch should not occur on the dates from Nominal Launch + 4 Days through Nominal Launch + 7 Days, which are the dates highlighted in Table 5.8.7.

*5.8.5 Statistical ΔV Analysis*

After obtaining the deterministic ΔV for the trajectory, a simple analysis was performed to arrive at preliminary values for the statistical ΔV required for the mission. To estimate the amount of statistical ΔV needed, the trajectory from Earth Parking orbit through Lissajous orbit about EL2 was divided into three parts for simpler analysis. The first investigation calculates the statistical ΔV needed to clean up the errors associated with the third-stage motor injecting the spacecraft into the transfer orbit. The second analysis picks up after the launch injection clean-up maneuvers and includes the transfer to EL2 and insertion into the Lissajous orbit. Lastly, the statistical ΔV needed for station-keeping during the Lissajous orbit is calculated.

Since this analysis occurs in the early-stages of development, the budgeted amount of statistical ΔV for the Genesis mission will provide a basis of comparison for the EPIC mission. By assuming that the second-stage places the spacecraft in the parking orbit accurately, the resulting error associated with injecting into the transfer trajectory is assumed to be due solely to the third-stage pointing error and specific impulse error. Table 5.8.8 lists the differences in launch injection errors between the EPIC and Genesis third-stage motors. During the analysis, both the pointing and specific impulse errors were assumed to have a normal distribution.

**Table 5.8.8. Third-Stage Error Characteristics**

| | Genesis | EPIC |
|---|---|---|
| 3rd Stage Motor | Star - 37 | Star - 48B |
| Specific Impulse Error | 0.50% | 0.34% |
| Pointing Error | 2.0° | 1.5° |

According to the Delta II Payload Planner's Guide, October 2000, the injection pointing accuracy is modeled as pitch and yaw angle. Furthermore, the Planner's Guide lists the specific impulse error as 0.75% while the Genesis Navigation Plan uses a value of 0.5%.

The method used to estimate the statistical ΔV for launch injection cleanup consists of combining the ΔV for two maneuvers. A diagram of this method is shown in Fig. 5.8.11.

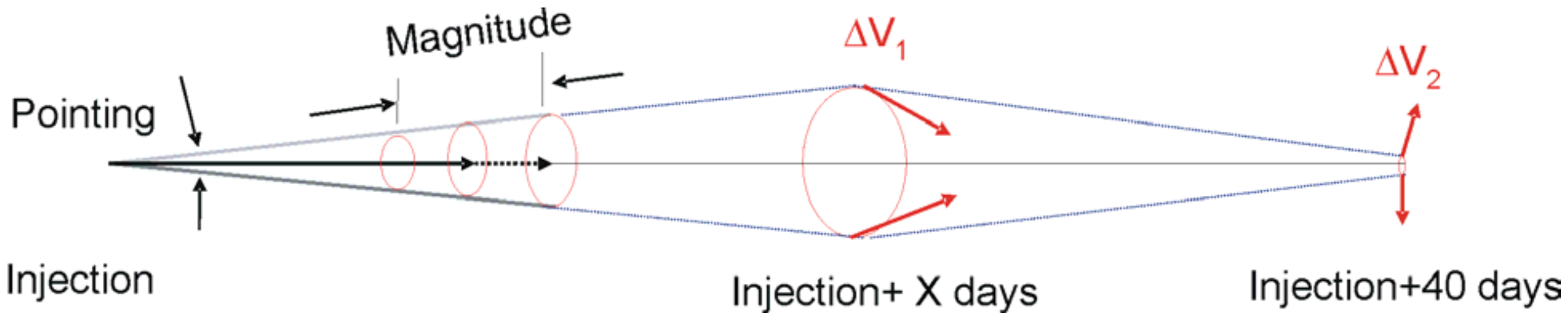

Figure 5.8.11. Position Error Growth After Injection.

As the spacecraft continues its orbit after injection, the pointing and specific impulse errors cause the spacecraft to deviate from the nominal trajectory in both position and velocity. Without any intervention, the spacecraft will not arrive at its destination. Therefore, the first correction maneuver is performed at some specified time after injection (the timing is discussed below) and is used to get the spacecraft back on the nominal trajectory at some later time. After the spacecraft is close to the nominal trajectory, another maneuver is performed to instantaneously correct the spacecraft's velocity to match the nominal trajectory velocity. The magnitude sum of these two burns gives the total statistical ΔV for the launch injection correction. This method of first correcting the position error at then the velocity error in two separate maneuvers is simple and provides conservative estimates for the statistical ΔV costs. However, a more detailed analysis can reduce the statistical ΔV cost by optimizing these two burns or by implementing an alternate method.

The amount of ΔV needed for the first correction maneuver is heavily dependent on the time between injection and the execution of the correction burn. The dependence on time compels the analysis to provide statistical ΔV values for several different maneuver execution times for the first burn after injection. Based on the Genesis maneuver plan and input from the supervisor of the Flight Path Control Group, Chris Potts, three maneuver schemes were developed to correct the launch injection errors from the third-stage motor. These three schemes offer the EPIC team a wide trade-space when designing the mission. The three schemes are identified as very conservative, semi conservative and less conservative. All three are described in Table 5.8.9.

**Table 5.8.9. Comparison Between Maneuver Schemes**

| | |
|---|---|
| Very Conservative: | First correction maneuver occurs at injection + 10 Days. Second correction maneuver occurs at injection + 40 days |
| Semi Conservative: | First correction maneuver occurs at injection + 10 days unless the necessary delta-v is greater than 130 m/s. If the delta-v exceeds the 130 m/s, the first correction maneuver is moved to injection + 4 days. Second correction maneuver occurs at injection + 40 days |
| Less Conservative: | First correction maneuver occurs at injection + 4 days unless the necessary delta-v is greater than 80 m/s. If the delta-v exceeds the 80 m/s, the first correction maneuver is moved to injection + 2 days. Second correction maneuver occurs at injection + 30 days. |

Using 5000 samples, the statistical ΔV values were generated once the maneuver schemes were created. Table 5.8.10 shows the results of the Monte-Carlo runs. For the semi and less conservative maneuver schemes, the contingency number represents the percentage of the 5000 runs that exceeded the maximum allowable ΔV and moved the first correction maneuver to an earlier time.

**Table 5.8.10. Statistical ΔV for Maneuver Schemes**

| | Avg. ΔV (m/s) | 1-σ (m/s) | Contingency |
|---|---|---|---|
| Very Conservative | 64 | 45 | 0% |
| Semi Conservative | 58 | 34 | 9% |
| Less Conservative | 38 | 22 | 8% |

*5.8.6. Post-Injection to Lissajous Orbit Insertion*

After the second injection correction maneuver, there are still residual errors in both the position and velocity of the spacecraft with respect to the nominal trajectory. Therefore, during the cruise portion of the mission there needs to be intermediate trajectory correction maneuvers (TCM's) to reduce the build-up of these errors over time. The position and velocity uncertainties used for this part of the statistical ΔV analysis are a position uncertainty of 50 meters and a velocity uncertainty of 2 cm/s. Both were assumed to have a normal distribution.

The estimation of the statistical ΔV necessary to complete the TCM's follows a similar methodology to the determination of the statistical ΔV for correcting the third-stage injection errors. Given the position and velocity uncertainties at a point in time, $T_1$, the uncertainties are propagated to a later point in time, $T_2$. A maneuver is performed at $T_2$ to correct the spacecraft's position at a future time, in this case $T_2$ + 5 days. Upon the spacecraft's arrival at $T_2$ + 5 days, a second maneuver is performed to correct the velocity at that time. The vector magnitude sum of the two maneuvers gives an estimate of the total ΔV for the TCM. Figure 5.8.12 illustrates the methodology used to approximate the statistical ΔV.

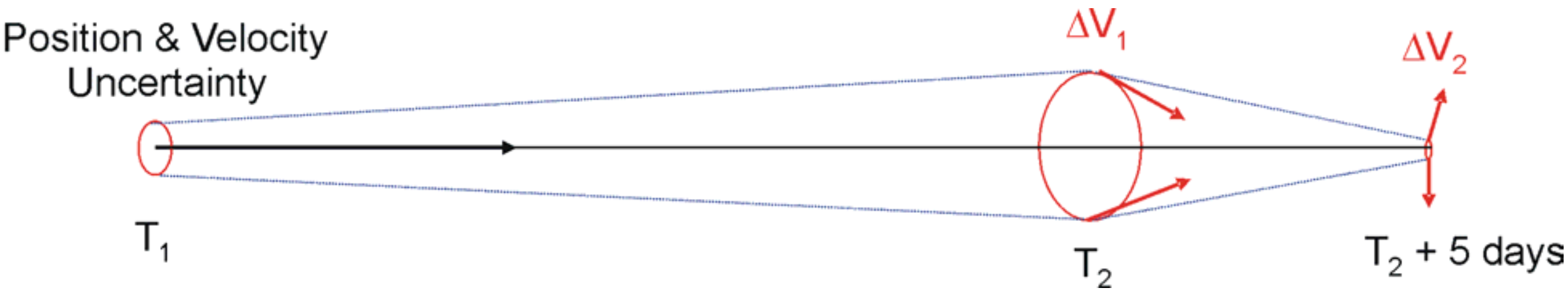

Figure 5.8.12. Maneuver Scheme for Transfer to Lissajous.

Four TCM's were modeled starting with the residual launch injection errors through insertion into the Lissajous orbit. The timing of the second TCM burns is shown in Figure 5.8.13. As with the launch injection analysis, the uncertainty values and the four pairs of TCM's were inputs into a Monte-Carlo simulation taking 5000 samples to produce an average statistical ΔV with a 1-sigma distribution. The result of the Monte-Carlo gives an average TCM ΔV of 7 m/s with a 3 m/s 1-sigma value. Considering that the nominal transfer trajectory has no deterministic TCM's the statistical ΔV value is equal to the total ΔV for transfer.

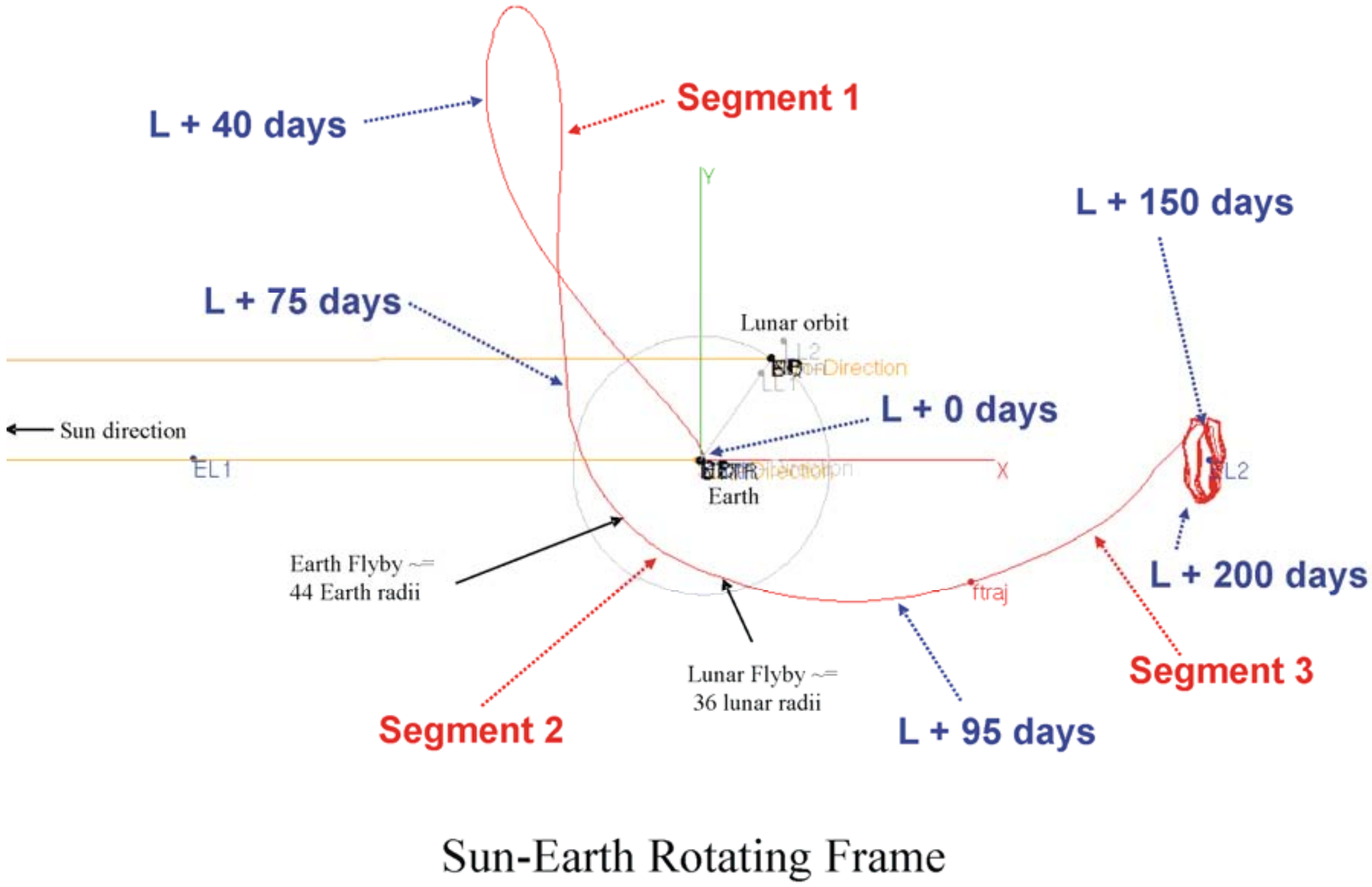

Figure 5.8.13. Transfer Analysis Diagram.

*5.8.7. Lissajous Station-Keeping Maneuvers*

Using a similar method to the two previously discussed, the statistical ΔV needed to maintain the Lissajous orbit was calculated using a position uncertainty of 50 m and a velocity uncertainty of 2 cm/s with normal distributions. Again, a 5000 sample Monte-Carlo simulation was used to generate the mean and one-sigma statistical ΔV values of 4 m/s and 1m/s, respectively.

*5.8.8. Statistical ΔV Summary*

By combining the three individual studies, an overall statistical ΔV budget can be generated. As a result of the multiple maneuver schemes tested to clean up the launch injection errors, there are three ΔV budget estimates for the nominal EPIC trajectory. Table 5.8.11 presents the collective results of the statistical ΔV analyses. Based on these preliminary results, the EPIC spacecraft will need to provide enough propellant to produce between 134 and 215 m/s.

As a reminder, the methods used to generate the values in Table 5.8.11 are only rough estimates. The analyses did not contain any maneuver optimization. However, with further time and analysis, these numbers may be reduced.

**Table 5.8.11. Summary of EPIC Statistical ΔV Budget**

| Event / Function | Very Conservative | | Semi Conservative | | Less Conservative (Genesis Compatible) | |
|---|---|---|---|---|---|---|
| | Mean | Sigma | Mean | Sigma | Mean | Sigma |
| Injection Cleanup | 64 | 45 | 58 | 34 | 38 | 22 |
| Post-Injection up to Lissajous Insertion (L + 200 days) | 7 | 3 | 7 | 3 | 7 | 3 |
| Sub Total | 72 | 45 | 65 | 34 | 45 | 21 |
| | | | | | | |
| Lissajous per Rev | 4 | 1 | 4 | 1 | 4 | 1 |
| 4-Years Sub Total | 31 | 10 | 31 | 10 | 31 | 10 |
| | | | | | | |
| Grand Total | 103 | 46 | 96 | 35 | 76 | 24 |
| Total 95% Probability | 195 | | 166 | | 124 | |
| Margin | 10 | | 10 | | 10 | |
| Total Budget | 215 | | 176 | | 134 | |

Note that the budget in Table 5.8.11 is based on a 4-year mission life. EPIC-LC, designed for 2-years, saves a modest ~16 m/s due to reduced station-keeping requirements.

*5.8.9. Future Work*

As stated before, the results from this analysis provide an adequate first-cut at the deterministic and statistical ΔV budget for the EPIC mission. As the mission design progresses, a more exhaustive analysis must be performed to increase the ΔV budget accuracy. This section provides a few suggestions on what should be considered in those future studies.

Given some characteristics of the spacecraft, i.e. solar-array size, spacecraft mass, etc., the effect of the solar radiation pressure on the spacecraft could be modeled and incorporated into the trajectory simulation. Although the perturbation force due to the solar radiation pressure is small compared to the gravitational forces of the Earth and Sun, it may be enough to force deterministic station-keeping maneuvers at the Lissajous orbit.

Currently, no other mission has gone into such a small Lissajous orbit as EPIC is proposing. Therefore, it would be prudent to have someone investigate these small orbits to uncover any potential unforeseen difficulties. In particular Roby Wilson has some background in Lissajous orbit stability.

As stated previously, the statistical ΔV values discussed were rough estimates and were not optimized to reduce the amount of ΔV. Furthermore, no intermediate deterministic TCM’s were included in the calculations, i.e. the TCM’s to extend the launch period. A maneuver specialist should be tasked to incorporate the TCM’s and optimize the maneuvers to provide better values for the statistical ΔV budget.

## 5.9 Mission Parameters and Spacecraft Definition

### *5.9.1 Scientific Operations*

EPIC carries out scientific observations from an L2 halo orbit. We reach L2 approximately 170 days after launch by means of a transfer orbit using lunar assist. The delta-V budget of 215 km/s includes 72 +/- 45 km/s for injection errors, a conservative trajectory correction strategy, and 4 years of orbit correction at L2. We take 95% probability on all maneuver errors add then include an additional 10% overall margin. The sunshield is deployed and the aperture lids are ejected early in the mission in order to reduce the heat load on the cryostat en route to L2.

**Table 5.9.1. Mission Design Summary**

| | |
|---|---|
| **Orbit** | L2 Halo |
| **Mission Life** | 1 year at L2 (required), 2 years at L2 (design) |
| **Maximum Eclipse Period** | 0 |
| **Spacecraft dry bus mass and contingency** | 713 kg, includes 43% contingency |
| **Spacecraft propellant mass and contingency** | 172 kg (ΔV budget and contingency shown in Table 5.8.11) |
| **Launch vehicle** | Atlas V 401, Delta IV 4040, option for Delta 2925H-9.5 Star 48 |
| **Launch vehicle mass margin** | 1702 kg (95%), 990 kg (56%) |

Once at L2, the instrument executes a single observing mode which consists of a spinning/precessing scan strategy (see Fig. 5.1.3 in foldout). This strategy provides uniform and redundant coverage of the sky and efficiently rotates the telescope direction on all regions of the sky. Data are transmitted to earth once per day via a toroidal-beam antenna, which enables downlink during observations without the use of a counter-rotating antenna. The single-shot adiabatic demagnetization refrigerator is cycled at regular intervals of 48 hours. The half wave plates in front of each telescope are stepped every 24 hours to remove the systematic effects of main beam asymmetries. With this repeated sequence of events and continuous observing, operations are rather simple. The sequence of operations are summarized in Tables 5.9.2 and 5.9.3.

**Table 5.9.2. Science Observations Operations**

| Mission Operation | Rate |
|---|---|
| Spin Spacecraft | Continuous, 0.1 - 3 rpm |
| Precess Spin Axis | Continuous, 1 rph |
| Step Wave plate | Once every 24 hours |
| Cycle ADR | Once every 48 hours |
| Downlink | Once every 24 hours |
| Maintain Orbit | Small maneuvers ~4 times per year |

**Table 5.9.3. Mission Operations and Ground Data Systems**

| Down link Information | Value, units |
|---|---|
| Number of Data Dumps per Day | 1 (baseline)<br>2 (TES option) |
| Downlink Frequency Band | 8.425 GHz (Near-Earth X-Band) |
| Telemetry Data Rate | 500 kbps (baseline)<br>4000 kbps (TES option) |
| S/C Transmitting Antenna Type(s) and Gain(s) | Toroidal-beam antenna, 9.0 dBi |
| Spacecraft transmitter peak power | 191 W (total power) |
| Downlink Receiving Antenna Gain | 58.7 dBi (baseline, 12-m DSN)<br>68.3 dBi (TES option, 34-m DSN) |
| Transmitting Power Amplifier Output | 100 W (RF power) |
| **Uplink Information** | **Value, units** |
| Number of Uplinks per Day | 1 |
| Uplink Frequency Band | 7.17 GHz |
| Telecommand Data Rate | 1 kbps at 45˚ |
| S/C Receiving Antenna Type(s) and Gain(s) | Low-gain omnis, 7.7 dBi boresight |

### *5.9.2 Payload and Spacecraft Resources*

A summary of the payload and spacecraft masses is listed in Table 5.1.9 in the foldout section. The total mass of the payload, including the deployable sunshield, support struts, X-band antenna, and LHe is 898 kg, which includes 43% contingency on all masses except LHe. A summary of the payload and spacecraft power requirements is listed in Table 5.9.4 below. The total payload power required is 272 W, including 43% contingency.

EPIC is sized for a Delta-II 2925-H-9.5 launch vehicle with a 3-m shroud, as shown in Fig. 5.1.1 in the foldout section. It now appears unlikely that this vehicle will still be available at the time of a launch opportunity, and even if it is available the future cost of a Delta-II appears to be comparable to an Atlas V 401 or Delta-IV 4040. Therefore we have based our masses, contingencies, and costs assuming an Atlas V 401 launch. We are equally compatible with a Delta-IV 4040. If a Delta-II 2925-H-9.5 is indeed a viable alternative, we can study a mission implementation for this vehicle. Our mass and volume requirements allow us to consider the possibilities of co-launch options or foreign launch vehicles.

**Table 5.9.4 Power Summary**

| Item | Power (CBE) [W] | Contingency [%] | Allocated [W] |
|---|---|---|---|
| Bolometer Electronics | 150 | 43 | 215 |
| ADR Electronics | 40 | 43 | 57 |
| **Subtotal Payload** | **190** | **43** | **272** |
| Attitude Control | 264 | 43 | 378 |
| C&DH | 69 | 43 | 99 |
| Power | 106 | 43 | 152 |
| Propulsion | 25 | 43 | 36 |
| Telecom (transmit mode) | 191 | 43 | 273 |
| Thermal | 31 | 43 | 44 |
| **Subtotal Spacecraft** | **686** | **43** | **981** |
| **Total Power** | **876** | **43** | **1253** |
| **GaAs Triple Junction Solar Panels** | | | |
| **Panel Area** | **Power [W]** | **Margin [%]** | **Margin [W]** |
| **4.0 m$^2$ Fixed at 45˚ Incidence** | 710 | | |
| **3.8 m$^2$ Deployed at 45˚ Incidence** | 670 | | |
| **Total** | **1380** | **10** | **127** |

### *5.9.3 Spacecraft Components*

We assume EPIC will operate with a custom-built commercial spacecraft bus. The spacecraft itself requires no new technology. A custom-designed X-band antenna producing a toroidal beam is baselined. This item would be provided equipment to the spacecraft vendor. EPIC requires a bus-mounted solar panel plus 6 hinged deployed panels on the sun-facing side of the bus. The deployable sunshield would be a provided payload element and is not part of the spacecraft.

We note that the requirements are close to the capabilities of a modified 'off-the-shelf' commercial bus. As an example, we show below the specifications of the Spectrum Astro-200HP spacecraft, the capabilities of which (from the RSDO catalog), are close to our specifications. For specificity, we compare our requirements to a modified SA-200HP, but at this stage in the project we have not selected an industrial partner and many options for a spacecraft are available.

**Table 5.9.5. Spacecraft Requirements and Capabilities**

| | RSDO Summary Capability | Units | Spectrum Astro SA-200HP | EPIC-LC Requirement |
|---|---|---|---|---|
| Compatibility | Payload Power (OAV) (EOL) | W | 650 | 272<br>(incl. 43% cont.) |
| | Payload Mass Limit of Bus | kg | 666 | 898<br>(incl. 43% cont.) |
| | Bus Dry Mass (w/o Payload) | kg | 354 | |
| | Science Data Downlink Capability | kbps | 50,000 (X-band) | 500 (baseline)<br>4,000 (TES option) |
| | Science Data Storage Capability | Gbit | 100 | 16 (baseline)<br>215 (TES option) |
| | Pointing Knowledge | arcsec | 0.5 | 30 |
| | Pointing Control | arcsec | 16 | 3600 |

| RSDO Summary Capability | | Units | Spectrum Astro SA-200HP | EPIC-LC Requirement |
|---|---|---|---|---|
| | Pointing stability (jitter) | arcsec/s | 0.1 | 20 |
| | Slewrate | deg/min | 120 | 360 (baseline)<br>1080 (TES option) |
| | Mission Design Life | yrs | 4 | 2 |
| | Compatible LVs | | Taurus, Athena I, Athena II, Delta II, Titan II, Atlas | Atlas V 401, Delta IV 4040, Delta II |
| | Types of Orbit Available | | LEO circular (nominal), many other orbits available | Earth-Sun L2 |
| | Internal Volume Available for Payload | | 100 cm dia. x 75 cm tall | Sufficient for warm electronics |
| **Description** | Attitude Control System | | 3-axis zero momentum bias/thruster based management | 3-axis momentum compensated |
| | Batteries | type/Ah | Two $NiH_2$ 50 Ah each | Two at 24 Ah each |
| | Arrays | Type/ area | Triple junction GaInP/GaAs/Ge 10.32 $m^2$ | Triple junction GaAs<br>4.0 $m^2$ body mounted<br>3.8 $m^2$ deployed |
| | Nominal Voltage | V | 28 | 28 |
| | C&CH Bus Architecture | | VME-based 32-bit RISC | 422 or 1553 |
| | Downlink Formats | | CCSDS: STDN/DSN | CCSDS |
| | Downlink Band | | X-band and S-band | X-band |
| | Structure | | Octagonal, Al space frame construction with honeycomb | Al or composite |
| | Propulsion | | Blowdown hydrazine | Hydrazine |
| | Propellant Capacity | kg | 67 | 172 |
| | Mass Delta-V | m/s | 131 | 215 |
| **Programmatic** | Heritage Missions | | New Millennium Deep Space 1 | |
| | Nominal Schedule | months | 36 | 36 |
| | Contract Options | | Full Redundancy | Replace S/C telecom with toroidal antenna |
| | | | Deep Space Configuration | Body mounted and deployed solar panels |
| | | | Ground Segment Integration Support | Add momentum wheel in spin axis |
| | | | | Modify propulsion tanks |
| | | | | Modify mechanical support |

We carried out a team-X study to assess and cost spacecraft components. This study assumes a commercially custom-built spacecraft bus. All the components required are space-proven technologies, either entirely off-the-shelf or with minor modifications. A summary of the component requirements is given in Table 5.9.6. An estimate of subsystem masses and power requirements are given in Table 5.9.7. Note that we have applied a conservative 43% contingency even to all spacecraft components.

**Table 5.9.6. Spacecraft Characteristics**

| | Spacecraft bus | Value/ Summary, units |
|---|---|---|
| **Structure** | Structures material | Aluminium or composite |
| | Number of articulated structures | None |
| | Number of deployed structures | 4 deployed solar panels |
| **T/C** | Type of thermal control used | Passive |
| **Propulsion** | Estimated delta-V budget | 215 m/s |
| | Propulsion type(s) and associated propellant(s)/oxidizer(s) | Hydrazine |
| | Number of thrusters and tanks | One 25 N Main Thruster<br>Twelve 0.9 N RCS Thrusters<br>One tank |
| | Specific impulse of each propulsion mode | 220 s |
| **Attitude Control** | Control method | 3-axis, momentum compensated |
| | Control reference | Inertial |
| | Attitude control capability | 1.0 deg |
| | Attitude knowledge limit | 30 arcsec |
| | Agility requirements | None |
| | Articulation/#–axes | None |
| | SENSORS:<br>Sun Sensors (8)<br>Star Trackers (2)<br>IMU (1)<br><br>ACTUATORS:<br>Reaction Wheels (4)<br>Momentum Wheels (4) | <br><br>1 arcsec accuracy<br>0.003 deg/hr stability<br><br><br>20 Nms momentum, 0.1 Nm torque<br>60 Nms momentum, 0.14 Nm torque |
| **C & DH** | Spacecraft housekeeping data rate | 10 kbps |
| | Data storage capacity | 16 Gbits (baseline)<br>215 Gbits (TES option) |
| | Maximum storage record rate | 98 kbps (baseline)<br>1270 kbps (TES option) |
| | Maximum storage playback rate | 500 kbps (baseline)<br>4000 kbps (TES option) |
| **Power** | Type of array structure | 4.0 $m^2$ body-mounted solar panels<br>3.8 $m^2$ hinged solar panels |
| | Array size, meters x meters | 7.8 $m^2$ |
| | Solar cell type | Triple-junction Ga-As |
| | Expected power generation | 1511 W BOL; 1380 W EOL |
| | On-orbit average power consumption | 981 W (incl. 43% contingency) |
| | Battery type | Li-Ion (two) |
| | Battery storage capacity | 50 Ah |

NOTE: the values supplied in this table are the EPIC requirements -- not the specifications for any particular implementation. The vendor for the spacecraft bus for this mission has not yet been selected.

**Table 5.9.7. Spacecraft Sub-System Characteristics**

| S/C Subsystem | Mass [kg, CBE] | Mass Ctgcy. [%] | Power [W, CBE] | Power Ctgcy. [%] |
|---|---|---|---|---|
| Attitude Control System | 81.9 | 43 | 264 | 43 |
| C&DH | 24.1 | 43 | 69 | 43 |
| Power | 52.6 | 43 | 106 | 43 |
| Propulsion (dry) | 22.1 | 43 | 25 | 43 |
| Structures and mechanisms | 212.9 | 43 | | |
| Launch adapter | 14.3 | 43 | | |
| Cabling | 46.4 | 43 | | |
| Telecom + X-band Antenna | 18.7 | 43 | 191 | 43 |
| Thermal | 25.5 | 43 | 31 | 43 |
| Propellant [$\Delta V$ = 215 m/s] | 172.0 | N/A | | |

## 5.10 Telemetry

We have estimated the telemetry band requirements for a range of options. These requirements can be met with a fixed low-gain X-band toroidal-beam antenna in conjunction with a 12-m or 30-m ground station. Downlink time per day is set by the currently available downlink bandwidth of 4 Mbps.

### *5.10.1 Input Data Rates*

We calculate data rates for two cases for EPIC-LC, with and without continuous waveplate modulation. The baseline case is scan modulation, where we step the waveplate every day only to mitigate beam effects, not for signal modulation. We investigate a variety of spin rates appropriate to the noise stability of the focal plane detectors. In the case of NTD Ge, noise stability has been demonstrated to a 1/f knee < 16 mHz for Planck, so a spin rate of 1 rpm is acceptable. For TES detectors, less is known about 1/f noise in real systems, although 40 mHz has been demonstrated in the lab (see Fig. 5.5.8). We use a maximum spin rate of 3 rpm in this case, which is starting to have an impact on the telemetry rate (a 30 m ground station is required) and choice of momentum wheels (exceeding commercial off-the-shelf wheels although much larger custom wheels are available). These are not hard constraints, and they could be overcome if it were absolutely necessary to spin faster than 3 rpm. Continuous modulation requires technological development, but offers an advantage in providing an additional level of signal modulation, thereby reducing 1/f stability requirements for the focal plane detectors and readouts. In this case we reduce the scan speed to 0.1 rpm to minimize the wave-plate spin rate requirements, and assume noise stability post waveplate demodulation to 1.6 mHz, the time scale for a spin to complete. The final choice of spin rate must be decided on the basis of detector noise stability with and without modulation.

<u>Case 1: Scan Modulated.</u> Assume we sample each detector at

$$\nu_s = 4\,[d\theta/dt \,/\, \theta_{FWHM}],$$

which corresponds to 2x the Nyquist sampling of the characteristic low-pass 3 dB frequency needed to avoid significant beam smearing, $[d\theta/dt \,/\, \theta_{FWHM}]$. The detectors must be fast enough such that

$\tau_{req} < 1/2\pi\ [\theta_{FWHM}\ /\ d\theta/dt]$.

The angular scan rate is given by

$d\theta/dt = 360 \sin(\theta_s)\ [\omega_s/1\ \text{rpm}]\ [\text{arcmin/s}]$,

where $\theta_s$ = spin cone angle = 55˚. Furthermore assume we sample each detector at 4 bits per sample, the same compression used on Planck. We assume spin rates of 1 rpm for the NTD Ge option and 3 rpm for the TES option in order to minimize effects from detector 1/f noise. At 3 rpm, the detector 1/f knee frequency would ideally be < 50 mHz.

**Table 5.10.1. Input Data Rate for NTD Ge Focal Plane**

| Freq [GHz] | Beam [arcmin] | Ndet [#] | $\tau_{req}$ [ms] | Sample Rate [Hz] | Data Rate [kbps] |
|---|---|---|---|---|---|
| 30 | 155 | 8 | 84 | 8 | 0.2 |
| 40 | 116 | 54 | 63 | 10 | 2 |
| 60 | 77 | 128 | 42 | 15 | 8 |
| 90 | 52 | 256 | 28 | 23 | 23 |
| 135 | 34 | 256 | 19 | 34 | 35 |
| 200 | 23 | 64 | 12 | 51 | 13 |
| 300 | 16 | 64 | 8 | 76 | 20 |
| **Total** | | **830** | | | **100** |

[1]Assumed spin rate of 1 rpm.
[2]Requires a 1/f knee < 16 mHz.
[3]Assumes 4 bits per sample per detector, and 2x Nyquist sampling.

**Table 5.10.2. Input Data Rate for TES Focal Plane**

| Freq [GHz] | Beam [arcmin] | Ndet [#] | $\tau_{req}$ [ms] | Sample Rate [Hz] | Data Rate [kbps] |
|---|---|---|---|---|---|
| 30 | 155 | 8 | 28 | 23 | 1 |
| 40 | 116 | 54 | 21 | 31 | 7 |
| 60 | 77 | 128 | 14 | 46 | 23 |
| 90 | 52 | 512 | 9 | 69 | 141 |
| 135 | 34 | 512 | 6 | 103 | 211 |
| 200 | 23 | 576 | 4 | 150 | 351 |
| 300 | 16 | 576 | 3 | 230 | 527 |
| **Total** | | **2366** | | | **1260** |

[1]Assumed spin rate of 3 rpm.
[2]Requires a 1/f knee < 50 mHz.
[3]Assumes 4 bits per sample per detector, and 2x Nyquist sampling.

Case 2: Waveplate Modulated. Assume in this case that we have N modulations of the polarization signal per beam crossing time. Conservatively we take N = 10, to avoid possible effects of mismatched beams, but this needs further simulation. The polarization signal is at a characteristic frequency $\nu_{pol} = N\,[d\theta/dt\ /\ \theta_{FWHM}]$, and we assume we sample this at 2x Nyquist.

The bolometers must be fast enough to respond to the polarization signal, taken as

$\tau_{req} < (1/4\pi\nu_{pol})$.

Furthermore assume we sample each detector at 4 bits per sample. This is the compressed sampling rate used on Planck, but needs to be checked for EPIC. As shown in Table 5.10.1, detector time constants are not an issue for TES bolometers. The waveplate rotates at $\nu_{pol}/4$. Because the waveplate modulates the signals at high frequencies, it stabilizes the detector system by classical switching, and the spacecraft spin rate can be reduced. We assume a spin rate of 0.1 rpm. Because 200 and 300 GHz detectors share common optics, we quote the rates based on the more demanding requirement at 300 GHz.

**Table 5.10.3. Data Rate with TES Detectors and Waveplate Modulation**

| Freq [GHz] | Beam [arcmin] | Ndet [#] | $\nu_{pol}$ [Hz] | $\tau_{req}$ [ms] | HWP R/R [rpm] | Sample Rate [Hz] | Data Rate [kbps] |
|---|---|---|---|---|---|---|---|
| 30 | 155 | 8 | 2.5 | 63 | 38 | 10 | 0.3 |
| 40 | 116 | 54 | | | | | 2 |
| 60 | 77 | 128 | 3.8 | 42 | 60 | 15 | 8 |
| 90 | 52 | 512 | 5.7 | 28 | 90 | 23 | 47 |
| 135 | 34 | 512 | 8.6 | 19 | 130 | 34 | 70 |
| 200 | 23 | 576 | 19.1 | 8.3 | 290 | 76 | 176 |
| 300 | 16 | 576 | | | | | 176 |
| **Total** | | **2366** | | | | | **480** |

[1]Assumed spin rate of 0.1 rpm, and 10 polarization cycles per beam crossing.
[2]Requires a 1/f knee < 2.5 Hz pre-demodulation.
[3]Assumes 4 bits per sample per detector, and 2x Nyquist sampling.

The downlink requirements, calculated in section 5.10.2 below, are summarized in Table 5.10.4 for all of the options. The baseline NTD Ge case could be accommodated with a 12-m ground station; all of the other options require a 34-m antenna. In the case of the 34-m antenna, the downlink budget is sized to fill the currently available maximum downlink rate of 4 Mbps. If this rate goes up, then the downlink budget can be increased and the downlink time per day can be reduced.

*5.10.2 Downlink Requirements*

We calculated the downlink requirements for the various cases above assuming the link budget calculations described in appendix D. These calculations were undertaken for the toroidal-beam antenna described in section 5.10.3 with the orbital parameters described in section 5.8. Appendix D carries out additional calculations for the full range of antennas, transmitters, and available bands for an earlier L2 orbit with a larger halo diameter. The antenna described in section 5.10.3 was specifically designed for a smaller halo. The downlink calculations are summarized in Table 5.10.4.

**Table 5.10.4. Telemetry and Downlink Requirements**

| Option | Spin rate [rpm] | Wave plate spin rate | Input rate[1] [kbps] | Downlink time per day [hrs] | |
|---|---|---|---|---|---|
| | | | | 12-m DSN | 34-m DSN |
| **Baseline** Scan-modulated NTD bolos[2] | 1.0 | step 22.5˚ per day | 100 | 4.8 | 0.6 |
| **Option** Wave plate-modulated TES bolos[3] | 0.1 | 40 – 300 rpm | 480 | - | 2.8 |
| **Option** Scan-modulated TES bolos[4] | 3.0 | step 22.5˚ per day | 1260 | - | 7.4 |

Notes:
[1]Assumes 4 bits per sample per detector (Planck compression ratio) with Nyquist sampling, plus 100% contingency.
[2]Requires a 1/f knee < 16 mHz (already demonstrated for NTD bolometers).
[3]Assumes 10 polarization cycles per beam crossing for each band. Requires 1/f knee < 2.5 Hz.
[4]Requires a 1/f knee < 50 mHz (near state-of-the-art for TES bolometers).

### *5.10.3 Low-Gain Torroidal-Beam X-band Antenna*

The downlink requirements in Tables 5.10.1 – 5.10.3 can be easily accommodated with a small gimballed antenna, which must counter spin continuously to counteract the spinning motion of the spacecraft. It must also be able to slowly steer in elevation to take out variations on the earth-spacecraft angle due to the size of the L2 halo. In order to eliminate any possible risk from a counter-spinning antenna, we have instead developed a low-gain X-band antenna which provides a toroidal beam shape with an opening angle of 45˚ and a beam width of 2˚. For the L2 halo orbit described in section 5.8, the earth always remains in the beam of this antenna. In this configuration, the antenna may be simply fixed and hard mounted to the back of the spacecraft.

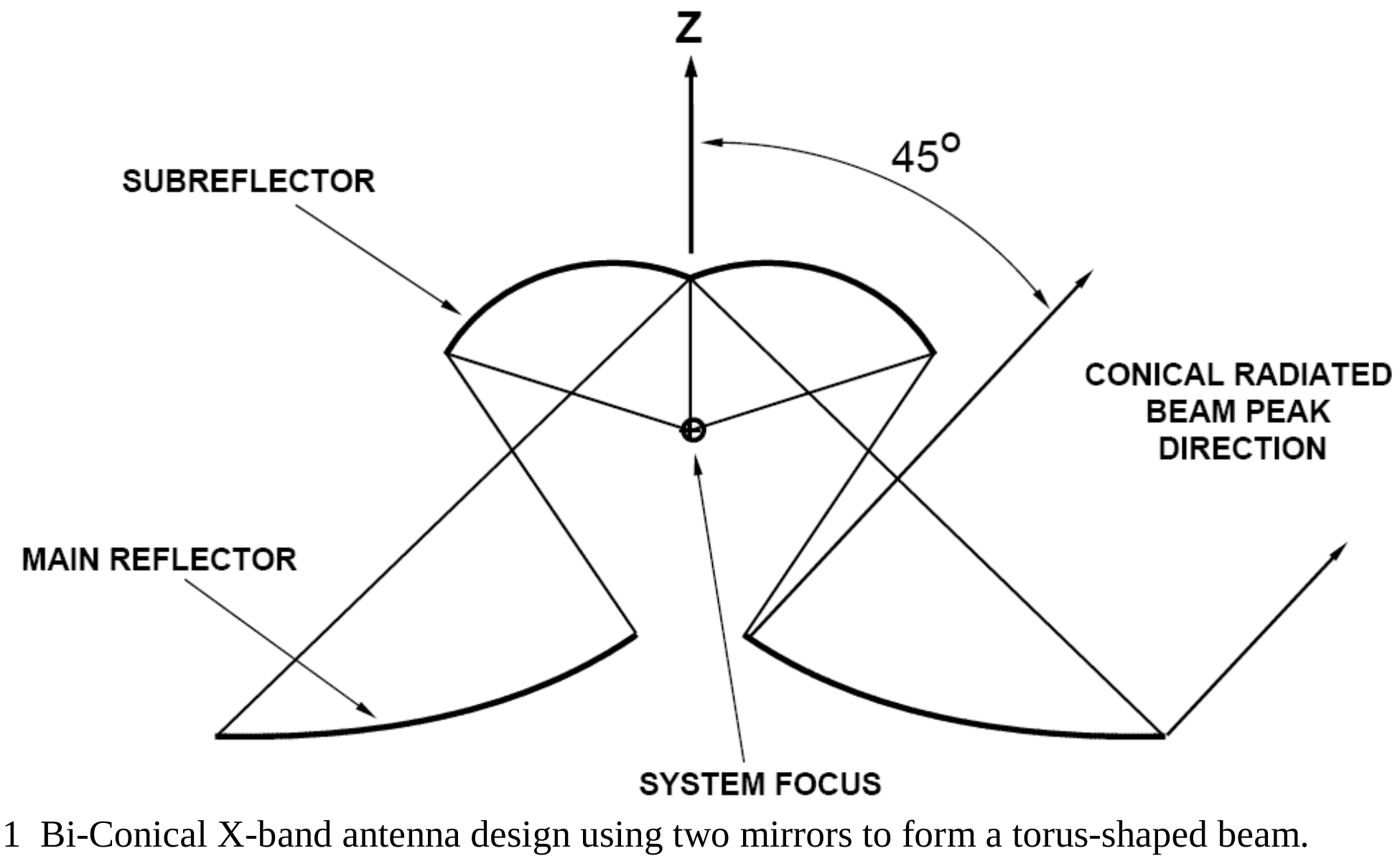

Figure 5.10.1 Bi-Conical X-band antenna design using two mirrors to form a torus-shaped beam.

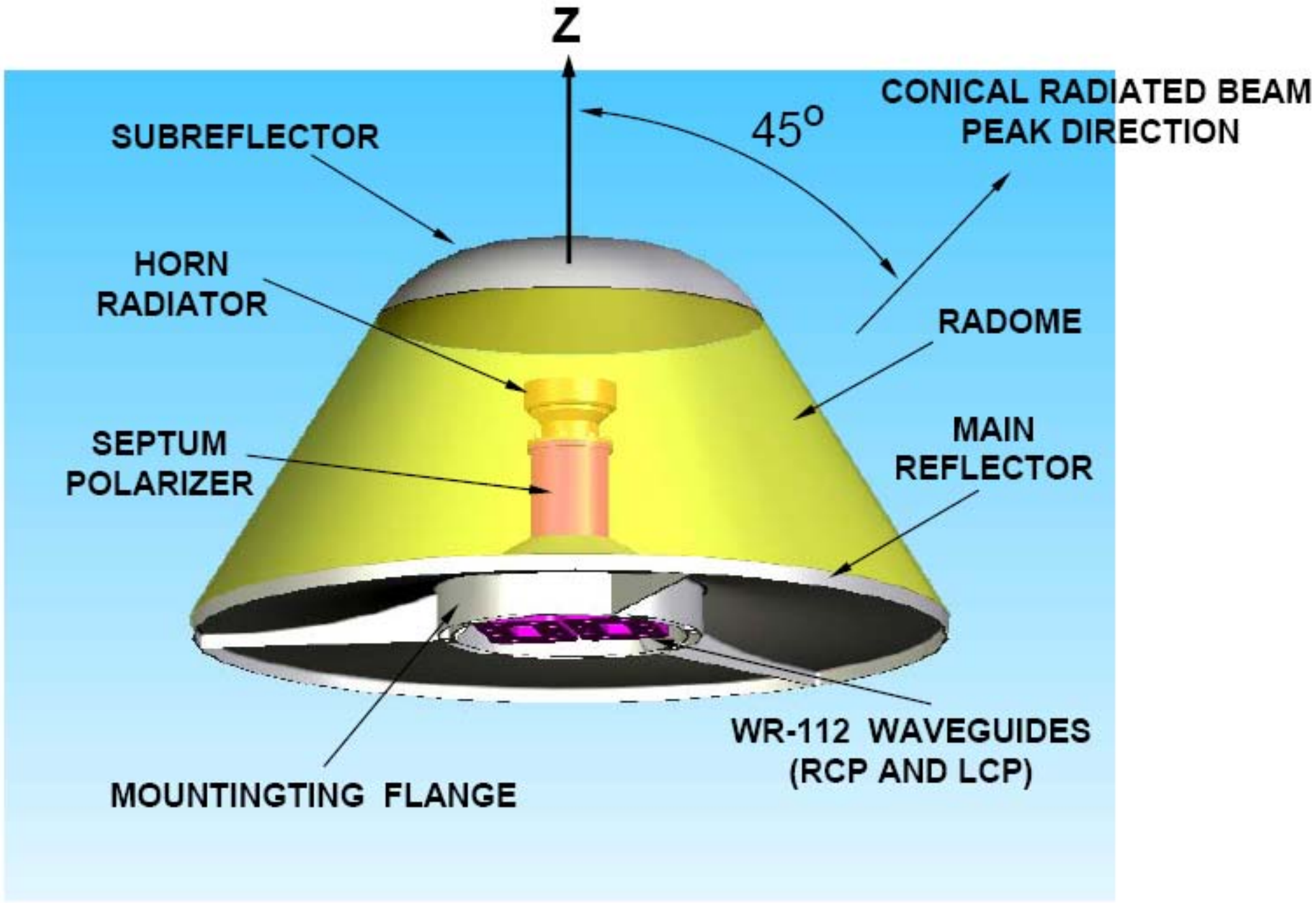

Figure 5.10.2. Mechanical configuration of antenna. RF power is channeled into standard waveguide flanges on the bottom mounting flange. The subreflector is supported by a radome.

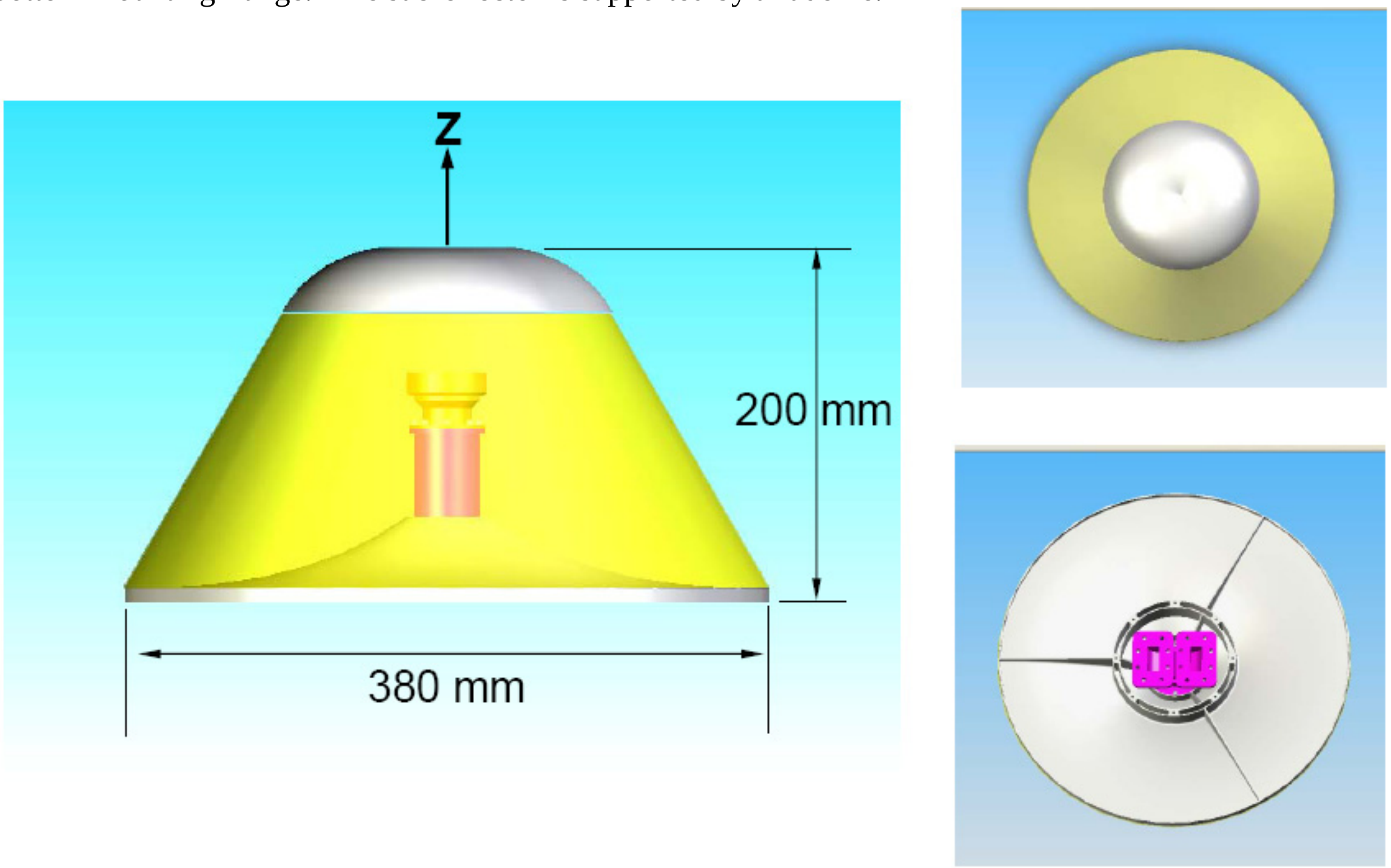

Figure 5.10.3. Dimensions and views of the antenna.

The radiation pattern associated with this antenna is shown in Figure 5.10.. The gain accounts for the losses associated with mismatch loss, ohmic loss, polarizer loss and mechanical errors.

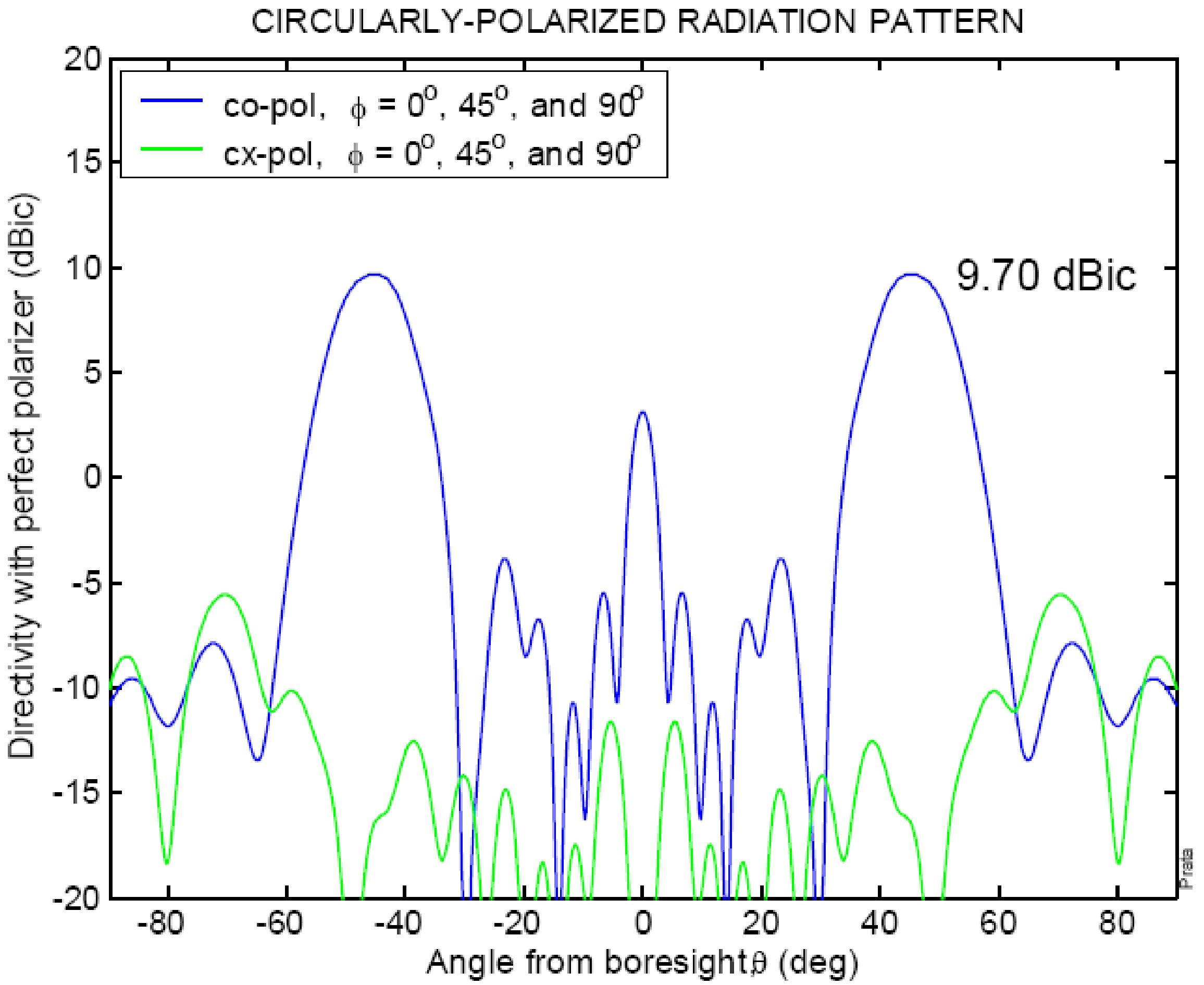

Figure 5.10.4. X-band bi-conical antenna radiation pattern.

**Table 5.10.5. Summary of Antenna Gain and Losses**

| | |
|---|---|
| Directivity at 45˚ | 9.697 dB |
| Radome losses | 0.050 dB |
| Surface losses (for 125 μm rms) | 0.004 dB |
| Feed Joule losses | 0.150 dB |
| Return loss (16 dB) | 0.110 dB |
| Polarization loss ($AR_{TX}$ = $AR_{RX}$ = 1.7 dB)* | 0.100 dB |
| Mechanical alignment | 0.050 dB |
| CP Gain at 45˚ | 9.23 dB |
| CP Gain at 43˚ | 9.0 dB |
| CP Gain at 47˚ | 9.1 dB |

* The total maximum loss due to the 1.7 dB imperfect axial ratio is 0.2 dB, but only half of it is included (the other half is left to the ground antenna).

The toroidal-beam antenna (with 12m or 34m ground antenna) requires a 100 W amplifier on the spacecraft. This amplifier would be similar to 100 W TWTA used on the Mars Reconnaissance Orbiter (MRO). Figure 5.10.5 shows the TWTA used on MRO. The mass of this TWTA is about 3 kg with about 50% efficiency thus requiring input power of 200 W.

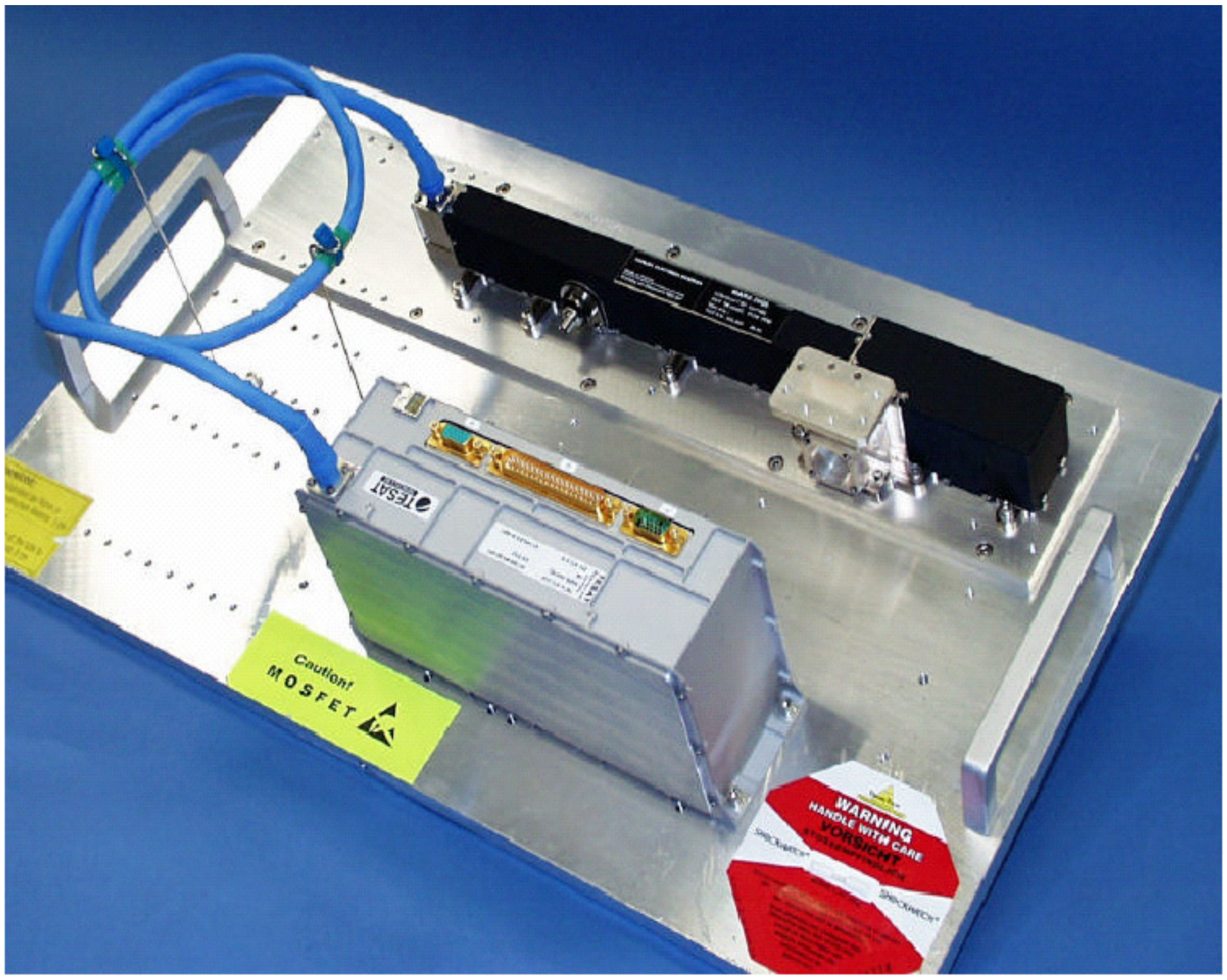

Figure 5.10.5. X-band 100W TWTA used on Mars Reconnaissance Orbiter. This unit requires minor modifications to convert from Deep Space X-band frequency to Near-Earth X-band frequency.

## 5.11 Cost Analysis

We have carried out a preliminary cost estimate and schedule. As it is beyond the resources of this study to provide a full grass-roots schedule and cost analysis, we estimate these parameters in analogy with similar missions such as *Spitzer*.

### *5.11.1 Project Schedule*

Our development schedule for the instrument is 18 months for phase A, 12 months for phase B, and 48 months for phase C/D. The spacecraft phase C/D is assumed to be 37 months, and decoupled from the instrument phase C/D. The phase C/D duration is longer than a typical mission, appropriate due to the cryogenic nature of the instrumentation. This schedule was adopted in analogy with phase C/D plans for similar missions, *WISE* and *Spitzer*, and is longer than the planned phase C/D for either mission, but shorter than the actual phase C/D of *Spitzer* (notable delays having to do with funding and launch vehicle availability).

<u>*Cryogenic System*</u>: Our planning is chosen to avoid costly and time-consuming cryogenic system tests at a high level of integration. EPIC-LC's design minimizes thermal complexity and maximizes parallel development. The telescope and detector subsystems are developed and tested outside of the flight cryostat. The flight cryostat is thermally tested at a representative shell temperature prior to integration with the telescope assemblies. In the instrument integration program, the flight cryostat is operated with the shell at room temperature. The passive thermal system of the v-groove radiators and sunshield is part of the instrument. The thermal interface to the spacecraft is simple:  the base of the bipod supports.  There are no heat pipes or cooling systems integral with the spacecraft.  All thermal interfaces are under the direct hardware control of the instrument team.

The thermal system is verified by a combination of testing and modeling. There is sufficient margin on the cryostat lifetime to compensate for the uncertainties in the modeling and for parameters that will not be verified in test.  Missions like *Spitzer* demonstrate that radiatively cooled cryogenic systems are successful when built with adequate margins combined with analysis and a limited test program.

Following this approach, the passive cooling system is tested independently of the spacecraft or cryostat, greatly simplifying tests at final integration.  To verify the performance of the passive cooling system, we will measure the infrared and thermal properties of the materials and carry out a thermal balance test on a scale model of the radiators and sunshield.

As the spacecraft and payload are integrated and tested separately, the cryogenic payload does not impose unusual or demanding requirements on the spacecraft bus during development and testing.  After the payload and spacecraft have been integrated, a thermal balance test of the spacecraft will be conducted, since the payload and sunshield thermal performance will be already verified separately.  This final test will be comparable in scope and complexity to that which would be performed with a non-cryogenic payload.

*Sunshade*:  The EPIC-LC sunshield is a deployed 3-layer 8-m shield with a hinged strut system. This system is significantly simpler and smaller than the 5-layer 22-m shield being developed for *JWST*.  We will use a gravitational offload system to comprehensively test the deployment of the structures and kapton membranes.  Gravitational offload techniques have been used to carry out testing of numerous structures of this size, mostly deployable antennas, with a high rate of success.  Several aerospace companies currently have this test capability.  Kapton membranes of the flight design and folding arrangement are integral to the deployment test.  Offloading the weight of the membranes is not a significant concern.  We will also test venting of the folded membranes in stowed configuration.

It is important to note that the sunshield only plays a minor role in passive cooling.  The sunshield does no thermal ‘work’, since it does not conduct or dissipate heat from the supports or wiring, which are the dominant heat paths to the cryostat.  Radiated power from the inner sunshield contributes a negligible fraction (1%) of the total heat load on the cryostat shell. Hence the cryostat shell temperature and the lifetime are both insensitive to the temperature of the sunshield.  The sunshield’s function is to block radiation from the sun and warmer sunshield layers from viewing the cryostat and optics, and to simply reflect radiated thermal power from the internal V-groove coolers to space.  Thus thermal tests of the sunshield can be limited in scope, a scale model to test thermal properties.

*Spacecraft*:  The spacecraft bus requirements are not demanding, and can be accommodated using a modified commercial bus.  Modifications include the addition of larger momentum wheels, a strengthened bus structure, body-mounted solar panels, and the JPL-provided toroidal antenna.  All spacecraft components (except the antenna) are off-the-shelf, flight-proven commercial hardware.  The 37-month spacecraft development schedule was provided by JPL's TeamX, which assumed a custom built bus using off-the- shelf commercial components for all subsystems.

*5.11.2 Cost Estimate*

Costs were generated by JPL's Advanced Concurrent Engineering Design Team (Team X), which includes experts in science, mission design, instruments, programmatics, ground system, and every spacecraft subsystem. Team members synthesize their own expertise and discipline-specific models to generate complete mission studies including cost details. JPL has used Team X to generate well over 600 project studies.

The Parametric Mission Cost Model (PMCM) is widely used for estimating project costs. It is comprised of a series of cost estimating relationships (CERs) that represent the cost of each project WBS element. The CERs were derived by multiple regression techniques from about 150 (Team X) studies. CERs take into account the key engineering technical drivers that affect mission cost. PMCM has been validated against the costs of actual missions flown by JPL.

Prior to the team-x session, the instrument costs for the deployable sunshade, antenna, cryostat, telescopes, focal plane detector arrays, and warm and cold readout electronics were calculated based on a grassroots basis by the team members involved in their design. These costs were scaled from actual costs on similar hardware delivered for Planck and Herschel where applicable. The grassroots cost for the instrument was $145M (FY07), so we instead used the larger team-X model-based instrument cost of $158M (FY07) in the above table.

**Table 5.11.1 Total Mission Cost Funding Profile**

| Item | FY09 | FY10 | FY11 | FY12 | FY13 | FY14 | FY15 | FY16 | Total (RY) | Total (FY07) |
|---|---|---|---|---|---|---|---|---|---|---|
| Phase | A | A-B | B-C/D | C/D | C/D | C/D | C/D-E | E | | |
| Concept Study | 0.1 | 2.7 | 1.3 | - | - | - | - | - | 4.1 | 3.7 |
| Science | 0.0 | 0.1 | 0.6 | 2.4 | 3.5 | 3.6 | 6.0 | 8.4 | 24.7 | 19.6 |
| Instrument | 0.1 | 1.2 | 9.4 | 37.2 | 55.3 | 57.1 | 31.1 | - | 191.4 | 157.9 |
| Spacecraft | 0.1 | 1.0 | 8.1 | 31.9 | 47.4 | 48.9 | 26.6 | - | 164.0 | 135.3 |
| Ground Data System Dev | 0.0 | 0.1 | 1.2 | 4.6 | 6.8 | 7.0 | 3.8 | - | 23.4 | 19.3 |
| MSI&T [3] | 0.0 | 0.1 | 0.4 | 0.3 | 0.4 | 3.9 | 4.1 | - | 9.2 | 7.4 |
| Launch services | - | - | - | 28.4 | 52.6 | 54.2 | 29.5 | - | 164.7 | 135.0 |
| MO&DA [4] | - | - | - | - | - | - | 5.8 | 12.0 | 17.8 | 13.7 |
| Education/Outreach | 0.0 | 0.0 | 0.0 | 0.1 | 0.2 | 0.2 | 0.7 | 1.3 | 2.6 | 2.0 |
| Reserves | 0.0 | 0.9 | 7.0 | 27.8 | 41.4 | 42.6 | 24.9 | 3.5 | 148.3 | 122.1 |
| Project Management | 0.0 | 0.1 | 0.6 | 2.4 | 3.5 | 3.7 | 2.3 | 0.7 | 13.1 | 10.7 |
| Project System Engineering | 0.0 | 0.1 | 0.9 | 3.4 | 5.1 | 5.2 | 2.8 | - | 17.5 | 14.4 |
| Safety Mission Assurance | 0.0 | 0.1 | 1.0 | 3.8 | 5.6 | 5.8 | 3.2 | 0.1 | 19.5 | 16.1 |
| Total Cost | **0.3** | **6.3** | **30.4** | **142.3** | **221.7** | **232.0** | **146.9** | **26.1** | **800.2** | **657.4** |
| Total Contributions | - | - | - | - | - | - | - | - | - | - |
| | | | | | | | Total Mission Cost | | | **657.4** |

1 FY costs in Real Year Dollars, Totals in Real Year and 2007 Dollars
2 Costs should include all costs including any fee
3 MSI&T - Mission System Integration and Test and preparation for operations
4 MO&DA - Mission Operations and Data Analysis

NAFCOM In order to validate the costs for the EPIC Mission, the costs were cross checked using NAFCOM v.2006, build date 4/18/2006. The NAFCOM costs for this mission were estimated to be $706M (FY07) after applying a 30% reserve, in good agreement with the above cost table. The inputs to NAFCOM were based on the mass and power summaries assuming an unmanned, earth-orbiting, scientific mission category.

Analogy to *Spitzer*. In order to further cross-check our cost estimate, we carried out a comparison to the actual costs of a similar cryogenic mission. The best example available was *Spitzer*, an infrared great observatory with a suite of 3 science instruments launched in 2003 with a cost of $1075M (FY07) for phases A-E without extended operations. We applied the following reductions to the *Spitzer* actual costs: 1) change phase E from 30 to 18 months; 2) scale the instrument development to a 48 month phase C/D from a 66 month phase C/D; 3) reduce the instrument requirements (3 instruments with a near-infrared diffraction-limited Be telescope to a single 100 mK instrument with mm-wave optics); 4) scale the spacecraft based on the less demanding pointing, control and data rate requirements for EPIC; 5) reduce the flight software for a single operating mode; 6) reduce the science management costs from that of a great observatory. Then we made the following additions: 1) add deployable sunshade cost; 2) add custom antenna; and 3) add higher launch vehicle cost. The estimate based on these adjustments agrees within 10% of the above cost estimate, although we must emphasize that the adjustments are significant due to the dissimilarity of the two missions.

*Disclaimer: The total estimated mission cost provided here are for budgetary and planning purposes only and does not constitute a commitment on the part of Caltech/JPL.*

# 6. Comprehensive Science Mission Option

A larger 3-m telescope aperture allows access an order of magnitude higher cutoff in $\ell$, providing access to the science themes described in section 1 and Table 4.1.1 that are inaccessible with a 30-cm telescope. These themes, measuring the scalar E-mode polarization signal to cosmological limits into the damping tail, using the lensing polarization signal to probe neutrino masses and dark energy, studying SZ clusters, and mapping Galactic polarization through polarized dust emission, broaden the secondary science case for EPIC, which otherwise must rely on combining high-resolution ground-based data with low-resolution all-sky space-borne measurements. Higher resolution also allows us to substantially subtract the foreground lensing polarization signal through precision measurements of the temperature anisotropy, allowing us to probe to lower values of the tensor to scalar ratio. We designate this 3-m version the *comprehensive science* option, or *EPIC-CS*.

## 6.1 Comprehensive Science Mission Overview

### *6.1.1. Instrument Requirements*

EPIC-CS is designed to carry out 1) a deep search for IGW B-mode polarization after lensing subtraction, and 2) the full secondary science themes described in section 1. We calculate the sensitivity and resolution required for subtraction of the lensing B-mode foreground using higher-order statistics from temperature anisotropy, as shown in Fig. 6.1.1. With lensing subtraction, the depth probed in r is a function both of the sensitivity and resolution, and while the subtraction continues to improve with higher resolution and sensitivity, most of the improvement is realized for a sensitivity $w_p^{-1/2}$ = 1-2 μK-arcmin and a resolution of ~5 arcmin.

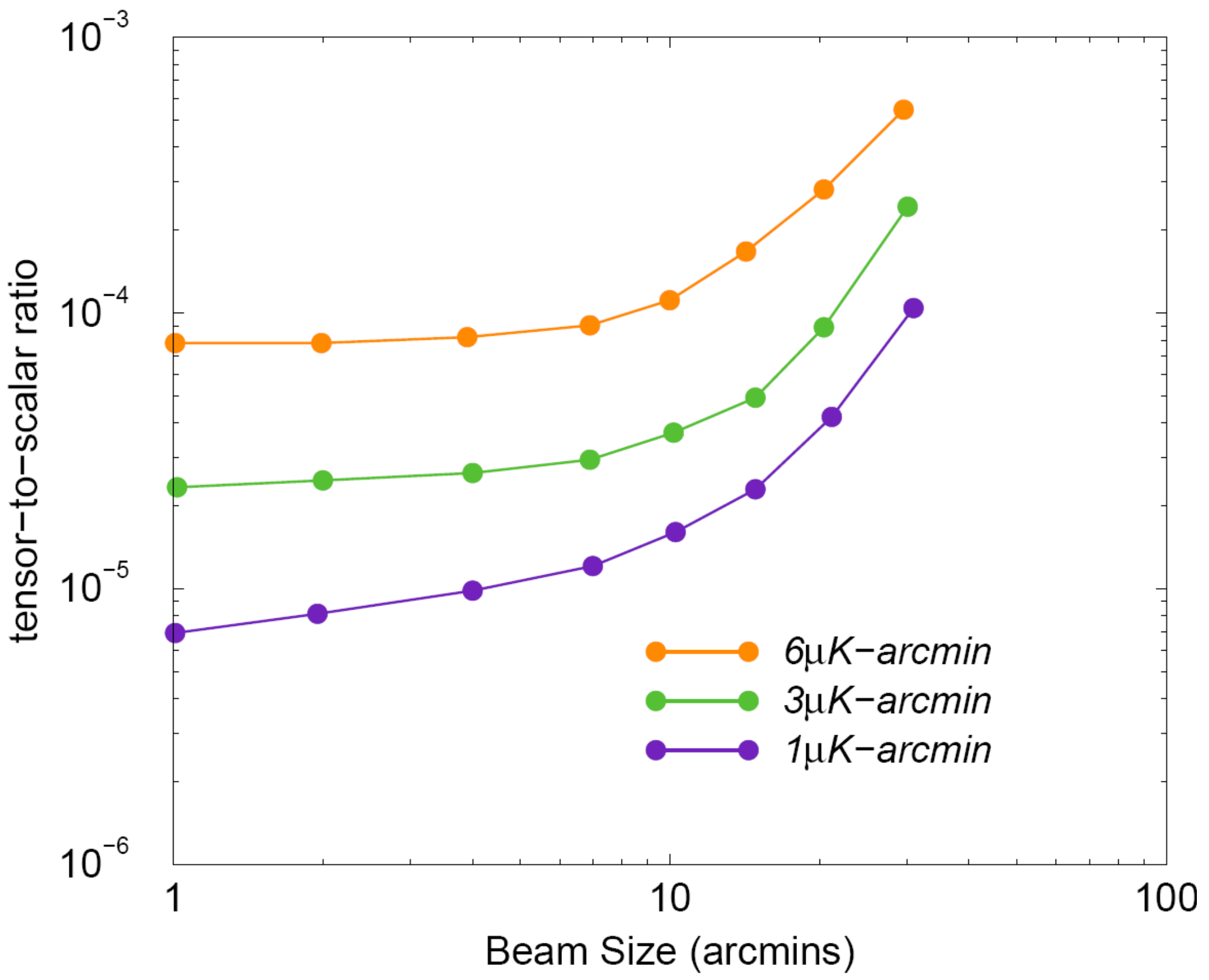

Fig. 6.1.1 The achievable tensor-to-scalar ratio as a function of the beam size. We show the expected tensor-to-scalar ratio for three noise level. The low tensor-to-ratio limit is achieved with "cleaning" of lensed B-modes using higher order statistics in the anisotropy maps using likelihood methods. For a given noise level, however, there is a limit on the beam size beyond which higher resolution maps no longer provide information related to lensing to properly reduce the confusion. For an experiment with a weight between 1 and 3 μK-arcmin, this resolution is between 4 to 7 arcmins.

For the secondary science goals, we take as the requirement measuring the E-mode and lensing B-mode polarization (which actually has non-Gaussian statistics) to cosmic variance into the Silk damping tail, in order to extract all of the available cosmological information. Once again, a sensitivity of 1-3 μK-arcmin and a resolution of 5 arcmin at 100 GHz meet these criteria, as can be seen in Fig. 1.3.1.

The requirements and goals from the criteria for lensing subtraction and secondary science are listed in Table 6.1.1, and serve to guide the definition of our study. Obviously sensitivity and resolution requirements drive mission lifetime and mirror aperture, and even a modest change in the requirements will have a significant impact on mission design, and ultimately cost. At present, the scientific requirements in Table 6.1.1 are somewhat idealized. A full analysis would include the extraction of cosmological parameters from scalar and lensing polarization signals, and include the effects of foreground contamination in carrying out the subtraction of lensing polarization. These calculations are challenging, but eventually will have to be carried out to fully justify the choice of mission parameters.

**Table 6.1.1 *Comprehensive Science* Mission Design Requirements and Goals**

| Instrument Criteria | Requirement | Design Goal |
|---|---|---|
| **High sensitivity** | $w_p^{-1/2} < 3.5$ μK-arcmin | $w_p^{-1/2} < 2$ μK-arcmin |
| **Subtract foreground signals to negligible levels** | Remove foregrounds to below r = 0.01 science goal | Optimize bands for foreground removal based on best knowledge |
| **Control systematic errors to negligible levels** | Suppress systematic errors to < 10% of r = 0.01 signal, after correction | Suppress raw systematic effects to less than 10% of statistical noise level |
| **Maintain sensitivity on large angular scales** | All-sky coverage with redundant interleaved scan strategy | |
| **Angular resolution** | < 5′ at 100 GHz | |

**Table 6.1.2 Baseline Instrument Parameters Summary Table**

| | |
|---|---|
| **Instruments** | Gregorian telescope (3-m effective aperture) |
| **Bands** | 30, 45, 70, 100, 150, 220, 340 & 500 GHz |
| **Detectors** | 1520 |
| **Sensitivity** | $w_p^{-1/2} = 3.5$ μK-arcmin (required), 1.8 μK-arcmin (design) |
| **Resolution** | 1 – 15 arcmin (FWHM), diffraction limit |
| **FOV** | 6 deg |
| **Pointing Knowledge** | 3" |
| **Focal Plane** | Transition-Edge Superconducting bolometers |
| **Read Out** | Multiplexed SQUID current amplifiers |
| **Pol. Modulation** | Half-wave plate or focal plane switched |
| **Optics** | 4 m x 6 m Off-axis Gregorian (3 m illuminated) |
| **Cryogenics** | Passive to 40 K / Mechanical cooler to 4 K / ADR to 0.1 K |
| **Mission Lifetime** | 2 years required / 4 years design |
| **Payload Mass** | 2735 kg including 43% contingency |
| **Payload Power** | 758 W including 43% contingency |
| **Average Data Rate** | 2300 kbps including 100% contingency |

### *6.1.2 Mission Description*

The EPIC-CS mission architecture consists of a passively cooled off-axis Gregorian Dragone with an effective aperture of 3 m. The telescope is housed inside a passively cooled radiation shield, isolated from the spacecraft by a bipod support system affixed to a support ring in the middle of the radiation shield. A 3-stage fixed V-groove provides passive cooling of the instrument and telescope optics, achieving a temperature of ~40 K for the radiation shield, and 35-40 K for the mirrors. Refracting foreoptics, with the possible inclusion of a waveplate, are cooled to 2-4 K by an active cooling system, either a liquid helium cryostat or a mechanical cooler. The bolometric focal plane is housed inside the fore-optics assembly, cooled to 100 mK. We have designed the receiver tube as a stand-alone unit inside a series of radiation shields and mounted off the base of the radiation shield. In the case of a cryostat, the liquid cryogens are contained in a separate vessel, and in flight the receiver is cooled to 2 K by operating cryogenic valves and a superfluid fountain effect system to pump superfluid liquid helium lines to the receiver. This configuration minimizes dewar mass and maximizes hold time. In the case of a mechanical cooler, the receiver design is essentially the same, but the cryostat is removed and replaced with a mechanical cooler which is coupled to the receiver and its radiation shields.

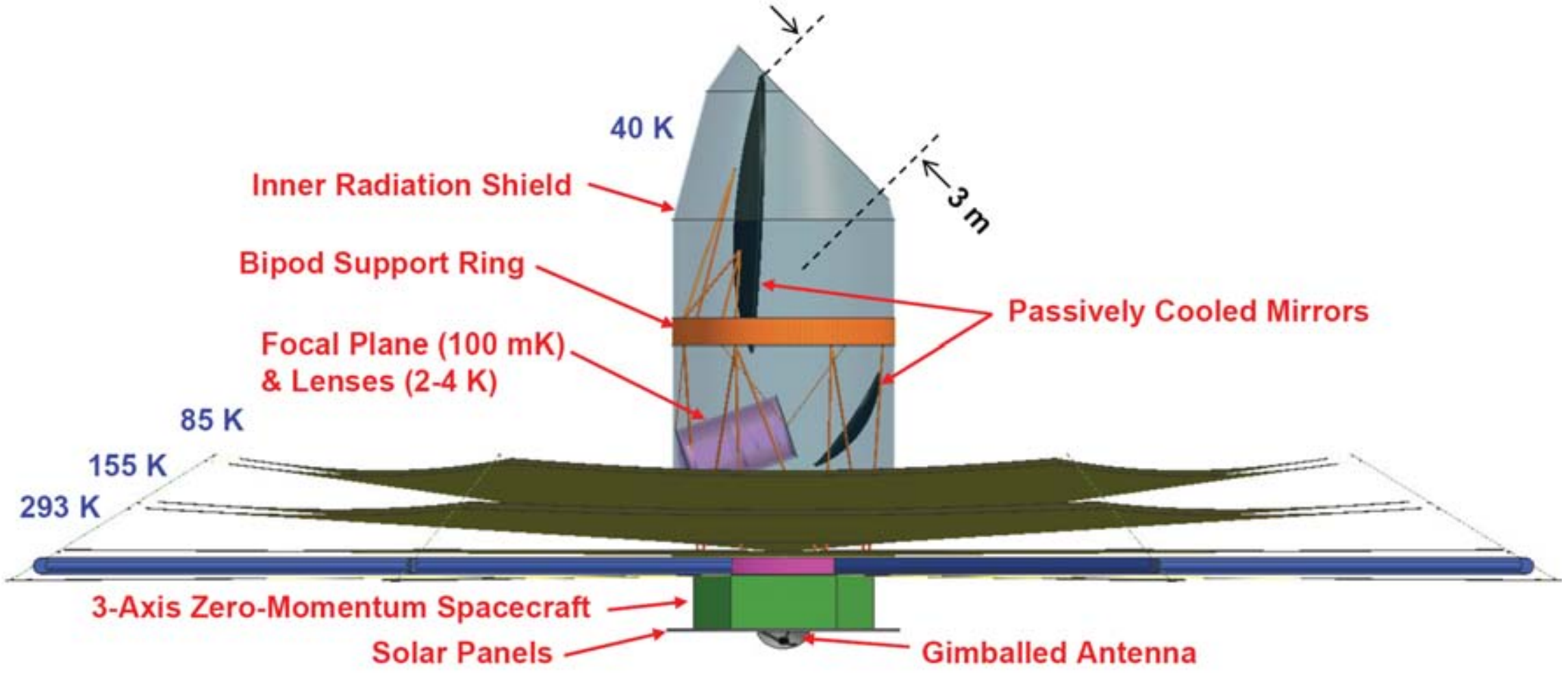

Fig. 6.1.2. Sideview of the EPIC-LC mission in flight configuration. The Gregorian Dragone 3-m reflection optics are passively cooled to < 40 K by a V-groove radiator (obscured by the deployed sunshade in this view). The receiver foreoptics are cooled to 2-4 K, and the focal plane is cooled to 100 mK. The same spinning/prcessing scan strategy employed in EPIC-LC is executed by a 3-axis zero-momentum spacecraft. Telemetry (due to the higher data rate) is provided through a counter-rotating gimbaled antenna.

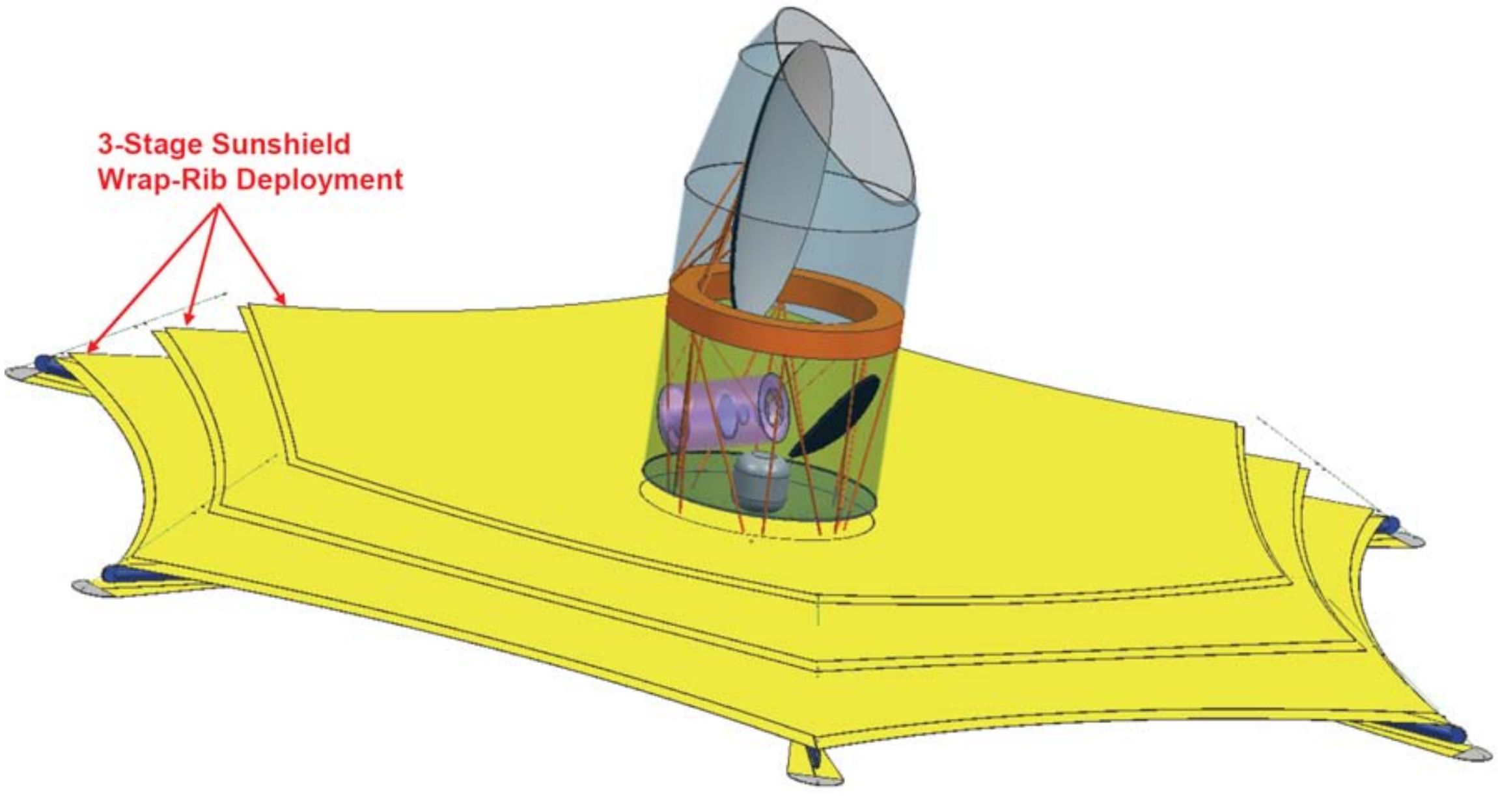

Fig. 6.1.3. Stereo view of the flight configuration showing the 3-stage deployed sunshield. The sunshield is designed to prevent sunshine from striking the inner 40 K radiation shield surrounding the telescope, and allow thermal radiation from the fixed v-groove radiator to escape to space. Note that the sunshield is 22 m diameter from scallop to scallop.

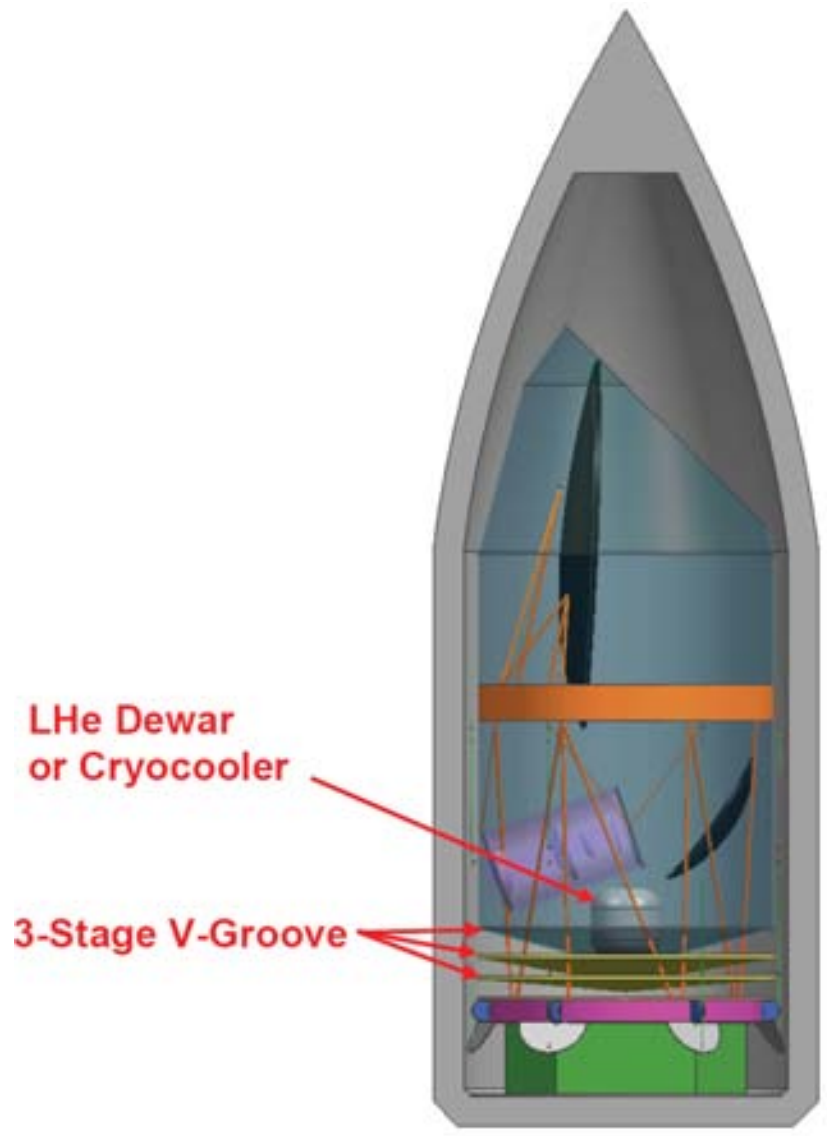

Fig. 6.1.2. The EPIC-CS mission in launch configuration. The external cooling system for the receiver foreoptics and focal plane is shown in this case as a cryostat. The sunshield deployment mechanism is located under the v-groove cooler, and the 3 sunshield films are each folded and stowed inside the 3-stage v-groove sections.

A large 22-m diameter (scallop to scallop dimension) 3-stage deployed sunshield keeps direct sunlight off the inner 40 K radiation shield. Three sections are need to reduce reradiated thermal power from the sunshield. Like the EPIC-LC sunshield, the EPIC-CS sunshield does not participate significantly in passive cooling, i.e. the sunshield does not conduct and radiate significant parasitic thermal power from the bipod support, and radiation from the sunshield is a small contributor to the total heat load on the 40 K stage. The optics are surrounded by a 40 K inner radiation shield. The shield minimizes radiant thermal inputs to the optics (from e.g. the moon or variations in the sunshield temperature), and can be used to provide baffling of stray light by coating the inside with mm-wave absorbing material. We are still assessing how much thermal and stray light baffling is needed, and if it is possible that it could be reduced or eliminated upon further study.

The scan pattern is essentially identical to the EPIC-LC scan pattern described in section 5.2.5. EPIC-CS has similar spin and precession angles as EPIC-LC, where the optical axis of the telescope is offset from the spin axis by 55˚ and the spin axis precesses in a 45˚ around the sun-spacecraft line. We have not carried out the same detailed analysis of the relative spin and precession rates needed to eliminate gaps in the scan pattern on short time scales, although we note this will be more of an issue with EPIC-CS's smaller field of view. Because the data rate is significantly higher, it is not possible to use a fixed toroidal-beam antenna. Instead we incorporate a counter-rotating gimbaled antenna to despin the scan pattern for purposes of telemetry.

## Table 6.1.3 Detailed Mass Summary

| Sub-Assembly | | Mass (CBE) [kg] | Contingency [%] | Allocated Mass [kg] |
|---|---|---|---|---|
| **Focal plane assembly** | | **12** | **43** | **17** |
| **Optics** | Primary mirror (3.6 x 5.3 m) | 105 | 43 | 150 |
| | Secondary mirror (1.3 x 1.8 m) | 10 | 43 | 14 |
| | Cooled Si fore lens (Φ1.0 m) | 88 | 43 | 126 |
| | Cooled Si aft lens (Φ0.73 m) | 42 | 43 | 60 |
| | **Total Optics Assembly** | **245** | **43** | **350** |
| **Active Cooling** | ADR | 8 | 43 | 11 |
| | 2-stage mechanical cooler | 60 | 43 | 86 |
| | Thermal strapping | 20 | 43 | 29 |
| | **Total Active Cooling Systems** | **88** | **43** | **126** |
| **Passive Cooling** | Inner thermal shield | (565) | (43) | (808) |
| | Receiver shielding | 31 | 43 | 44 |
| | V-grooves (3) | 187 | 43 | 267 |
| | **Total Passive Cooling Systems** | **218** | **43** | **311** |
| **Sunshade** | Kapton films | 268 | 43 | 383 |
| | Struts and supports | 178 | 43 | 255 |
| | Deployment hardware | 51 | 43 | 73 |
| | **Total Sunshade Mass** | **497** | **43** | **711** |
| **Structural Supports** | Primary mirror support struts | 50 | 43 | 72 |
| | Secondary mirror support struts | 5 | 43 | 7 |
| | Main support ring | 283 | 43 | 405 |
| | Ring support struts | 153 | 43 | 219 |
| | Cold optics struts | 121 | 43 | 173 |
| | **Total Structural Supports** | **612** | **43** | **875** |
| Cabling | | 65 | 43 | 93 |
| Warm Electronics | | 40 | 43 | 57 |
| Optics cover and mechanism | | 135 | 43 | 193 |
| **Subtotal for Payload** | | **1912** | **43** | **2735** |
| Attitude Control System | | 240 | 25 | 300 |
| C&DH | | 24 | 30 | 31 |
| Power | | 73 | 30 | 95 |
| Propulsion (dry) | | 47 | 27 | 60 |
| Structures and mechanisms | | 595 | 30 | 774 |
| Launch adapter | | 39 | 30 | 51 |
| Cabling | | 63 | 30 | 82 |
| Telecom + X-band Antenna | | 24 | 19 | 29 |
| Thermal | | 53 | 30 | 69 |
| Propellant [ΔV = 215 m/s] | | 437 | 0 | 437 |
| **Subtotal for Wet Spacecraft** | | **1595** | | **1928** |
| **Total Launch Mass** | | **3507** | | **4663** |
| **Launch Vehicle Maximum Payload Mass to L2 (C3 = -0.6)** | | | | |
| **Vehicle** | | **Pld Mass [kg]** | **Margin [%]** | **Margin [kg]** |
| **Atlas V 541** | | **5886** | **26** | **1223** |
| **Atlas V 551** | | **6401** | **37** | **1738** |

*Estimate excludes the mass of inner thermal shield, assumed here to be made of 5 kg/$m^2$ Al honeycomb sandwich. It is possible to lightweight the shield since it is not structural, if it does not have to be mm-wave absorbing on the inside, but this requires further study.

Much of the study results on EPIC-LC in section 5 apply to the larger EPIC-CS case. In the remainder of section 6 we highlight aspects that are different, and refer the reader back to appropriate information in section 5 to avoid unnecessary duplication. A summary of the attributes which are significantly different than the material presented for the EPIC-LC configuration is summarized below. Due to the limited scope of this study we were not able to estimate the cost of the EPIC-CS option.

**Systematic Error Mitigation (Section 6.2):** Main beam effects are strongly dependent on beam size, and thus significantly different for EPIC-CS. We recalculate these effects for the case of EPIC-CS and find that many of the requirements on main beam effects are reduced due to smaller beam sizes.

**Reflecting Optics (Section 6.3):** EPIC-CS uses an off-axis Gregorian Dragone telescope with an effective aperture of 3 m. This system provides large AΩ throughput, needed for large focal plane arrays, and high sensitivity. The optical system must accommodate a wide range of wavelengths simultaneously. The mirrors are passively cooled to < 40 K, but present modestly higher optical background on the focal plane detectors. Prototypes of this system are being developed for EBEX and Polarbear.

**Polarization Modulators (Section 6.4):** The main reason for polarization modulation in EPIC-LC is to remove polarization artifacts created by main beam imperfections. This need is significantly reduced due to the smaller beam sizes in EPIC-CS. However, since active modulation can still provide some benefits, we consider options that are appropriate to EPIC-CS, with its much wider instantaneous spectral band coverage.

**TES/SQUID Focal Plane (Section 6.5):** We estimate focal plane parameters for EPIC-CS, which are somewhat different due to emission from the warm mirrors, and the available throughput provided by the optics.

**Cooling to 100 mK (Section 6.6):** We calculate passive cooling for the EPIC-CS configuration. We explore the parameters of a mechanical cooler option for cooling the lenses to 2-4 K and providing a precooling stage for the 100 mK cooler.

**Deployed Sunshade (Section 6.7):** The larger telescope aperture requires a significantly larger 3-stage deployed sunshield. We have developed a different deployment design based on a wrap-rib concept to meet the requirements of the larger shield.

**"Off-the-shelf" Hardware (Section 6.9):** Spacecraft components are resized for the larger EPIC-CS experiment.

**Telemetry (Section 6.10):** Data rate requirements for EPIC-CS are recalculated, and we find a pointed downlink antenna is necessary due to the higher transmission rates.

### 6.2 Systematic Error Mitigation

The systematic error budget was calculated for EPIC-CS using the same methods described in section 5.2. Because the overall sensitivity of EPIC-CS is similar to the EPIC-LC

TES option, the focal plane sensitivity to temperature drifts is only slightly different, with the largest change being increased sensitivity to the mirror temperature. The most significant effects are in the response to beam effects, due to the significantly higher resolution of EPIC-CS.

**Table 6.2.1.** Systematic Error Goals and Requirements for EPIC-CS

| Systematic Error | Description | Suppression to Meet Goal | Knowledge to Meet Requirement |
|---|---|---|---|
| *Main Beam Effects[1] – Instrumental Polarization* | | | |
| **Δ Beam Size** | $FWHM_E \neq FWHM_H$ | $(\sigma_1-\sigma_2)/\sigma < 6 \times 10^{-3}$ | $(\sigma_1-\sigma_2)/\sigma < 2 \times 10^{-2}$ |
| **Δ Gain** | Mismatched gains | $(g_1-g_2)/g < 10^{-4}$ | $(g_1-g_2)/g < 3 \times 10^{-4}$ |
| | Mismatched AR coating | $\Delta n/n < 6 \times 10^{-4}$ | $\Delta n/n < 2 \times 10^{-3}$ |
| **Δ Beam Offset** | Pointing E ≠ H | $\Delta\theta$ < 0.14" raw scan<br>$\Delta\theta$ < 10" sym. scan | $\Delta\theta$ < 0.4" raw scan<br>$\Delta\theta$ < 30" sym. scan |
| **Δ Ellipticity** | $e_E \neq e_H$<br>$\Delta e = (e_1-e_2)/2$ | $\Delta e < 7 \times 10^{-2}$, $\psi = 0°$<br>$\Delta e < 8 \times 10^{-4}$, $\psi = 45°$ | $\Delta e < 2 \times 10^{-1}$, $\psi = 0°$<br>$\Delta e < 2 \times 10^{-3}$, $\psi = 45°$ |
| **Satellite Pointing** | Q and U beams offset | < 12" | < 36" |
| *Main Beam Effects – Cross Polarization* | | | |
| **Δ Rotation** | E & H not orthogonal | $\theta_1-\theta_2 < 4'$ | $\theta_1-\theta_2 < 12'$ |
| **Pixel Rotation** | E ⊥ H but rotated w.r.t. beam's major axis | < 2.4′ | < 7.2′ |
| **Optical Cross-Pol** | Birefringence | $n_e-n_o < 4 \times 10^{-5}$ | $n_e-n_o < 10^{-4}$ |
| *Scan Synchronous Signals* | | | |
| **Far Sidelobes** | Diffraction, scattering | $< 1\ nK_{CMB}$ | $< 3\ nK_{CMB}$ |
| **Thermal Variations** | Solar power variations | | |
| **Magnetic Pickup** | Susceptibility in readouts and detectors | | |
| *Thermal Stability[2]* | | | |
| **40 K Baffle[3,5]** | Varying optical power from thermal emission | 4 mK/√Hz; 10 μK s/s | 12 mK/√Hz; 30 μK s/s |
| **2 K Optics[3,6]** | | 400 μK/√Hz; 0.3 μK s/s | 1.2 mK/√Hz; 1 μK s/s |
| **0.1 K Focal Plane[4,7]** | Thermal signal induced in detectors | 200 nK/√Hz; 0.4 nK s/s | 600 nK/√Hz; 1.2 nK s/s |
| *Other* | | | |
| **1/f Noise** | Detector and readout drift | 0.016 Hz (1 rpm) | 0.2 Hz (1 rpm) |
| **Band Mismatch** | Variation in filters | $\Delta\nu_c/\nu_c < 1 \times 10^{-4}$ | $\Delta\nu_c/\nu_c < 1 \times 10^{-3}$ |
| **Gain Error** | Gain uncertainties between detectors | $< 10^{-4}$ | $< 3 \times 10^{-4}$ |

[1]Main beam effects calculated at 100 GHz, no averaging over the focal plane is assumed

[2]Calculated at 100 GHz, at signal modulation frequencies, expressed for instantaneous and scan-synchronous signals respectively.

[3]Assumes 1% matching to unpolarized optical power, calculated at 100 GHz to give 1 $nK_{CMB}$(rms).

[4]Assumes 5% matching to focal plane drifts, calculated at 100 GHz to give 1 $nK_{CMB}$(rms).

[5]Planck achieves < 30 μK/Hz at 4 K regulated on Sterling-cycle cooler stage

[6]Planck achieves < 5 μK/√Hz at 1.6 K regulated on open-cycle dilution refrigerator J-T stage

[7]Planck achieves < 40 nK/√Hz at 0.1 K regulated on focal plane with open-cycle dilution refrigerator

The required tolerances on instrumental defects for EPIC-CS were derived from the techniques described in section 5.2. We force the level of the systematic effect under consideration to produce a level of spurious B-mode power spectra equal to the goal level to be below the statistical noise, as described in Section 3. This calculation yields the nominal contamination level and, given the scaling behavior of these systematic effects, we analytically derive the required value of the defect to reach a given level of spurious B-mode polarization. This was done using the map-based calculation, and confirmed using the multi-pole space calculation. As was the case for EPIC-LC, the agreement between the two simulations is excellent.

In the following figures we illustrate the results of our map-based calculations. However, as discussed above, the multi-pole space calculation was used to appraise second-order beam effects neglected by the map-based calculation. In addition to the main beam effects described above, we simulated the effect of satellite pointing errors after reconstruction.

We stress that all of these effects were simulated for a single focal plane pixel at a single frequency (100 GHz). To the extent that parameters vary over the focal plane, these effects will partially average down to give a smaller residual signal, and therefore, our goals and requirement levels are conservative. Beam effects have different dependences on the beam width $\sigma$, which can be calculated analytically in advance. In power spectrum units ($\mu K^2$), differential gain and differential rotation are independent of beam size, but differential beam width and differential ellipticity scale as $\sigma^4$. Differential pointing scales in a complicated manner, but for our uniform scan strategy we found it scales as $\sigma^2$.

Satellite pointing errors produce a systematic effect in a complicated manner. With dual analyzers, we instantaneously extract a single linear Stokes parameter (Q or U) in each beam that is not susceptible to pointing error. To extract the second parameter (U or Q), we must wait for the beams to rotate on the sky. Satellite pointing errors thus displace the Q and U beams on the sky. Pointing error does not give a simple gradient effect with dipolar symmetry; and it has a different multipole dependence and smaller amplitude than the differential pointing effect. As seen in Fig. 6.2.1, the residual power spectrum from satellite pointing more closely resembles that of differential rotation.

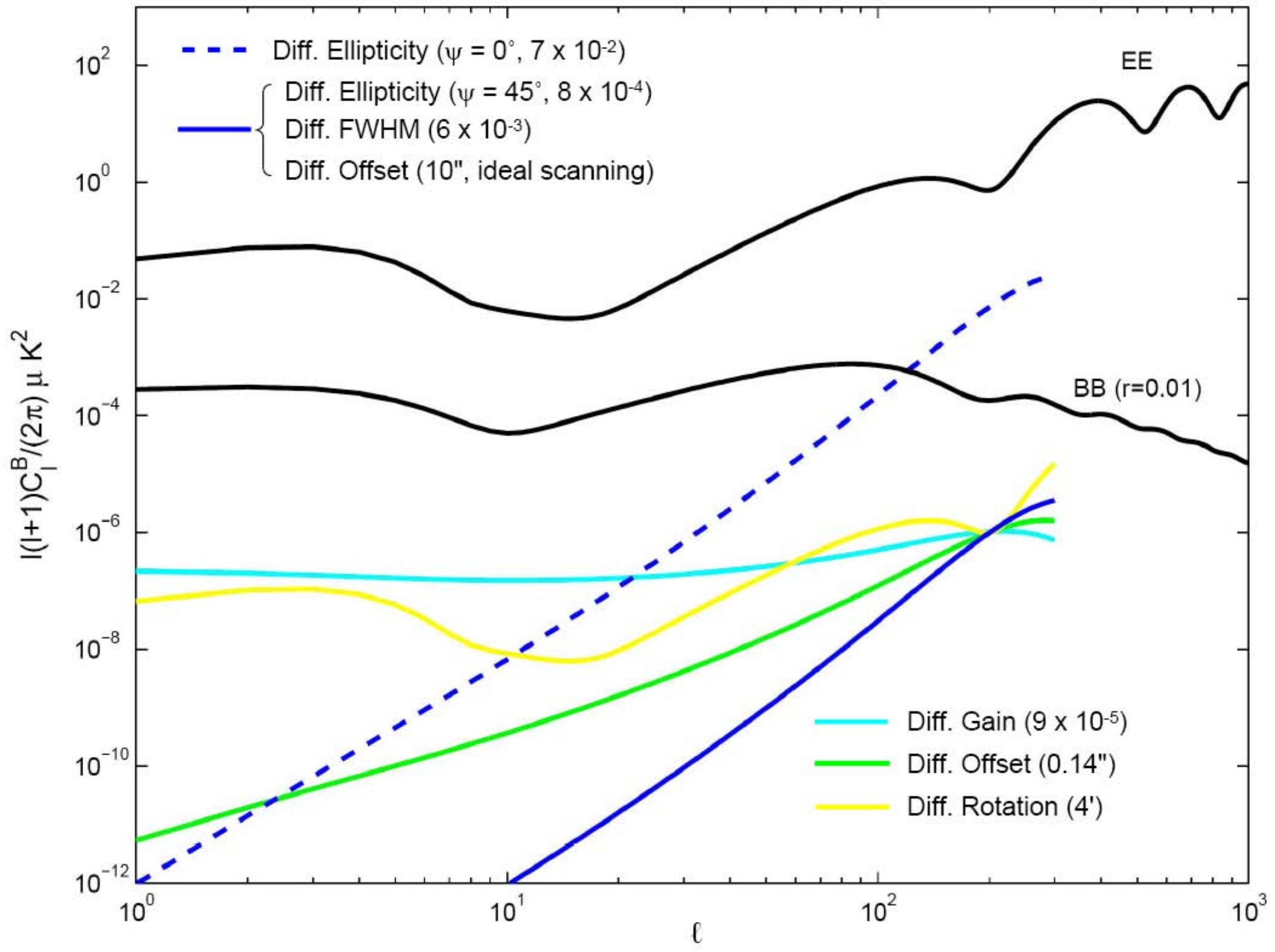

Fig. 6.2.1. Spurious B-mode power spectra for EPIC-CS with 5′ beams at 100 GHz. The amplitudes of these main-beam effects are all chosen to produce equivalent spurious B-mode power at ℓ = 200 at the goal level specified in section 3. The legend for each trace indicates the level of systematic of each type which produces spurious polarization signals. Note that the solid blue curve corresponds to three separate effects which have the same power spectrum. Differential ellipticity is shown for ψ = 0˚, which only produces E-mode polarization, and for ψ = 45˚, which only produces B-mode polarization. The result of a shift in the beam centroids, differential beam offset, is shown for two cases. One case is for the EPIC scan strategy, the other case is for an idealized scan pattern covering all scan angles uniformly over the entire sky. With the present scan pattern it may be possible to approximate the ideal scan pattern by mathematically weighting scans. These spectra indicate the level of the raw effect, and further reduction is possible given prior knowledge of the beam effects.

## 6.3 Gregorian Dragone Optics

Off-axis Gregorian telescopes have considerable heritage, both in space and with suborbital experiments. The WMAP satellite, the balloon-borne payloads BOOMERANG, MAXIMA, and Archeops, and the ground-based ACBAR instrument have all used the Offset Gregorian design. In the near future, the design will be deployed on ESA's Planck, the South Pole Telescope, the Atacama Cosmology Telescope and the EBEX balloon-borne payload. We have developed a wide-field design for EPIC to meet the optical throughput requirement.

### *6.3.1 Design*

The throughput requirements for EPIC-CS are given in Table 6.3.1. The optical system consists of a two-mirror Offset Gregorian telescope that couples radiation into a section of cold optics. The cold optics include 3 silicon lenses, an achromatic HWP, and the focal plane. The offset Gregorian telescope consists of a paraboloidal primary with a major axis of 3.5 m and an ellipsoidal secondary. The entrance aperture is 3 m in diameter and the overall f-number of the optical system is 1.7. The apertures of the mirrors are offset from the parent conics so as to produce a completely unobstructed light path in a compact design, which satisfies the Mizugushi-Drgaone condition [1,2] (see Fig. 6.3.1). The telescope and cryostat fit into an Atlas V shroud.

**Table 6.3.1** Parameters for EPIC-CS Optics

| Frequency [GHz] | Throughput[1] [$cm^2$ sr] | FOV[2] (deg) | Strehl Ratio[3] |
|---|---|---|---|
| 30 | 31 | 3.7 | 0.94 |
| 45 | 56 | 3.4 | 0.92 |
| 70 | 63 | 2.9 | 0.91 |
| 100 | 45 | 2.2 | 0.95 |
| 150 | 24 | 1.4 | 0.99 |
| 220 | 8.2 | 0.8 | 0.98 |
| 340 | 1.5 | 0.35 | 0.99 |
| 500 | 0.6 | 0.2 | 0.95 |

[1] The product of throughput per pixel ($\lambda^2$) and the total number of pixels at a given frequency. A pixel on the focal plane contains two polarization sensitive TES detectors.

[2] Pixels are arranged on a square grid with a circular boundary. We give the cumulative outer diameter of the FOV. The lower frequency pixels are arranged in annuli around the higher frequency ones.

[3] Ratio given at the outermost diameter of the frequency band.

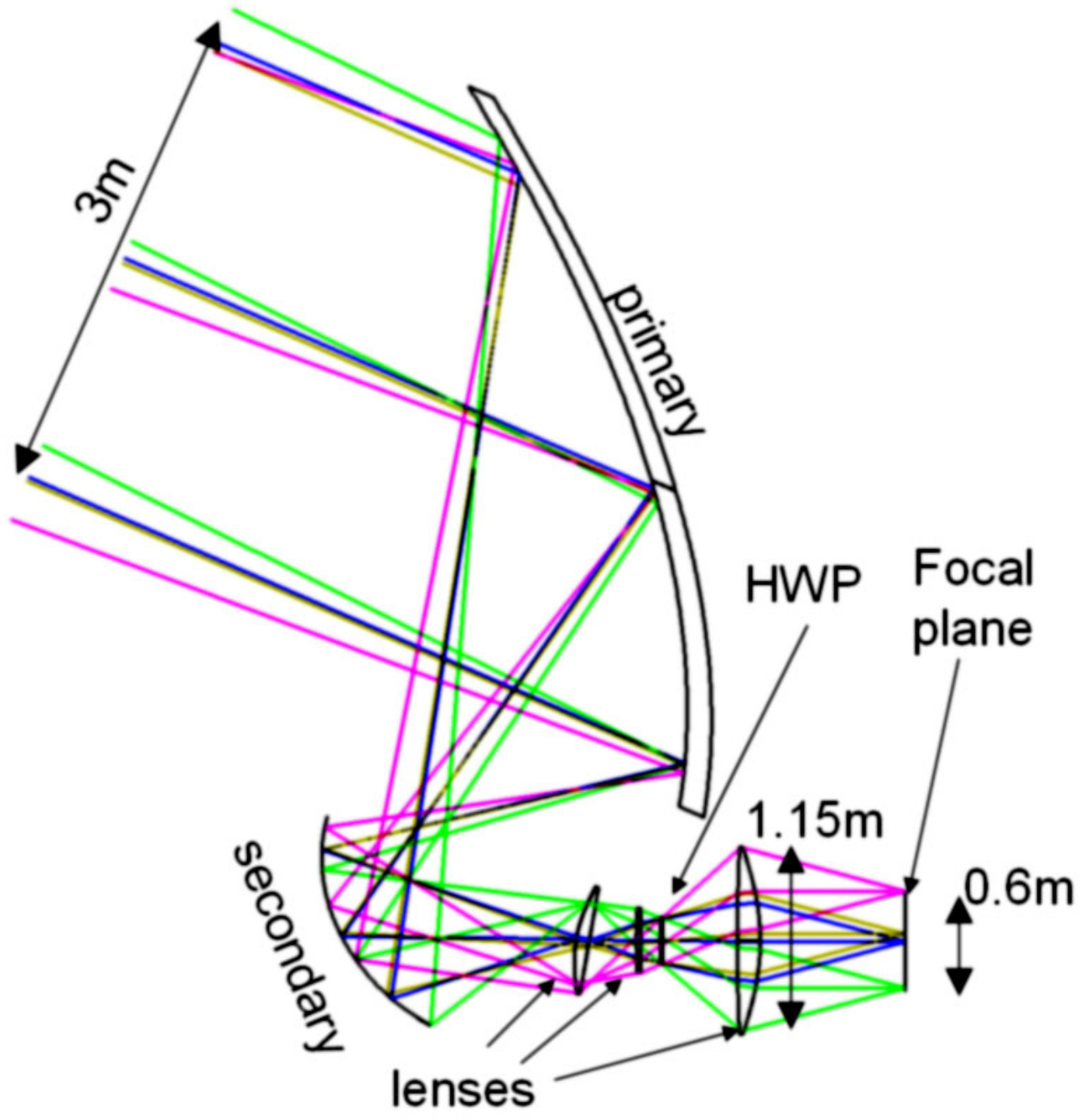

Figure 6.3.1. A raytracing diagram of the optical system for EPIC-CS. The largest dimension of the primary is 3.5 m, however the illumination, controlled by the Lyot stop, falls to -10dB within the central ~70% of the area and drops rapidly to the edge of the primary. The largest cold lens is 1.15 m in diameter, but could be decreased by adding one more lens to the design.

In contrast to the LC system design, EPIC-CS has only one focal plane. The detectors for the 500 GHz bands are at the center of the focal plane and successive lower frequency bands are arranged in annuli around the center. The optical design provides image quality over the entire FOV that is much better than the diffraction criterion of 0.8 Strehl ratio. The diffraction performance at the edge of the FOV of each frequency band is given in Table 6.3.1.

*6.3.2 Performance*

The Greogrian design shows values of instrumental polarization that are smaller than 0.5% even for non-ideal anti-reflection coatings. Table 6.3.2 gives the Muller matrix elements as calculated from CODE V and including only elements on the sky side of the half wave plate. Signals from these elements are modulated by the HWP whereas instrumental polarization from elements behind the HWP are not modulated. Note that with EPIC-LC the HWP is the first element in the optical path and therefore it modulates only the signals from the sky. Another contrast with the optics of EPIC-LC is the magnitude of the mixing between Q and U as encoded by the QU term of the Muller matrix. With EPIC-CS a rotation of about 3 degrees is induced on an incident polarization vector primarily by the curvature of the reflectors. This rotation, which is a function of the position on the focal plane, needs to be calibrated out using both ground and in-flight measurements.

We carried out a preliminary study using the geometric theory of diffraction of the antenna response of the EPIC-CS telescope, including the primary and secondary mirrors, PLANCK-style baffles around the mirrors, and the deployable sunshield. It was beyond the scope of this study to optimize the baffling structure for the EPIC-CS telescope. Figure 6.3.2 shows the geometry of the telescope and baffles and gives the definition of the coordinate system. In the simulations, which were done using GRASP9, the secondary mirror was illuminated by a Gaussian beam that produced an edge taper of -15 dB at the edge of the primary mirror. The baffle was assumed black. A cut in the antenna response in the xy plane is shown in Figure 6.3.3. For most of the range the response is below -40 dBi which appears adequate given the simulations shown in Figure 5.3.3. The features near -20 degrees with amplitude of zero dBi are due to diffraction at the edge of the primary. There are several important differences between our simulations and the anticipated implementation. The assumption of -15 dB edge taper at the edge of the primary is conservative because it does not include the effects of the cold aperture stop, which should decrease the edge taper substantially thereby reducing all far sidelobe levels. In particular, inclusion of the aperture stop should reduce the amplitude of the feature arising from diffraction at the edge of the primary near -20 degrees. Our simulations do not include any of the internal optics, which include lenses and the window. These optical elements could increase scattering and thus sidelobe levels.

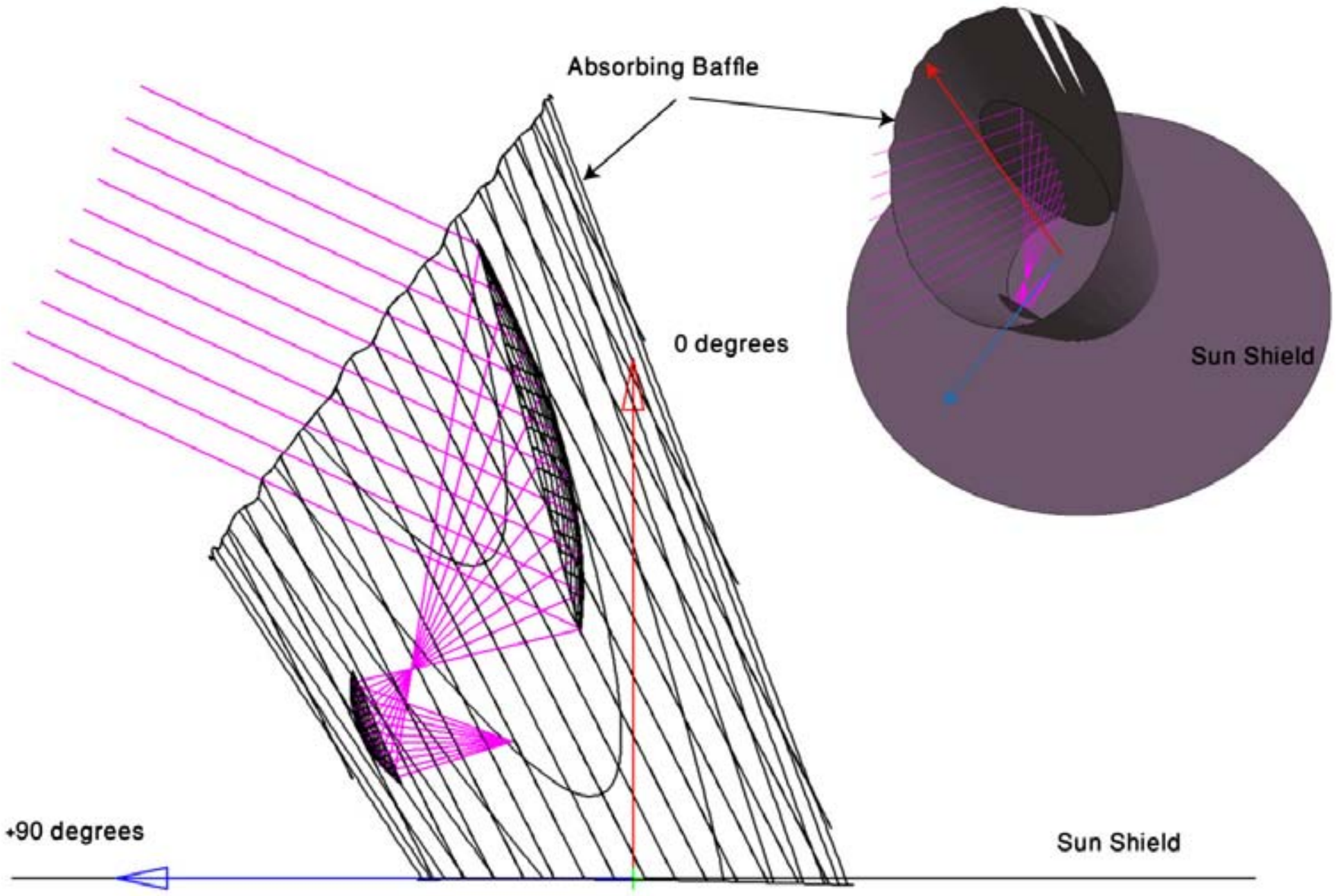

Figure 6.3.2. The geometry used for calculating the antenna response of EPIC-CS.

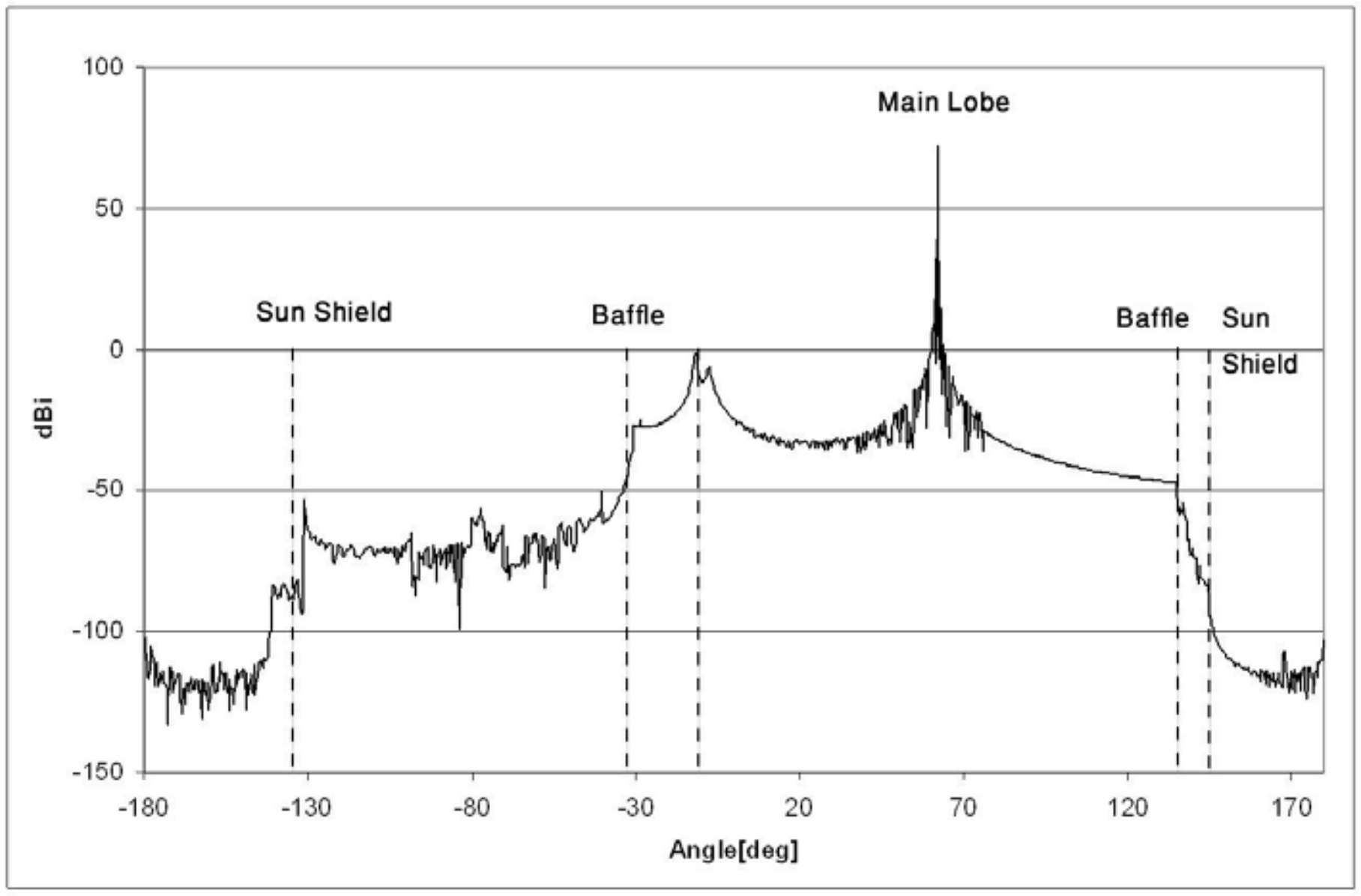

Figure 6.3.3. Antenna response of the EPIC-CS telescope assuming baffling that is similar to that of the PLANCK satellite. We assume -15 dB edge illumination at the edge of the primary, which neglects the attenuating effects of the cold aperture. Feature marked with dashed lines correspond to increased attenuation from the baffle and the sun-shield. The un-marked feature at ~-20 degrees corresponds to diffraction around the edge of the primary and would decrease for lower edge illumination.

There are a number of challenges in the implementation of the EPIC-CS optics. The current design employs large cold silicon lenses. A silicon lens with a diameter of 1.15 m has not been demonstrated. This lens could be made smaller at the expense of adding another lens to the system. Since there is a single multi-frequency focal plane, the lenses require broad band, cryogenic anti-reflection coating. Such coating is not yet available for silicon lenses, although a *narrow-band*, cryogenic coating has been recently developed [3]. The technical limitations introduced by the silicon lenses could be mitigated to some extent by using polyethylene lenses. However, because the index of polyethylene is about a factor of 2 smaller than silicon, it is likely that the number and size of lenses will increase. The design of a half-wave plate becomes problematic, since it must cover more than a decade in frequency. The EBEX experiment uses an achromatic HWP made of a stack of 5 plates. The stack provides modulation efficiency larger than 95% over a bandwidth of ~350 GHz. Designing a HWP that will simultaneously serve frequencies between 30 and 500 GHz may not be feasible, so in section 6.4 we explore the possibility of a focal plane modulator.

**Table 6.3.2 Muller Matrix Elements for the Gregorian Dragone**

| Matrix | Ideal ARC | ARC Index 10% high | ARC Thickness 10%Thin |
|---|---|---|---|
| IQ | $1.6 \times 10^{-4}$ | $3.6 \times 10^{-3}$ | $1.7 \times 10^{-3}$ |
| IU | $8 \times 10^{-5}$ | $1.9 \times 10^{-3}$ | $9.2 \times 10^{-4}$ |
| QU | 0.111 | 0.107 | 0.105 |

Table 6.3.2. Mueller matrix elements for the edge of the 150 GHz band of the EPIC-CS telescope. Only elements on the sky side of the half-wave plate are included in the calculation. The ideal ARC is a $\sqrt{n}$, λ/4 anti-reflection coating for 150 GHz. The columns labeled '10% high' and '10% thin' assume an index that is 10% higher or a coating 10% thinner compared to the ideal coating, respectively. A QU term with a value of 0.11 corresponds to a rotation of the polarization vector by an angle of about 3 degrees.

Following Tran *et al.* 2007 [4], we have also studied the tradeoffs between a Gregorian Dragone design and a crossed Dragone, also known as a compact range antenna. The crossed Dragone, see appendix B, presents generally better optical performance but lacks a cold Lyot stop. A challenge with the crossed Dragone is meeting the view angle requirements while packaging the system into the space available in a launch shroud.

### 6.4. Focal Plane Modulators

The EPIC-CS optics present a bandwidth challenge for a single polarization modulator. An alternative to a HWP, which modulates the polarization over the entire focal plane, is to modulate the polarization for each detecting element individually. Here the modulator is fabricated on the focal plane and is becoming part of each detector. It only needs to provide a limited bandwidth of $\Delta \nu / \nu \sim 30\%$. A focal plane modulator in this case will modulate whatever polarization is produced by the upstream optics, so its primary function is to modulate the signal band at a convenient audio frequency.

The approach for modulating polarized signals in the focal plane is to use RF switches between the antennas on the focal plane and the bolometer. For example, a dual-point dual-toggle switch allows two antennas and two bolometers to form Q or U Stokes parameters and to use all the radiation impinging on the pixel. This type of switching is common for coherent HEMT-based radiometers, but in that case switching comes after an amplifier allowing the switch to have dissipative loss. For bolometric detectors, a very low-loss and low-power-dissipation switch needs to be developed. There are two types of promising technologies to achieve the RF switching. One is based on Josephson junctions and the second is based on RF micro-machine membrane (MEMs) switches.

The Josephson junction switch is based on the fact that the RF impedance of a tunnel junction depends strongly on the bias of the junction. The JPL/CIT group has built a prototype pixel as shown in Fig 6.4.1. This pixel uses quadrature hybrids to efficiently switch polarizations given the RF impedance swing of the tunnel junction. The prototypes have achieved a high switching ratio, but further work is needed to optimize the switches, and then operate them with TES bolometers.

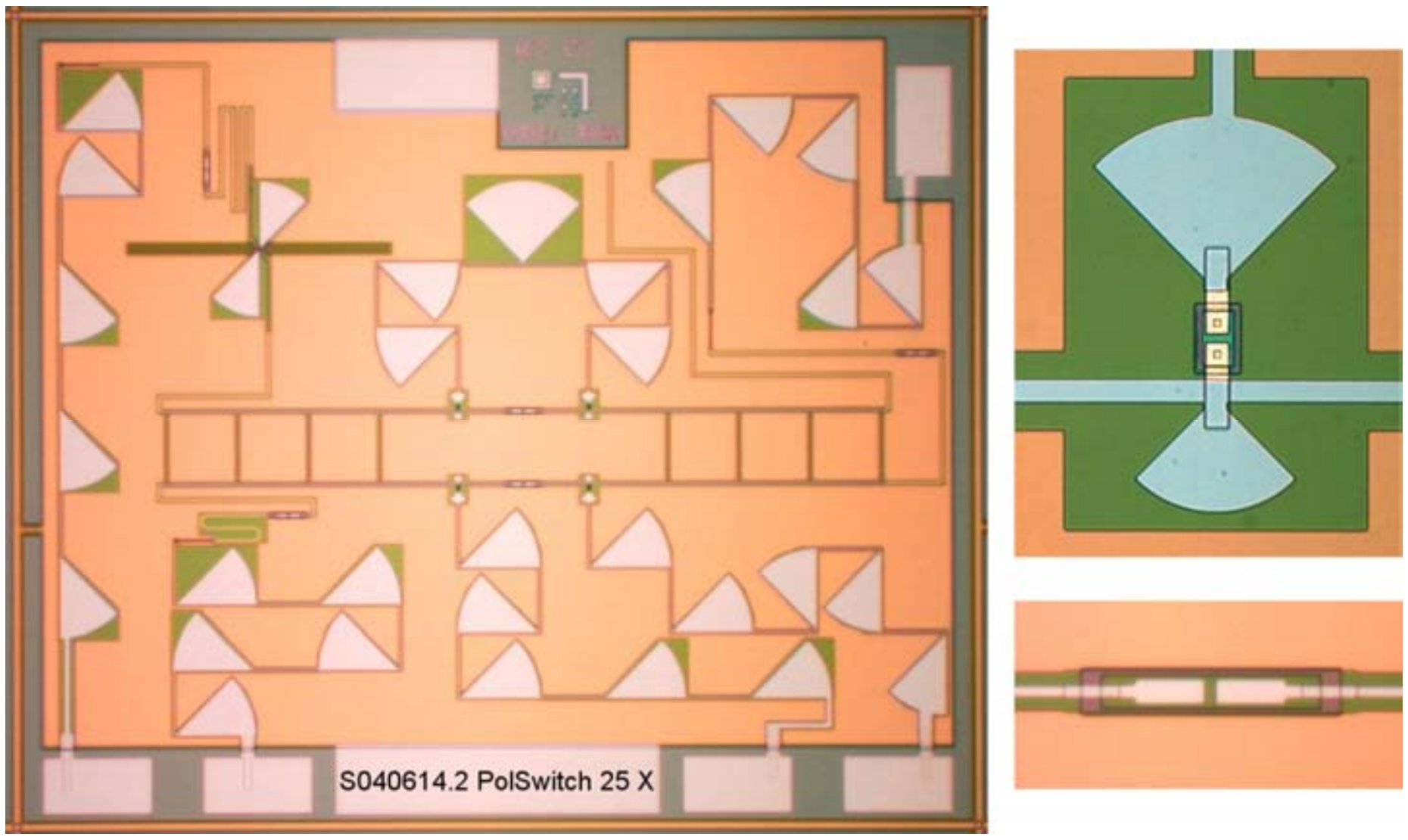

Figure 6.4.1. Junction-based polarization switch with two input arms and two output arms. The input arms are 1) a slot antenna, in order to introduce an optical signal, and 2) a termination resistor. The output arms go to junction detectors – these will be replaced by antenna-coupled bolometers in a full device. The signal passes through an arrangement of junction switches (see inset top right) which may be switched on and off by applying bias current. The DC drive current is electrically isolated from the detectors by 4 series capacitor sections in the transmission line (see inset bottom right).

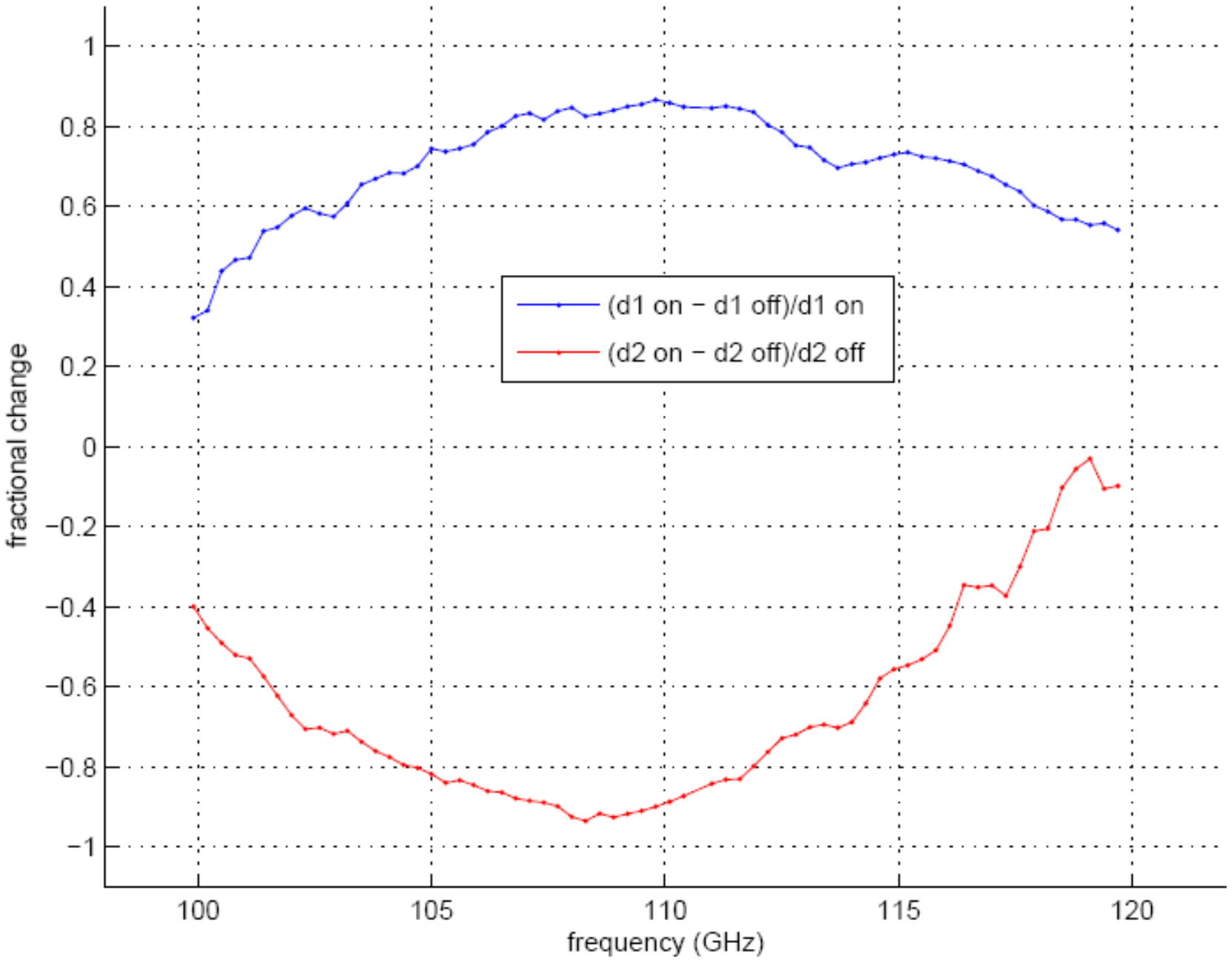

Fig. 6.4.2. Measured modulation of polarized signals by the junction-based switch, operating efficiently in a ~10% band centered at ~110 GHz.

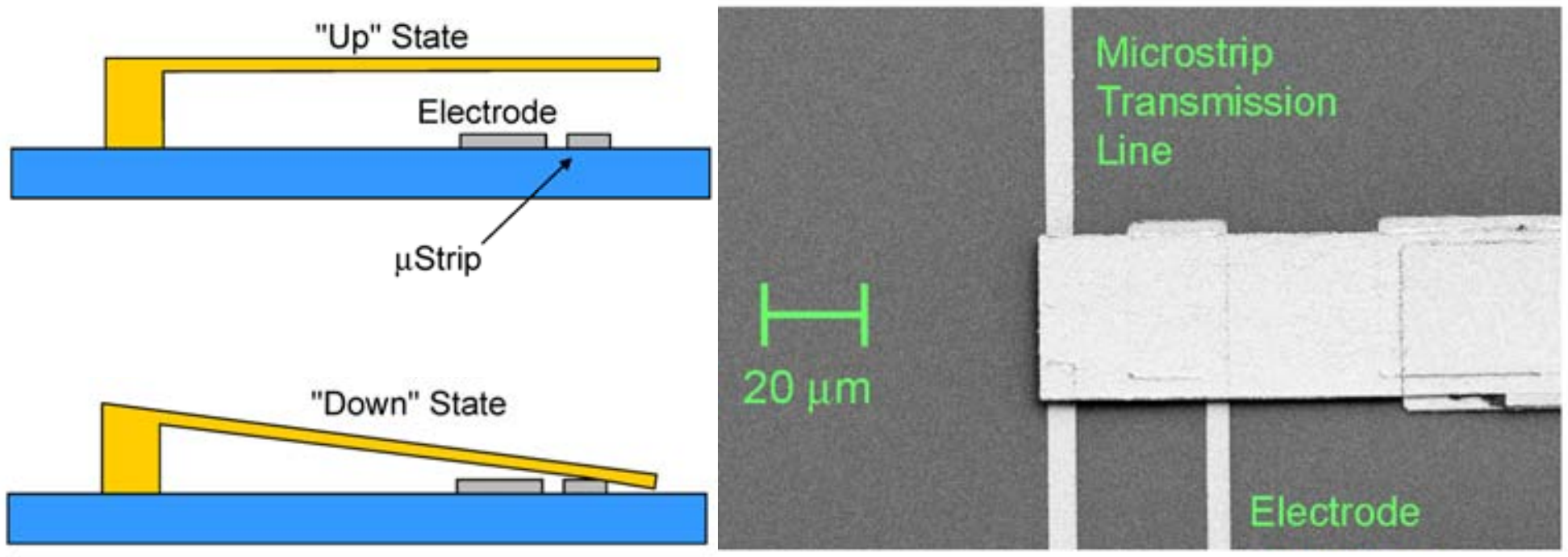

Fig. 6.4.3. A MEM switch developed by Kogut and his collaborators for the balloon borne PAPPA instrument.

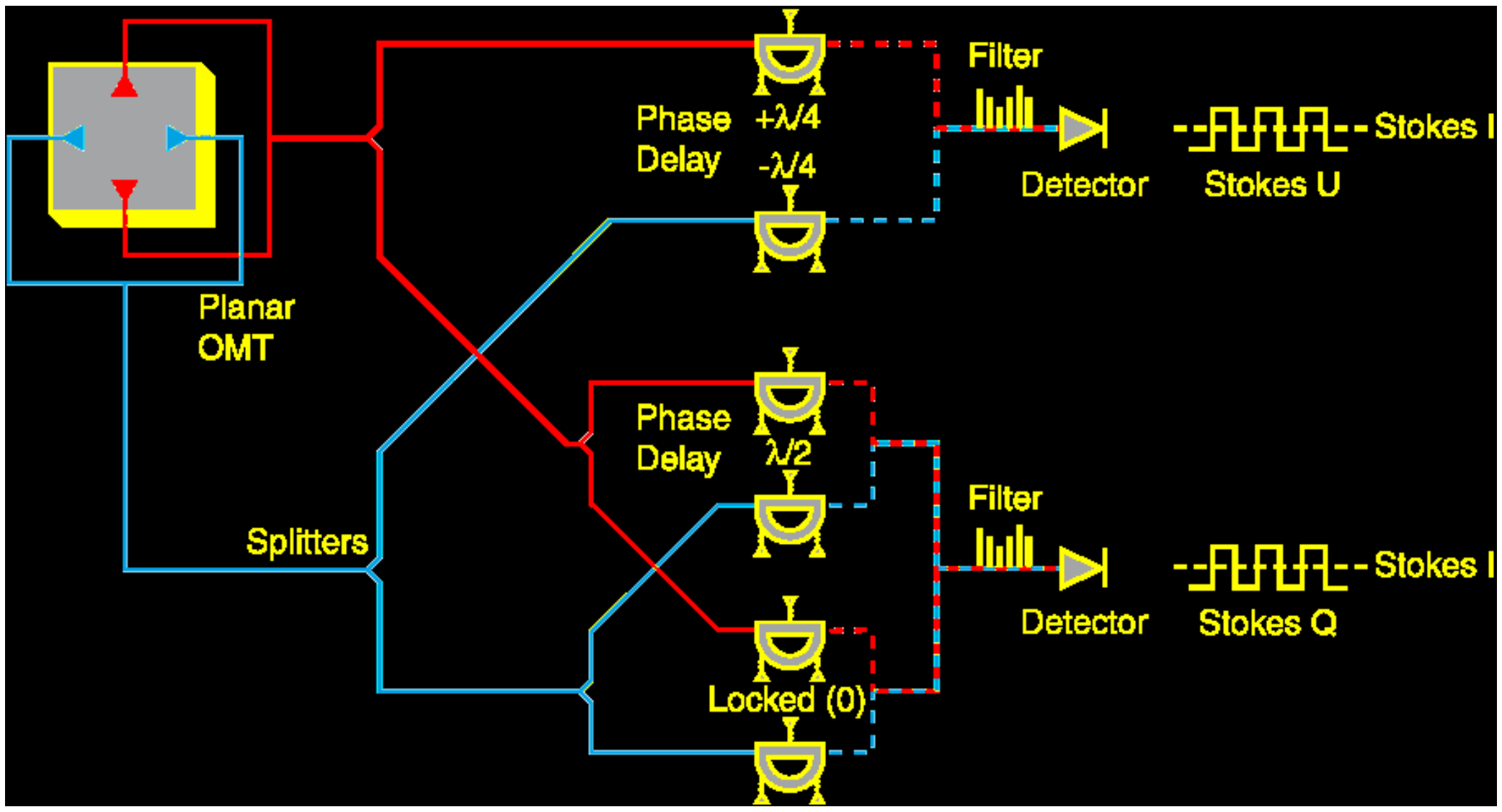

Fig. 6.4.4. The design for the focal plane polarization modulator of the PAPPA balloon borne instrument. Phase delays introduced by MEM switches on polarized signals are combined to produce output Q and U Stokes parameters (figure courtesy of Al Kogut).

## 6.5 Focal Plane Design

The focal-plane design for the EPIC-CS reflector is largely similar to that of the EPIC-LC refractor, except that the reflector has a single focal-plane with multiple frequency bands whereas there are multiple focal planes each operating in a single-band in the EPIC-LC refractor. The pixel designs for the two concepts can be identical. The reflector focal plane layout will have a "bullseye" design with annular rings of pixels of the same frequency. The highest frequency pixels will be at the center where aberrations are minimized.

### *6.5.1 Focal Plane Parameters*

Table 6.5.1 shows the design parameters for the reflector focal plane. The total field-of-view given by the optics is large enough to accommodate the required focal plane.

The small beam size places more demanding requirements on the detector speed of response. The spin rate is set to 1 rpm, basically the same rate as the refractor, due to the 1/f knee requirement. With approximately 10 times smaller beams, the beam crossing time is 10 times shorter, and the detector speed of response must be 10 times faster. Under these conditions, NTD Ge detectors are limited by speed of response, and require either increasing the thermal conductivity at a reduction in sensitivity, or slowing down the scan rate. Therefore we only consider TES bolometers for this mission option. The sensitivity for EPIC-CS is derived under the assumptions of a low-emissivity 60 K telescope, as described in Table 6.5.1. The emission from the telescope results in somewhat higher photon NEPs than the EPIC-LC refractor with cold optics.

**Table 6.5.1 Sensitivity Model Input Assumptions**

| | | | | | |
|---|---|---|---|---|---|
| Lens temperature | $T_{opt}$ | 4 K | Focal plane temperature | $T_0$ | 100 mK |
| Lens coupling[*] | $\varepsilon_{opt}$ | 10% | Optical efficiency[*] | $\eta$ | 40% |
| Wave plate temperature | $T_{wp}$ | 4 K | Fractional bandwidth[*] | $\Delta\nu/\nu$ | 30% |
| Wave plate coupling[*] | $\varepsilon_{wp}$ | 2% | TES heat capacity[*] | $C_0$ | 0.2 pW/K |
| Mirror temperature | $T_{baf}$ | 60 K | TES dln(R)/dln(T) | $\alpha$ | 100 |
| Mirror coupling at 300 GHz[*] | $\varepsilon$ | 1.0% | TES safety factor[†] | $P_{sat}/Q$ | 5 |

*Emissivity given by $\varepsilon = 0.01\ \mathrm{sqrt}(\nu/300\ \mathrm{GHz})$. See [1,2] for measured emissivities

**Parameter based on experimental measurement.

†Selectable design parameter.

**Table 6.5.2 Detailed Bands and Sensitivities for TES Option**

| Freq [GHz] | $\theta_{FWHM}$ [′] | $N_{bol}$[1] [#] | $\tau_{req}$[2] [ms] | $\tau$ [ms] | Required Sensitivity[3,4] | | | |
|---|---|---|---|---|---|---|---|---|
| | | | | | NET[5] [μK√s] | | $w_p^{-1/2}$ [μK-′][6] | δTpix[7] [nK] |
| | | | | | bolo | band | | |
| 30 | 15.5 | 20 | 9.7 | 1.2 | 85 | 19 | 41 | 240 |
| 45 | 10.3 | 80 | 6.4 | 0.9 | 72 | 10 | 22 | 130 |
| 70 | 6.6 | 220 | 4.1 | 0.7 | 62 | 4.2 | 9 | 54 |
| 100 | 4.6 | 320 | 2.9 | 0.6 | 58 | 3.2 | 7 | 41 |
| 150 | 3.1 | 380 | 1.9 | 0.6 | 61 | 3.1 | 7 | 40 |
| 220 | 2.1 | 280 | 1.3 | 0.6 | 88 | 5.2 | 11 | 67 |
| 340 | 1.4 | 120 | 0.9 | 0.6 | 270 | 25 | 53 | 320 |
| 500 | 0.9 | 100 | 0.6 | 0.3 | 2100 | 210 | 450 | 2700 |
| **Total[8]** | | **1520** | | | | 1.8 | 3.5 | 21 |

Notes:

1. Two bolometers per focal plane pixel
2. $\tau_{req} = (1/2\pi)\ \theta_{FWHM}/d\theta/dt$ at 1 rpm
3. Calculated sensitivity with 2-year mission life
4. Sensitivity margin of √2 applied to all NETs
5. Sensitivity of one bolometer in a focal plane pixel
6. $[8\pi\ \mathrm{NET}_{bolo}^2/(T_{mis}\ N_{bol})]^{1/2}(10800/\pi)$
7. Sensitivity δT in a 120′ x 120′ pixel
8. Combining all bands together

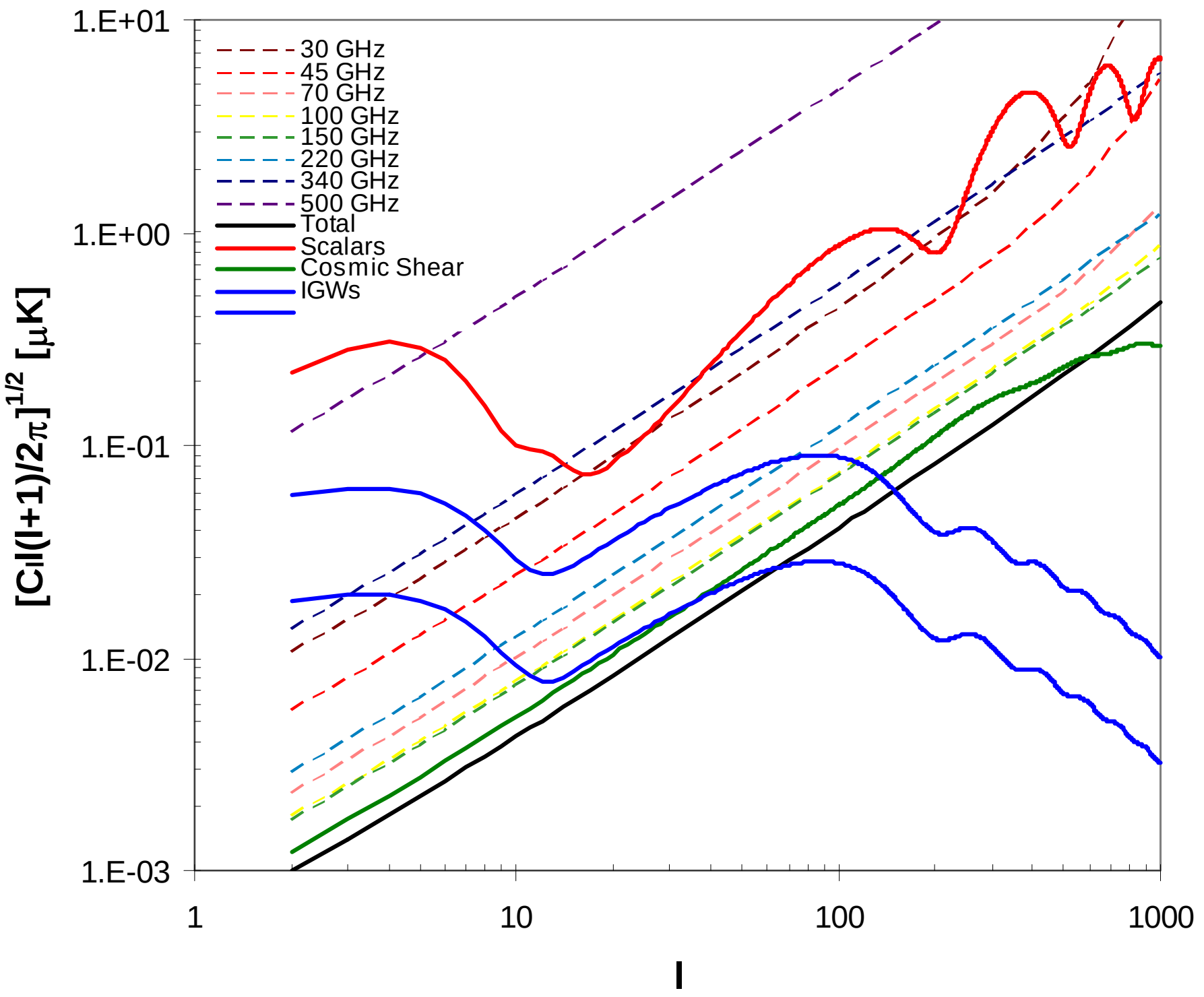

Fig. 6.5.1. Noise Cls calculated for TES bolometers for all of the bands (dashed), and combined (solid black) compared to scalar EE (red), cosmic shear (green), and IGW BB (blue at r = 0.1 and r = 0.01). Note that this option measures scalar EE to sample variance out to ℓ ~ 1000, and can measure lensing BB to "sample variance" (a convenient misnomer since the statistics are non-Gaussian) out to ℓ ~ 300.

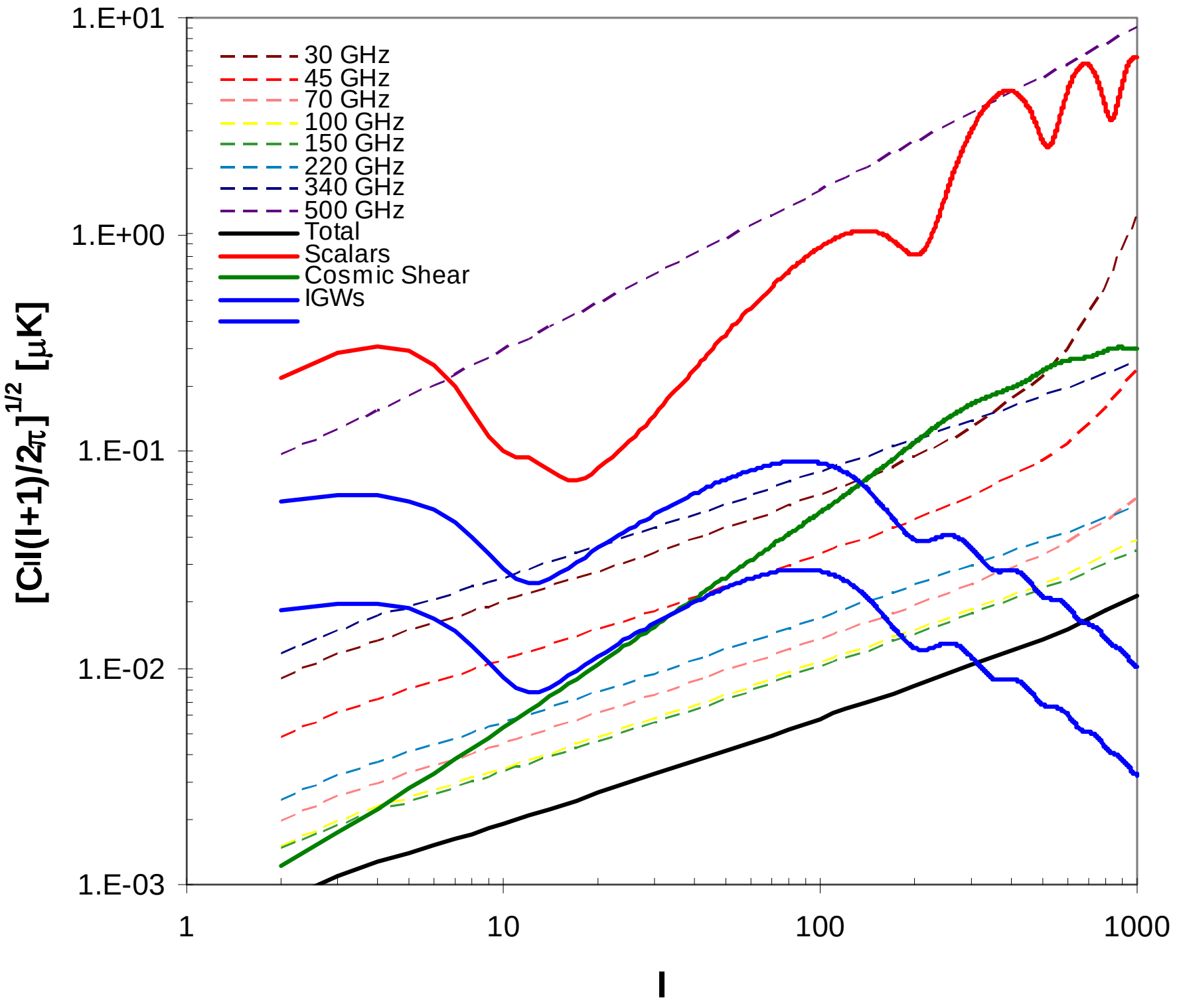

Fig. 6.5.2. Errors on Cls calculated for TES bolometers for all of the bands (dashed), and combined (solid black) compared to scalar EE (red), cosmic shear (green), and IGW BB (blue at r = 0.1 and r = 0.01). The calculation assumes fsky = 0.8, Δℓ/ℓ = 0.3 binning, and ignores sample variance.

*6.5.2 Q/U Analyzer*

It may be advantageous, for either the EPIC-CS or the EPIC-LC options, to improve the polarimetry capabilities of the focal plane. A dual polarization antenna offers simultaneous measurements of two linear polarization states with matched beams. With an accurate relative calibration, such polarimeters reject common-mode unpolarized sky signals. This feature suppresses the most serious systematic, temperature to polarization mixing, in a CMB polarization experiment. Similarly, a simultaneous measurement of both the Stokes Q and U parameters (rotated 45° with respect to each other) is advantageous in the fidelity of the reconstruction of the polarization vector, reducing another potential systematic, E-mode to B-mode mixing.

We have developed devices that measure power in two detectors, one in vertical polarization $E_x^2$ and one in horizontal polarization $E_y^2$ (see Figs. 5.5.9 and 5.5.11). This arrangement is simple, and maximizes the optical power on the fewest detectors. However, Stokes U must either be obtained in a different antenna pixel, rotated in the focal plane by 45° (the solution used in Planck), or, after a long time interval, rotating the instrument field of view on the sky by 45°. Either of these approaches may introduce systematic errors.

An alternative is to extract Stokes Q and U simultaneously in a single pixel, a 'Q/U analyzer', as shown in Fig. 6.5.3. The technique for doing this is simple, and consists of splitting half of the vertical and horizontal signal into detectors, producing $E_x^2/2$ and $E_y^2/2$, and the other half into a 180° hybrid followed by detectors, producing $(E_x+E_y)^2/4$ and $(E_x-E_y)^2/4$. If the detectors are photon noise limited, there is no sensitivity penalty in such an arrangement.

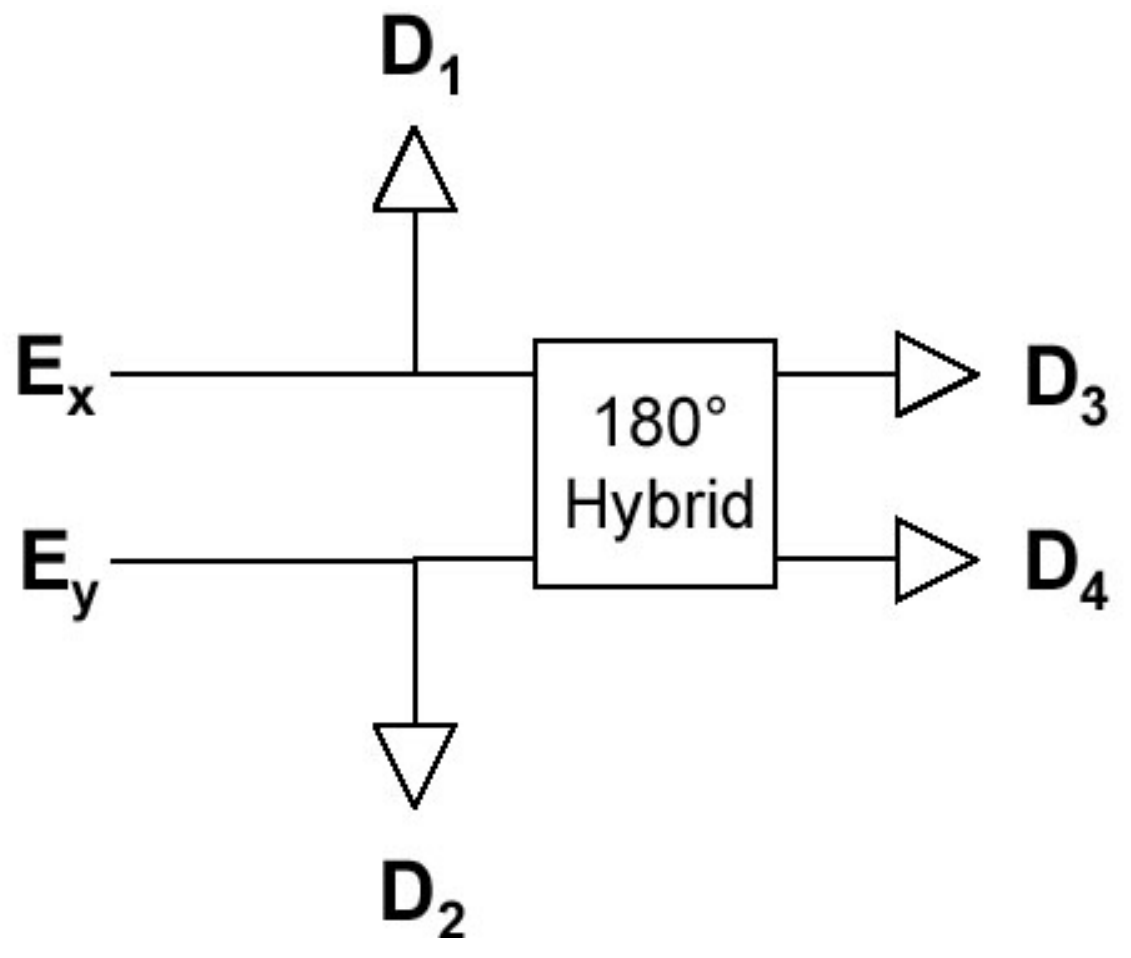

Figure 6.5.3. QU-analyzer for direct detectors. Electromagnetic radiation from a feed or antenna is transmitted to two power splitters. Half the signal from each splitter goes to a 180° hybrid, and half to a detector (D1 and D2). The output channels of the hybrid, which give the sum and difference of the inputs, then pass to detector pairs D3 and D4. The signals in the 4 detectors are a combination of 3 Stokes parameters: $S_1 = E_x^2 = (I + Q)$, $S_2 = E_y^2 = (I - Q)$, $S_3 = (E_x + E_y)^2 = (I + U)$, and $S_4 = (E_x - E_y)^2 = (I - U)$. All of the linear Stokes polarization information can be extracted by forming the pair differences and sums: $Q = 2(S_1 - S_2)$, $U = 2(S_3 - S_4)$, $I = 2(S_1 + S_2) = 2(S_3 + S_4)$.

A wideband microstrip 180° hybrid can accomplish the Q/U analysis and be integrated seamlessly with the antenna. As shown in Fig. 6.5.4 such a hybrid consists of a "rat-race" coupler, familiar to microwave engineers, combined with a broadband microstrip crossover, which rearranges the two input ports of the rat-race coupler to be on the same side. The wide design bandwidth is achieved by tuning the reflected/transmitted waves at the T-junctions, controlled by the impedance of each section.

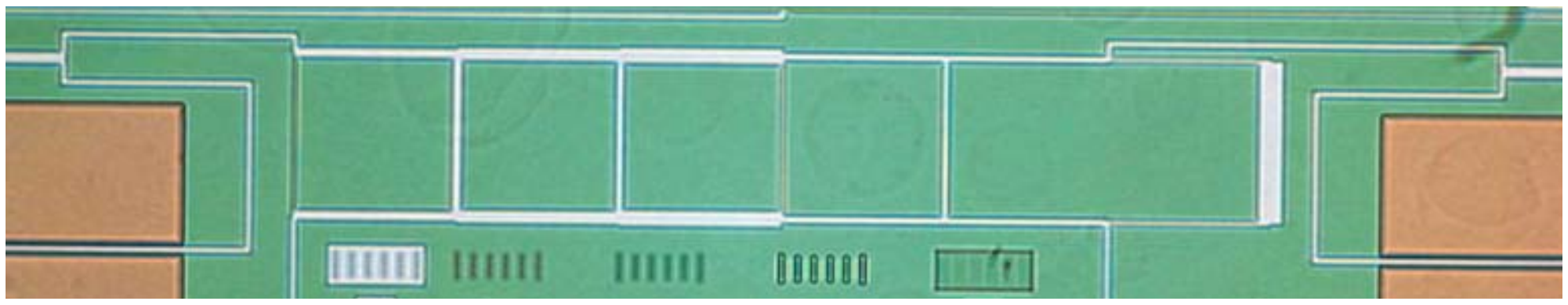

Fig. 6.5.4. Broad bandwidth 180˚ hybrid in microstrip.

## 6.6 Cooling System

We baseline a cryocooler for EPIC-LC, offering essentially unlimited mission lifetime, lower weight, and allowing for decontamination in flight. The cooler consumes 340W of electrical power (nominal). By contrast, a cryostat option offers (perhaps) a lower ultimate temperature, no vibration, and higher temperature stability, simpler interfaces, but has higher mass and a more challenging launch accommodation. A cold-launched design also minimizes the number of thermal cycles of the instruments and focal planes prior to launch.

The passive cooling calculation is similar to that described in section 5.6.2 for the EPIC-LC configuration. The major difference is a baffle tube surrounding the Gregorian telescope. The telescope baffle is modeled as one conductively floating node below the support ring, and one node above. The interior of the upper portion is black for stray light control. The exterior of the upper portion, as well as both sides of the lower portion, are specular and low emittance. The telescope support ring is low emittance except for the upper annular surface, which is black, as is the back of the primary mirror. The front of the primary is (of course) low emittance.

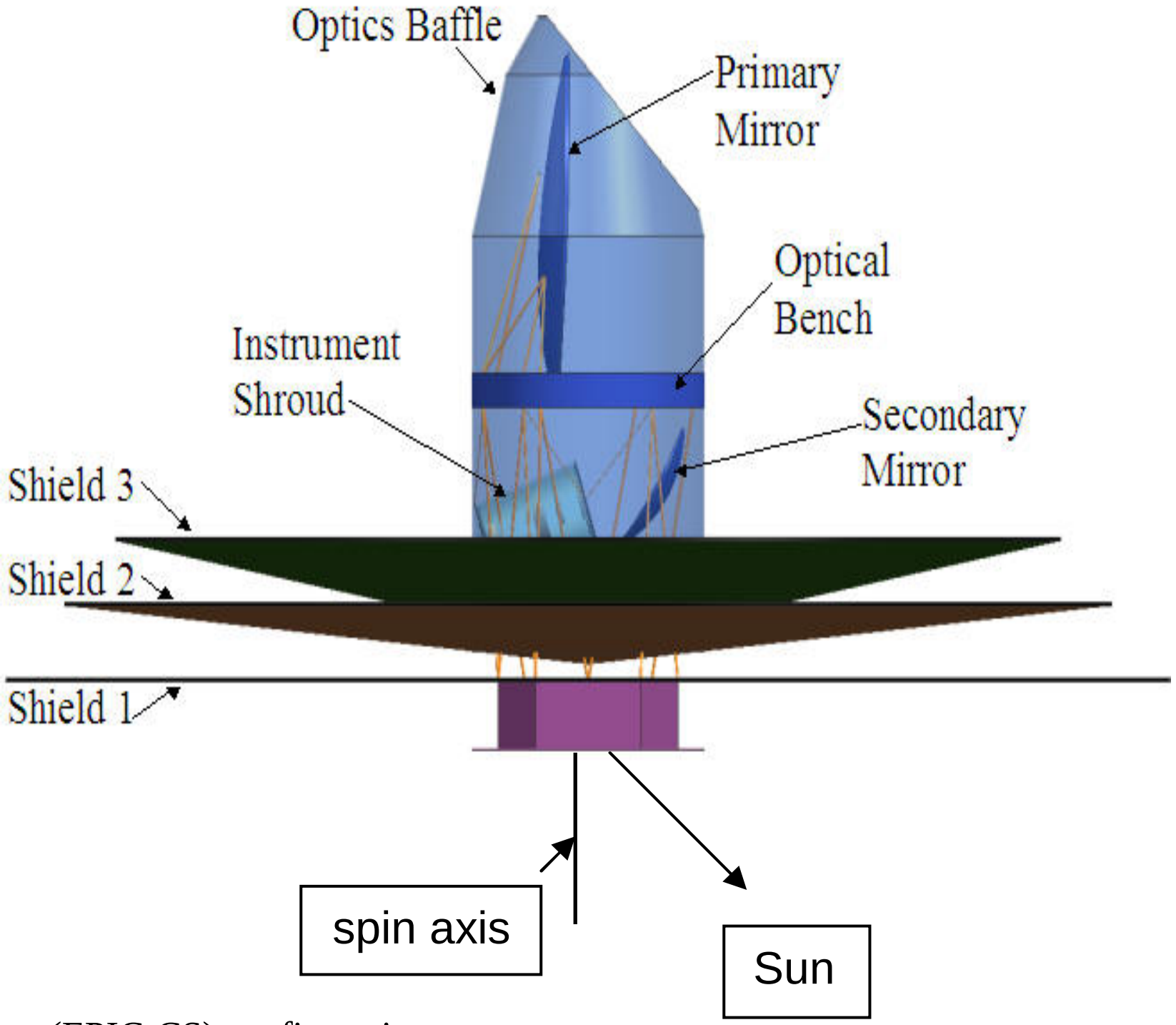

Figure 6.6.1. Dragone (EPIC-CS) configuration.

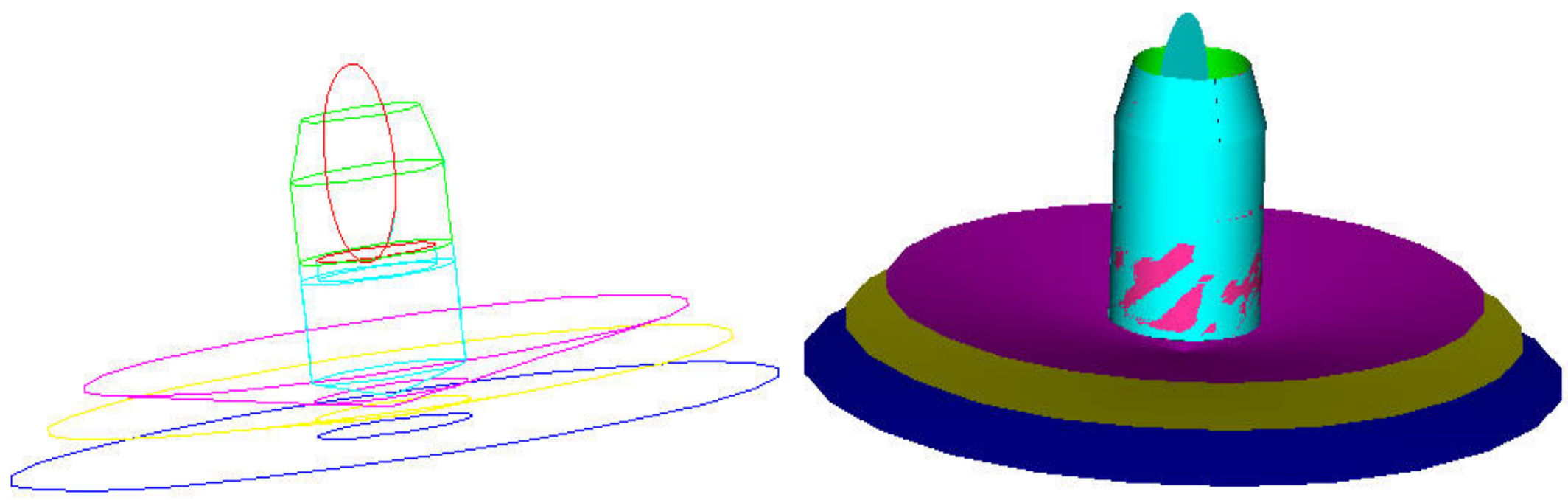

Figure 6.6.2. Wire frame and shaded views of the Dragone (EPIC-CS) radiative geometry model. The diameter of the largest deployed shield is 27.3 m tip-to-tip.

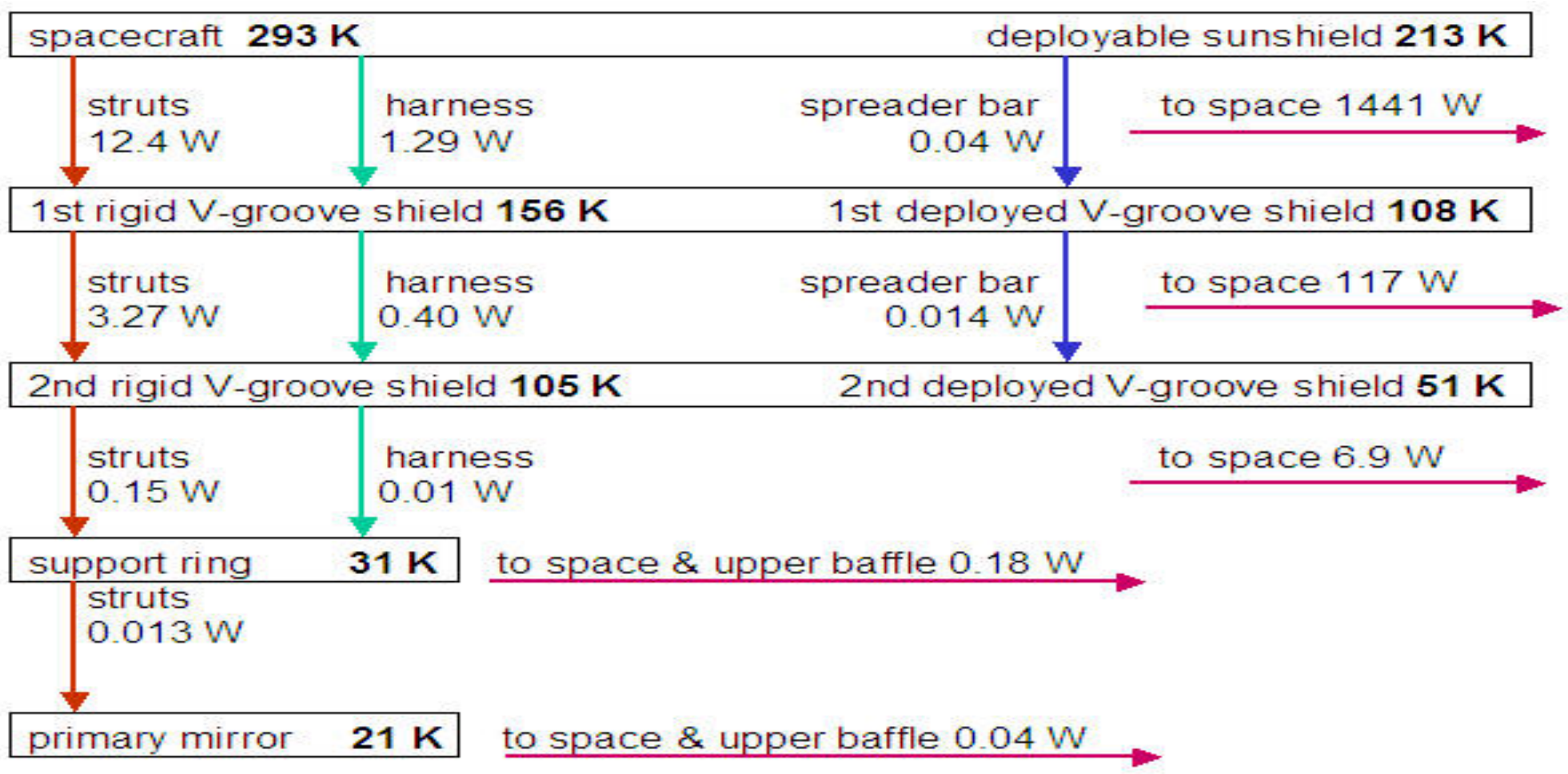

Figure 6.6.3. Dragone simplified heat flow and temperature map.

## 6.7 Large Aperture Sunshade

We have developed the design of a deployable, three-level sunshade for shielding the EPIC space telescope. The deployment is based on high-TRL lenticular wrapped ribs that meet the mechanical and thermal requirements for the EPIC-CS option.

### *6.7.1 Requirements*

The Dragone telescope is sized for launch in an Atlas V rocket with a 5-m diameter fairing. While in a halo orbit at L2, the spacecraft will rotate at 1 RPM about its central axis for the telescope, which has a line of sight tilted ~55° with respect to the spin axis, to scan a 90° conical region once per minute. The spin axis will precess at 1 RPH at a cone half-angle of ~45° to give a full hemispherical scan once per hour. To prevent mechanical disturbances resulting from this rotation, the lowest natural frequency of the sunshade in the plane of rotation should be at least 10 times the rotation rate. Thus, the first mode should occur above 0.167 Hz. In addition to this frequency requirement, the load carrying members of the sunshade must have an acceptable factor of safety against buckling and material strength failure. Because the three layers of the sunshade membrane are tensioned to minimize film wrinkling, the supporting struts to compressive loading. Furthermore, because the top two sunshield layers are additionally tensioned by the spreader bars to maintain their separation and support, the spreader bars themselves are subjected to bending stresses. Based on this mission description, a summary of the requirements for the deployable sunshade is given in Table 6.7.1.

**Table 6.7.1. Summary of EPIC Sunshade Requirements**

| Requirement | Dragone Telescope |
|---|---|
| Maximum shade diameter (tip to tip) | 27.3 m |
| Minimum shade diameter (scallop to scallop) | 21.7 m |
| Stowabe inside rocket fairing | Atlas V |
| Fairing diameter | 5 m |
| Fundamental frequency (in the plane of sunshade) | >0.167 Hz |
| Factor of safety on buckling of lenticular struts | >6.0 |
| Factor of safety on bending strength of spreader bars | >3.0 |
| Viewable area of spreader bar cross section | 1 $cm^2$ |
| Number of reinforced aluminized shade layers | 6 |
| Maximum mass for sunshade and deployment hardware | 500 kg |

*6.7.2 Technical Approach*

We studied the wrap-rib lenticular strut concept in detail because it provides a reliable, high-TRL level deployment system that has flight heritage (ATS-6[1,2] and hundreds of classified antennas) and is lightweight. The ATS-6 antenna is shown in Fig. 6.7.1. The EPIC sunshade design is directly analogous to ATS-6: radial ribs are wrapped around a central hub, and are released (or, motor driven) in space to deploy a lightweight mesh (or, film for EPIC).

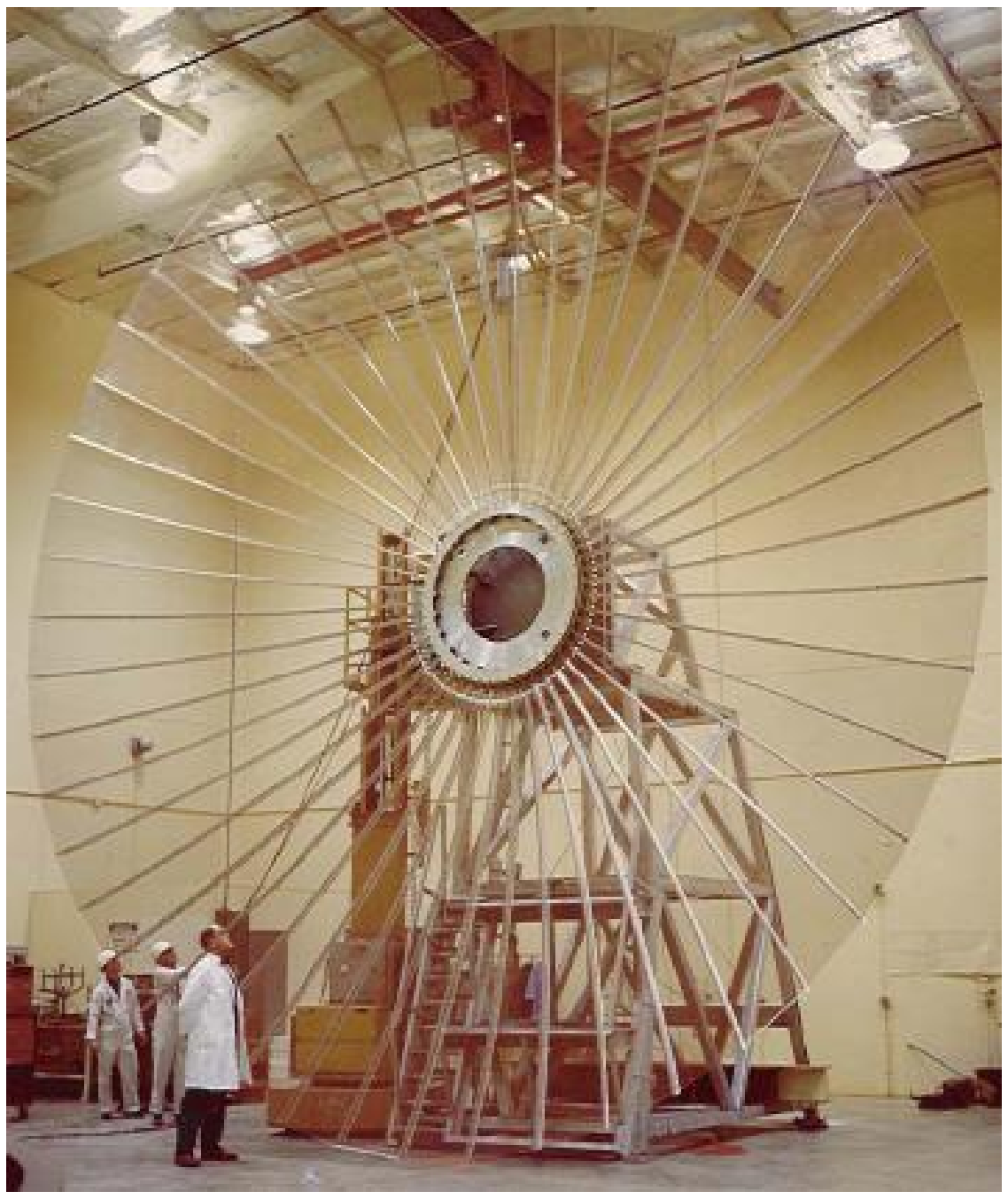

Figure 6.7.1. ATS-6 - 30-ft (9.1-m) Diameter Wrap-Rib Deployed Antenna, Lockheed Missile & Space Corporation, Inc.

The deployed EPIC sunshade, depicted in Fig. 6.7.2, is mounted to three conductive aluminum-faced honeycomb V-groove radiators (5-cm thick) spaced vertically 13.5-cm apart (center-to-center) at their center and 25-cm apart at their edges at the center of three dual-layers of 1-mil ($2.54x10^{-5}$-m) thick, reflective, reinforced membrane film. This thickness of film is chosen so that it can be easily folded and stored without becoming permanently creased. The reinforcement is to prevent ripping of the film due to pinholes caused by micrometeorites. The dual-layer configuration is used because the close spaces between the films radiate heat to cold space almost as a black body. These film layers are supported by six lenticular cross-section struts that extend radially from a central mounting hub. The ends of the three dual-layer films are supported and separated by a spreader bar attached at a 45° angle to the ends of the booms.

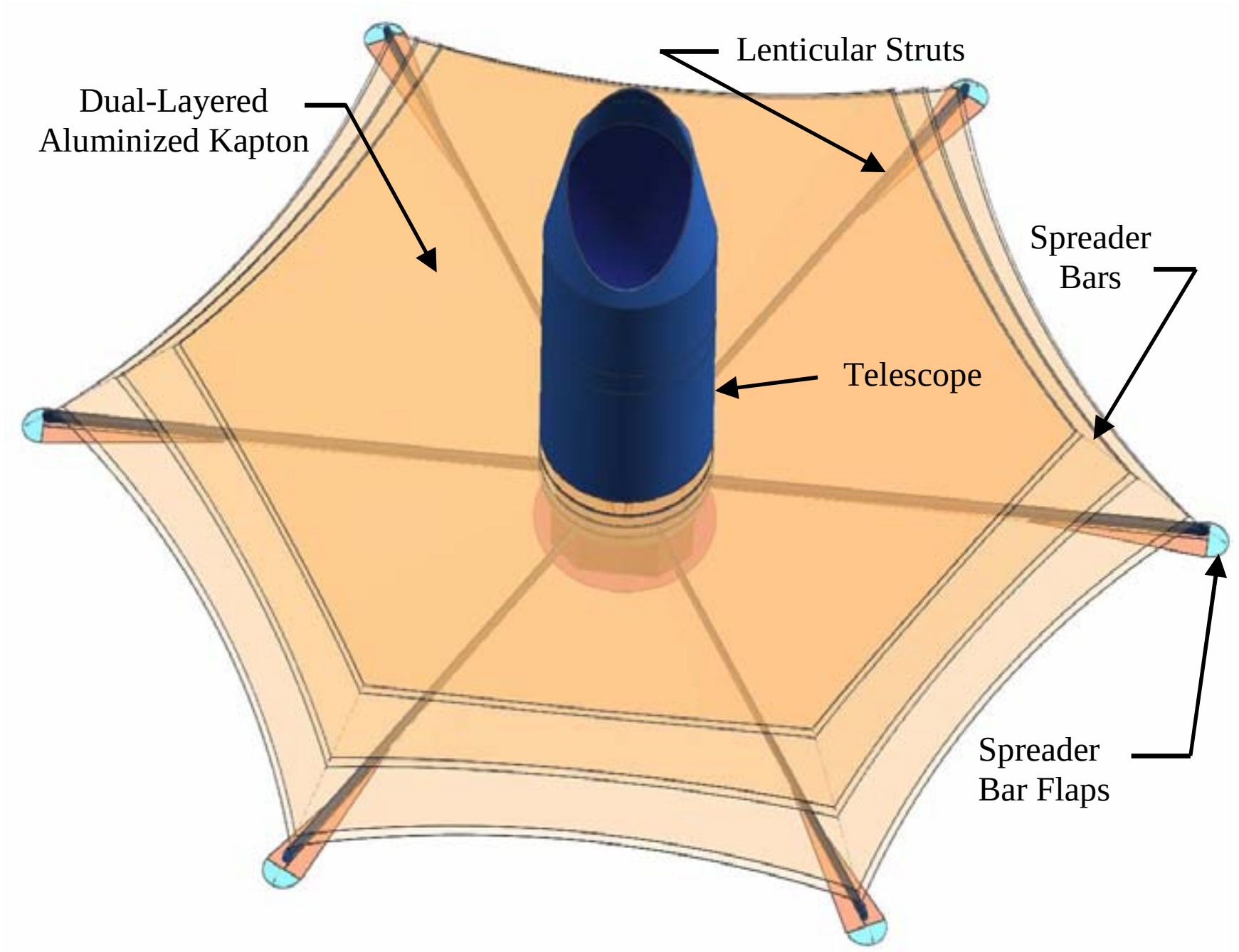

Figure 6.7.2. Sketch of Deployed EPIC Sunshade.

### *6.7.3 Wrap-Rib Heritage*

The lenticular wrap-rib concept was developed by Lockheed Missile and Space Corporation in the early 1960's for use with deployable antennas[3]. As seen in Fig. 6.7.3, this system is composed of a set of ribs having a lenticular cross-section (Fig. 6.7.4), cantilevered from a central, rigid hub. These ribs support, for the case of a sunshade, the thin, reflective film. For stowage during launch, these ribs are wrapped around the central hub. The nature of the lenticular cross-section allows the tubes to be collapsed elastically, thus storing a large amount of strain energy within the strut material. Upon release, this stored strain energy causes the collapsed, wrapped ribs to spring into their stiff, undeformed (deployed) shape. However, for larger wrap-rib systems, there is such a large amount of stored strain energy that a simple "release and deploy" approach would happen too quickly, likely causing damage to the struts, the film, and possibly the spacecraft or instrument. To limit the deployment rate, the EPIC design uses a motorized hub that attaches and supports the lenticular struts. This motorized deployment concept was used also by Lockheed, and a deploying antenna is shown in Fig. 6.7.5.

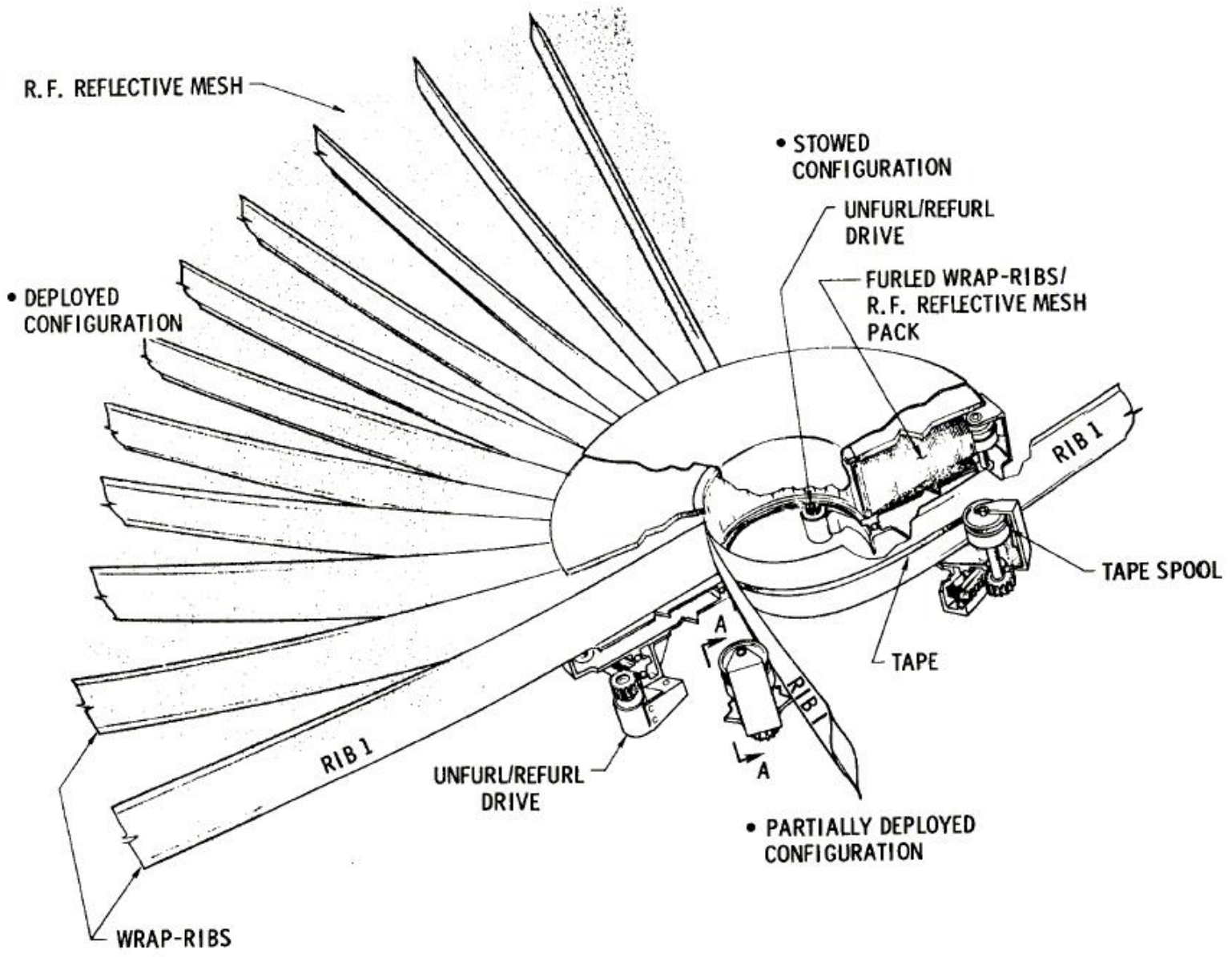

Figure 6.7.3. Wrap-ribs attached to central, motor driven hub [Chadwick and Jones, 1975].

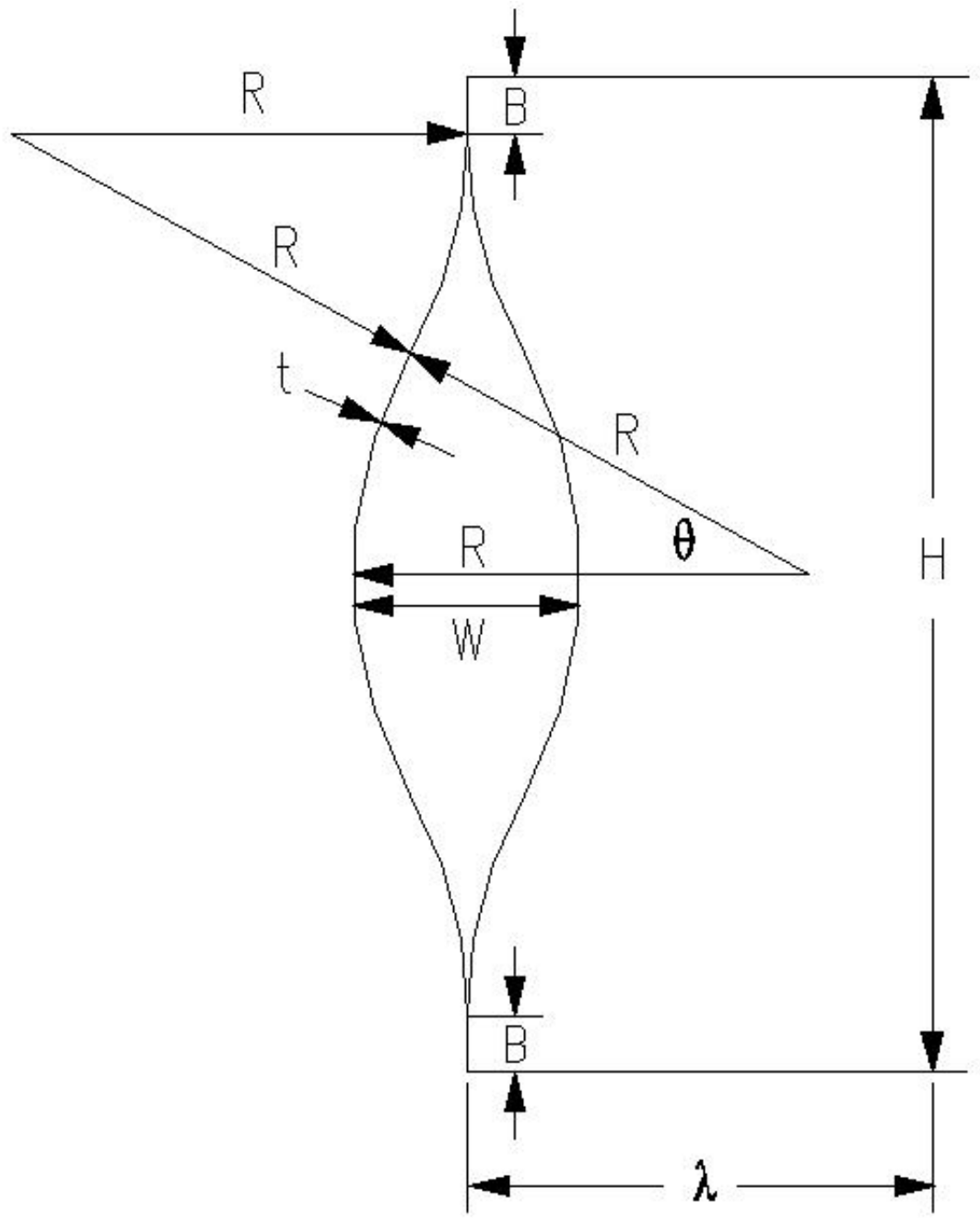

Figure 6.7.4. Cross-section of a single lenticular wrap-rib.

Lenticular wrap-rips have been successfully used to deploy spacecraft antennas, such as the 30-ft (9.1-m) diameter X-band reflector for the ATS-6 mission,[1,2] seen in Fig. 6.7.1, and hundreds of classified missions of various sizes over the past 30 years. Lockheed Missiles & Space Company have also designed, built and flight qualified 3 (0.91)-, 6 (1.82)-, 10 (3)-, 20 (6.1)- and 30 (9.1)-ft (m) diameter deployable wrap-rib reflectors, as well as partially constructed a 55-m diameter antenna for NASA in 1984[4]. All of these antennas were ground tested in a 1-g environment.

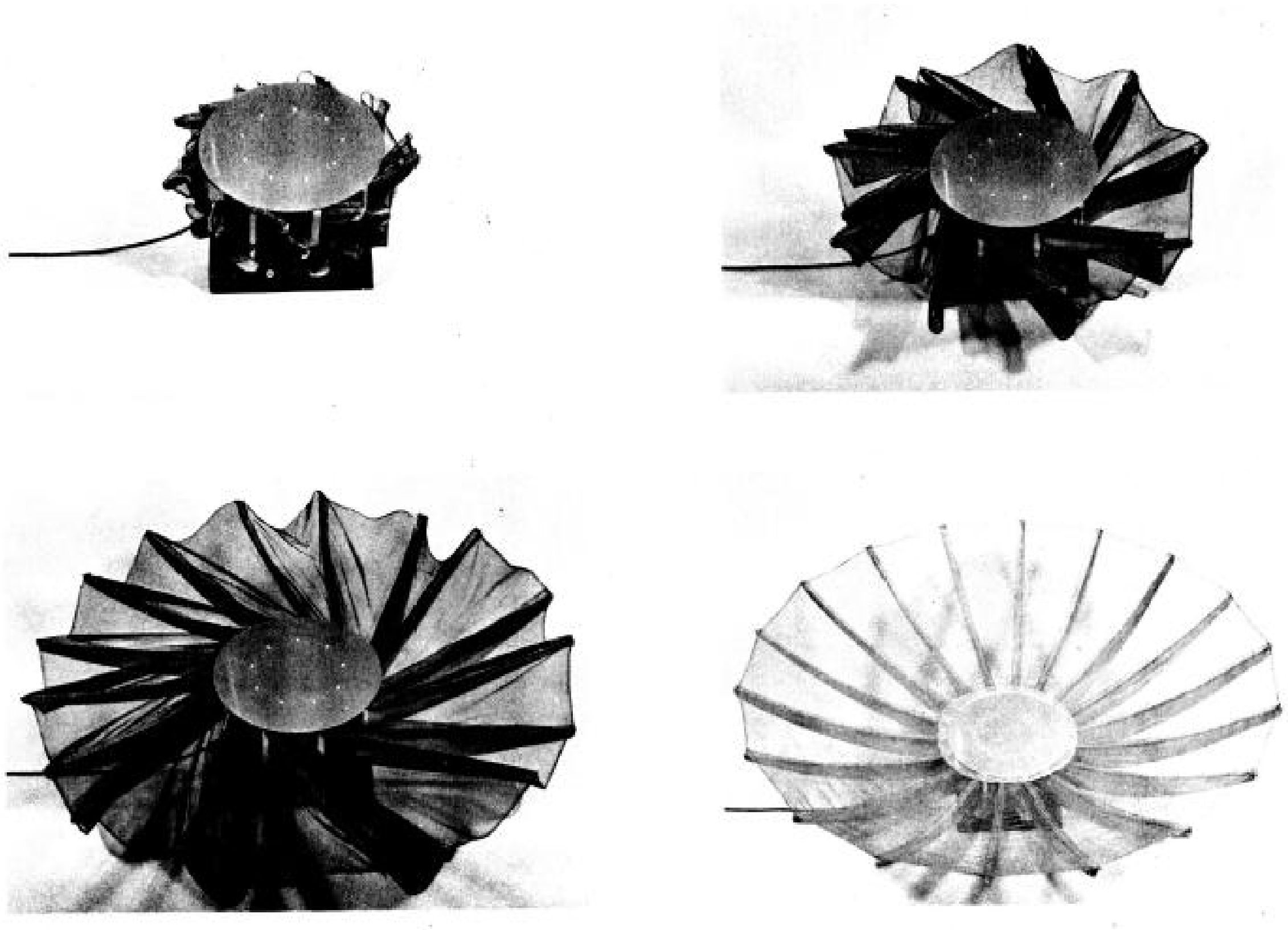

Fig. 6.7.5. Deployment using a Motor Driven Hub [Chadwick and Jones, 1975].

Lenticular struts have also flown successfully in space and have been tested extensively in the laboratory. The European Space Agency used an 8-m long lenticular collapsible tube mast as a stand-alone monopole antenna on its ULYSSES spacecraft[5, 6]. It was also wrapped around a central hub for launch and deployed using the centrifugal force of the spacecraft. The deployment process was designed to be irreversible, that is, it was never retracted during the mission. A retractable version was later developed by the Spanish company SENER[7]. SENER and British company Harwell[5] have both studied the manufacturing, modeling, and deployment of lenticular booms for space applications. Their concept, shown in Figure 6, involves each strut being wrapped on its own mandrel and independently deployed.

A 14-m long continuous tube mast was also developed by the German Aerospace Centre (DLR) for use with solar sails[8]. In 1965, one of the earliest studies of the mechanics of lenticular struts was performed at NASA Lewis Research Center. Three struts were fabricated and tested to determine the stress and strain behavior of the material when collapsed and wrapped around a central hub. They found the lenticular design to be a practical solution for a strut that is foldable for launch, reliably deployed in space, and provides a great deal of structural rigidity after deployment[9]. The early 1980's saw a three phase study between NASA Langley and Howard University investigated manufacturing and testing of lenticular struts with a large number of various cross-sectional geometries. Their work focused on full-deployment and buckling stability of the struts[10]. They found the lenticular geometry offers more resistance to buckling than a comparable circular cylinder.

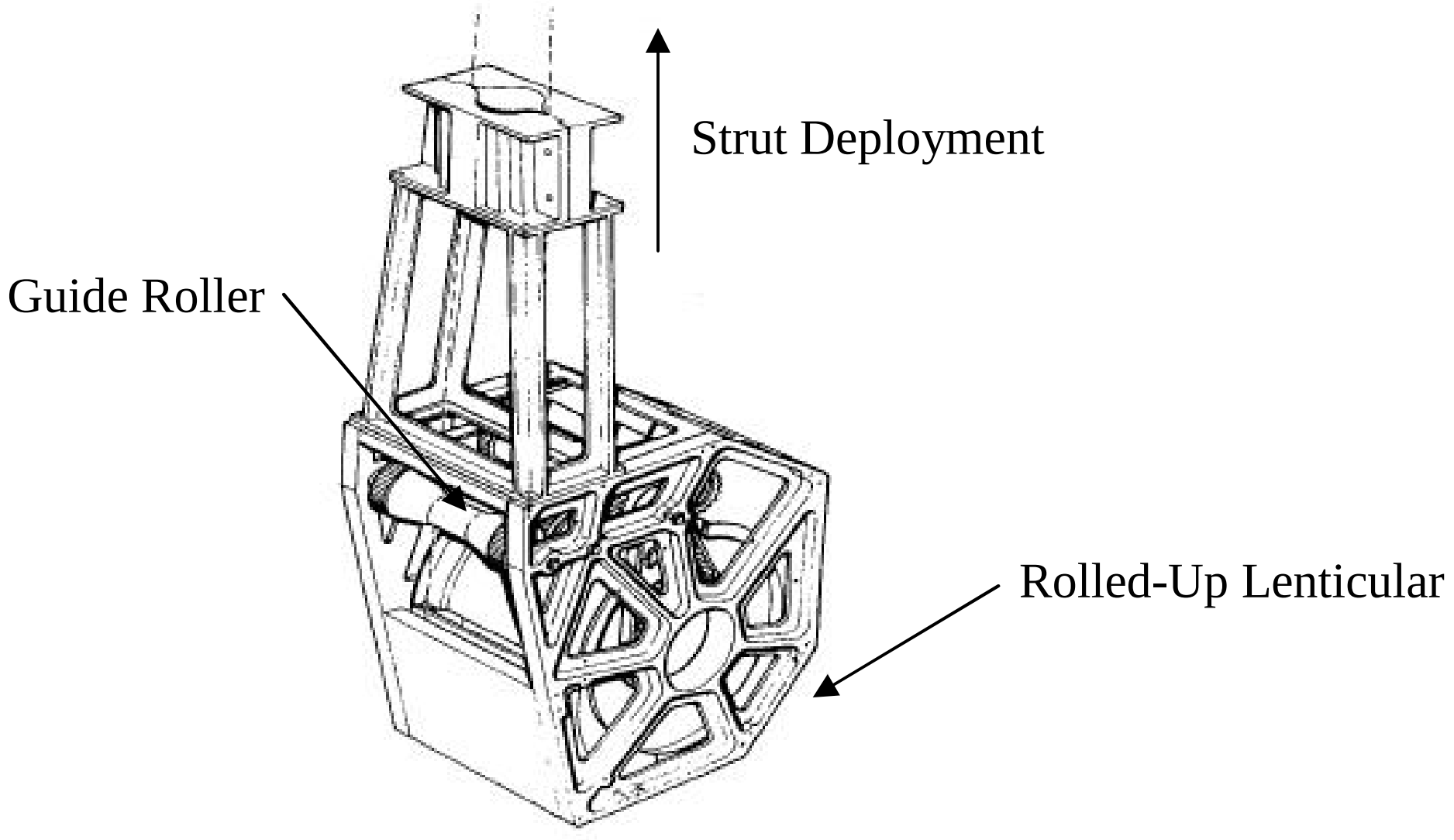

Figure 6.7.6. Continuous Tube Mast Stowage and Deployment System [Aguirre et al. 1985].

Currently in the United States, there are at least three vendors for this type of technology: Lockheed Missiles & Space Company, Composite Technology Development Inc., and Composite Optics, recently acquired by ATK Space Systems. These are in addition to the European companies discussed above. The next phase of the EPIC study will likely necessitate partnering with one or more of these companies to determine their manufacturing capabilities for graphite-epoxy lenticular struts

*6.7.3 Storage and Deployment of Sunshade*

Based upon the flight heritage of ATS-6 and the other deployable technologies described above, this section describes the storage and deployment of the EPIC sunshade shown in Fig. 6.7.2. The stowed configuration of the lenticular struts is presented in Fig. 6.7.7, where the folded membranes are omitted for clarity.

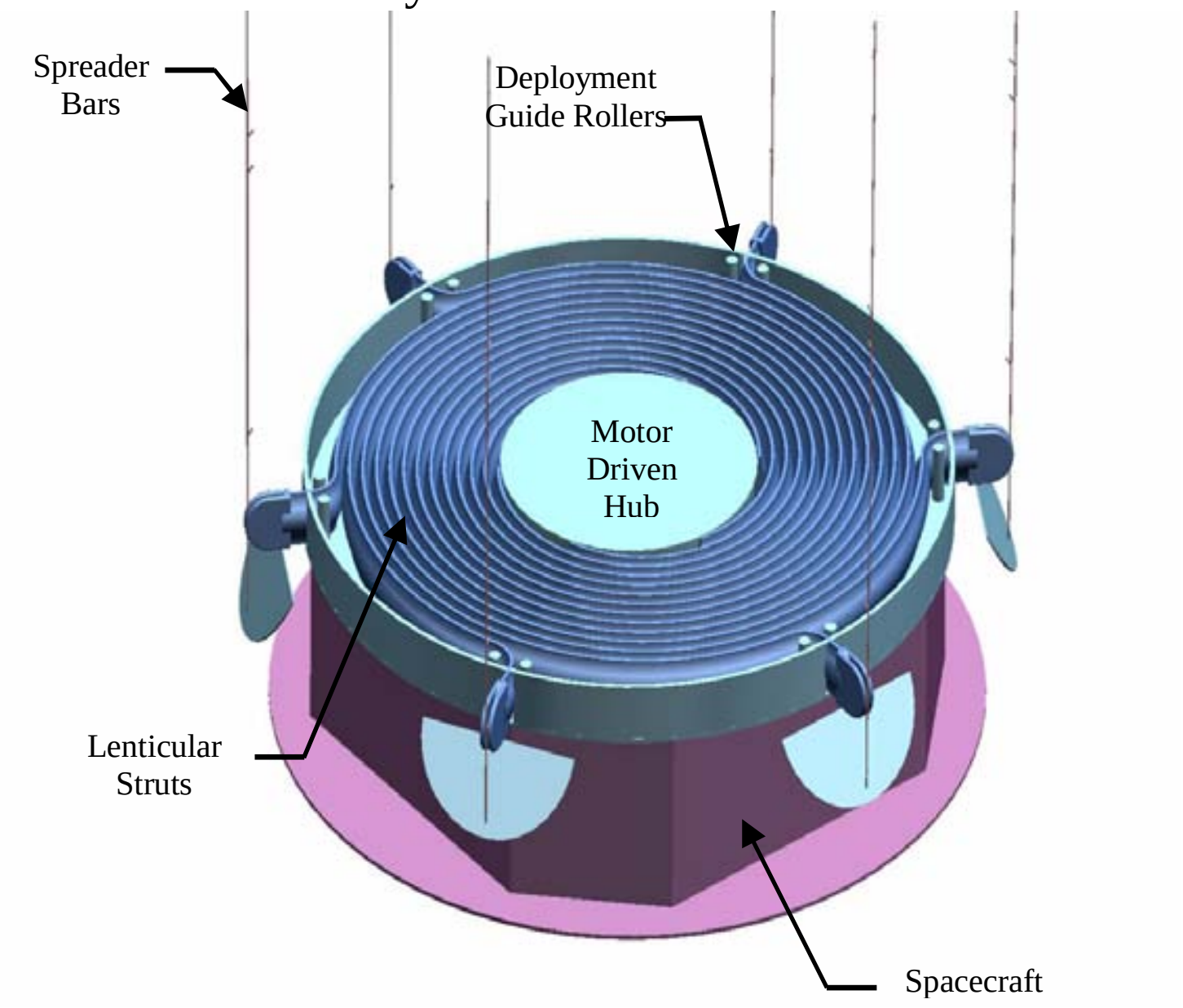

Figure 6.7.7. Stowed Configuration of EPIC Sunshade.

The spreader bars are stowed along side the telescope during launch. The lenticular struts are hinged to and wrap around the motor driven hub, located beneath the bottom V-groove radiator and above the spacecraft. Next, the reflective film is folded according to the pattern shown in Fig. 6.7.8.

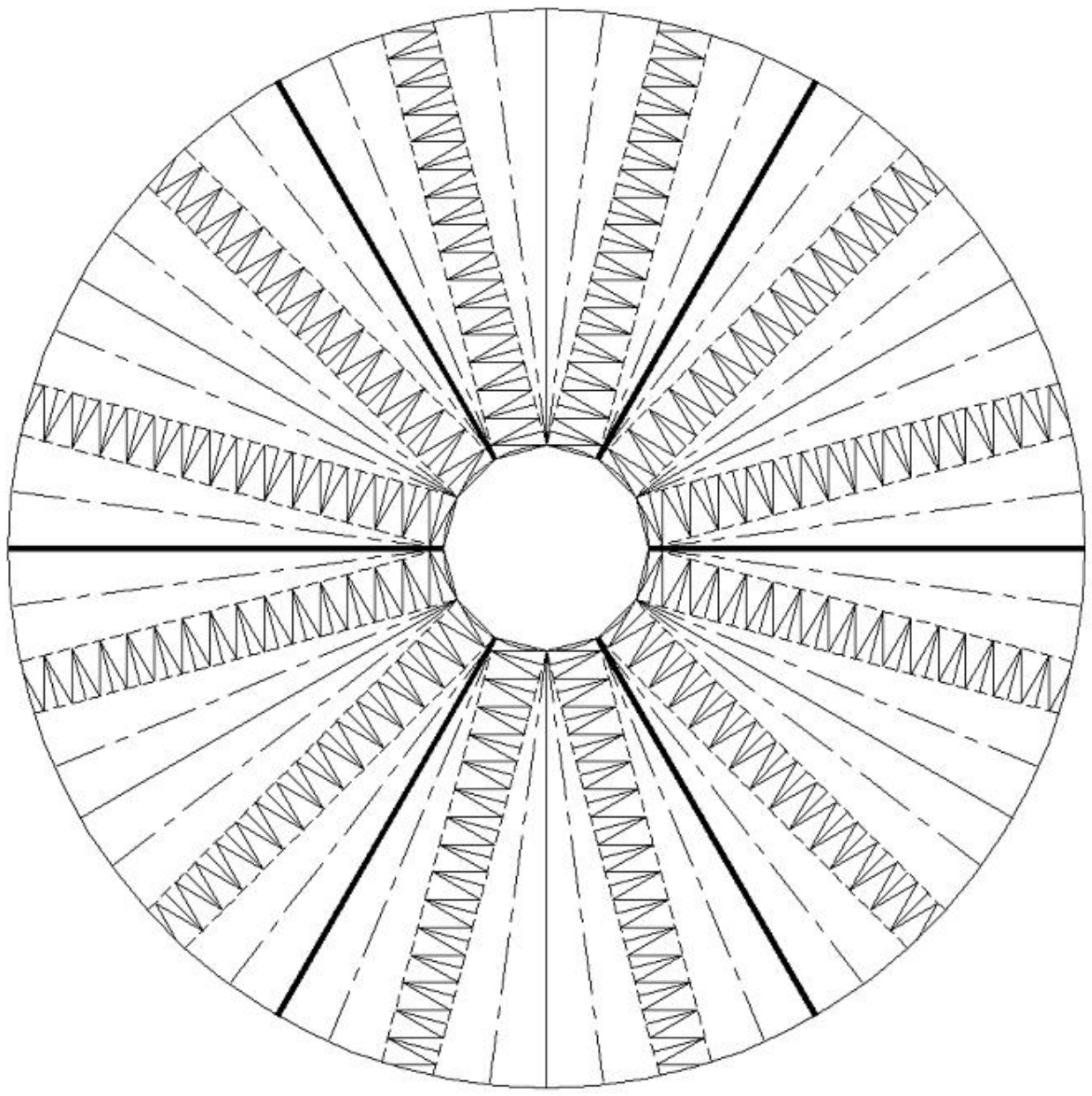

Figure 6.7.8. Folding Pattern for EPIC Sunshade Film.

This folding pattern is designed as if the sunshade had 12-sides/struts to make the width of the folded sunshade equal to 0.3-m, which is the amount of space allotted for membrane storage. The folding pattern is merely a geometry issue; it does not matter that the EPIC sunshade only has 6-sides/struts. The bold lines in Fig. 6.7.8 represent the location of the struts (beneath and not touching the sunshade), as well as define the 6 sections of the film that will be folded individually and then seamed together during packaging. The two types of dashed radial lines are folded first so that the "zig-zag" regions touch to form a diamond pattern. The diamond lines are folded next, and finally the long axes of the diamonds are folded. As the folding progresses, the sunshade will evolve into an accordion-like column of film that is readily deployed in the radial direction. Then, the folded membrane is stored around the perimeter of the three V-groove radiators, as shown in Fig. 6.7.9.

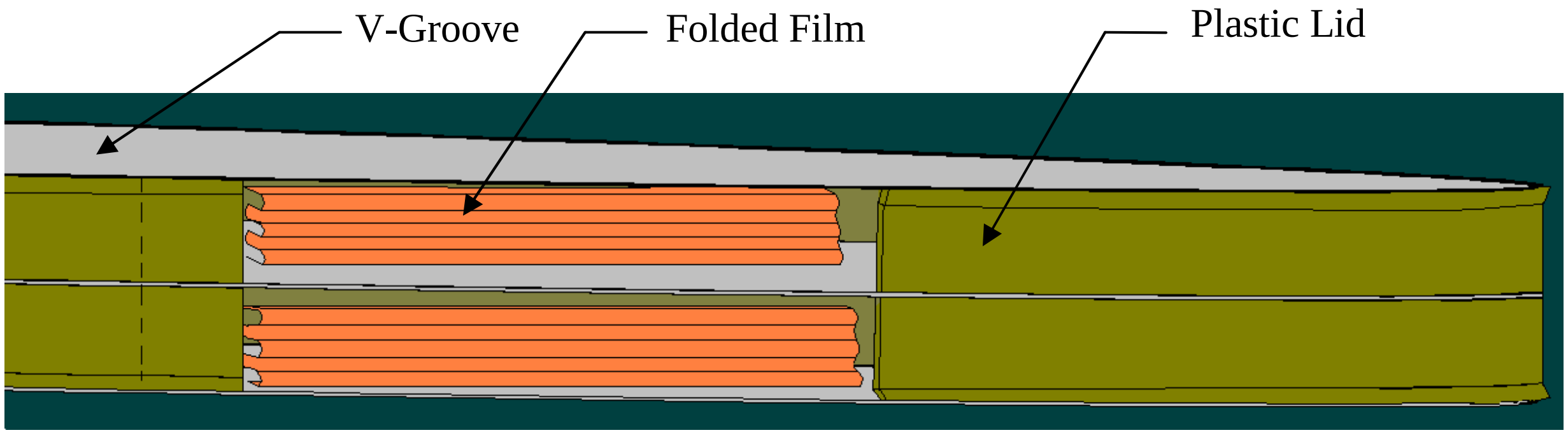

Fig. 6.7.9. Sketch of the Folded Sunshade Film, Storage Container and Lid.

The lightweight plastic lids are designed to allow any trapped air to vent during launch to prevent damage to the folded sunshade. During deployment, these lids allow only one fold to deploy at a time and maintain tension in the film during deployment to prevent tangling or damage to the membrane. Table 2 shows the calculated volume required to store the folded membrane inside the outer rim of the V-groove radiators as shown in Fig. 6.7.9. For this calculation, the film is presumed to have a thickness of 0.002-in (.05-mm) to conservatively account for the reinforcement thickness, and a packing factor of 3 is assumed. However, this increased thickness is not used for the stress analysis of the sunshade. Clearly, all three folded dual-layers of the sunshade will easily fit into the storage boxes inside the V-groove radiators. In space, the motor driven hub is rotated and the sunshade deploys as depicted in Fig. 6.7.10.

The sunshade begins in (a) in the stowed configuration, inside the edges of the V-groove radiators, as detailed above in Fig. 6.7.9. When the motor is engaged, the strain energy stored within the struts will drive them outwards in the radial direction through the guide rollers (b). The tension of the membrane unfolding from the storage boxes will cause the spreader bars to rotate from their vertical position slightly (c), and then to their approximately 45° inclination (d), where a hard-stop inside the spreader bar pivot will hold them in place for the remainder of the deployment phase (e). During this time, the deployment of the lenticular struts can be monitored using either a low-resolution video feed or an encoder-type position sensor along the guide rollers or within the motorized deployment hub. Also, the spreader bar deployment cables will be feeding out to keep pace with the ends of the struts, but their monitoring their tension is not important at this point because the spreader bars are held in the correct position by the hard-stop within the spreader bar pivot. The spreader bar cables need only be tight enough not to get tangled during deployment. Should the deployment become stuck for some reason at this point, the motor can be reversed to slightly retract the struts in order to correct the potential sticking point. Once the struts are fully deployed, the spreader bar deployment cables are tightened using a motorized spool to the proper tension to hold the spreader bars at the correct angle for the duration of the mission (d). This tension should be monitored in order to ensure proper positioning of the spreader bars while in transit and upon arrival at L2. Constant force springs stored inside the spreader bar pivots will tension the membranes directly and accurately by way of Kevlar cords.

**Table 6.7.2. Packaging Volumes for Dragone Sunshade**

| Layer Number | Film Volume, $m^3$ | Packing Factor | Packaged Volume, $m^3$ | Volume Required in V-Groove Storage Box, $m^3$ | V-Groove Storage Box Volume, $m^3$ |
|---|---|---|---|---|---|
| 1 (Warmest) | 0.02147 | 3 | 0.06440 | 0.1271 | 0.1838 $m^3$ |
| 2 | 0.02090 | 3 | 0.06271 | | |
| 3 | 0.01734 | 3 | 0.05203 | 0.1028 | 0.1838 $m^3$ |
| 4 | 0.01693 | 3 | 0.05079 | | |
| 5 | 0.01433 | 3 | 0.04299 | 0.0850 | 0.1838 $m^3$ |
| 6 (Coldest) | 0.01401 | 3 | 0.04203 | | |

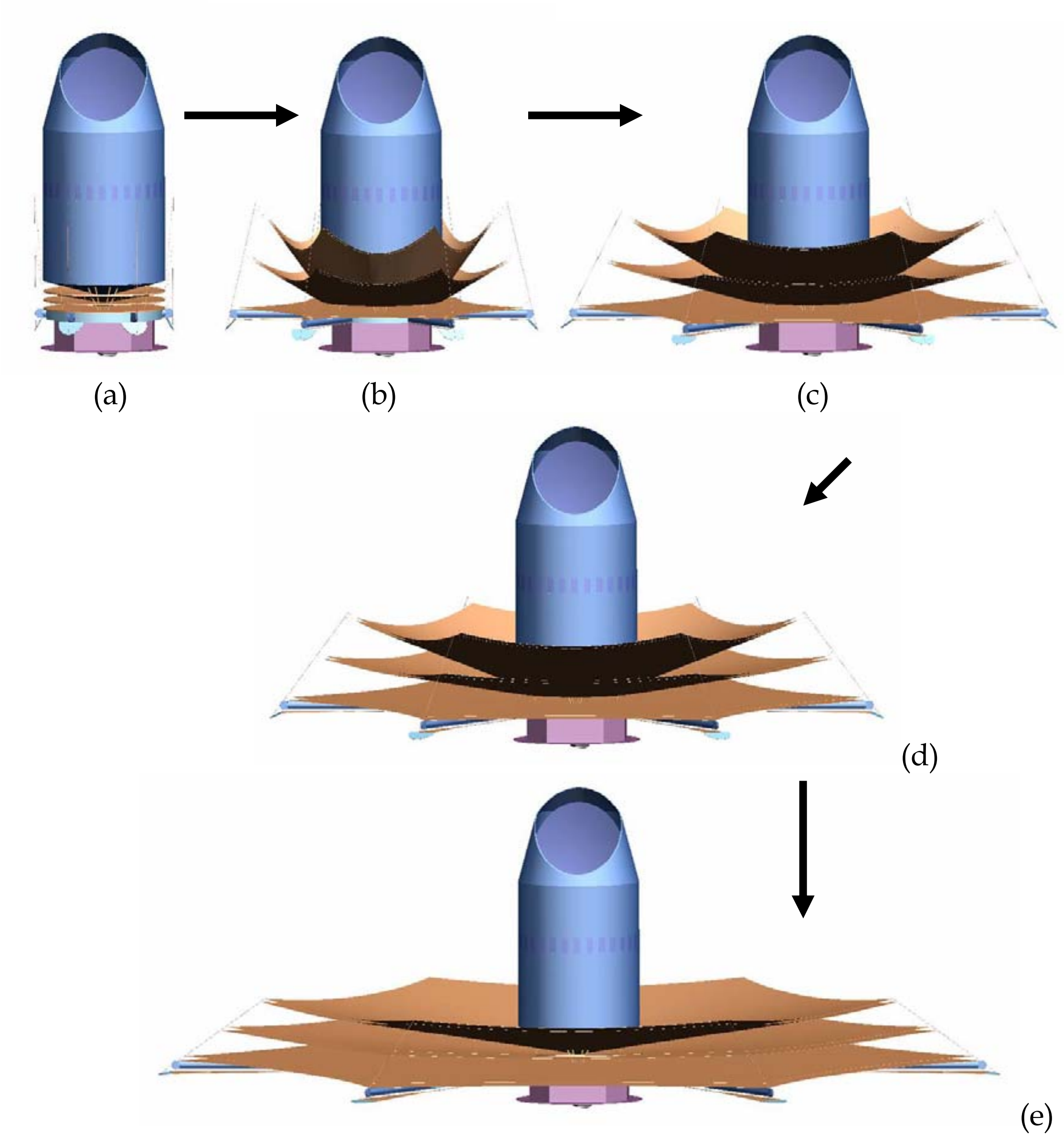

Figure 6.7.10. Deployment of the EPIC Sunshade.

While deployment can happen very quickly, the motor driven hub is estimated to deploy the EPIC sunshade in a matter of a few minutes. The spinning motion of the telescope will slightly complicate the deployment because it only rotates at 1 RPM, so one may consider deploying the sunshade before spin-up in space.

### *6.7.4 Deployed Configuration of Sunshade*

Given the architecture for stowing and deploying the sunshade, the details of the deployed configuration are now discussed. The fully deployed sunshade is shown in Fig. 6.7.2, and Fig. 6.7.11 shows a detailed view of a spreader bar region. The spreader bar is attached to the end of the strut using a pivoting mechanism with a built-in hard-stop that allows the spreader bar to be stowed upright during launch, and then to rotate as necessary during deployment. When the struts reach their fully deployed position, the spreader bar deployment cable tension is adjusted using the spreader bar deployment motor, which properly positions the three dual-layers of reflective film. To eliminate stray light reflected into the telescope optics, the exact position of the films is controlled by adjusting the tension using constant force springs and Kevlar cables that are contained inside the lenticular strut or spreader bar pivot. These constant force springs will be designed to maintain the proper tension in all of the layers at all times, even in the presence of thermally-induced deformations, such as the Kevlar cords expanding or the Kapton films contracting when they are cooled. The spreader bar flap attaches to the bottom of the bottom film to keep the spreader bar in the shade at all times, again to eliminate radiation heat transfer into the telescope. As described earlier, the telescope is spinning at 1 RPM. This rotation causes a small amount of tensile force in the lenticular struts, which would tend to slightly increase the fundamental natural frequency. Therefore, the present analysis, which does not include this rotation, presents a conservative design. During the life of the mission, various components of the sunshade will see different temperatures. The lower temperatures will slightly decrease the damping in the composite struts. However, high damping is not required as the system is designed to have its resonance much higher than the 1 RPM excitation frequency. Also, since the lenticular struts are always in the sun and the sunshade is deployed while warm (early in the mission), no major effects are expected due to temperature dependent changes in material properties.

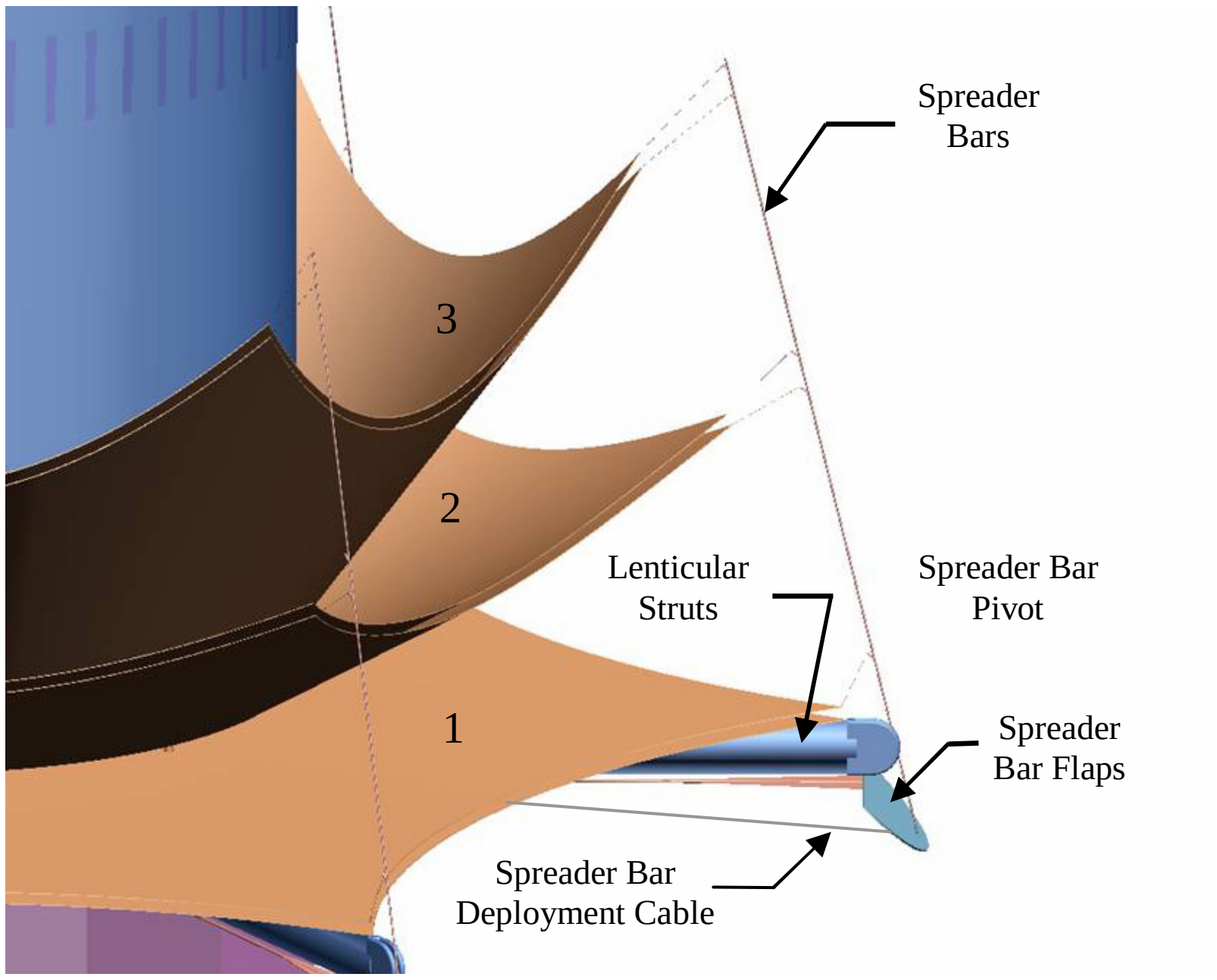

Figure 6.7.11. Detailed View of Spreader Bar Region.

### *6.7.5 Ground Testing*

As discussed above, previous wrap-rib antennas have been successfully ground tested prior to launch. The EPIC sunshade can also be tested on the ground, which offers advantages in terms of verification testing and system reliability. This section shows a simple mechanics of materials model to show that ground testing of the EPIC-CS sunshade is possible without buckling the lenticular struts. Since the g-level during burn maneuvers in space are much smaller than 1g, this design will easily survive the orbital insertion maneuvers near L2. A single strut is modeled as a cantilevered beam in a 1g environment with both a tip load (the spreader bar pivot and the equivalent mass of the sunshade it supports) and a distributed load (the weight of the strut) as depicted in Fig. 6.7.12.

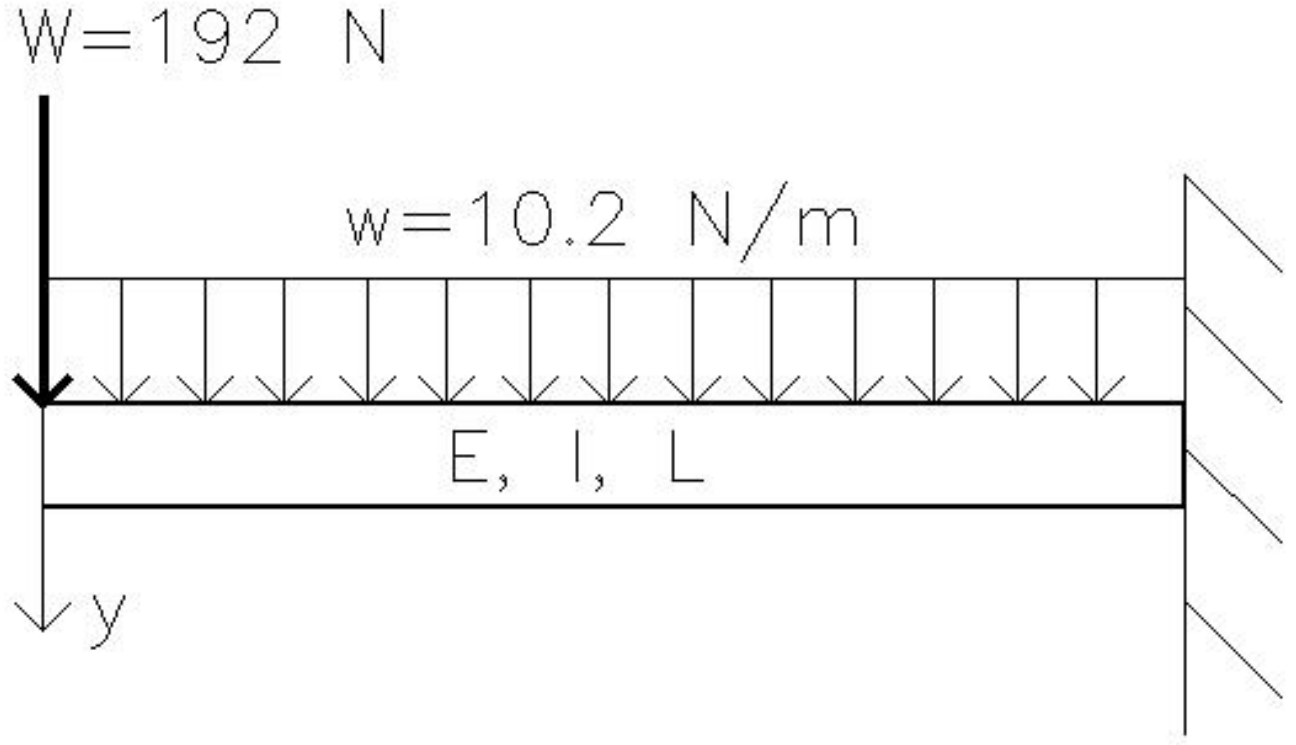

Figure 6.7.12. Model of Lenticular Strut from EPIC-CS Sunshade (Length = 11.76-m) with Tip and Distributed Load in 1g.

For this loading scenario, the maximum deflection is calculated to be 36-cm for the 11.76-m long strut. This deflection represents about 3% of the length, which is reasonably small, but of enough concern to check the local wall buckling of the strut. Considering the lenticular cross section, the maximum compressive stress, which occurs at the bottom of the beam, is 85 MPa. Since wall-buckling is a localized effect, the beam is considered as a tube with radius equal to the radius of curvature of the lenticular. For this geometry, the wall-buckling load is 489 MPa, thus there is a factor of safety of 5.75 on local wall-buckling of the struts for the EPIC-CS sunshade during ground testing in 1g. However, this calculation does not include the dynamic loads that may be experienced during deployment. Therefore, it is recommended that the tips of the struts be gravity-offloaded during deployment testing in 1g. In this case, the gravity offload system will be designed to offload the tip mass, equivalent load of the sunshade material and lenticular strut so that no deflection occurs at the tip. The only deflection would occur along the span of the beam due to its own weight, with a maximum deflection of 3-cm just to the left of center. This amount of deflection is very small for a beam of this length, thus the strut will clearly not buckle if it is gravity offloaded during deployment testing in 1g. The details of this analysis are in Appendix C.

### *6.7.6 Material Selection*

With the deployed configuration given above, the materials for the key components are selected. Table 6.7.3 lists the key components and the corresponding type of material. The lenticular struts are made from graphite epoxy composite to minimize weight, while the V-grooves are aluminum-honeycomb to conduct heat on their faces while being as light as possible. The inter-hub struts (bipods) and spreader bars are gamma alumina and S-glass epoxy composite, respectively, to reduce conduction and radiation heat transfer from the warmer shields into the telescope. The pulleys, motors, motorized hub, constant force springs, and spreader bar deployment cables can be metal as they are on the warm side of the sunshade and are not coupled thermally to the colder telescope components.

Table 6.7.3. Key Structural Components and Selected Material

| Item | Material |
|---|---|
| Lenticular Struts | Graphite/Epoxy composite |
| V-Grooves | Aluminum-faced honeycomb |
| 2 m-Hub attachements/hinges | Aluminum |
| Ring support struts | Gamma alumina |
| Sunshade membrane | Aluminum coated, reinforced Kapton |
| Spreader bars | S-glass/epoxy composite |
| Spreader bar pivot | Aluminum |
| Constant force springs | Spring steel |
| Membrane attachements | Kevlar cord |
| Spreader bar deployment cables | Steel |
| Pulleys, motors, bearings, guide rollers | Aluminum |
| Motor-driven hub and bottom plate | Aluminum-faced honeycomb |

### *6.7.7 Deployable Technologies*

As clearly described above, the EPIC sunshade design requires the capability for compact stowage coupled with the capability to gracefully deploy a large, multi-layer membrane to the proper shape precision. The deployment scheme selected for EPIC involves simultaneously extending six struts using an articulating boom technology that pushes out the spreader bars and the film layers attached to them. This deployment motion unfolds and properly tensions the membrane films. There are several types of articulating, deployable boom concepts available for consideration. ATK-ABLE has had success in space with the articulating ADAM MAST (STS-99 STRM, IPEX, and WSOA missions[11]). However, these are high strength, high precision structures that are not required for EPIC sunshades, as they would likely be cost prohibitive, as well as present stowage problems between the V-groove radiators during launch. ATK-ABLE also makes coilable longeron booms, GR1 and GR2 (ST8 mission[11]), which are a mature technology, but the screw-driven deployment canister hardware mass has yet to be investigated. The coilable deployment also involves a twisting motion that must be counteracted using additional hardware that has not been investigated for this project. ILC Dover has developed the Space Inflatable Ultraboom[11], which is an uncured composite isogrid structure that is unrolled on-orbit using an internal air bladder, and then rigidized (cured) in space. Again, the deployment/inflation and curing hardware is thought to be prohibitively heavy. A good overview of these and other related deployable structures is given by Tibert[12]. While other deployable boom options could be investigated in much more detail, the concept selected for analysis in this report is the lenticular wrap-rib, as it offers a fairly simple, reliable, and lightweight deployment system.

The architecture of the James Webb Space Telescope (JWST), shown in Fig. 6.7.13, is similar to that of EPIC in that it has a large, deployable sunshade that passively cools a telescope.

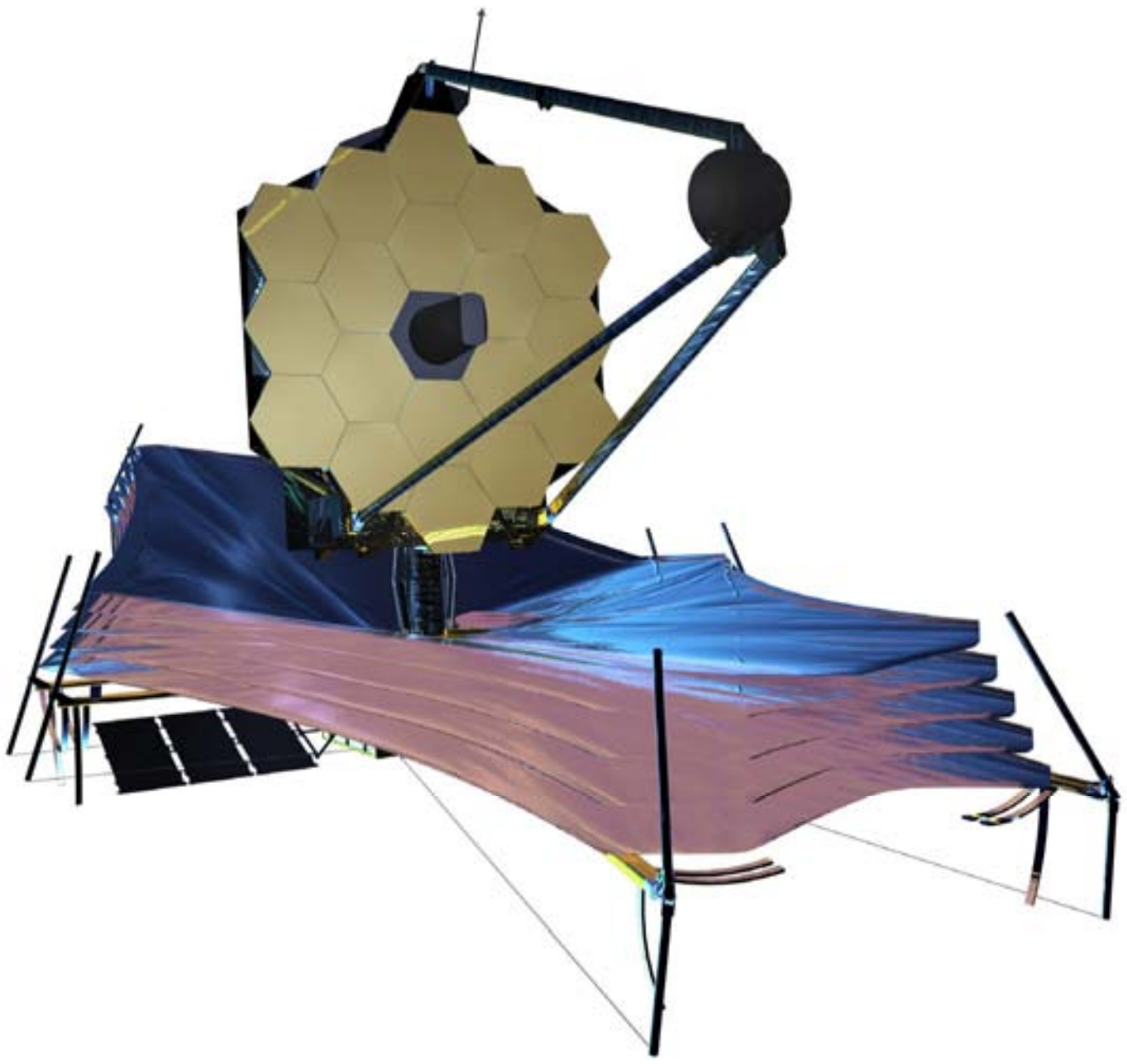

Figure 6.7.13: James Webb Space Telescope (JWST)

While JWST is farther along in its design, this EPIC design is simpler. First, the JWST sunshade has five film layers instead of three for EPIC. These layers are essentially folded only one time in a four-petal-like fashion, and stowed along side the telescope while in transit to L2. JWST's sunshade then unfolds like a flower, and a more complicated set of seven spreader bars undergo a complex rotational sequence to tension and separate the five sunshade layers. This motion is also controlled by a spreader bar deployment cable. While there are no booms to deploy on JWST, the large scale, independent motion of the four petals and spreader bars is more risky because of the increased number of autonomous components. The failure of any of these deployable sections could jeopardize the success of the mission by not properly cooling the telescope. In contrast, the simultaneous deployment of the EPIC shields using only the rotation of a single motorized hub is simpler. There are other complex issues that JWST is still addressing for its sunshade, including complicated folding patterns, air entrapment/venting and chafing during launch and solar radiation, electrical charging, and micrometeorite damage on orbit. Presumably, these hurdles in technology will be addressed by JWST in time for the solutions to be utilized by EPIC.

### *6.7.8 Specifications*

With the deployed configuration given above, a structural analysis was performed in order to design the structure to meet the requirements given in Table 1. While the details of this analysis are omitted here for simplicity and presented fully in Appendix C, they allow the geometry of the sunshade to be completely designed and the total system mass estimated. Fig. 6.7.14 shows the lenticular strut root cross-section design for the sunshade as well as the first in-plane mode shape of the sunshade.

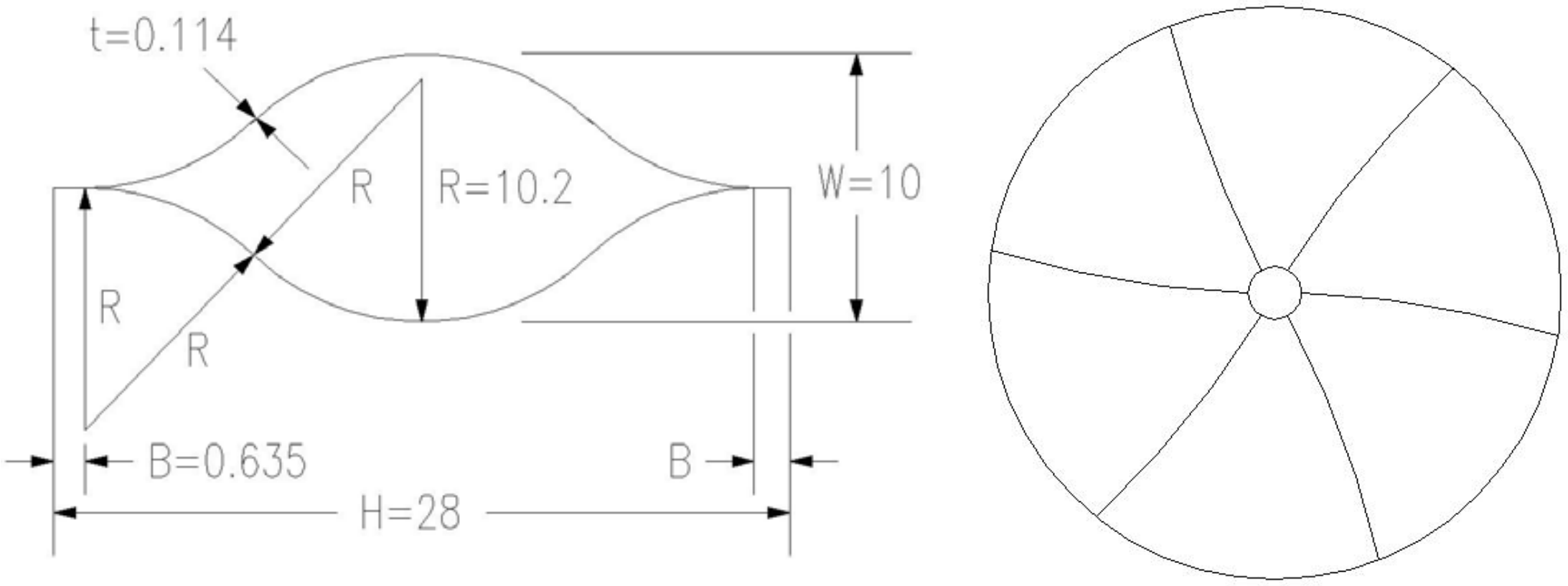

Moderately Flat: f = 0.51 (dim. in cm)

First in-plane mode:
~0.6 Hz

Figure 6.7.14: a) Lenticular cross section for the EPIC-CS sunshade and b) First in-plane mode frequency.

Tables 6.7.4 and 6.7.5 present the geometry of the three layers for both sunshades. Geometric and material properties are given in Table 6.7.6.

Based on the designed deployed geometry and the properties of the selected materials, the following sunshade mass estimates are given in Table 6.7.7. These mass estimates include only the sunshade film and the structural support and deployment hardware. The masses of the rigid, central aluminum-honeycomb V-groove radiators are not included. However, Table 6.7.8 presents the sunshade masses compared to the corresponding V-groove radiator mass.

Table 6.7.4. Sunshade Geometry for the EPIC-CS Sunshade.

| Bottom shield (i=1) | | | Middle Shield (i=2) | | | Top Shield (i=3) | | |
|---|---|---|---|---|---|---|---|---|
| L= | 13.86 | m | L= | 12.82 | m | L= | 12.00 | m |
| R1= | 11.00 | m | R= | 10.10 | m | R= | 9.39 | m |
| h= | 1.00 | m | h= | 1.00 | m | h= | 1.00 | m |
| R2= | 24.51 | m | R2= | 21.04 | m | R2= | 18.50 | m |
| theta= | 0.57 | radians | theta= | 0.62 | radians | theta= | 0.66 | radians |
| theta(degree) | 32.84 | degrees | theta(degree) | 35.47 | degrees | theta(degree) | 37.85 | degrees |
| shade area | 443.3 | $m^2$ | shade area | 375.4 | $m^2$ | shade area | 325.8 | $m^2$ |
| Total Area (6 shades) | 2288.9 | $m^2$ | | | | | | |

Table 6.7.5. Sunshade Geometry for EPIC-LC Sunshade.

| Bottom shield (i=1) | | | Middle Shield (i=2) | | | Top Shield (i=3) | | |
|---|---|---|---|---|---|---|---|---|
| L= | 4.91 | m | L= | 4.51 | m | L= | 4.18 | m |
| R= | 4.00 | m | R= | 3.65 | m | R= | 3.23 | m |
| h= | 0.25 | m | h= | 0.25 | m | h= | 0.25 | m |
| R2= | 12.17 | m | R2= | 10.27 | m | R2= | 8.86 | m |
| theta= | 0.41 | radians | theta= | 0.44 | radians | theta= | 0.48 | radians |
| theta(degree) | 23.27 | degrees | theta(degree) | 25.33 | degrees | theta(degree) | 27.28 | degrees |
| shade area | 57.65 | $m^2$ | shade area | 48.19 | $m^2$ | shade area | 39.45 | $m^2$ |
| Total area (6 shades) | 290.58 | $m^2$ | | | | | | |

Table 6.7.6. Strut, Film, and Material Properties.

| Item | Symbol | Value | Units |
|---|---|---|---|
| V-groove spacing (center) | $d$ | 0.135 | m |
| V-groove spacing (edge) | $d_{edge}$ | 0.25 | m |
| Film stress | $\sigma$ | 20,684 | Pa |
| Film thickness | $t$ | 2.54E-05 | m |
| Film density | $\rho_{film}$ | 0.09 | kg/m$^2$ |
| Spreader bar thickness | $t_{SB}$ | 0.001524 | m |
| Strut modulus | $E_{lenticular}$ | 72.8 | GPa |
| Strut density | $\rho_{lenticular}$ | 1522 | kg/m$^3$ |
| Strut thickness (22-m) | $t_{lenticular}$ | 9.53E-04 | m |
| Strut thickness (8-m) | $t_{lenticular}$ | 7.62E-04 | m |

Table 6.7.7. Mass Estimate for the EPIC-CS Sunshade

| **Dragone Sunshade** | | | | |
|---|---|---|---|---|
| **Item** | **Mass each (kg)** | **Qty** | **Substructure mass (kg)** | **Comments** |
| Lenticular Struts | 12.18 | 6 | 73.06 | Give f=0.60 Hz, FS>6 |
| 2-m-Hub attachements/hinges | 2.44 | 6 | 14.61 | 20% of strut mass |
| Spreader bar pivot | 2.44 | 6 | 14.61 | 20% of strut mass |
| Spreader bars | 0.39 | 6 | 2.32 | FS=3.9 |
| Pulleys | 0.01 | 30 | 0.42 | |
| Constant Force Spring | 0.05 | 18 | 0.81 | |
| Connectors | 0.005 | 18 | 0.09 | |
| Spreader bar deployment cable | 0.55 | 1 | 0.55 | |
| Kevlar cord | 0.14 | 1 | 0.14 | |
| Aluminized-Kapton Film (6 shades) | 267.80 | 1 | 267.80 | 90 g/m$^2$, 0.001" thick, 30% for seams |
| Spreader bar flaps | 0.38 | 6 | 2.25 | 1.63-m dia, semicircle |
| Support rod | 0.27 | 6 | 1.61 | 0.5-cm dia |
| Ring support strut to Vgroove attachments | 1.91 | 36 | 68.85 | 25% of strut mass |
| Deployment system | - | - | - | |
| Motor driven hub (2 m) | 8.48 | 1 | 8.48 | 2-m dia, circ. cyl., 5-cm thick, 30-cm tall, attaches lenticular struts |
| Deployment guide rollers (pair) | 2.00 | 6 | 12.00 | |
| Botton plate | 12.72 | 1 | 12.72 | 3-m dia, 2-cm thick |
| Motor, mount and gearing | 10.00 | 1 | 10.00 | |
| Bearings | 5.00 | 1 | 5.00 | |
| Spreader Bar Motor | 2.00 | 1 | 2.00 | |
| | | | 497.33 | **Total Sunshade Support/Deployment Hardware Mass** |

Table 6.7.8. A Comparison between Sunshade Mass and V-Groove Mass

| Sunshade Configuration | Sunshade Mass (kg) | V-Groove Mass (kg) |
|---|---|---|
| EPIC-CS | 497 | 187 |

From these mass estimates, a few important conclusions can be made. Having the three sunshade layers supported individually requires the positioning of struts between the sunshield layers, thereby reducing the radiative efficiency of the V-grooves. Next, the largest mass drive for the sunshades is the aluminized Kapton film. 54% of the sunshade mass comes from having a dual-layer membrane attached to each of the three V-groove radiators. As discussed earlier, the dual-layer configuration is used because the small gaps radiate heat to cold space much like a black body, thus more efficiently cool the telescope. Clearly, while this configuration offers increased thermal performance, it carries a large mass penalty. Lastly, the aluminum-honeycomb V-grooves have significant mass compared to the sunshade, and their design should be optimized in the future.

### *6.7.9 Future Work*

While this effort has outlined a preliminary design for a deployable, lightweight sunshade that meets or exceeds the requirements for the EPIC telescope, much work is needed in the future as this project moves forward towards possible launch. First, only a conceptual design of the motor driven hub deployment system was provided. Previous experience with motorized hubs controlling the deployment of rigidizable, inflatable struts indicates that precautions must be taken in order to successfully deploy all types of struts. A more detailed analysis of the power required to safely deploy the system and of the required strengths of the hub, bottom plate, and mounting hinges for the lenticular struts must be performed in order to reduce the overall system mass. The next step in a more detailed, preliminary design would also focus on the spreader bar pivots, ring support strut-to-V-groove connectors, as well investigate the best set of material properties (tailorable for composite lenticular struts). More experience is also needed in the folding of such large, sectioned membranes and the effects that creases will have on the thermal performance of the sunshade system. Also, a more detailed analysis could investigate a tensioned cord around the perimeter of the sunshade, which would increase the natural frequency by inducing clamped-pinned-type mode shapes. Likewise, the structural model ignores the small shear stiffness contribution of the membrane film. Either modification to the model would result in lighter lenticular struts. Thus far, no analysis has been performed to determine if the proposed design would survive the mechanical and acoustical conditions imposed on the stowed sunshade during launch. There are several more design issues to be addressed before a large scale testing could begin, however, a small, proof-of-deployment-concept study for the membrane folding, storage, and wrap-rips could be performed in the near-term.

## 6.8 EL2 Halo Orbit

The orbital study described in Section 5.8 was designed for 4-years of observations without eclipses, and applies equally to EPIC-CS. See Section 5.8 for a full description. We note that the requirement for a small halo orbit was largely driven by the telemetry solution of the EPIC-LC configuration, and could be relaxed for EPIC-CS. The system implication of a larger halo is that the shields must increase to accommodate the smaller off-axis angles of the earth and moon.

## 6.9 Standard Spacecraft Components

### *6.9.1 Scientific Operations*

Operations for EPIC-CS are the same as described in section 5.9.1, with the exception that the downlink is accomplished using a counter-spinning antenna rather than a toroidal beam antenna. Mission parameters are summarized in the tables below. Also see the mass summary table 6.1.3.

**Table 6.9.1. Mission Design Summary**

| | |
|---|---|
| **Orbit** | L2 Halo |
| **Mission Life** | 2 years at L2 (required), 4 years at L2 (design) |
| **Maximum Eclipse Period** | 0 |
| **Spacecraft dry bus mass and contingency** | 1491 kg, includes 29% average contingency |
| **Spacecraft propellant mass and contingency** | 437 kg (ΔV budget and contingency shown in Table 5.8.11) |
| **Launch vehicle** | Atlas V 541, Atlas V 551 |
| **Launch vehicle mass margin** | 1223 kg (26%), 1738 kg (37%) |

**Table 6.9.2. Science Observations Operations**

| **Mission Operation** | **Rate** |
|---|---|
| Spin Spacecraft | Continuous, ~1 rpm |
| Precess Spin Axis | Continuous, ~1 rph |
| Cycle ADR | Continuous operation |
| Downlink | Once every 24 hours |
| Maintain Orbit | Small maneuvers ~4 times per year |

**Table 6.9.3. Mission Operations and Ground Data Systems**

| **Down link Information** | **Value, units** |
|---|---|
| Number of Data Dumps per Day | 2 (X-band), 1 (Ka-band) |
| Downlink Frequency Band | 8.425 GHz (Near-Earth X-Band)<br>25.5 - 27 GHz (Near-Earth Ka Band) |
| Average Telemetry Data Rate | 2300 kbps |
| S/C Transmitting Antenna Type(s) and Gain(s) | 0.4 m, 28.4 dBi (X-band) or<br>0.4 m, 38.1 dBi (Ka-band) |
| Spacecraft transmitter peak power | < 80 W (total power) |
| Downlink Receiving Antenna Gain | 58.7 / 68.4 dBi (12-m DSN X / Ka)<br>68.3 / 76.0 dBi (34-m DSN X / Ka) |
| Transmitting Power Amplifier Output | < 40 W (RF power) |
| **Uplink Information** | **Value, units** |
| Number of Uplinks per Day | 1 |
| Uplink Frequency Band | 7.17 GHz |
| Telecommand Data Rate | 1 kbps at 45˚ |
| S/C Receiving Antenna Type(s) and Gain(s) | Low-gain omnis, 7.7 dBi boresight |

*6.9.2 Payload and Spacecraft Resources*

A summary of the payload and spacecraft masses is listed in Table 6.1.3. The total mass of the payload, including the deployable sunshield, support struts, antenna, and cooler is 2735 kg, which includes 43% contingency on all masses. A summary of the payload and spacecraft power requirements is listed in Table 6.9.4 below. The total payload power required is 758 W, including 43% contingency.

**Table 6.9.4 Power Summary**

| Item | Power (CBE) [W] | Contingency [%] | Allocated [W] |
|---|---|---|---|
| Bolometer Electronics | 150 | 43 | 215 |
| Mechanical cooler | 340 | 43 | 486 |
| ADR Electronics | 40 | 43 | 57 |
| **Subtotal Payload** | **530** | **43** | **758** |
| Attitude Control | 148 | 30 | 192 |
| C&DH | 122 | 30 | 159 |
| Power | 75 | 30 | 98 |
| Propulsion | 1 | 30 | 1 |
| Telecom (transmit mode) | 55 | 30 | 72 |
| Thermal | 99 | 30 | 129 |
| **Subtotal Spacecraft** | **500** | **30** | **650** |
| **Total Power** | **1030** | **37** | **1408** |
| **GaAs Triple Junction Solar Panels** | | | |
| **Panel Area** | **Power [W]** | **Margin [%]** | **Margin [W]** |
| **10.0 m$^2$ Fixed at 45° Incidence** | **1775** | **26** | **367** |

*6.9.3 Spacecraft Components*

We assume EPIC-CS will operate with a custom-built commercial spacecraft bus. The spacecraft itself requires no new technology. EPIC-CS requires a bus-mounted solar panel on the sun-facing side of the bus. The deployable sunshield would be a provided payload element and is not part of the spacecraft. The downlink antenna must be gimbaled and continuously rotate (see section 6.10.2).

We carried out a team-X study to assess the spacecraft components. The ACS requires 2000 - 4000 Nms momentum wheels, which exceeds the current capability of commercial wheels. However, this is within the range of second-generation wheels flown on a defense satellite program. Otherwise, all the components are space-proven technologies, either entirely off-the-shelf or with minor modifications. A summary of the component requirements is given in Table 6.9.5. An estimate of subsystem masses and power requirements are given in Table 6.9.6.

**Table 6.9.5. Spacecraft Characteristics**

| | Spacecraft bus | Value/ Summary, units |
|---|---|---|
| **Structure** | Structures material | Aluminium or composite |
| | Number of articulated structures | None |
| | Number of deployed structures | None |

| | Spacecraft bus | Value/ Summary, units |
|---|---|---|
| T/C | Type of thermal control used | Passive |
| Propulsion | Estimated delta-V budget | 215 m/s |
| | Propulsion type(s) and associated propellant(s)/oxidizer(s) | Hydrazine |
| | Number of thrusters and tanks | One 35 N Main Thruster<br>Twelve 1 N RCS Thrusters<br>One tank |
| Attitude Control | Control method | 3-axis, momentum compensated |
| | Control reference | Inertial |
| | Attitude control capability | 40 arcsec |
| | Attitude knowledge limit | 2 arcsec (3$\sigma$) |
| | Agility requirements | None |
| | Articulation/#–axes | None |
| | <u>SENSORS</u>:<br>Sun Sensors (14)<br>Star Trackers (2)<br>IMU (1)<br><br><u>ACTUATORS</u>:<br>Reaction Wheels (4)<br>Momentum Wheels (4) | <br><br>1 arcsec accuracy<br>0.003 deg/hr stability<br><br><br>150 Nms momentum, 0.1 - 0.2 Nm torque<br>2400 Nms momentum |
| C&DH | Spacecraft housekeeping data rate | 10 kbps |
| | Data storage capacity | 600 Gbits |
| | Maximum storage record rate | 2300 kbps |
| | Maximum storage playback rate | 20 Mbps |
| Power | Type of array structure | 4.0 $m^2$ body-mounted solar panels<br>3.8 $m^2$ hinged solar panels |
| | Array size, meters x meters | 10.0 $m^2$ |
| | Solar cell type | Triple-junction Ga-As |
| | Expected power generation | 1940 W BOL; 1770 W EOL |
| | On-orbit average power consumption | 1408 W (incl. 37% contingency) |
| | Battery type | Li-Ion (two) |
| | Battery storage capacity | 50 Ah |

**Table 6.9.6. Spacecraft Sub-System Characteristics**

| S/C Subsystem | Mass [kg, CBE] | Mass Ctgcy. [%] | Power [W, CBE] | Power Ctgcy. [%] |
|---|---|---|---|---|
| Attitude Control System | 240 | 25 | 148 | 30 |
| C&DH | 24 | 30 | 122 | 30 |
| Power | 73 | 30 | 75 | 30 |
| Propulsion (dry) | 47 | 27 | 1 | 30 |
| Structures and mechanisms | 595 | 30 | | |
| Launch adapter | 39 | 30 | | |
| Cabling | 63 | 30 | | |
| Telecom + X-band Antenna | 24 | 19 | 55 | 30 |
| Thermal | 53 | 30 | 99 | 30 |
| Propellant [$\Delta V$ = 215 m/s] | 437 | N/A | | |

## 6.10 Telemetry

### *6.10.1 Telemetry Rate Requirements*

The input data rate for EPIC-CS is summarized in Table 6.10.1 assuming scan-modulated TES bolometers.

**Table 6.10.1. Input Data Rate for TES Focal Plane**

| Freq [GHz] | Beam [arcmin] | $N_{det}$ [#] | $\tau_{req}$ [ms] | Sample Rate [Hz] | Data Rate [kbps] |
|---|---|---|---|---|---|
| 30 | 16 | 20 | 10 | 66 | 5 |
| 45 | 10 | 80 | 6 | 100 | 32 |
| 70 | 7 | 220 | 4 | 150 | 140 |
| 100 | 5 | 320 | 3 | 220 | 280 |
| 150 | 3 | 380 | 2 | 330 | 500 |
| 220 | 2 | 280 | 1.3 | 480 | 540 |
| 340 | 1.4 | 120 | 0.9 | 750 | 360 |
| 500 | 0.9 | 100 | 0.6 | 1100 | 440 |
| **Total** | | **1520** | | | **2300** |

We calculated the downlink requirements for the various cases above assuming the link budget calculations described in appendix D. The data rate is too large to allow the use of the toroidal beam antenna, so downlink must be accomplished by a gimbaled and continuously rotating 0.4 m high-gain antenna. While this adds the complexity of a mechanism, it provides significantly higher data rate with lower transmitter power. At X-band, there calculations assume that the downlink obtains the current maximum available bandwidth of 4 Mbps. For Ka-band, we assume target a maximum downlink rate of 20 Mbps. Sufficient memory storage must be included in this case due to greater weather dependency. Appendix D carries out additional calculations for the full range of antennas, transmitters, and available bands.

**Table 6.10.2. Input Telemetry Rates**

| Option | Spin rate [rpm] | Modulator rate [Hz] | Input rate[1] [kbps] |
|---|---|---|---|
| **Baseline**<br>Scan-modulated TES bolos[2] | 1.0 | N/A | 2300 |
| **Option**<br>Wave plate-modulated TES bolos[3] | 0.1 | 16 - 275 | 2300 |

Notes:
[1]4 bits per sample per detector (Planck compression) with Nyquist sampling, plus 100% contingency.
[2]Requires a 1/f knee < 16 mHz.
[3]Assumes 10 polarization cycles per beam crossing for each band. Requires 1/f knee < 2.5 Hz.

**Table 6.10.3. Downlink Requirements**

| Band | S/C Dish [m] | S/C Gain [dB] | RF Power [W] | Ground station [m] | Ground gain [dB] | Downlink rate [Mbps] | Downlink /day [h] |
|---|---|---|---|---|---|---|---|
| X | 0.4 | 28.4 | 10 | 12 | 58.7 | 4.0 | 13.8 |
| | | | 1 | 34 | 68.3 | | |
| Ka | 0.4 | 38.1 | 40 | 12 | 68.4 | 20.0 | 2.8 |
| | | | 7 | 34 | 76 | | |

Notes: Downlink requirements calculated for 2300 kbps input data rate. Weather dropouts not accounted.

## *6.10.2 Gimbaled Downlink Antenna*

EPIC will be placed into a halo orbit around the Earth-Sun Lagrange point L2. While the baseline orbit at L2 gives a small halo, this requirement can be relaxed for EPIC-CS. Therefore we allow for an angular radius of this as large as 9 degrees. The spacecraft is spinning about its longitudinal axis one revolution per minute, and the spin axis is "coning" with a 45 degree half-angle at a rate of one revolution per hour. The complex spinning/coning scan motion of the EPIC spacecraft, combined with its "halo" orbit around the L2 point, make pointing the telecom antenna for high-data rate downlink communications a challenge. This geometry is illustrated in Fig. 6.10.1.

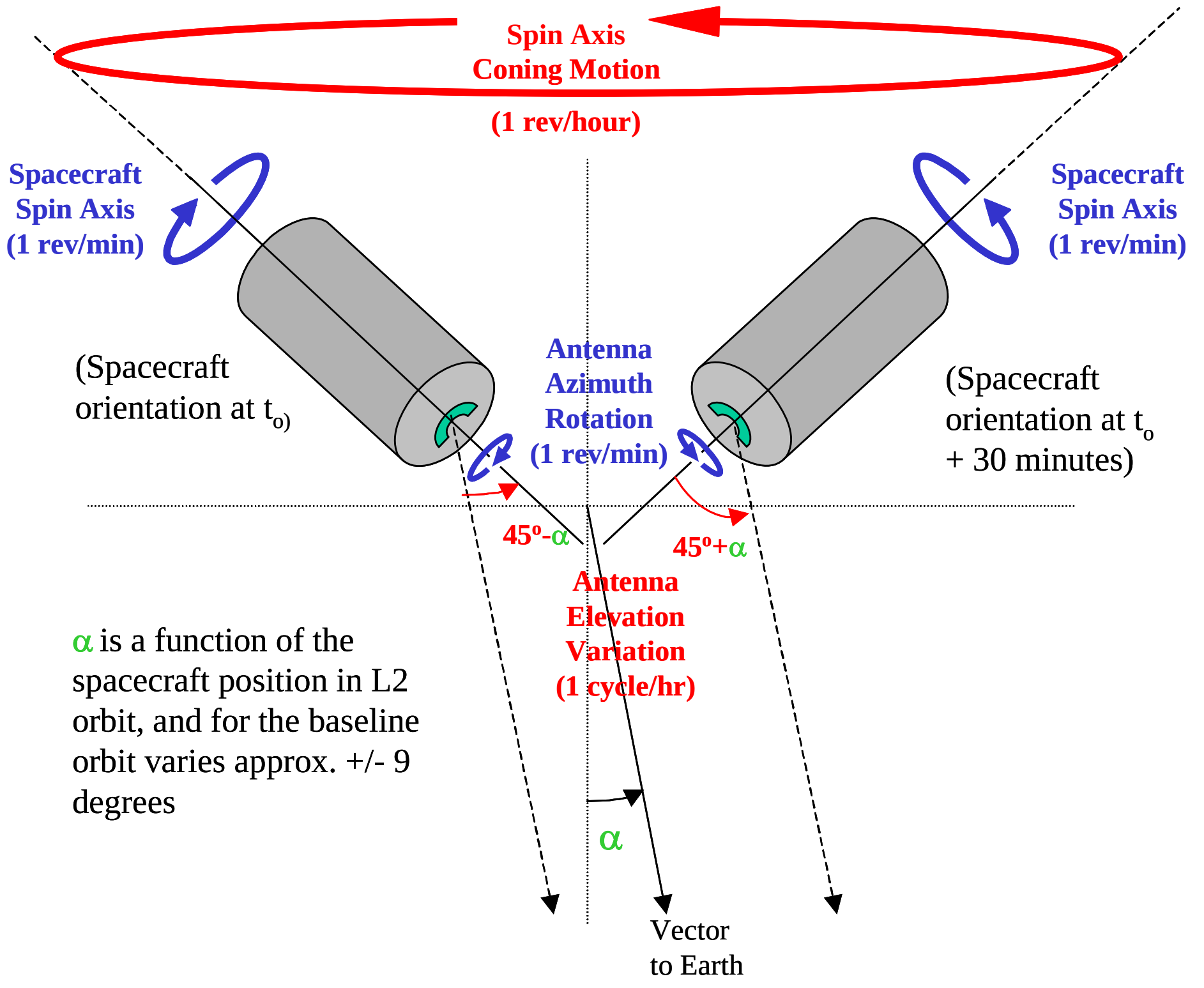

Fig. 6.10.1. Geometry for the downlink antenna due to the spinning/precessing scan pattern.

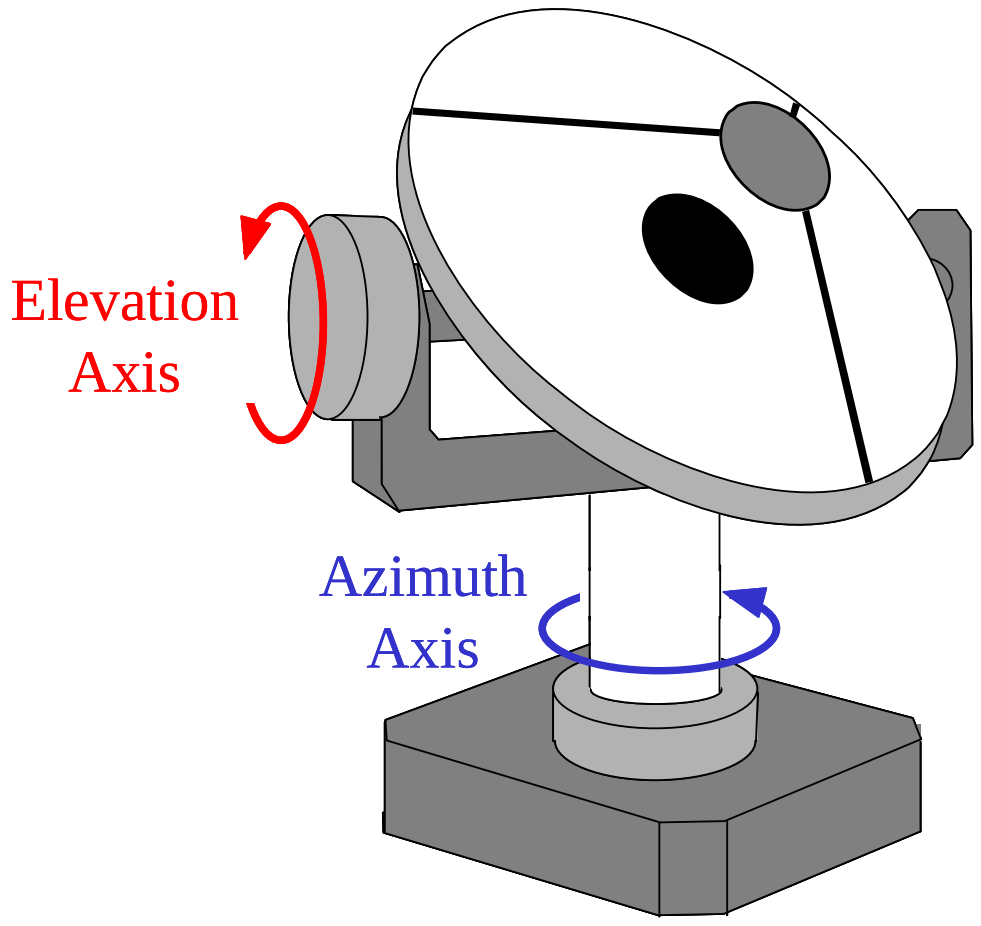

**Fig. 6.10.2.** Two-axis gimbal geometry. The antenna spins continuous in azimuth and can be slowly driven in elevation.

The telemetry rate requirements can be realized with a small (0.4 m) steered antenna, continuously counter-rotating as in Fig. 6.10.2. Both azimuth and elevation axes should pass through the center-of-mass of the supported hardware to minimize reaction disturbances to the spacecraft. Power for the elevation axis drive will be transmitted via a slipring. The RF signal to the antenna passes through a spinning waveguide interface. The requirements on the gimbaled mechanism are listed in Table 6.10.1. The 0.4˚ pointing accuracy requirement does not appear to be too demanding, based on current industry capability.

**Table 6.10.1 Gimbaled Mechanism Requirements**

| PARAMETER | VALUE | COMMENTS |
|---|---|---|
| Azimuth Scan Rate | 1 rev/minute | Driven by spacecraft spin rate |
| Elevation Scanning Angle | +/- 10 degrees about a nominal 45 degree offset | A function of L2 orbit parameters |
| Elevation Scanning Rate | one cycle per hour | Driven by coning rate |
| Supported Antenna Mass | ~1 kg for the 0.4 m antenna | |
| Operational Lifetime | 4 years | |
| Antenna Pointing Accuracy | ~0.4˚ for Ka-band | End-to-end |

Current industry capability for single- and two-axis gimbals is represented by the two leading vendors: Ball Aerospace and MOOG Schaeffer Magnetics Division [1,2]. Both have extensive flight heritage in single-axis and continuously-scanning gimbal mechanisms, and some hardware heritage for two-axis (azimuth/elevation) gimbals, though none of these is an exact match to the EPIC requirements. JPL has some limited experience building and flying a two-axis gimbal, the Bearing and Power Transfer Assembly (BAPTA), which was part of the Special Sensor Microwave/Imager (SSM/I) instrument flown on the Defense Meteorological Satellite Program (DMSP) Block 5D-2 F8 spacecraft.

**Table 6.10.2. Summary of Ball Aerospace Two-Axis Gimbal Heritage**

| Program/Customer | Equipment Description | Technical Information | Status |
|---|---|---|---|
| Mars Exploration Rovers (MER)<br>JPL | High Gain Antenna Gimbal<br>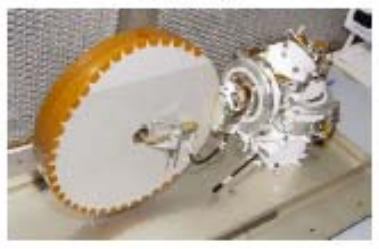 | • HGAG is a two-axis gimbal mounted on the rover equipment deck<br>  • Points the high gain antenna back to Earth<br>• The PMA provides the *eyes* for the rover<br>  • Two sets of panoramic cameras<br>    • Navigation<br>    • Stereo imaging<br>  • Four PMA mechanisms<br>    • Two mirrors<br>    • Calibration target<br>    • Structure support | • Flight deliveries – August 2002<br>• Launch Mid-2003 |
| Mars Exploration Rovers (MER)<br>JPL | PanCam Mast Assembly<br>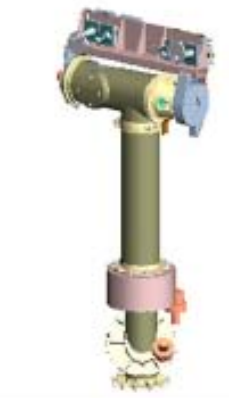 | • The PMA provides the *eyes* for the rover<br>  • Two sets of panoramic cameras<br>    • Navigation<br>    • Stereo imaging<br>  • Four PMA mechanisms<br>    • Two mirrors<br>    • Calibration target<br>    • Structure support | • First flight delivery – August 2002<br>• Mid-2003 launch |
| YSB<br>Sandia Labs | YSB<br>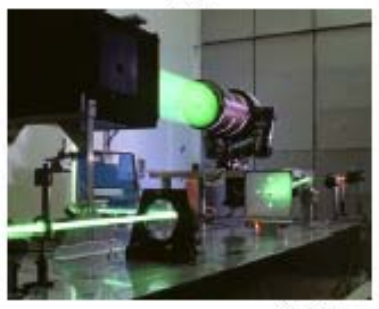<br>83-478c | • 305 mm yoke, all-beryllium two-axis gimbal<br>• 18.6 kg gimbal weight<br>• 2.5 μrad accuracy<br>• 50 Hz resonance<br>• ≤0.01 deg pointing error<br>• 0.001 to 2 deg/sec slew rate<br>• Space proven<br>• Over 10 years on-orbit performance by two units | • Three delivered<br>• Two flown late 1980s<br>• One unit refurbished 1995-1996 |
| Crosslink<br>Eastman Kodak | Crosslink<br>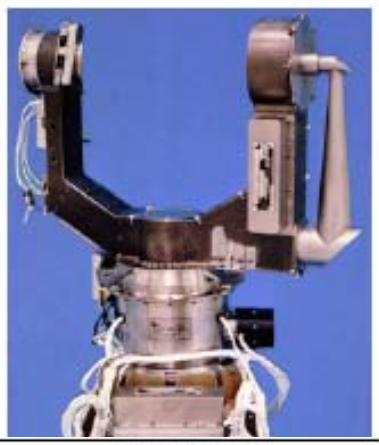 | • 216 mm yoke titanium two-axis gimbal mechanism<br>• Integrated electronics and Coudé optics path<br>• 13 kg gimbal weight<br>• 4.5 kg payload capacity<br>• 100 Hz resonance<br>• Continuous azimuth travel<br>• 30 deg elevation travel<br>• 96 μrad position accuracy<br>• Space proven | • 14 deliveries through 1992<br>• Program cancelled before first launch |
| Pointing Tracking System (PTS)<br>High Resolution Doppler Imager<br>University of Michigan | PTS-HRDI<br>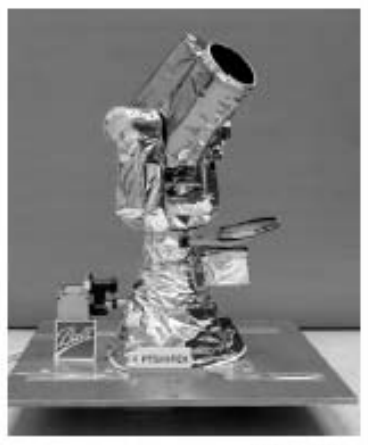<br>87-1263c | • 762 x 914 mm, two-axis aluminum gimbal mechanism<br>• 40.9 kg gimbal weight<br>• 50 Hz resonance<br>• 90 deg elevation; 350 deg azimuth<br>• 3 arcmin position accuracy<br>• Space proven | • Launched in 1992 on UARS<br>• Successful on orbit operation |
| Optical Airborne Measurement Program (OAMP)<br>aka COBRA EYE<br>MIT LL | 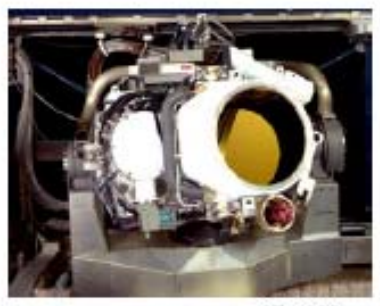<br>87-1605c | • 25-in. telescope aperture<br>• Cryogenic system<br>• Cable handling mechanism<br>• Rugged aircraft use<br>• 2,000 lb telescope payload | • Successful aircraft flight mission 1987-1993 |
| SRS<br>McDonnell Douglas and Martin Marietta | SRSTAG/Comlink<br>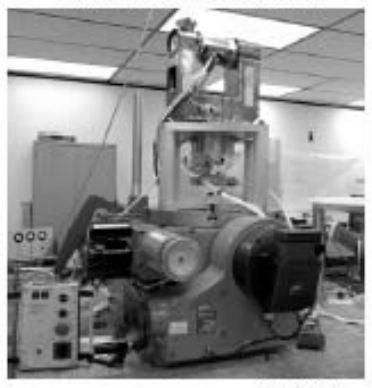<br>84-3323c | • 170 mm yoke span, two-axis titanium gimbal<br>• 11.8 kg gimbal weight<br>• 10.5 kg payload capacity<br>• 60 Hz resonance<br>• ±30 deg elevation; ±90 deg azimuth<br>• 96 μrad position accuracy<br>• 100 deg/sec slew rate<br>• 4 rad/sec$^2$ acceleration<br>• Space proven | • Six deliveries |

The specific design implementation for the EPIC consists of a continuously scanning azimuth axis (at 1 rev. per minute, synchronous with the spacecraft spin), and an elevation axis which continuously scans up to +/-10 degrees at a rate of one cycle per hour (synchronous with the spacecraft coning/precession; the actual angle dependent on the L2 orbit parameters). It appears that the most straightforward approach to implementing a two-axis gimbal for EPIC will be a custom-build using flight-proven commercial hardware and design heritage, built either by JPL using commercial parts, or by an industrial contractor after a competitive bid.

### 6.11 Cost Analysis

Cost analysis of the EPIC-CS option was beyond the resources of this study.

## 7. Technology Roadmap

The development of EPIC requires bringing several laboratory technologies to space readiness. As has been the case for COBE, WMAP, and Planck, ground and balloon-based experiments will be critical pathfinders to prove the new technologies. The main technologies that require development and test are the antenna-coupled TES bolometer arrays with their multiplexed SQUID readouts and the polarization modulators, such as the half-wave plate rotator. Support for development of the basic technologies and sub-orbital experiments will be critical. The current NRA-based effort has allowed the community to develop the basic technology ideas, but to bring the required technology to maturity for an mission selection, a higher level of resources is required. Fig. 7.1 shows a timeline for the technology development for EPIC. The timeline assumes work from the current EPIC team, but the team is open contribution and collaboration from the entire community.

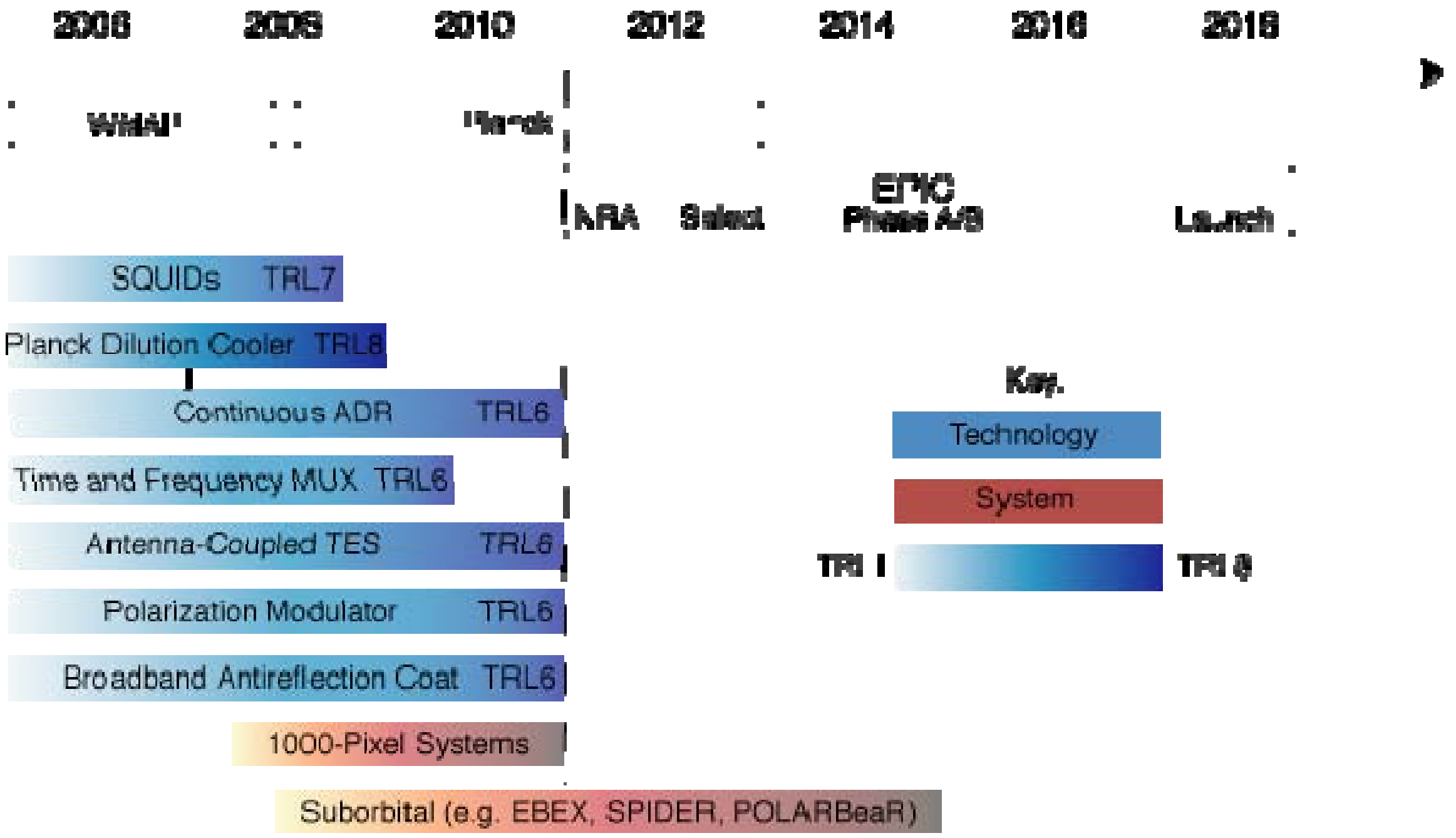

Figure 7.1. Timeline for EPIC technology development.

**Antenna-coupled TES detectors:** The development of the focal-plane technology is one of the key areas where progress is required to design and build EPIC. As discussed in the earlier detector section, the basic concepts for the focal-plane exist. It will take a concerted effort to develop each of the ideas to the point where the tradeoffs between different concepts becomes clear. Fabrication of large arrays systems and tests of these systems in ground and balloon systems is required. Also, development specific to space missions where mass, vibration, and lifetime requirements are more stringent will be essential and will require dedicated funding. The established groups at JPL/CIT, NIST, UCB/LBNL, and GSFC have the necessary expertise, but a dedicated line of funding is required to push beyond the "mid-TRL hump" in order to obtain fully functional arrays from the current mix of demonstrated component technologies.

**SQUID Multiplexers:** DC SQUIDs have been flown on the GPB mission. The SQUIDs required for EPIC are similar in construction and materials, but the type of multiplexed TES readout will determine the exact SQUID configurations required. SQUIDs for the CMB community are almost all built by NIST, and therefore it is important that NIST has sustained support. The time-domain MUX is being developed by NIST and the frequency domain MUX is being developed by UCB/LBNL. Ground- and balloon-based experiments will test these readout technologies with 1000 element arrays. For space, further work is required on power dissipation, both cryogenic and ambient. Such work will not likely be done for suborbital experiments, since resource requirements are less stringent.

**Coolers:** The Planck dilution cooler will be flight tested in 2008. It has a 100 nW of cooling power at 100 mK for the detector system after a much larger portion of the cooling power is allocated for the cooler supports. This cooler would be suitable for EPIC, but a continuous ADR which can have more cooling power would allow more flexibility in design of the multiplexed readout. Continuous ADRs are being developed by GSFC and JPL.

**Polarization Modulators:** The status and outlook for polarization modulators is similar to that for the focal-plane technologies. There are several plausible concepts including rotating half-wave plates, Faraday modulators, and microstrip RF switches. Each of these will be tested in ground and balloon tests, but dedicated funding would be required to bring them to a high TRL level before 2011, the NRA date planned by the Weiss committee.

**Broadband antireflection coatings:** For EPIC-LC, each aperture has only a single frequency band and therefore the anti-reflection coating requirements are simple. Suitable coatings already exist. For EPIC-CS, broadband antireflection coatings are required and these have not yet been demonstrated. This is another area that requires funding, both at the NRA level and also dedicated funding toward CMBPOL.

**Suborbital Pathfinders:** The members of our collaboration are working on existing, planned, and proposed CMB polarization experiments on the ground and in balloons. These are staged with the future experiments having increasing capability. BiCEP and QUAD are observing now at the South Pole with ~100 NTD detectors. EBEX is a funded balloon experiment that will use ~1000 TES detectors in an LDB flight in 2008. SPIDER (LDB balloon) and POLARBeaR (ground) will use 1000 element arrays of planar-coupled TES detectors. These experiments will be able to test several focal-plane technologies, polarization modulators, and observation strategies. The members of the EPIC team will collaborate to compare the resulting lessons to refine the design of EPIC.

## Appendix A. Formalism for Main-Beam Systematics

Here we summarize calculations which were used to simulate the effects of main-beam systematic distortions. We have found that previous results, derived in real space, can be formally generalized by including infinitely many higher-order corrections which can be summed up and represented as analytic functions. In practice, to reduce computation time, we truncate the expansion at second-order, as summarized below, but our analytic expressions allow us to bound the effects of higher-order residuals. These results were subsequently compared to the full analytic expressions, where the simulations were performed in multipole-space, and confirmed the analytic approach with good agreement. A primary benefit of the real-space simulation is that it accounts for the *exact* scan strategy employed, allowing for optimization as a method of mitigating the main-beam systematic effects.

We approximate the signal observed by a single polarimeter pixel by Taylor expanding the underlying field on the sky. We expand the real-space temperature and polarization fields on the sky up to the second order. Here, *p* represents the angular coordinates of the center of a given pixel, and *r* is the exact direction of observation. The observed temperature signal is:

$$T(r) \approx T(p) + \nabla T(p)(r-p) + \frac{1}{2}(r-p)^T D^2 T(p)(r-p).$$

Identical definitions apply for the two linear polarization fields Q(r) and U(r). $D^2T(p)$ represents the second derivatives of the temperature field near the point *r*.

The signal, *s*(*r*), measured by a detector is the convolution of the underlying T,Q and U signals with the pixel's antenna response or "beam pattern". Here we consider the effects of the main beam only. The signal is:

$$s(r) = \int B(r-r',\beta,\theta)\left[T(r') + Q(r')\cos 2\alpha + U(r')\sin 2\alpha\right]dr'$$

where β is the angle between the scan axis and the local meridian, and θ is the angle between the beam's major axis and the scan axis. The angle between the polarization sensitivity direction of the detector and the local meridian is α. The following condition holds for the angular variables: $\alpha = \beta + \theta + \psi$, where ψ is the possible rotation of the polarization sensitivity direction of the detector with respect to the major axis of the (potentially elliptical) main-beam. The virtue of the Taylor expansion is that it allows us to compute the integral as a function of the idealized beam parameters and T, Q, U fields, which is much faster than a real space convolution of the *full* beam and field expressions. After some math, one can show that *s*(*r*) becomes:

$$s(r) = \frac{1}{2}\Big[\tilde{T} + Z_T + cos2(\alpha-\psi)W_T + sin2(\alpha+\psi)X_T + cos2\alpha\left(\tilde{Q} + Z_Q + cos2(\alpha-\psi)W_Q + sin2(\alpha+\psi)X_Q\right)$$
$$+ sin2\alpha\left(\tilde{U} + Z_U + cos2(\alpha-\psi)W_U + sin2(\alpha+\psi)X_U\right)\Big]$$

where

$$\tilde{T} = T(p) + \nabla T(p)(r-p) + \frac{1}{2}(r-p)^T D^2 T(p)(r-p)$$

and

$$Z_T = \frac{\sigma_x^2 + \sigma_y^2}{4}(T_{11} + T_{22}),\ W_T = \frac{\sigma_x^2 - \sigma_y^2}{4}(T_{11} - T_{22}),\ X_T = \frac{\sigma_x^2 - \sigma_y^2}{4}(T_{12} + T_{21}),$$

in which x and y are the axes of the beam and $T_{ij}$ are the second spatial derivatives of *T*. The Matrices *Z, W*, and *X* encapsulate the systematic distortions from an ideal, circularly-symmetric Gaussian beam. The various symmetries (e.g. dipolar, quadrupolar) of the beam distortions, as described in Section 3 couple to first and second derivatives of the underlying fields. For example, the monopole symmetric systematic related to differential gain can produce a non-vanishing spurious polarization signal even if the underlying Q or U signal in that pixel is zero. Similar definitions hold for Q and U. First and second derivatives are obtained from simulated maps, and the simulations can be marginalized over many realizations to isolate the intrinsic effects of the systematic distortions.

Given a realization of the underlying sky, simulations of the expected detector time-ordered-data (TOD) streams are produced. Once the data streams are computed using the above equations, we project them onto maps using HEALPIX. For a given pixel, the set of the *n* samples that fall into this pixel is formed into a Stokes vector, called s. We then have:

$$s = A\begin{pmatrix} I \\ Q \\ U \end{pmatrix} \text{ with } A = \frac{1}{2}\begin{pmatrix} 1 & \cos 2\alpha_1 & \sin 2\alpha_1 \\ 1 & \cos 2\alpha_2 & \sin 2\alpha_2 \\ \vdots & \vdots & \vdots \\ 1 & \cos 2\alpha_n & \sin 2\alpha_n \end{pmatrix}$$

In the above, the matrix *A* is the "pointing matrix" which maps time-ordered data to map pixels. To reconstruct I, Q and U, we invert the above expression by performing a least square minimization, using standard matrix techniques, yielding three synthetic maps $(I,Q,U) = \left(A^T A\right)^{-1} A^T \vec{s}$ . These maps can then be used to synthesize maps of E and B-mode polarization, from which power spectra are produced. To study EPIC's susceptibility to main-beam distortions the process is repeated for varying systematic effect levels.

## Appendix B. Alternative Optical Designs

The crossed-Dragone design (also known as a compact range antenna) offers a number of advantages when compared to its Gregorian counterpart. A crossed-Dragone design is a Cassegrain telescope with a decentered entrance aperture that satisfies the Mizuguchi-Dragone condition. An example designed for EPIC-CS is shown in Figure B.1. The system consists of two reflectors that feed the radiation directly into the focal plane. A HWP is placed just in front of the focal plane (not shown in the figure).

In the crossed-Dragone the radii of curvature of the mirrors are much less severe than for a Gregorian system of comparable size and f/#. This reduces both instrumental and cross-polarization systematic effects, and also diminishes the effect of aberrations; the aberration performance and polarization properties of the system are given in Table B.1 and B.2. Furthermore, the focal plane of the crossed-Dragone is nearly flat (focal plane radius of curvature ~ 32 meters) and telecentric (deviations of ~ 1° from telecentricity at the edge of the field of view), thus eliminating the need for refractive re-imaging optics that would otherwise be needed to meet these conditions. As a result, it is possible to achieve a single, large deeply diffraction limited field-of-view *without any additional lenses*. This is a substantial advantage relative to either the Gregorian system, which requires a number of relay lenses, or the refractor design, which is monochromatic.

One major tradeoff in a fully reflective crossed-Dragone design is that the primary mirror itself is the only natural stop in the system. This necessitates that the perimeter of the primary mirror be surrounded by a black surface that is as cold as possible, and that the focal plane be more sparsely populated in order to control spillover without a cold stop. In the crossed design we use a focal plane spacing of $d = 3.25(f/\#)\lambda$ to achieve a similar spillover as in the $2(f/\#)\lambda$ spacing of the Gregorian or refractor designs that have cold stops. This spacing, along with an oversized primary, limits the entrance aperture size that will fit into an Atlas-5 launch vehicle. The crossed entrance aperture is 2.0 m, compared to 2.8 m for the Gregorian design. Because of the potential advantages of the crossed Dragone design it is worthwhile to study carefully the trade-offs between aperture size, focal plane spacing of detectors, beam spill-over, beam size, scientific return and overall technical challenge. The crossed-Dragone provides substantial simplicity in implementation relative to the Gregorian design, but with somewhat coarser angular resolution.

A second trade-off with the crossed Dragone design is the proximity of the incident beam to the secondary and to the focal plane. If such a design is chosen as candidate for a future mission careful attention should be given to these constraints and to diffraction they may cause.

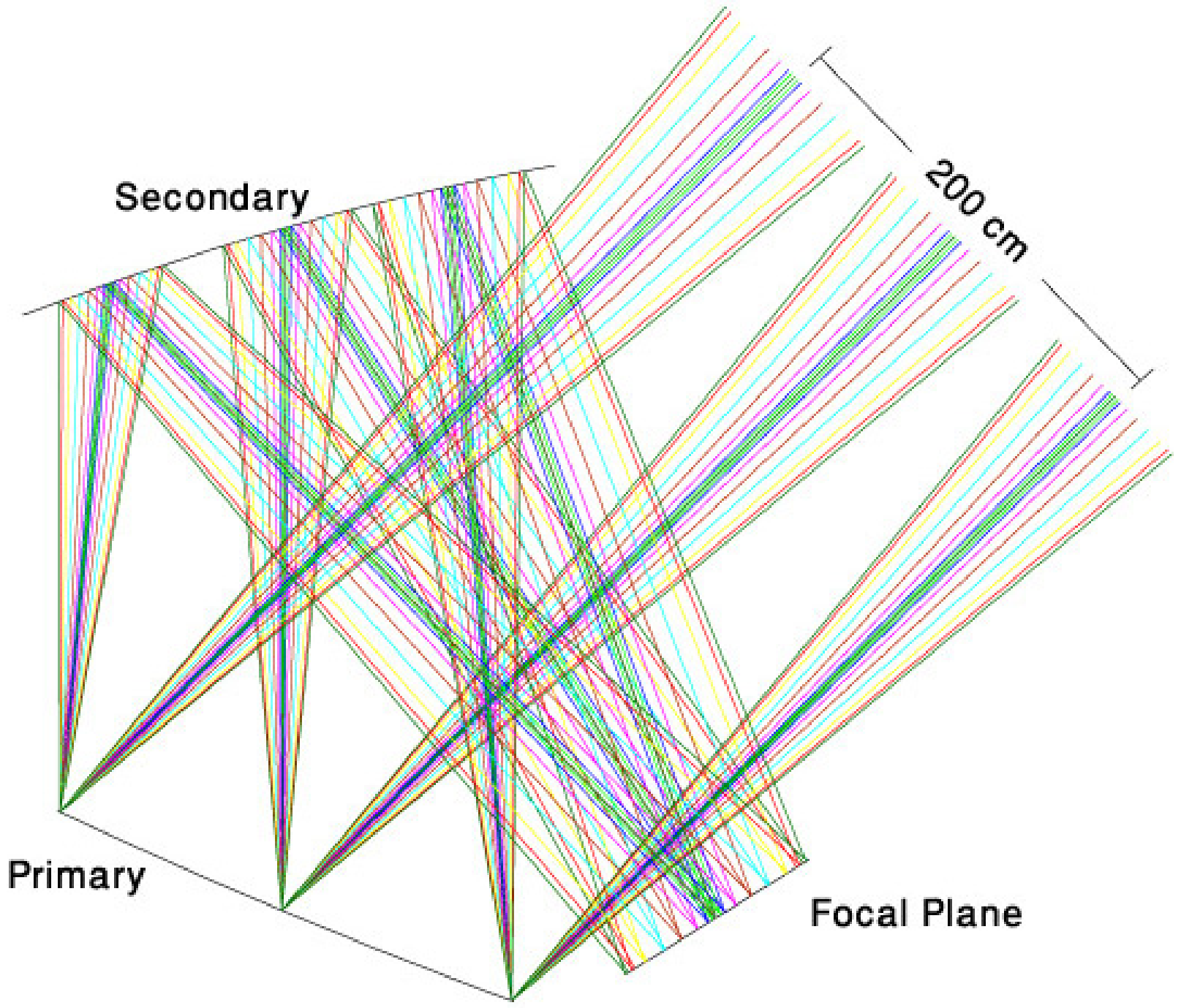

Figure B.1: A crossed-Dragone system with 200 cm open aperture for EPIC-CS. This system can fit inside an Atlas V shroud, but a larger aperture system would not fit. This system provides an achromatic, nearly telecentric, and diffraction-limited field of view with low instrumental and cross-polarization without any lenses.

**Table B.1 Parameters for EPIC Crossed-Dragone Optics**

| Frequency [GHz] | Throughput[1] [$cm^2$ sr] | FOV[2] (deg) | Strehl Ratio[3] |
|---|---|---|---|
| 30 | 31 | 10.2 | 0.99 |
| 45 | 56 | 8.4 | 0.98 |
| 70 | 63 | 7.1 | 0.96 |
| 100 | 45 | 5.3 | 0.97 |
| 150 | 24 | 3.5 | 0.97 |
| 220 | 8.2 | 1.9 | 0.99 |
| 340 | 1.5 | 0.9 | 0.99 |
| 500 | 0.6 | 0.5 | 0.99 |

[1] Defined as the product of throughput per pixel ($\lambda^2$) and the total number of pixels at a given frequency. A pixel on the focal plane contains two polarization sensitive TES detectors.
[2] Pixels are arranged on a square grid with a circular boundary. We give the cumulative outer diameter of the FOV. The lower frequency pixels are arranged in annuli around the higher frequency ones.
[3]Ratio given at the outermost diameter of the frequency band.

Table B.1: Lowest Strehl ratios provided by the Crossed-Dragone telescope at the edge of the FOV for each of the frequency bands. Strehl ratios larger than 0.8 are considered diffraction limited. Note that the required number of detectors and associated detective throughput is the same as the EPIC-CS Gregorian Dragone summarized in Table 6.3.1. Compared to the Gregorian design, the Crossed design has much larger FOV, due in part to a smaller primary but mostly due to the larger spacing between pixels, 3.25 f$\lambda$ instead of 2 f$\lambda$.

**Table B.2. Polarization Properties of the Crossed-Dragone 150 GHz Band**

| **Matrix Element** | **Level** |
|---|---|
| IQ | $1.5 \times 10^{-4}$ |
| IU | $< 1 \times 10^{-5}$ |
| QU | 0.00563 |

Table B.2: Mueller matrix elements for the edge of the field of view of the 150 GHz band of the 2 meter aperture EPIC Crossed-Dragone telescope. A mixing of QU at the level shown would rotate an incident polarization vector by 0.16 degrees. The finite conductivity of the surfaces is included in the calculation.

# Appendix C. Mechanical Calculations for Deployed Sunshield

Given the deployed configuration of the sunshade, analytical tools were used to design the structure to meet the given requirements in Table 6.7.1. This section will present the developed analytical tools used to design this sunshade. Resulting specifications are given in Sections 5.7.6 and 6.7.8. The general design methodology is to first design the structure to meet the natural frequency requirements, then design the spreader bars, and finally check the buckling behavior of the lenticular struts. If the factor of safety requirement on buckling load is met, then no further work is required. However, if the buckling requirement is not met, then the strut must be redesigned not to buckle. While this redesign will surely meet the dynamic requirement, a calculation should be made to ensure that the fundamental frequency requirement of the buckling-re-designed strut is met.

## C.1 Lenticular Geometry Analysis

The geometry of the lenticular strut shown in Fig. 6.7.4 has many parameters, but for a given thickness, *t*, and tab length, *B*, a cross section is specified uniquely by any two of *H, W,* or *R*. The relationships between these parameters are

$$\lambda = R - \frac{W}{2} \tag{C.1}$$

$$H = 2\sqrt{4R^2 - (R+\lambda)^2} \tag{C.2}$$

It is also useful to define the flatness ratio, *f*, for the lenticular geometry:

$$f = \frac{\lambda}{R} \tag{C.3}$$

Deployable lenticular struts should be moderately flat, which implies an *f* value between about 0.4 and 0.6[10]. The last parameter of interest is the angle $\theta$, given as

$$\theta = \sin^{-1}\left(\frac{\sqrt{4-(1+f)^2}}{2}\right) \tag{C.4}$$

For the present EPIC study, the lenticular cross-section has a linearly tapered height, *H*, which is specified along with the constant radius, *R*, as the height of the strut will be limited by the spacing between the V-groove radiators and the radius by mission requirements. The remaining parameters can then be calculated and used in the subsequent mechanical analysis. The nonlinear nature of Equations C.2 and C.4 shows that while *H* tapers linearly, *W* and $\theta$ do not. Such relationships are used subsequently when the density and moment of inertia are calculated along the length of the strut.

## C.2 Mechanical Analysis

With the relationships between the lenticular strut cross section design variables, the next step in the structural design is an analysis of the natural frequency to size the cross section of the lenticular struts so that the sunshade meets the fundamental frequency requirement. A Rayleigh-Ritz procedure[13] is used, which involves selecting a shape function for the in-plane displacement of the lenticular strut cantilevered from the central hub that meets the geometric boundary conditions at the fixed end (displacement and slope equal to zero). The selected shape function is

$$u = \sum_{i=1}^{N} a_i(t)\left(1 - \cos\frac{i\pi\xi}{2}\right) \tag{C.5}$$

where $\xi = x/L$ is the normalized spatial variable along the length of the strut with length *L*, $a_i(t)$ are time-dependent scaling parameters, and *N* is the number of terms to include in the shape function. Increasing the value of *N* increases the number of calculated natural frequencies as well as the accuracy of the predicted values. Using this equation along with Hamilton's Principle results in the standard eigenvalue problem for structural resonances:

$$\left(M-\omega^2 K\right)\{a\}=0 \tag{C.6}$$

Here, $M$ is the mass matrix and $K$ is the stiffness matrix. The determinate of the term in parentheses in Equation C.6 provide the natural frequencies, ω, of the system, while the nonzero vector *{a}* gives the linear combinations of the shape functions from Equation C.5 that approximate the true eigenvectors. The mass matrix for a single sunshade is

$$M_{ij}=L\int_0^1 \rho(\xi)\left(1-\cos\frac{i\pi\xi}{2}\right)\left(1-\cos\frac{j\pi\xi}{2}\right)d\xi \tag{C.7}$$

where

$$\rho(\xi)=4\rho t H_0\left(\frac{B}{H_0}+2\left(\frac{R}{H_0}\right)\theta(\xi)\right)+\frac{\rho_m L 2\pi\xi}{n}+m_{tip} \tag{C.8}$$

is the density per unit length of the strut, $\rho$ is the strut material density, $\rho_m$ is the density of the reflective film, $m_{tip}$ is the tip mass of the strut (20% of the beam mass plus the mass of the spreader bar, as calculated later), $n$ is the number of lenticular struts (6 for EPIC) and a subscript $0$ refers to a cross-section parameter at the root/base of the strut. This model assumes that the film mass is distributed evenly along the length of the strut. The stiffness matrix for in-plane motion of the sunshade is

$$K_{ij}=L\int_0^1 \frac{EI(\xi)}{L^4} i^2 j^2 \frac{\pi^4}{16}\cos\frac{i\pi\xi}{2}\cos\frac{j\pi\xi}{2}d\xi \tag{C.9}$$

where the moment of inertia about the bending axis is

$$I(\xi)=4tR^3\left\{\left[\left(\frac{W(\xi)}{2R}-1\right)^2+2\right]\theta_0+2\left(\frac{W(\xi)}{2R}-2\right)\sin\theta_0+\frac{\sin 2\theta_0}{2}\right\} \tag{C.10}$$

and

$$\theta_0=\sin^{-1}\left(\frac{0.5H_0-B}{2R}\right) \tag{C.11}$$

A computer program was written, where the material properties and selected lenticular geometry was input, and the resulting natural frequency of the sunshade calculated. The program is used iteratively, varying the geometry until the required natural frequency is obtained.

### C.3 Stress Analysis of Spreader Bar

The loads imparted to the spreader bars are due to the tension in the membrane film layers. From Fig. 6.7.6, a free body diagram of these loads is constructed in Fig. C.1. Since the bottom membrane attaches close to the spreader bar pivot at the end of the lenticular strut, its load contribution creates a small moment, and therefore has not been included in this analysis.

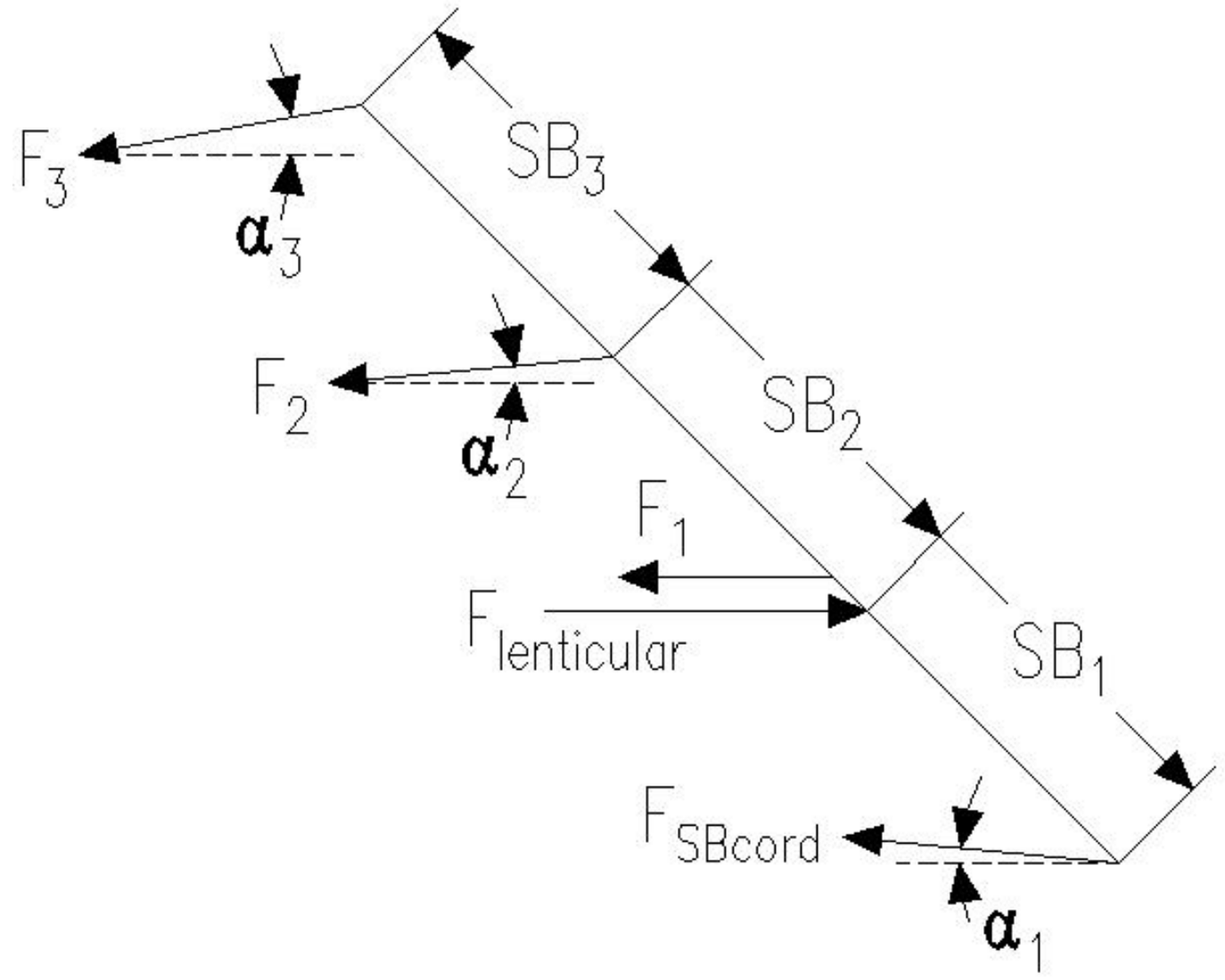

Figure C.1: Free Body Diagram of the Spreader Bar.

For EPIC, the values of $\alpha_2$ and $\alpha_3$ are 5° and 10°, respectively. To determine $\alpha_1$ the loads, the geometry of the sunshade, shown in Fig. 6.7.8, must first be determined[14,15].

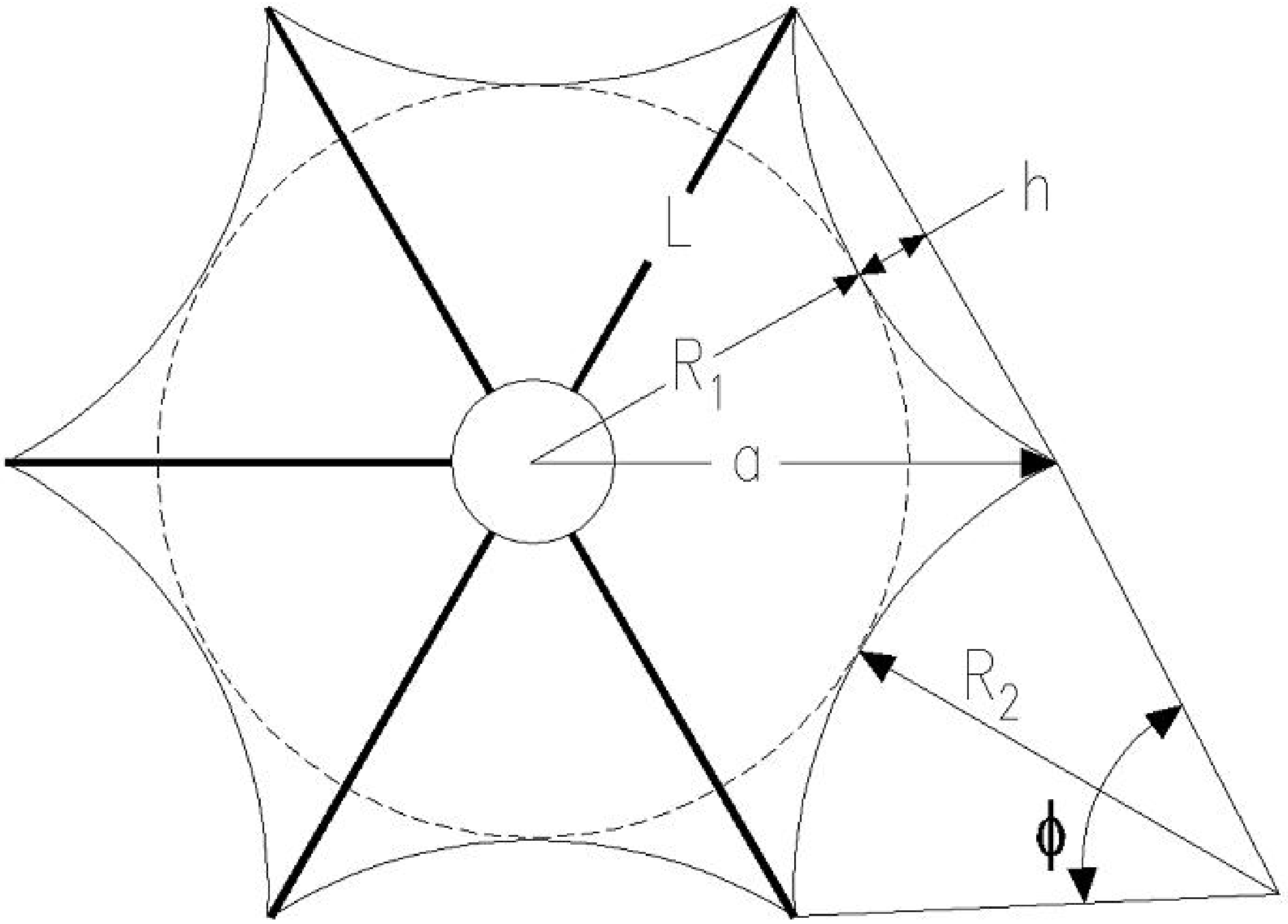

Figure C.2: Sunshade Geometry for Bottom Sunshade Layer.

Given a uniform stress in the membrane film, $\sigma$, the force imparted to each spreader bar by the film is

$$F_i = \sigma t R_{1i} \quad i = 1, 2, 3 \tag{C.12}$$

Here, $i$ denotes the sunshade layer number, as specified in Figure 11. From Figure C.2, $R_{11}$ is known from the required diameter of the sunshade; however, the length of the lenticular struts is *not* equal to $R_{11}$. Likewise, $R_{1i}$ is the distance from the center of the V-groove to the closest scalloped-edge point for the $i^{th}$ sunshade layer. The distance from the center of the V-groove to the tip of the lenticular strut is

$$a_1 = \frac{\sqrt{2}}{3} R_{11} + \frac{2h}{\sqrt{3}} \tag{C.13}$$

while the actual length of the strut is shorter by the radius of the V-groove, or

$$L_1 = a_1 - \frac{d_{Vgroove}}{2} \tag{C.14}$$

where $d_{Vgroove}$ is the diameter of the V-groove radiator. For the middle and top layers,

$$L_2 = \sin^{-1}\left( \frac{\sqrt{2}}{2} \frac{\sin(L_1 - d)}{\sin(135^\circ - \alpha_2)} \right) \tag{C.15}$$

$$L_3 = \sin^{-1}\left( \frac{\sqrt{2}}{2} \frac{\sin(L_1 - 2d)}{\sin(135^\circ - \alpha_3)} \right) \tag{C.16}$$

where $d$ is the center-to-center spacing of the V-groove radiators. $a_2$ and $a_3$ are found by adding the radius of the V-groove radiators to $L_2$ and $L_3$, respectively. Also, for the top and middle layers,

$$R_{12} = \frac{\sqrt{3}}{2} \left( a_2 - \frac{2h}{\sqrt{3}} \right) \tag{C.17}$$

$$R_{13} = \frac{\sqrt{3}}{2} \left( a_3 - \frac{2h}{\sqrt{3}} \right) \tag{C.18}$$

At this point, Equation C.12 can be used to find the force resultants for the membrane layers. The force required by the spreader bar cable is

$$F_{SBcable} = \frac{F_3 \sin\alpha_3 + F_2 \sin\alpha_2}{\sin\alpha_1} \tag{C.19}$$

where

$$\alpha_1 = \sin^{-1}\left(\frac{\sin SB_1 \sin 135^\circ}{\sin L_{SBcable}}\right) \tag{C.20}$$

and the length of the spreader bar cable is

$$L_{SBcable} = \sqrt{SB_1^2 + L_1^2 - 2SB_1 L_1 \cos 135^\circ} \tag{C.21}$$

where $SB_1$, $SB_2$, and $SB_3$ are segment lengths of the spreader bar defined in Fig. 6.7.7. The geometry of the sunshade specifies that the length of the spreader bar segments be

$$SB_2 = \frac{2\left(L_2 \sin(\alpha_2) + d\right)}{\sqrt{2}} \tag{C.22}$$

$$SB_3 = \frac{SB_2\left(2d + L_3 \sin(\alpha_3) + d\right)}{d + L_2 \sin\alpha_2} - SB_2 \tag{C.23}$$

The length of $SB_1$ must be selected to ensure proper stowage along side the telescope during launch, and so that the stress level in the spreader bar does not exceed the factor of safety requirement on bending strength. For the 22-m diameter sunshade (scallop-to-scallop diameter), $SB_1$ is chosen to be 0.5-m.

With the known forces and distances from Fig. C.1, one can calculate the moments about the spreader bar pivot point:

$$M_3 = F_3 \cos(45 - \alpha_3)(SB_2 + SB_3) \tag{C.24}$$

$$M_2 = F_2 \cos(45 - \alpha_2) SB_2 \tag{C.25}$$

$$M_{SBcable} = F_{SBcable} \cos(45 + \alpha_1) SB_1 \tag{C.26}$$

The bending stress in the spreader bar due to these moments is

$$\sigma_{SB} = \frac{\left(M_2 + M_3 + M_{SBcable}\right) r_{SB}}{\pi r_{SB}^3 t_{SB}} \tag{C.27}$$

The radius of the spreader bar is determined from the maximum viewable cross sectional area requirement in Table 6.7.1. The thickness is chosen to meet the factor of safety on strength requirement, which is

$$F.S._{strength} = \frac{\sigma_{allowable}}{\sigma_{SB}} \tag{C.28}$$

where the allowable stress for S-glass is 1.7-GPa. S-glass is chosen because its low conductivity will minimize heat transfer from the warmest to the coldest shield.

**C.3 Sunshade Area**

Most of the geometry of the sunshade has been presented, except, the edge scallop of the sunshade between the struts is found by solving for $R_{2i}$ and $\phi_i$ simultaneously[14] from

$$L_i = 2R_{2i}\sin\frac{\phi_i}{2} \quad and \quad h = R_{2i}\left(1-\cos\frac{\phi_i}{2}\right) \tag{C.29}$$

At this point, the area of the three sunshade layers can be calculated as

$$Area = \sum_{i=1}^{3} 3\left(a_i\left(R_{1i}+h\right) - R_{2i}^2\left(\phi_i - \sin\phi_i\right)\right) \tag{C.30}$$

**C.4 Buckling Analysis**

From the free body diagram of the spreader bar in Fig. C.1, the compressive load on the lenticular strut is

$$F_{Lenticular} = F_3\cos\alpha_3 + F_2\cos\alpha_2 + F_{SBcable}\cos\alpha_1 + \sigma\, t\, R_{11} \tag{C.31}$$

The Euler buckling load for a cantilevered beam is

$$F_{Euler} = \frac{\pi^2 EI}{4L^2} \tag{C.32}$$

For a calculated compressive load in the strut, the factor of safety against buckling is

$$F.S._{buckling} = \frac{F_{Euler}}{F_{lenticular}} \tag{C.33}$$

If this factor of safety does not meet the requirement, then the necessary moment of inertia, *I*, for the beam is calculated by applying the required factor of safety to the buckling load in Equation C.32. This new moment of inertia is then used to determine the new required lenticular geometry, and the mechanical analysis is performed again to ensure that the fundamental frequency requirement is met. For EPIC, the struts tended to be driven by the buckling load, not the frequency requirement.

### C.5 Stress Analysis of Lenticular Strut in 1-g

This section of the appendix gives the equations used to determine if the 22-m diameter EPIC sunshade can be deployed and tested on the ground in a 1-g environment. For the geometry in Fig. 6.7.12, the maximum deflection of the strut occurs at the free end. The deflection due to the distributed load is

$$y_{distributed} = -\frac{wL^4}{8EI} \tag{C.34}$$

while the deflection due to the tip load is

$$y_{tipload} = -\frac{WL^3}{3EI} \tag{C.35}$$

These tip deflections add together to give the total deflection for the loading scenario in Fig. 6.7.12. The compressive stress at the bottom of the lenticular strut is

$$\sigma_{applied} = -\frac{(WL + 0.5wL)H}{2I} \tag{C.36}$$

where $H$ is the height of the lenticular cross-section. The local wall-buckling stress for a thin-walled tube of radius $R$ and wall thickness $t$ with no imperfections is

$$\sigma_{local} = -\frac{0.6Et}{R} \tag{C.37}$$

The factor of safety against local wall-buckling is then simply

$$F.S._{wall-buckling} = \frac{\sigma_{local}}{\sigma_{applied}} \tag{C.38}$$

For the case where the struts are gravity-offloaded, the beam deflects in a pinned-cantilevered manner, and the maximum deflection occurs at 0.4215L from the pinned end with a value of

$$y_{offload} = -0.0054\frac{wL^4}{EI} \tag{C.39}$$

## Appendix D. Telemetry Link Budget Calculations

The focus of the Telecom study was to provide spacecraft antenna options along with possible ground options. As the study progressed, an option to reduce antenna complexity with low data rate was also studied. The telecom study included telescope options listed below:

a. An EPIC-LC mission (X-band downlink @ 4Mbps with gimballed antenna)
b. An EPIC-CS mission (Ka-band downlink @ 20Mbps)

c. An EPIC-LC mission with low data rate (X-band downlink with bi-conical antenna). This report provides Telecom inputs and summarizes various X-band and Ka-band data downlink options.

## D.1 Telecom Requirements

The EPIC spacecraft orbits around the L2 Lagrange point. The spacecraft's spin axis is 45 degrees tilted from Earth-Sun line. The spin occurs at ~1 rpm and the system precesses at ~1 revolution per hour. The spacecraft orientation and spin is shown in Fig. D.1.

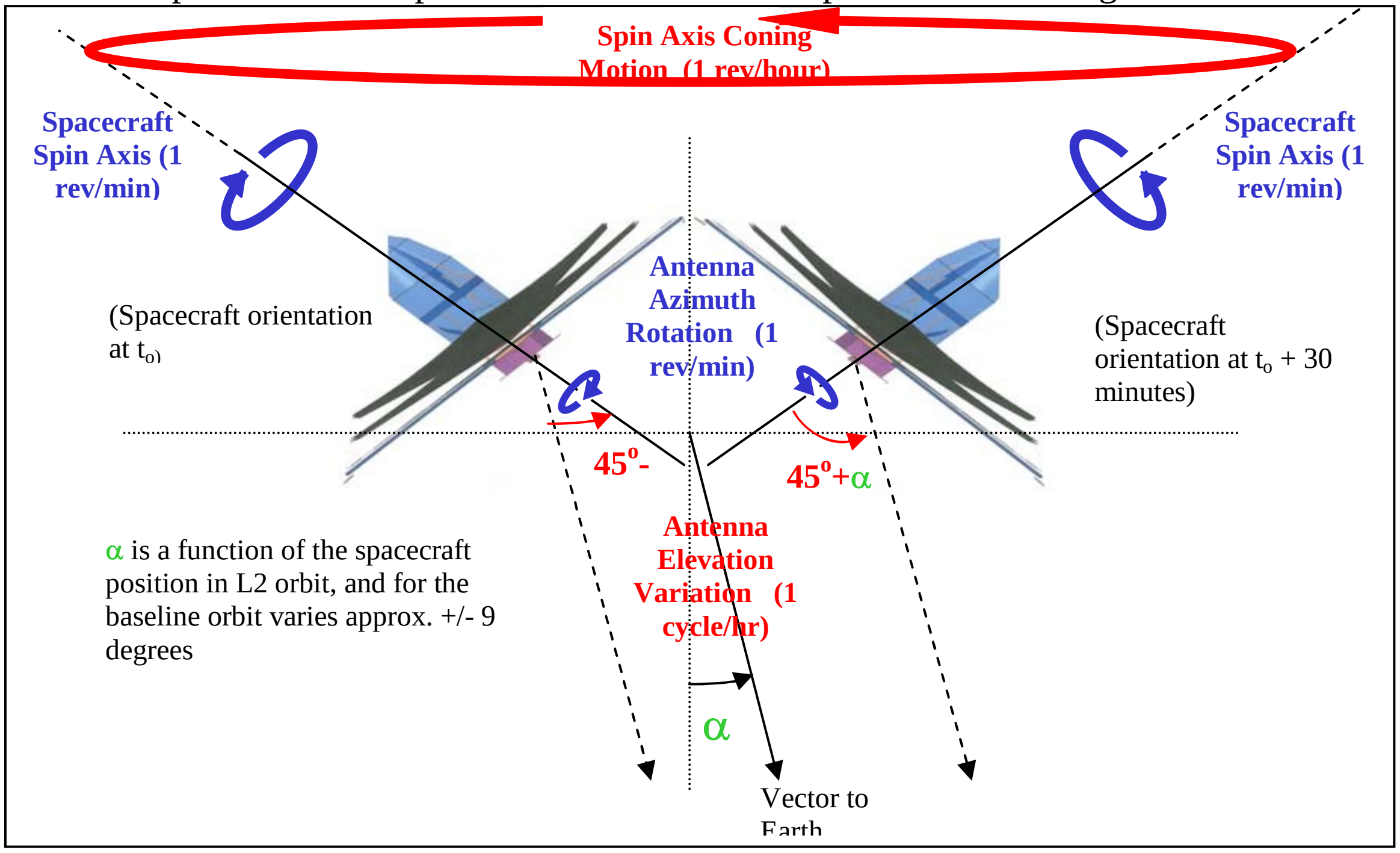

Figure D.1. Spacecraft Orientation, spin and precession angles.

The telecom requirements for EPIC based on an earlier TeamX Study were 4 Mbps downlink @ X-band for EPIC-LC and 20 Mbps downlink @ Ka-band for EPIC-CS. The goal for the Telecom effort was to study antenna pointing requirements and determine various spacecraft antenna and ground network options. This was achieved by performing simulations for antenna pointing requirements followed by various link analyses by changing various parameters in the link such as space craft antenna, spacecraft amplifier, and ground station to determine possible downlink data rate.

The frequencies used for EPIC mission will be near earth frequencies and not DSN frequencies since range for the EPIC mission is less that $2x10^6$ km. The X-band frequency is limited to 10 MHz, limiting the downlink data rate to 4 Mbps thus the need to use Ka-band for any option requiring a downlink data rate greater than 4 Mbps. The near earth X-band or Ka-band frequencies are not channelized (per Frequency Allocation group at JPL) as in the case for Deep Space frequencies, thus use of these frequencies will require coordination with other programs.

With a standard antenna, EPIC needs a 2-axis gimbaled drive due to the spinning and precession of the spacecraft. There was a need to study the antenna pointing requirements and determine if there were other antenna options which can possibly eliminate the gimbals. Also, based on various telecom options a link analysis was performed to determine possible downlink data rates.

To determine the antenna pointing requirement a simulation model for EPIC spacecraft trajectory with spinning and precession was built on the Satellite Orbit Analysis Program (SOAP) tool. Three ground stations, with one at each DSN site were modeled. The spacecraft communication antenna was modeled and was pointed towards Earth Nadir. A simulation was executed to determine the antenna pointing requirements. The angle between antenna pointing axis and the spacecraft spin axis was recorded. The results show the effect of the spacecraft spinning, precession and the Halo orbit.

Fig. D.2 shows the angles between the spacecraft spin axis and antenna pointing axis as generated by SOAP. The plot on the left labeled as 702AnglePrecToEarthNadir, shows the angle variation along the elevation axis. What is seen is that antenna pointing is 45 deg off the Spacecraft spin axis and goes thru a variation with +/-10 deg (worst case) due to the effect of precession and halo orbit (note this is with a larger halo orbit, but has now been reduced to +/- 2 deg). The shaded region in the plot shows the effect of the precession and its details are seen in Fig. D.3, including the de-spin affect on the azimuth axis.

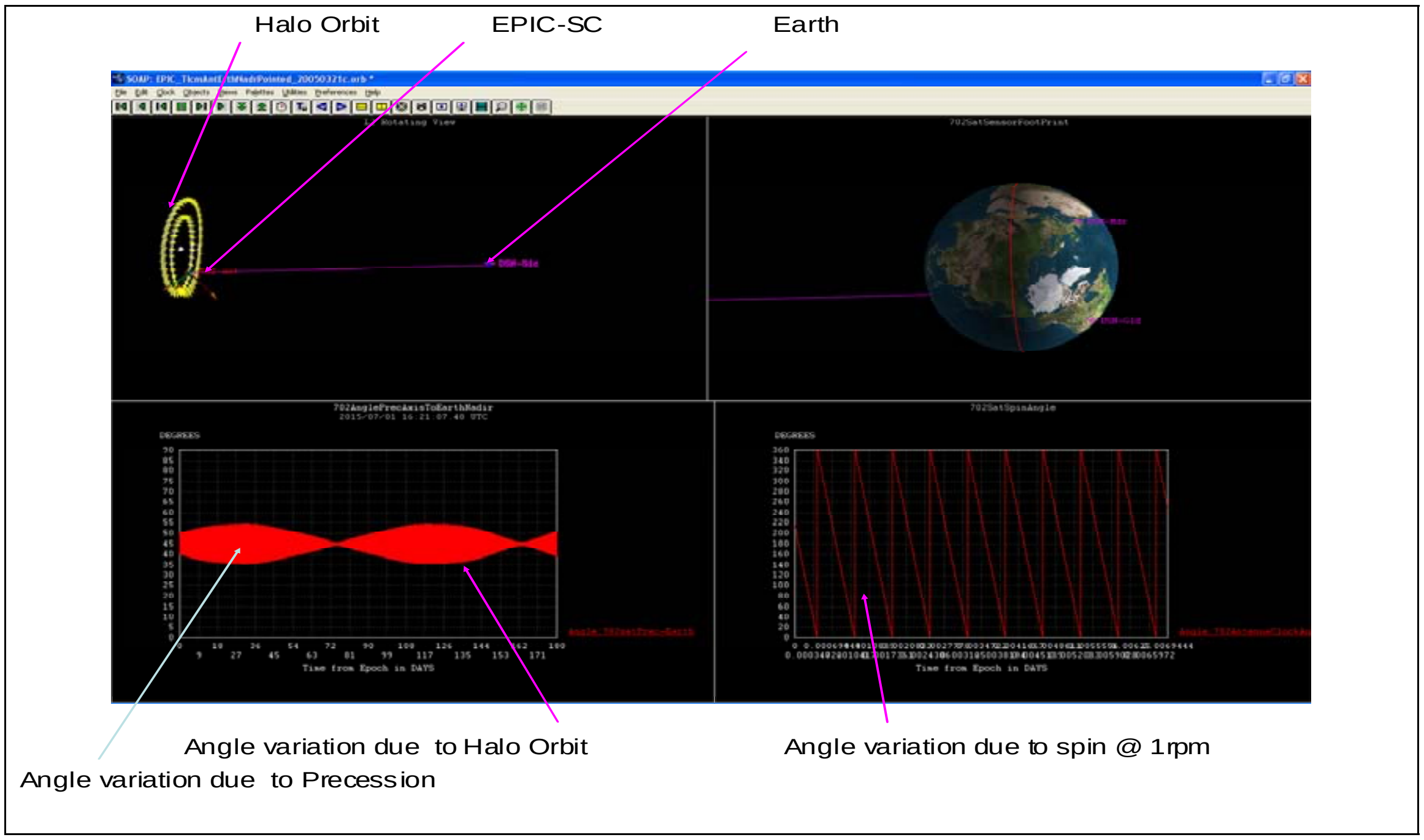

Figure D.2: SOAP antenna pointing results (effects of halo orbit, precession and spin). Note that the angle variation is computed for an earlier orbit with a significantly larger halo diameter.

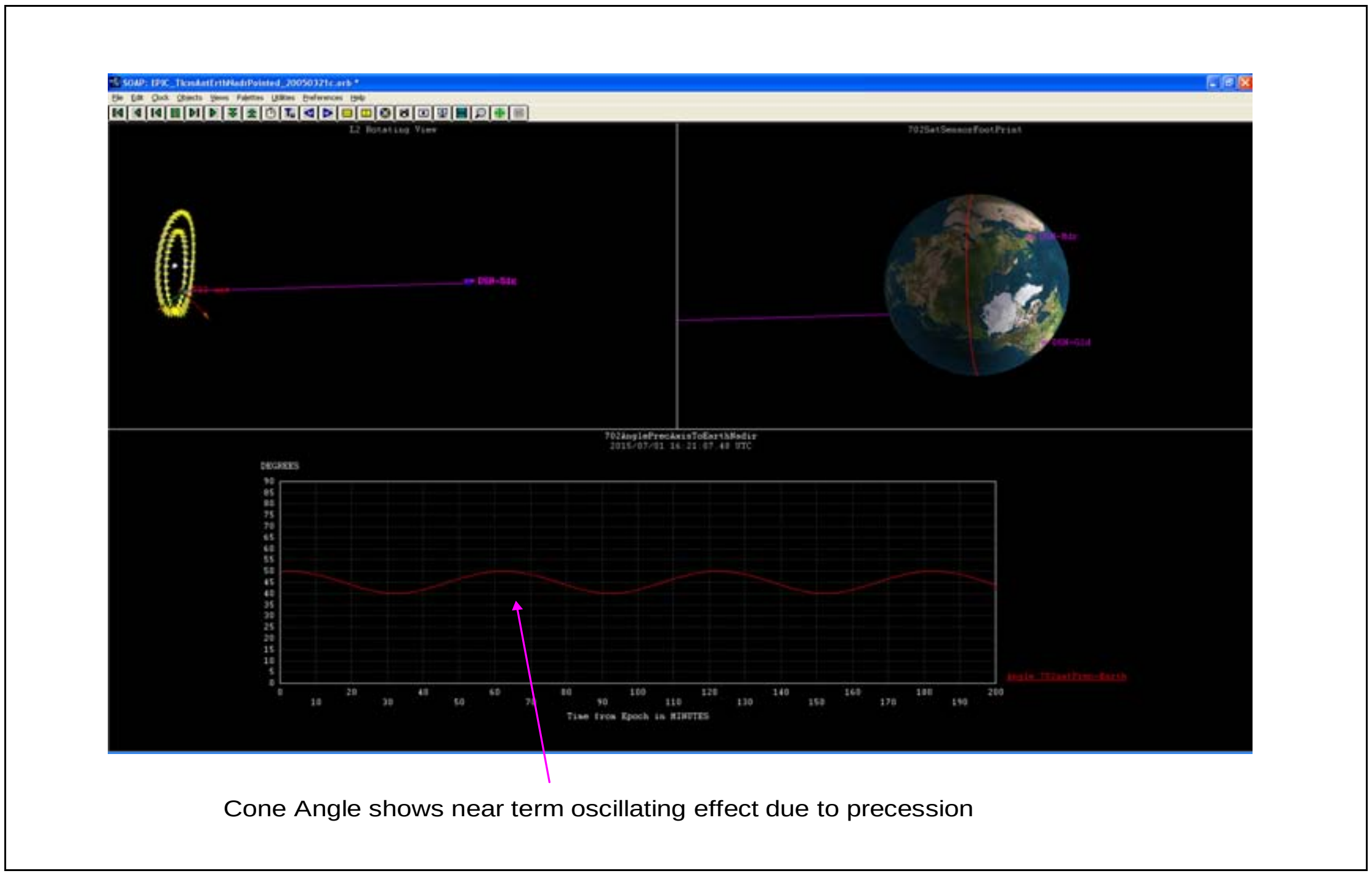

Figure D.3. SOAP antenna pointing result (effect of precession).

The SOAP results indicate that there are 3 options for the spacecraft antenna which are as follows:

a. Earth Nadir pointed HGA requiring continuous 2-axis control
   - Axis-1 to counteract the spin-affect of the spacecraft at a rate of 360deg/min.
   - Axis-2 to continuously point antenna to Earth Nadir at a rate of 0.33deg/min
b. Earth Nadir Pointed antenna with wider beamwidth of 21.4 deg requiring continuous1-axis control for to de-spin. This option is applicable to EPIC-LC at low data rates.
c. A bi-conical antenna pointing in the direction of spin axis. Antenna pattern starts 35 deg from spin axis to 55 deg from spin axis. This option is also applicable for the EPIC-LC low data rate case. (Note that this option has now been revised based on a smaller orbit, and a higher antenna gain).

## D.2 Specifications

After the SOAP simulations, link analysis was performed along with various spacecraft antenna, amplifier and ground network options to determine the feasibility of the link and achievable data rates. Table D.1 shows the link analysis results performed for the X-band downlink

## Table D.1. X-band Downlink Performance Summary

| Pointing Option | AntType | AntSize (m) | Mass (kgms) | AntBW (deg) | Ant Gain (dB) | SC-Amp (W) | GndStation | Gnd-Gain (dB) | DataRate (Mbps) | Comments |
|---|---|---|---|---|---|---|---|---|---|---|
| 2-axis Control | Dish | 0.4 | 6.9 | 7.47612426 | 28.3825333 | 13 | 12m | 58.65 | 4 | TeamX results |
| | Dish | 0.5 | 6.6 | 5.98089941 | 30.3207335 | 5 | 34m | 68.29 | 4 | TeamX results |
| | | | | | | | DSN-Array | | 4 | Don't need DSN array |
| 1-axis Control | Dish | 0.14 | 9.1 | 21.360355 | 19.2638941 | 70 | 12m | 58.65 | 4 | TeamX Results |
| | | 0.14 | 7.1 | 21.360355 | 19.2638941 | 10 | 34m | 68.29 | 4 | 34m option studied and is feasible |
| | | | | | | | DSN-Array | | | Don't need DSN array |
| Annular Pattern (18 element antenna network) | Dish | 0.14 | 24.1 | 21.360355 | 19.2638941 | 70 | 12m | 58.65 | 4 | Use horn for 1G option with total of 18 elements to cover 360deg. There will be switching network in the backend to switch between antennas. This will require switching algorithm and knowledge of the earth with inputs provided from ACS. This option requires a complex switching network with estimated mass of about 15kgms. |
| | | 0.14 | 24.1 | 21.360355 | 19.2638941 | 10 | 34m | 68.29 | 4 | Same as above |
| | | | | | | | DSN-Array | | 4 | Don't need DSN array |
| Single Annular Antenna | Horn | 289x175mm | 9.7 | 20.5 | 5 | 100 | 12m | 58.65 | 0.197 | Biconical antenna as designed by Aluizio with 20.5 deg HPBW provides 5dB antenna gain |
| | | 289x175mm | 9.7 | 20.5 | 5 | 100 | 34m | 68.29 | 1.83 | Biconical antenna as designed by Aluizio with 20.5 deg HPBW provides 5dB antenna gain |
| | | | | | | | DSN-Array | | | Don't need DSN array |

| | | | | | |
|---|---|---|---|---|---|
| TeamX Study | Other Options | LowRate Option | DSN-Array Option | Not Needed | Not Feasible |

**NOTES:**

1. Mass estimate includes Antenna, Amplifier, Radio, Switching network (if applicable) and other misc components as stated in TeamX. The mass estimates do not include the mass for the gimbals.
2. DSN-array is currently not a project. The rule of thumb for DSN-array is to use < 50% of total elements which amounts to 200.
3. Antenna gain for bi-conical antenna is based on theoretical estimates provided from antenna team and requires further analysis to provide more accurate inputs.

A 12m ground antenna suffices for X-band downlink at 4 Mbps with a 2-axis or 1-axis antenna pointing mechanism. In the case where there is a need to avoid any pointing requirement on the spacecraft antenna, a single bi-conical antenna with a 12 m ground antenna will require a reduction in the data rate to about 197 kbps. The other option is to use a 34 m ground station which supports 1.83 Mbps downlink as seen above.

## Table D.2 Ka-band Downlink Performance Summary

| Pointing Option | AntType | AntSize (m) | Mass (kgms) | AntBW (deg) | Ant Gain (dB) | SC-Amp (W) | GndStation | Gnd-Gain (dB) | DataRate (Mbps) | Comments |
|---|---|---|---|---|---|---|---|---|---|---|
| | | | | | | | | | | |
| 2-axis Control | HGADish | 0.8 | 8.9 | 1.21487019 | 44.165466 | 10 | 12m | 68.4 | 20 | TeamX results |
| | | 0.8 | 5.7 | 1.21487019 | 44.165466 | 5 | 34m | 76 | 20 | 34m option studied and is feasible |
| | | 0.8 | 8.7 | 1.21487019 | 44.165466 | 5 | 34m | 76 | 60 | Other 34m options to reduce tracking time |
| | | 0.8 | 8.9 | 1.21487019 | 44.165466 | 10 | 34m | 76 | 121 | Other 34m options to reduce tracking time |
| | | | | | | | DSN-Array | | 20 | Don't need DSN array for 2-gimbal option |
| | | | | | | | | | | |
| 1-axis Control | Horn | 0.04 | | 24.2974038 | 18.144866 | 4000 | 12m | 68.4 | 20 | Option not feasible, since the maximum amplifier assumption is 100W (project input) |
| | Horn | 0.04 | 8.4 | 24.2974038 | 18.144866 | 100 | 12m | 68.4 | 0.5 | |
| | Horn | 0.04 | | 24.2974038 | 18.144866 | 800 | 34m | 76 | 20 | Option not feasible, since the maximum amplifier assumption is 100W (project input) |
| | Horn | 0.04 | 8.4 | 24.2974038 | 18.144866 | 100 | 34m | 76 | 3.4 | |
| | Horn | 0.04 | | 24.2974038 | 18.144866 | 10 | DSN-Array | 95.4 | 20 | Option not feasible since it requires DSN-array with 661 elements --> Max elements is 400 |
| | Horn | 0.04 | 8.4 | 24.2974038 | 18.144866 | 100 | DSN-Array | 85.4 | 20 | Option feasible w/ 70-element DSN array. Important to remember that DSN-array is not a project and a backup option will be required if Array is not ready by the time of this project |
| | | | | | | | | | | |
| Annular Pattern (20 element antenna network) | Horn | 0.04 | | 24.2974038 | 18.144866 | 4000 | 12m | 68.4 | 20 | Not feasible |
| | Horn | 0.04 | 23.9 | 24.2974038 | 18.144866 | 100 | 12m | 68.4 | 0.5 | Low Rate Option |
| | Horn | 0.04 | | 24.2974038 | 18.144866 | 800 | 34m | 76 | 20 | Not feasible |
| | Horn | 0.04 | 23.9 | 24.2974038 | 18.144866 | 100 | 34m | 76 | 3.4 | Low Rate Option |
| | Horn | 0.04 | 23.9 | 24.2974038 | 18.144866 | 100 | DSN-Array | 85.4 | 20 | Use horn for 1G option with total of 20 elements to cover 360deg. There will be switching network in the backend to switch between antennas. This will require switching algorithm and knowledge of the earth with inputs provided from ACS. This option requires a complex switching network with estimated mass of about 15kgms. |
| | | | | | | | | | | |
| Single Annular Antenna | Horn | 335x231mm | 10.3 | 20 | 10 | 100 | 12m | 68.4 | 0.078 | Conical antenna with 20deg BW can provide about 10dB antenna gain. |
| | | 335x231mm | 10.3 | 20 | 10 | 100 | 34m | 76 | 0.5 | Conical antenna with 20deg BW can provide about 10dB antenna gain. |
| | | 335x231mm | 10.3 | 20 | 10 | 100 | DSN-Array | 85.4 | 3.9 | This option requires 70 element array of the DSN-array |
| | | 335x231mm | 10.3 | 20 | 10 | 100 | DSN-Array | 92.5 | 20 | This option requires 300 element array of the DSN-array |

| | | | | | |
|---|---|---|---|---|---|
| TeamX Study | Other Options | LowRate Option | DSN-Array Option | Not Needed | Not Feasible |

**NOTES:**

1. Mass estimate includes Antenna, Amplifier, Radio, Switching network (if applicable) and other misc components as stated in TeamX. The mass estimates do not include the mass for the gimbals.
2. DSN-array is currently not a project. The rule of thumb for DSN-array is to use < 50% of total elements which amounts to 200
3. Antenna gain for biconical antenna is based on theoretical estimates provided from antenna team and requires further analysis to provide more accurate inputs.

Ka-band downlink requires 2-axis control to support a 20 Mbps downlink. There are other data rates that can be supported by 2-gimbal spacecraft antenna by using a 34 m ground station. 20 Mbps can be supported with a 5 W amplifier and a 34 m ground station with about 7 dB link margin. The data rate can be increased to 60 Mbps with the same uplink and downlink scenario but reducing the link margin to 2.2 dB. The final scenario is to increase the amplifier to 10 W and double the data rate to 121 Mbps. These data rates are feasible at Ka-band since Ka-band is not band-limited like X-band. The other options for Ka-band link are to use either 1-axis spacecraft antenna or a Ka-band bi-conical antenna but either option requires DSN-array ground network. A DSN-array is currently not a project and future plans are not well understood.

# References

**Section 0**

**Section 1**

**Section 5.7**

**Section 6.3**

**Section 6.4**

**Section 6.5**

**Section 6.7**

**Section 6.10**

**Appendix D**